\documentclass{article}
\usepackage{epubtk}
\usepackage{epsf}
\usepackage{amsmath}
\usepackage{amssymb}

\showlistoftablesfalse

\newcommand{\J}{\mbox{\usefont{U}{rsfs}{m}{n} I}}
\DeclareMathOperator{\tr}{tr}
\DeclareMathOperator{\real}{Re}
\DeclareMathOperator{\imag}{Im}
\DeclareMathOperator{\Ric}{Ric}

\makeatletter
  \let\hangafter\@hangfrom
\makeatother

\newcounter{penumerate}
\newenvironment{penumerate}
  {\vspace{0.8 em}
   \refstepcounter{penumerate}
   \hangafter{\hspace{1 em}\thepenumerate~}}
  {

  }
\renewcommand{\thepenumerate}{(P\arabic{penumerate})}

\newcounter{cenumerate}
\newenvironment{cenumerate}
  {\vspace{0.8 em}
   \refstepcounter{cenumerate}
   \hangafter{\hspace{1 em}\thecenumerate~}}
  {

  }
\renewcommand{\thecenumerate}{(C\arabic{cenumerate})}

\newcounter{aenumerate}
\newenvironment{aenumerate}
  {\vspace{0.8 em}
   \refstepcounter{aenumerate}
   \hangafter{\hspace{1 em}\theaenumerate~}}
  {

  }
\renewcommand{\theaenumerate}{(A\arabic{aenumerate})}

\newcounter{henumerate}
\newenvironment{henumerate}
  {\vspace{0.8 em}
   \refstepcounter{henumerate}
   \hangafter{\hspace{1 em}\thehenumerate~}}
  {

  }
\renewcommand{\thehenumerate}{(H\arabic{henumerate})}

\newcounter{nenumerate}
\newenvironment{nenumerate}
  {\vspace{0.8 em}
   \refstepcounter{nenumerate}
   \hangafter{\hspace{1 em}\thenenumerate~}}
  {

  }
\renewcommand{\thenenumerate}{(N\arabic{nenumerate})}

\newcounter{menumerate}
\newenvironment{menumerate}
  {\vspace{0.8 em}
   \refstepcounter{menumerate}
   \hangafter{\hspace{1 em}\themenumerate~}}
  {

  }
\renewcommand{\themenumerate}{(M\arabic{menumerate})}

\newcounter{renumerate}
\newenvironment{renumerate}
  {\vspace{0.8 em}
   \refstepcounter{renumerate}
   \hangafter{\hspace{1 em}\therenumerate~}}
  {

  }
\renewcommand{\therenumerate}{(R\arabic{renumerate})}

\newcounter{senumerate}
\newenvironment{senumerate}
  {\vspace{0.8 em}
   \refstepcounter{senumerate}
   \hangafter{\hspace{1 em}\thesenumerate~}}
  {

  }
\renewcommand{\thesenumerate}{(S\arabic{senumerate})}

\begin{document}

\title{Gravitational Lensing from a Spacetime Perspective}

\author{\epubtkAuthorData{Volker Perlick}
        {Physics Department\\
         Lancaster University \\
         Lancaster LA1 4YB \\
         United Kingdom}
        {vperlick@lancaster.ac.uk}
        {}
       }

\date{}
\maketitle


\begin{abstract}
  The theory of gravitational lensing is reviewed from a spacetime
  perspective, without quasi-Newtonian approximations. More precisely,
  the review covers all aspects of gravitational lensing where
  light propagation is described in terms of lightlike geodesics of
  a metric of Lorentzian signature. It includes the basic equations
  and the relevant techniques for calculating the position, the shape, and
  the brightness of images in an arbitrary general-relativistic
  spacetime. It also includes general theorems on the classification
  of caustics, on criteria for multiple imaging, and on the possible
  number of images. The general results are illustrated with examples
  of spacetimes where the lensing features can be explicitly calculated,
  including the Schwarzschild spacetime, the Kerr spacetime, the
  spacetime of a straight string, plane gravitational waves, and others.
\end{abstract}

\epubtkKeywords{gravitational lenses, differential geometry}

\newpage


\section{Introduction}
\label{sec:introduction}

In its most general sense, \emph{gravitational lensing} is a collective term
for all effects of a gravitational field on the propagation of electromagnetic
radiation, with the latter usually described in terms of rays. According to
general relativity, the gravitational field is coded in a metric of Lorentzian
signature on the 4-dimensional spacetime manifold, and the light rays are the
lightlike geodesics of this spacetime metric. From a mathematical point of view,
the theory of gravitational lensing is thus the theory of lightlike geodesics
in a 4-dimensional manifold with a Lorentzian metric.

The first observation of a `gravitational lensing' effect was made when the
deflection of star light by our Sun was verified during a Solar eclipse in 1919.
Today, the list of observed phenomena includes the following: \\

\noindent
{\bf Multiple quasars.} \\
\noindent
The gravitational field of a galaxy (or a cluster of galaxies) bends the light
from a distant quasar in such a way that the observer on Earth sees two or
more images of the quasar. \\

\noindent
{\bf Rings.} \\
\noindent
An extended light source, like a galaxy or a lobe
of a galaxy, is distorted into a closed or almost closed ring
by the gravitational field of an intervening galaxy. This phenomenon
occurs in situations where the gravitational field is almost
rotationally symmetric, with observer and light source close to the
axis of symmetry. It is observed primarily, but not exclusively,
in the radio range. \\

\noindent
{\bf Arcs.} \\
\noindent
Distant galaxies are distorted into arcs by the gravitational
field of an intervening cluster of galaxies. Here the situation is less
symmetric than in the case of rings. The effect is observed in the
optical range and may produce ``giant luminous arcs'', typically
of a characteristic blue color. \\

\noindent
{\bf Microlensing.} \\
\noindent
When a light source passes behind a compact mass,
the focusing effect on the light leads to a temporal change in brightness
(energy flux). This microlensing effect is routinely observed since
the early 1990s by monitoring a large number of stars in the bulge of
our Galaxy, in the Magellanic Clouds and in the Andromeda
galaxy. Microlensing has also been observed on quasars. \\

\noindent
{\bf Image distortion by weak lensing.} \\
\noindent
In cases where the distortion
effect on galaxies is too weak for producing rings or arcs, it can be
verified with statistical methods. By evaluating the shape of a large
number of background galaxies in the field of a galaxy cluster, one
can determine the surface mass density of the cluster. By evaluating
fields without a foreground cluster one gets information about the
large-scale mass distribution. \\

Observational aspects of gravitational lensing and methods of how to use
lensing as a tool in astrophysics are the subject of the
Living Review by Wambsganss~\cite{wambsganss-98}. There the reader may
also find some notes on the history of lensing.

The present review is meant as complementary to the review by Wambsganss. While all
the theoretical methods reviewed in~\cite{wambsganss-98} rely on quasi-Newtonian
approximations, the present review is devoted to the theory of gravitational
lensing from a spaectime perspective, without such approximations. Here
the terminology is as follows: ``Lensing from a spacetime perspective''
means that light propagation is described in terms of lightlike geodesics
of a general-relativistic spacetime metric, without further approximations.
(The term ``non-perturbative lensing'' is sometimes used in the same sense.)
``Quasi-Newtonian approximation'' means that the general-relativistic
spacetime formalism is reduced by approximative assumptions to essentially
Newtonian terms (Newtonian space, Newtonian time, Newtonian gravitational field).
The quasi-Newtonian approximation formalism of lensing comes in several
variants, and the relation to the exact formalism is not always evident
because sometimes plausibility and ad-hoc assumptions are implicitly made.
A common feature of all variants is that they are ``weak-field approximations''
in the sense that the spacetime metric is decomposed
into a background (``spacetime without the lens'') and a small perturbation
of this background (``gravitational field of the lens''). For the background
one usually chooses either Minkowski spacetime (isolated lens) or a spatially
flat Robertson--Walker spacetime (lens embedded in a cosmological model).
The background then defines a Euclidean 3-space, similar to Newtonian space,
and the gravitational field of the lens is similar to a Newtonian
gravitational field on this Euclidean 3-space. Treating the lens as a small
perturbation of the background means that the gravitational field of the
lens is weak and causes only a small deviation of the light rays from the
straight lines in Euclidean 3-space. In its most traditional version, the
formalism assumes in addition that the lens is ``thin'', and that the lens
and the light sources are at rest in Euclidean 3-space, but there are also
variants for ``thick'' and moving lenses. Also, modifications for a spatially
curved Robertson--Walker background exist, but in all variants a non-trivial
topological or causal structure of spacetime is (explicitly or implicitly)
excluded. At the center of the quasi-Newtonian formalism is a ``lens
equation'' or ``lens map'',
which relates the position of a ``lensed image'' to the position of the
corresponding ``unlensed image''. In the most traditional version one
considers a thin lens at rest, modeled by a Newtonian gravitational potential
given on a plane in Euclidean 3-space (``lens plane''). The light rays are
taken to be straight lines in Euclidean 3-space except for a sharp bend at
the lens plane. For a fixed observer and light sources distributed on a plane
parallel to the lens plane (``source plane''), the lens map is then a
map from the lens plane to the source plane. In this way, the geometric
spacetime setting of general relativity is completely covered behind a curtain
of approximations, and one is left simply with a map from a plane to a
plane. Details of the quasi-Newtonian approximation formalism can be found
not only in the above-mentioned Living Review~\cite{wambsganss-98}, but
also in the monographs of Schneider, Ehlers, and
Falco~\cite{schneider-ehlers-falco-92} and Petters, Levine, and
Wambsganss~\cite{petters-levine-wambsganss-2001}.

The quasi-Newtonian approximation formalism has proven very successful for
using gravitational lensing as a tool in astrophysics. This is impressively
demonstrated by the work reviewed in~\cite{wambsganss-98}. On the other hand,
studying lensing from a spacetime perspective is of relevance under
three aspects: \\

\noindent
{\bf Didactical.} \\
\noindent
The theoretical foundations of lensing can be properly formulated
only in terms of the full formalism of general relativity. Working
out examples with strong curvature and with non-trivial causal or
topological structure demonstrates that, in principle, lensing situations
can be much more complicated than suggested by the quasi-Newtonian
formalism. \\

\noindent
{\bf Methodological.} \\
\noindent
General theorems on lensing (e.g., criteria for multiple imaging,
characterizations of caustics, etc.) should be formulated within the exact
spacetime setting of general relativity, if possible, to make sure that they
are not just an artifact of approximative assumptions. For those results which
do not hold in arbitrary spacetimes, one should try to find the precise
conditions on the spacetime under which they are true. \\

\noindent
{\bf Practical.} \\
\noindent
There are some situations of astrophysical interest to which the
quasi-Newtonian formalism does not apply. For instance, near
a black hole light rays are so strongly bent that, in principle,
they can make arbitrarily many turns around the hole. Clearly, in
this situation it is impossible to use the quasi-Newtonian formalism
which would treat these light rays as small perturbations of straight
lines. \\

The present review tries to elucidate all three aspects. More precisely,
the following subjects will be covered:
\begin{itemize}
\item
  The basic equations and all relevant techniques
  that are needed for calculating the position, the shape, and the
  brightness of images in an arbitrary general-relativistic spacetime
  are reviewed.
  Part of this material is well-established since decades, like
  the Sachs equations for the optical scalars (Section~\ref{ssec:Sachs}),
  which are of crucial relevance for calculating distance measures
  (Section~\ref{ssec:distance}), image distortion
  (Section~\ref{ssec:distortion}), and the brightness of images
  (Section~\ref{ssec:brightness}). It is included here
  to keep the review self-contained. Other parts refer to more recent
  developments which are far from being fully explored, like the exact
  lens map (Section~\ref{ssec:cone}) and variational techniques
  (Section~\ref{ssec:fermat}). Specifications and simplifications are
  possible for spacetimes with symmetries. The case of spherically symmetric
  and static spacetimes is treated in greater detail
  (Section~\ref{ssec:ss}).
\item
  General theorems on lensing in arbitrary spacetimes, or in certain
  classes of spacetimes, are reviewed. Some of these results are of
  a local character, like the classification of locally stable caustics
  (Section~\ref{ssec:front}). Others are related to global aspects, like
  the criteria for multiple imaging in terms of conjugate points
  and cut points (Sections~\ref{ssec:cut} and~\ref{ssec:crit}). The
  global theorems can be considerably strengthened if one
  restricts to globally hyperbolic spacetimes (Section~\ref{ssec:hypcrit})
  or, more specifically, to asymptotically simple and empty spacetimes
  (Section~\ref{ssec:asy}). The latter may be viewed as spacetime
  models for isolated transparent lenses. Also, in globally hyperbolic
  spacetimes Morse theory can be used for investigating whether the total
  number of images is finite or infinite, even or odd (Section~\ref{ssec:morse}).
  In a spherically symmetric and static spacetime, the occurrence of an
  infinite sequence of images is related to the occurrence
  of a ``light sphere'' (circular lightlike geodesics), like in the
  Schwarzschild spacetime at $r=3m$ (Section~\ref{ssec:ss}).
\item
  Several examples of spacetimes are considered, where the lightlike
  geodesics and, thus, the lensing features can be calculated explicitly.
  The examples are chosen such that they illustrate the general results.
  Therefore, in many parts of the review the reader will find suggestions
  to look at pictures in the example section. The best known and
  astrophysically most relevant examples are the Schwarzschild
  spacetime (Section~\ref{ssec:schw}), the Kerr spacetime
  (Section~\ref{ssec:kerr}) and the spacetime of a straight string
  (Section~\ref{ssec:str}). Schwarzschild black hole lensing and
  Kerr black hole lensing was intensively investigated already in the
  1960s, 1970s, and 1980s, with astrophysical applications concentrating
  on observable features of accretion disks. More recently,
  the increasing evidence that there is a black hole at the center
  of our Galaxy (and probably at the center of most galaxies) has
  led to renewed and intensified interest in black hole lensing
  (see Sections~\ref{ssec:schw} and~\ref{ssec:kerr}). This is a
  major reason for the increasing number of articles on lensing
  beyond the quasi-Newtonian approximation. 
\end{itemize}

This introduction ends with some notes on subjects \emph{not}
covered in this review: \\

\noindent
{\bf Wave optics.} \\
\noindent
In the electromagnetic theory, light is described by wavelike
solutions to Maxwell's equations. The ray-optical treatment used
throughout this review is the standard high-frequency approximation
(geometric optics approximation) of the electromagnetic theory for
light propagation in vacuum on a general-relativistic spacetime (see,
e.g., \cite{misner-thorne-wheeler-73}, \S~22.5
or~\cite{schneider-ehlers-falco-92}, Section~3.2). (Other notions of
vacuum light rays, based on a different approximation procedure, have
been occasionally suggested~\cite{mashhoon-87}, but will not be
considered here. Also, results specific to spacetime dimensions other than
four or to gravitational theories other than Einstein's are not covered.)
For most applications to lensing the ray-optical
treatment is valid and appropriate. An exception, where wave-optical
corrections are necessary, is the calculation of the brightness of
images if a light source comes very close to the caustic of the observer's
light cone (see Section~\ref{ssec:brightness}). \\

\noindent
{\bf Light propagation in matter.} \\
\noindent
If light is directly influenced
by a medium, the light rays are no longer the lightlike geodesics of
the spacetime metric. For an isotropic non-dispersive medium, they
are the lightlike geodesics of another metric which is again of
Lorentzian signature. (This ``optical metric'' was introduced by
Gordon~\cite{gordon-23}. For a rigourous derivation, starting from
Maxwell's equation in an isotropic non-dispersive medium, see
Ehlers~\cite{ehlers-67}.) Hence, the formalism used throughout this
review still applies to this situation after an appropriate re-interpretation
of the metric. In anisotropic or dispersive media, however,
the light rays are not the lightlike geodesics of a Lorentzian metric.
There are some lensing situations where the influence of matter
has to be taken into account. For instance., for the deflection of radio
signals by our Sun the influence of the plasma in the Solar corona
(to be treated as a dispersive medium) is very well measurable.
However, such situations will not be considered in this review.
For light propagation in media on a general-relativistic spacetime,
see~\cite{perlick-2000b} and references cited therein. \\

\noindent
{\bf Kinetic theory.} \\
\noindent
As an alternative to the (geometric optics approximation of)
electromagnetic theory, light can be treated as a photon gas,
using the formalism of kinetic theory. This has relevance, e.g.,
for the cosmic background radiation. For basic notions of
general-relativistic kinetic theory see, e.g., \cite{ehlers-73}.
Apart from some occasional remarks, kinetic theory will not be
considered in this review. \\

\noindent
{\bf Derivation of the quasi-Newtonian formalism.} \\
\noindent
It is not satisfacory if the quasi-Newtonian formalism of lensing is set up
with the help of ad-hoc assumptions, even if the latter look plausible. From
a methodological point of view, it is more desirable to start from the exact
spacetime setting of general relativity and to derive the quasi-Newtonian
lens equation by a well-defined approximation procedure. In comparison to
earlier such derivations~\cite{schneider-ehlers-falco-92, sasaki-93,
  seitz-schneider-ehlers-94} later effort has led to considerable
improvements. 
For lenses embedded in a cosmological model, see Pyne and
Birkinshaw~\cite{pyne-birkinshaw-96} who consider lenses that need not be
thin and may be moving on a Robertson--Walker background (with positive, negative,
or zero spatial curvature). For the non-cosmological situation, a Lorentz
covariant approximation formalism was derived by Kopeikin and
Sch{\"a}fer~\cite{kopeikin-schaefer-99}. Here Minkowski spacetime is
taken as the background, and again the lenses need not be thin and
may be moving.

\newpage


\section{Lensing in Arbitrary Spacetimes}
\label{sec:general}

By a \emph{spacetime} we mean a 4-dimensional manifold $\mathcal{M}$ with
a ($C^{\infty}$, if not otherwise stated) metric tensor field $g$ of signature
$(+,+,+,-)$ that is time-oriented. The latter means that the non-spacelike
vectors make up two connected components in the entire tangent bundle,
one of which is called ``future-pointing'' and the other one ``past-pointing''.
Throughout this review we restrict to the case that the
light rays are freely propagating in vacuum, i.e., are not influenced by
mirrors, refractive media, or any other impediments. The light rays
are then the lightlike geodesics of the spacetime metric. We first
summarize results on the lightlike geodesics that hold in arbitrary spacetimes.
In Section~\ref{sec:hyp} these results will be specified for spacetimes with
conditions on the causal structure and in Section~\ref{sec:symmetry} for
spacetimes with symmetries.


\subsection{Light cone and exact lens map}
\label{ssec:cone}

In an arbitrary spacetime $(\mathcal{M},g)$, what an observer at an event $p_\mathrm{O}$
can see is determined by the lightlike geodesics that issue from $p_\mathrm{O}$ into
the past. Their union gives the \emph{past light cone} of $p_\mathrm{O}$. This is the
central geometric object for lensing from the spacetime perspective. For a
point source with worldline $\gamma_\mathrm{S}$, each past-oriented lightlike geodesic
$\lambda$ from $p_\mathrm{O}$ to $\gamma_\mathrm{S}$ gives rise to an image of $\gamma_\mathrm{S}$ on
the observer's sky. One should view any such $\lambda$ as the central ray of
a thin bundle that is focused by the observer's eye lens onto the observer's
retina (or by a telescope onto a photographic plate). Hence, the intersection
of the past light cone with the world-line of a point source (or with the
world-tube of an extended source) determines the visual appearance of
the latter on the observer's sky.

In mathematical terms, the observer's \emph{sky} or \emph{celestial sphere}
$\mathcal{S}_\mathrm{O}$ can be viewed as the set of all lightlike directions at
$p_\mathrm{O}$. Every such direction defines a unique (up to parametrization) lightlike
geodesic through $p_\mathrm{O}$, so $\mathcal{S}_\mathrm{O}$ may also be viewed as a subset of
the space of all lightlike geodesics in $(\mathcal{M},g)$ (cf.~\cite{low-93}).
One may choose at $p_\mathrm{O}$ a future-pointing vector $U_\mathrm{O}$ with $g(U_\mathrm{O},U_\mathrm{O})=-1$,
to be interpreted as the 4-velocity of the observer. This allows identifying
the observer's sky $\mathcal{S}_\mathrm{O}$ with a subset of the tangent space
$T_{p_\mathrm{O}} \mathcal{M}$,
\begin{equation}
  \label{eq:skyO}
  \mathcal{S}_\mathrm{O} \simeq
  \left\{ w \in T_{p_\mathrm{O}} \mathcal{M} \,\big|\, g(w,w)=0
  \mbox{~~and~~} g(w,U_\mathrm{O})=1 \right\}.
\end{equation}
If $U_\mathrm{O}$ is changed, this representation changes according to the
standard aberration formula of special relativity.
By definition of the \emph{exponential map} $\exp$, every affinely
parametrized geodesic $ s \mapsto \lambda (s)$ satisfies $\lambda (s) =
\exp \bigl( s \, \dot{\lambda} (0) \bigr)$. Thus, the
past light cone of $p_\mathrm{O}$ is the image of the map
\begin{equation}
  \label{eq:cone}
  (s,w) \longmapsto \exp (s w),
\end{equation}
which is defined on a subset of $] 0, \infty [ \times \mathcal{S}_\mathrm{O} $. If we
restrict to values of $s$ sufficiently close to 0, the map~(\ref{eq:cone})
is an embedding, i.e., this truncated light cone is an embedded submanifold; this
follows from the well-known fact that $\exp$ maps a
neighborhood of the origin, in each tangent space, diffeomorphically
into the manifold. However, if we extend the map~(\ref{eq:cone}) to
larger values of $s$, it is in general neither injective nor an
immersion; it may form folds, cusps, and other forms of \emph{caustics},
or transverse self-intersections. This observation is of crucial
importance in view of lensing. There are some lensing phenomena, such
as multiple imaging and image distortion of (point) sources into
(1-dimensional) rings, which can occur only if the light cone fails
to be an embedded submanifold (see Section~\ref{ssec:crit}). Such lensing
phenomena are summarized under the name \emph{strong lensing} effects.
As long as the light cone is an embedded submanifold, the effects exerted
by the gravitational field on the apparent shape and on the apparent
brightness of light sources are called \emph{weak lensing} effects. For
examples of light cones with caustics and/or transverse self-intersections,
see Figures~\ref{fig:schwcon}, \ref{fig:strcon1}, and~\ref{fig:strcon2}.
These pictures show light cones in spacetimes with symmetries, so
their structure is rather regular. A realistic model of our own light
cone, in the real world, would have to take into account numerous
irregularly distributed inhomogeneities (``clumps'') that bend light rays in
their neighborhood. Ellis, Bassett, and Dunsby~\cite{ellis-bassett-dunsby-98}
estimate that such a light cone would have at least $10^{22}$ caustics
which are hierarchically structured in a way that reminds of fractals.

For calculations it is recommendable to introduce coordinates
on the observer's past light cone. This can be done by choosing an
orthonormal tetrad $(e_0,e_1,e_2,e_3)$ with $e_0=-U_\mathrm{O}$ at the observation
event $p_\mathrm{O}$. This parametrizes the points of the observer's celestial sphere
by spherical coordinates $(\Psi, \Theta)$,
\begin{equation}
  \label{eq:ThetaPsi}
  w = \sin \Theta \, \cos \Psi \, e_1 + \sin \Theta \, \sin \Psi \, e_2 +
  \cos \Theta \, e_3 + e_0.
\end{equation}
In this representation, map~(\ref{eq:cone}) maps each $(s, \Psi, \Theta)$ to
a spacetime point. Letting the observation event float along the observer's
worldline, parametrized by proper time $\tau$, gives a map that assigns to each
$(s, \Psi, \Theta, \tau)$ a spacetime point. In terms of coordinates $x =
(x^0,x^1,x^2,x^3)$ on the spacetime manifold, this map is of the form
\begin{equation}
  \label{eq:obsco}
  x^i = F^i (s, \Psi, \Theta, \tau),
  \qquad
  i=0,1,2,3.
\end{equation}
It can be viewed as a map from the world as it appears to the
observer (via optical observations) to the world as it is. The
coordinates $(s, \Psi, \Theta, \tau)$ were called \emph{optical 
coordinates} by Temple~\cite{temple-38} and 
\emph{observational coordinates} by Ellis~\cite{ellis-80}. 
A detailed discussion of 
their properties can be found in~\cite{ellis-nel-maartens-stoeger-whitman-85}.
They are particularly useful in cosmology but can be introduced for
\emph{any} observer in \emph{any} spacetime. It is useful to consider
observables, such as distance measures (see Section~\ref{ssec:distance}) or
the ellipticity that describes image distortion (see Section~\ref{ssec:distortion})
as functions of the observational coordinates. Some observables, e.g., the
redshift and the luminosity distance, are not determined by
the spacetime geometry and the observer alone, but also depend on the
4-velocities of the light sources. If a vector field $U$ with
$g(U,U)=-1$ has been fixed, one may restrict to an observer and
to light sources which are integral curves of $U$. The above-mentioned
observables, like redshift and luminosity distance, are then uniquely
determined as functions of the observational coordinates. In applications
to cosmology one chooses $U$ as tracing the mean flow of luminous matter
(``Hubble flow'') or as the rest system of the cosmic background radiation;
present observations are compatible with the assumption that these two
distinguished observer fields coincide~\cite{blake-wall-2002, crawford-2009,
itoh-yahata-takada-2010}.

Writing map~(\ref{eq:obsco}) explicitly requires solving the lightlike geodesic
equation. This is usually done, using standard index notation, in the Lagrangian
formalism, with the Lagrangian $\mathcal{L} = \frac{1}{2} g_{ij}(x) \dot{x}^i
\dot{x}^j$, or in the Hamiltonian formalism, with the Hamiltonian $\mathcal{H} =
\frac{1}{2} g^{ij}(x) p_i p_j$. A non-trivial example where the solutions can
be explicitly written in terms of elementary functions is the string spacetime of
Section~\ref{ssec:str}. Somewhat more general, although still very special,
is the situation that the lightlike geodesic equation admits three
independent constants of motion in addition to the obvious one
$g^{ij}(x) p_i p_j =0$. If, for any pair of the four
constants of motion, the Poisson bracket vanishes (``complete integrability''),
the lightlike geodesic equation can be reduced to first-order form, i.e., the
light cone can be written in terms of integrals over the metric coefficients.
This is true, e.g., in spherically symmetric and static spacetimes
(see Section~\ref{ssec:ss}).

Having parametrized the past light cone of the observation event $p_\mathrm{O}$
in terms of $(s,w)$, or more specifically in terms of $(s, \Psi, \Theta)$,
one may set up an \emph{exact lens map}. This exact lens map is analogous
to the lens map of the quasi-Newtonian approximation formalism, as far as
possible, but it is valid in an arbitrary spacetime without approximation.
In the quasi-Newtonian formalism for thin lenses at rest, the lens map assigns
to each point in the \emph{lens plane} a point in the \emph{source
  plane} (see, e.g., \cite{schneider-ehlers-falco-92, petters-levine-wambsganss-2001,
wambsganss-98}). When working in an arbitrary
spacetime without approximations, the observer's sky $\mathcal{S}_\mathrm{O}$ is
an obvious substitute for the lens plane. As a substitute for the source
plane we choose a 3-dimensional submanifold $\mathcal{T}$ with a
prescribed ruling by timelike curves. We assume that $\mathcal{T}$ is globally
of the form $\mathcal{Q} \times \mathbb{R}$, where the points of the
2-manifold $\mathcal{Q}$ label the timelike curves by which $\mathcal{T}$
is ruled. These timelike curves are to be interpreted as the worldlines
of light sources. We call any such $\mathcal{T}$ a \emph{source surface}.
In a nutshell, choosing a source surface means choosing a two-parameter
family of light sources.

The exact lens map is a map from $\mathcal{S}_\mathrm{O}$ to $\mathcal{Q}$. It is
defined by following, for each $w \in \mathcal{S}_\mathrm{O}$, the past-pointing
geodesic with initial vector $w$ until it meets $\mathcal{T}$ and then
projecting to $\mathcal{Q}$ (see Figure~\ref{fig:map}). In other words, the
exact lens map says, for each point on the observer's celestial sphere,
which of the chosen light sources is seen at this point. Clearly,
non-invertibility of the lens map indicates multiple imaging. What
one chooses for $\mathcal{T}$ depends on the situation. In applications
to cosmology, one may choose galaxies at a fixed redshift $z=z_\mathrm{S}$ around
the observer. In a spherically-symmetric and static spacetime one may choose
static light sources at a fixed radius value $r=r_\mathrm{S}$. Also, the surface of
an extended light source is a possible choice for $\mathcal{T}$.

\epubtkImage{figure01.png}
{\begin{figure}[hptb]
   \def\epsfsize#1#2{0.5#1}
   \centerline{\epsfbox{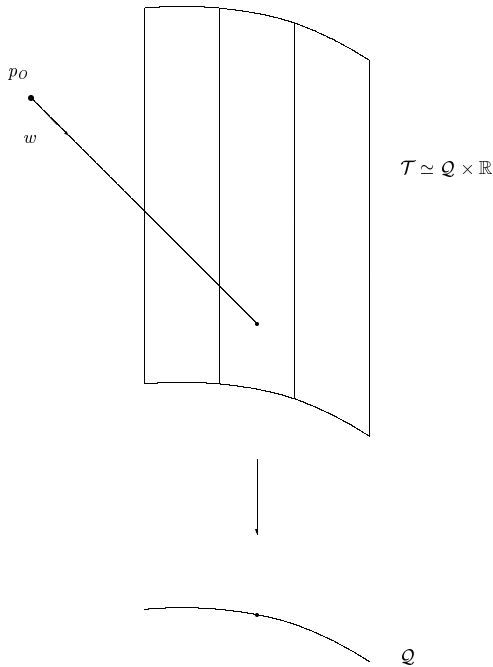}}
   \caption{Illustration of the exact lens map. $p_\mathrm{O}$ is
     the chosen observation event, $\mathcal{T}$ is the chosen
     source surface. $\mathcal{T}$ is a hypersurface ruled by
     timelike curves (worldlines of light sources) which are
     labeled by the points of a 2-dimensional manifold
     $\mathcal{Q}$. The lens map is defined on the observer's
     celestial sphere $\mathcal{S}_\mathrm{O}$, given by
     Equation~(\ref{eq:skyO}), and takes values in $\mathcal{Q}$. For
     each $w \in \mathcal{S}_\mathrm{O}$, one follows the lightlike
     geodesic with this initial direction until it meets $\mathcal{T}$
     and then projects to $\mathcal{Q}$. For illustrating the exact
     lens map, it is an instructive exercise to intersect the light
     cones of Figures~\ref{fig:schwcon}, \ref{fig:strcon1},
     \ref{fig:strcon2}, and~\ref{fig:wave1} with various source
     surfaces $\mathcal{T}$.}
   \label{fig:map}
 \end{figure}
}

The exact lens map was introduced by Frittelli and
Newman~\cite{frittelli-newman-99} and further discussed
in~\cite{ehlers-frittelli-newman-2003, ehlers-2000}.
The following global aspects of the exact lens map were investigated
in~\cite{perlick-2001}. First, in general the lens map is not
defined on all of $\mathcal{S}_\mathrm{O}$ because not all past-oriented lightlike
geodesics that start at $p_\mathrm{O}$ necessarily meet $\mathcal{T}$. Second, in
general the lens map is multi-valued because a lightlike geodesic might meet
$\mathcal{T}$ several times. Third, the lens map need not be
differentiable and not even continuous because a lightlike geodesic
might meet $\mathcal{T}$ tangentially. In~\cite{perlick-2001}, the
notion of a \emph{simple lensing neighborhood} is introduced which
translates the statement that a deflector is transparent into precise
mathematical language. It is shown that the lens map is globally
well-defined and differentiable if the source surface is the boundary
of such a simple lensing neighborhood, and that for each light source that does
not meet the caustic of the observer's past light cone the number of
images is finite and odd. This result applies, as a special case,
to asymptotically simple and empty spacetimes (see Section~\ref{ssec:asy}).

For expressing the exact lens map in coordinate language, it is recommendable
to choose coordinates $(x^0,x^1,x^2,x^3)$ such that the source surface
$\mathcal{T}$ is given by the equation $x^3 = x^3_\mathrm{S}$, with a constant
$x^3_\mathrm{S}$, and that the worldlines of the light sources are
$x^0$-lines. In this situation the remaining coordinates $x^1$ and
$x^2$ label the light sources and the exact lens map takes the form
\begin{equation}
  \label{eq:exactlens}
  (\Psi, \Theta) \longmapsto (x^1,x^2).
\end{equation}
It is given by eliminating the two variables $s$ and $x^0$ from the
four equations~(\ref{eq:obsco}) with $F^3(s, \Psi, \Theta, \tau)
=x^3_\mathrm{S}$ and fixed $\tau$. This is the way in which the lens map was
written in the original paper by Frittelli and Newman; see Equation~(6)
in~\cite{frittelli-newman-99}. (They used complex coordinates
$(\eta, \overline{\eta})$ for the observer's celestial sphere
that are related to our spherical coordinates $(\Psi, \Theta)$
by stereographic projection.) In this explicit coordinate version,
the exact lens map can be succesfully applied, in particular, to
spherically symmetric and static spacetimes, with $x^0=t$,
$x^1=\varphi$, $x^2=\vartheta$, and $x^3=r$ (see Section~\ref{ssec:ss}
and the Schwarzschild example in Section~\ref{ssec:schw}). The exact
lens map can also be used for testing the reliability of approximation
techniques. In~\cite{kling-newman-perez-2000} the authors find that
the standard quasi-Newtonian approximation formalism may lead to
significant errors for lensing configurations with two lenses.


\subsection{Wave fronts}
\label{ssec:front}

Wave fronts are related to light rays as solutions of the Hamilton--Jacobi
equation are related to solutions of Hamilton's equations in classical
mechanics. For the case at hand (i.e., vacuum light
propagation in an arbitrary spacetime, corresponding to the Hamiltonian
$\mathcal{H}=\frac{1}{2} g^{ij}(x)p_ip_j$), a wave front is a subset of the spacetime
that can be constructed in the following way:
\begin{enumerate}
\item Choose a spacelike 2-surface $\mathcal{S}$ that is orientable.
\item At each point of $\mathcal{S}$, choose a lightlike direction
  orthogonal to $\mathcal{S}$ that depends smoothly on the
  foot-point. (You have to choose between two possibilities.)
\item Take all lightlike geodesics that are tangent to the
chosen directions. These lightlike geodesics are called the
\emph{generators} of the wave front, and the wave front is the union
of all generators.
\end{enumerate}
Clearly, a light cone is a special case of a
wave front. One gets this special case by choosing for $\mathcal{S}$
an appropriate (small) sphere. Any wave front is the envelope of
all light cones with vertices on the wave front. In this sense,
general-relativistic wave fronts can be constructed according
to the \emph{Huygens principle}.

In the context of general relativity the notion of wave fronts was introduced
by Kermack, McCrea, and Whittaker~\cite{kermack-mccrea-whittaker-32}. For a
modern review article see, e.g., Ehlers and Newman~\cite{ehlers-newman-2000}.

A coordinate representation
for a wave front can be given with the help of (local) coordinates
$(u^1,u^2)$ on $\mathcal{S}$. One chooses a parameter value $s_0$
and parametrizes each generator $\lambda$ affinely such that
$\lambda (s_0) \in \mathcal{S}$ and $\dot{\lambda} (s_0)$ depends
smoothly on the foot-point in $\mathcal{S}$. This gives the wave
front as the image of a map
\begin{equation}
  \label{eq:front}
  (s, u^1, u^2) \longmapsto F^i (s, u^1, u^2),
  \qquad
  i=0,1,2,3.
\end{equation}
For light cones we may choose spherical coordinates, $(u^1=\Psi, u^2=\Theta)$,
(cf.\ Equation~(\ref{eq:obsco}) with fixed $\tau$). Near $s=s_0$, map~(\ref{eq:front})
is an embedding, i.e., the wave front is a submanifold. Orthogonality
to $\mathcal{S}$ of the initial vectors $\dot{\lambda} (s_0)$ assures
that this submanifold is lightlike. Farther away from $\mathcal{S}$,
however, the wave front need not be a submanifold. The \emph{caustic} of
the wave front is the set of all points where the map~(\ref{eq:front})
is not an immersion, i.e., where its differential has rank $<3$.
As the derivative with respect to $s$ is always non-zero, the
rank can be $3-1$ (caustic point of \emph{multiplicity} one,
\emph{astigmatic} focusing) or $3-2$ (caustic point of \emph{multiplicity}
two, \emph{anastigmatic} focusing). In the first case, the
cross-section of an ``infinitesimally thin'' bundle of generators
collapses to a line, in the second case to a point (see
Section~\ref{ssec:Sachs}). For the case that the wave front is a
light cone with vertex $p_\mathrm{O}$, caustic points are said to be
\emph{conjugate} to $p_\mathrm{O}$ along the respective generator. For
an arbitrary wave front, one says that a caustic point
is \emph{conjugate} to any spacelike 2-surface in the wave front.
In this sense, the terms ``conjugate point'' and ``caustic point''
are synonymous. Along each generator, caustic points are isolated
(see Section~\ref{ssec:Sachs}) and thus denumerable. Hence, one may
speak of the first caustic, the second caustic, and so on. At all
points where the caustic is a manifold, it is either spacelike
or lightlike. For instance, the caustic of the
Schwarzschild light cone in Figure~\ref{fig:schwcon} is a spacelike
curve; in the spacetime of a transparent string, the caustic of
the light cone consists of two lightlike 2-manifolds that meet
in a spacelike curve (see Figure~\ref{fig:strcon2}).

Near a non-caustic point, a wave front is a hypersurface $S=
\mbox{constant}$ where $S$ satisfies the Hamilton--Jacobi equation
\begin{equation}
  \label{eq:eikonal}
  g^{ij} (x) \, \partial_i S (x) \, \partial_j S (x) = 0.
\end{equation}
In the terminology of optics, Equation~(\ref{eq:eikonal}) is called the
\emph{eikonal equation}.

At caustic points, a wave front typically forms cuspidal edges or
vertices whose geometry might be arbitrarily complicated, even locally.
If one restricts to caustics which are \emph{stable} against perturbations
in a certain sense, then a local classification of caustics is possible
with the help of Arnold's singularity theory of Lagrangian or
Legendrian maps. Full details of this theory can be found
in~\cite{arnold-gusein-varchenko-85}. For a readable review of
Arnold's results and its applications to wave fronts in general
relativity, we refer again to~\cite{ehlers-newman-2000}.
In order to apply Arnold's theory to wave fronts, one
associates each wave front with a Legendrian submanifold
in the projective cotangent bundle over $\mathcal{M}$ (or
with a Lagrangian submanifold in an appropriately reduced
bundle). A caustic point of the wave front corresponds to a point
where the differential of the projection from the Legendrian submanifold
to $\mathcal{M}$ has non-maximal rank. For the case
$\dim (\mathcal{M})=4$, which is of interest here,
Arnold has shown that there are only
five types of caustic points that are stable with respect to
perturbations within the class of all Legendrian submanifolds.
They are known as \emph{fold}, \emph{cusp}, \emph{swallow-tail},
\emph{pyramid}, and \emph{purse} (see Figure~\ref{fig:caustics}).
Any other type of caustic is unstable in the sense that it
changes non-diffeomorphically if it is perturbed within the
class of Legendrian submanifolds.

\epubtkImage{figure02.png}
{\begin{figure}[hptb]
   \def\epsfsize#1#2{0.8#1}
   \centerline{\epsfbox{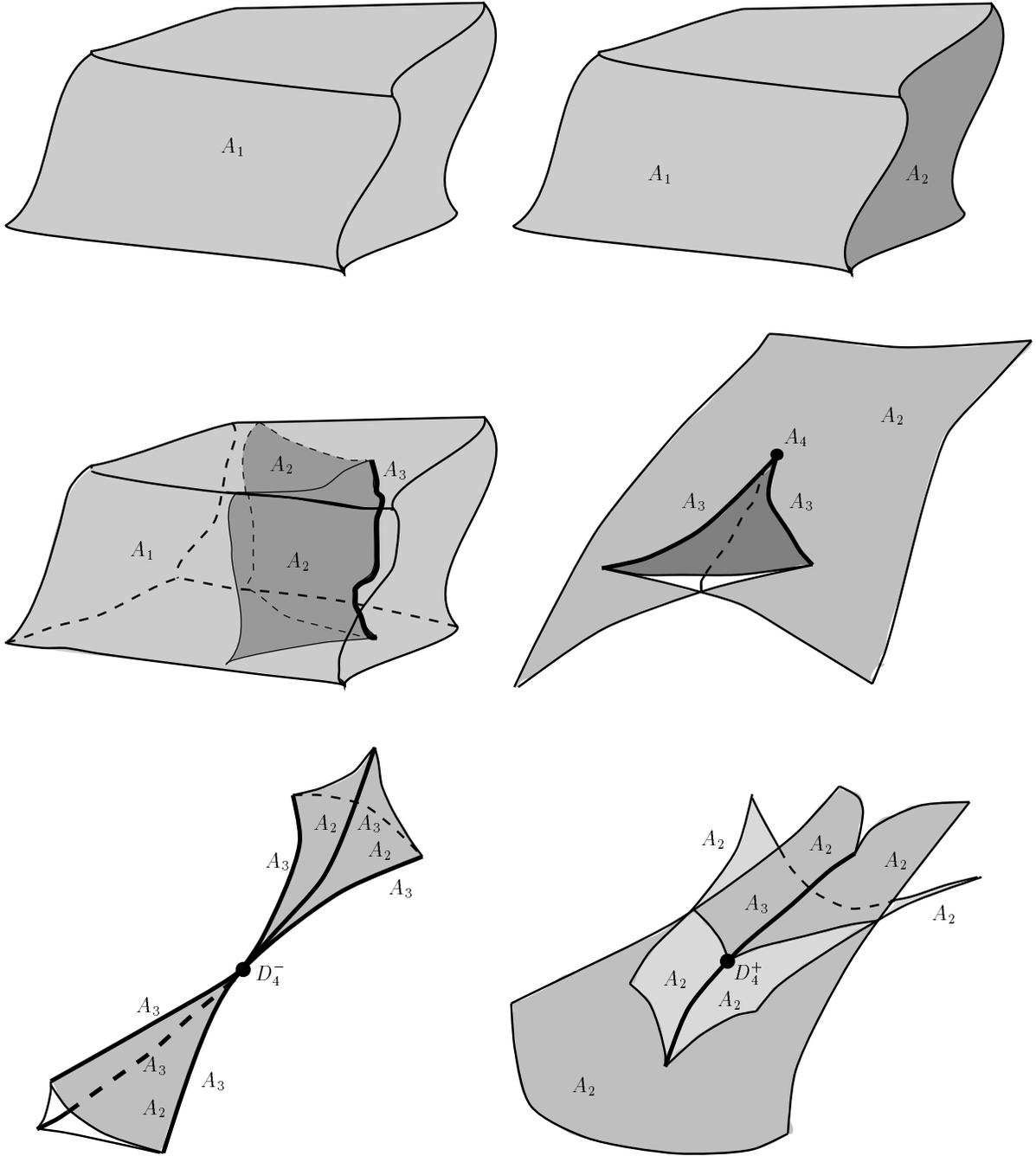}}
   \caption{Wave fronts that are locally stable in the sense
     of Arnold. Each picture shows the projection into 3-space of a
     wave-front, locally near a caustic point. The projection is made
     along the integral curves of a timelike vector field. The
     qualitative features are independent of which timelike vector field is
     chosen. In addition to regular, i.e., non-caustic, points $(A_1)$, there
     are five kinds of stable points, known as fold $(A_2)$, cusp $(A_3)$,
     swallow-tail $(A_4)$, pyramid $(D_4^-)$, and purse $(D_4^+)$.
     The $A_k$ and $D_k$ notation refers to a relation to exceptional
     groups (see~\cite{arnold-gusein-varchenko-85}). The picture
     is taken from~\cite{hasse-kriele-perlick-96}.}
   \label{fig:caustics}
 \end{figure}
}

Fold singularities of a wave front form a lightlike 2-manifold in
spacetime, on a sufficiently small neighborhood of any fold caustic
point. The second picture in Figure~\ref{fig:caustics} shows such a
``fold surface'', projected to 3-space along the integral curves of a
timelike vector field. This projected fold surface separates a region
covered twice by the wave front from a region not covered at all. If the
wave front is the past light cone of an observation event, and if one
restricts to light sources with worldlines in a sufficiently small
neighborhood of a fold caustic point, there are two images for light
sources on one side and no images for light sources on the
other side of the fold surface. Cusp singularities of a wave front form
a spacelike curve in spacetime, again locally near any cusp caustic point.
Such a curve is often called a ``cusp ridge''. Along a cusp ridge,
two fold surfaces meet tangentially. The third picture in
Figure~\ref{fig:caustics} shows the situation projected to 3-space.
Near a cusp singularity of a past light cone, there is local triple-imaging
for light sources in the wedge between the two fold surfaces
and local single-imaging for light sources outside this wedge.
Swallow-tail, pyramid, and purse singularities are points where
two or more cusp ridges meet with a common tangent, as illustrated
by the last three pictures in Figure~\ref{fig:caustics}.

Friedrich and Stewart~\cite{friedrich-stewart-83} have demonstrated
that all caustic types that are stable in the sense of Arnold can be
realized by wave fronts in Minkowski spacetime. Moreover, they stated
without proof that, quite generally, one gets the same stable caustic
types if one allows for perturbations only within the class of wave fronts
(rather than within the larger class of Legendrian submanifolds). A proof
of this statement was claimed to be given in~\cite{hasse-kriele-perlick-96}
where the Lagrangian rather than the Legendrian formalism was used.
However, the main result of this paper (Theorem~4.4
of~\cite{hasse-kriele-perlick-96}) is actually too weak to justify
this claim. A different version of the desired
stability result was indeed proven by another approach. In this approach
one concentrates on an \emph{instantaneous wave front}, i.e., on the
intersection of a wave front with a spacelike hypersurface $\mathcal{C}$.
As an alternative terminology, one calls the intersection of a (``big'')
wave front with a hypersurface $\mathcal{C}$ that is transverse to all generators a
``small wave front''. Instantaneous wave fronts are special cases of small
wave fronts. The caustic of a small wave front is the set of
all points where the small wave front fails to be an immersed 2-dimensional
submanifold of $\mathcal{C}$. If the spacetime is foliated by spacelike
hypersurfaces, the caustic of a wave front is the union of the caustics
of its small (=\,instantaneous) wave fronts. Such a foliation can always
be achieved locally, and in several spacetimes of interest even globally.
If one identifies different slices with the help of a timelike vector field,
one can visualize a wave front, and in particular a light cone, as a motion
of small (=\,instantaneous) wave fronts in 3-space. Examples are shown
in Figures~\ref{fig:schwfrt}, \ref{fig:monofrt1}, \ref{fig:monofrt2},
\ref{fig:strfrt1}, and~\ref{fig:strfrt2}. Mathematically, the same can
be done for non-spacelike slices as long as they are transverse to the
generators of the considered wave front (see Figure~\ref{fig:wave2}
for an example). Turning from (big) wave fronts to small wave fronts
reduces the dimension by one. The only caustic points of a small wave
front that are stable in the sense of Arnold are cusps and swallow-tails.
What one wants to prove is that all other caustic points are unstable with
respect to perturbations of the wave front \emph{within the class of
wave fronts}, keeping the metric and the slicing fixed. For spacelike
slicings (i.e., for instantaneous wave fronts), this was indeed
demonstrated by Low~\cite{low-98}. In this article, the author views
wave fronts as subsets of the space $\mathcal{N}$ of all lightlike
geodesics in $(\mathcal{M},g)$. General properties of this space
$\mathcal{N}$ are derived in earlier articles by Low~\cite{low-89,
low-93} (also see Penrose and Rindler~\cite{penrose-rindler-86},
volume~II, where the space $\mathcal{N}$ is treated in twistor language).
Low considers, in particular, the case of a globally
hyperbolic spacetime~\cite{low-98}; he demonstrates the desired stability result for
the intersections of a (big) wave front with Cauchy hypersurfaces
(see Section~\ref{ssec:hypfront}). As every point in an arbitrary spacetime
admits a globally hyperbolic neighborhood, this local stability result
is universal. Figure~\ref{fig:strfrt2} shows an instantaneous
wave front with cusps and a swallow-tail point. Figure~\ref{fig:schwfrt}
shows instantaneous wave fronts with caustic points that are neither cusps
nor swallow-tails; hence, they must be unstable with respect to perturbations
of the wave front within the class of wave fronts.

It is to be emphasized that Low's work allows to classify the stable
caustics of small wave fronts, but not directly of (big) wave fronts.
Clearly, a (big) wave front is a one-parameter family of small wave
fronts. A qualitative change of a small wave front, in dependence of
a parameter, is called a ``metamorphosis'' in the English literature
and a ``perestroika'' in the Russian literature. Combining Low's
results with the theory of metamorphoses, or perestroikas, could lead
to a classsification of the stable caustics of (big) wave fronts. However,
this has not been worked out until now.

Wave fronts in general relativity have been studied in a long series of
articles by Newman, Frittelli, and collaborators. For some aspects of their
work see Sections~\ref{ssec:fermat} and~\ref{ssec:asy}. In the
quasi-Newtonian approximation formalism of lensing, the classification
of caustics is treated in great detail in the book by Petters, Levine,
and Wambsganss~\cite{petters-levine-wambsganss-2001}. Interesting related
mateial can also be found in Blandford and Narayan~\cite{blandford-narayan-86}.
For a nice exposition of caustics in ordinary optics see Berry and
Upstill~\cite{berry-upstill-80}.

A light source that comes close to the caustic of the observer's past
light cone is seen strongly magnified. For a point source whose
worldline passes exactly through the caustic, the ray-optical treatment
even gives an infinite brightness (see Section~\ref{ssec:brightness}).
If a light source passes behind a compact deflecting mass, its brightness
increases and decreases in the course of time, with a maximum at the
moment of closest approach to the caustic.
Such \emph{microlensing} events are routinely observed by monitoring
a large number of stars in the bulge of our Galaxy, in the Magellanic
Clouds, and in the Andromeda Galaxy (see, e.g., \cite{mollerach-roulet-2002}
for an overview). In his millennium essay on
future perspectives of gravitational lensing, Blandford~\cite{blandford-2001}
mentioned the possibility of observing a chosen light source strongly
magnified over a period of time with the help of a space-born telescope.
The idea is to guide the spacecraft such that the worldline of the
light source remains in (or close to) the one-parameter family of
caustics of past light cones of the spacecraft over a period of time.
This futuristic idea of ``caustic surfing'' was mathematically further
discussed by Frittelli and Petters~\cite{frittelli-petters-2002}.


\subsection{Optical scalars and Sachs equations}
\label{ssec:Sachs}

For the calculation of distance measures, of image distortion, and of
the brightness of images one
has to study the \emph{Jacobi equation} (=\,equation of geodesic
deviation) along lightlike geodesics. This is usually done in terms
of the \emph{optical scalars} which were introduced by Sachs et
al.~\cite{jordan-ehlers-sachs-61, sachs-61}. Related background material
on lightlike geodesic congruences can be found in many text-books
(see, e.g., Wald~\cite{wald-84}, Section 9.2). In view of applications
to lensing, a particularly useful exposition was given by Seitz,
Schneider and Ehlers~\cite{seitz-schneider-ehlers-94}. In the
following the basic notions and results will be summarized. \\

\noindent
{\bf Infinitesimally thin bundles.} \\
\noindent
Let $s \longmapsto \lambda (s)$ be an affinely parametrized
lightlike geodesic with tangent vector field $K = \dot{\lambda}$.
We assume that $\lambda$ is past-oriented, because in applications
to lensing one usually considers rays from the observer to the source.
We use the summation convention for capital indices $A,B,\dots$
taking the values 1 and 2. An \emph{infinitesimally thin bundle} (with
elliptical cross-section) along $\lambda$ is a set
\begin{equation}
  \label{eq:B}
  \mathcal{B} =
  \left\{ c^A Y_A \,\big|\, c^1, c^2 \in \mathbb{R},
  ~~ \delta_{AB} \, c^A c^B \le 1 \right\}.
\end{equation}
Here $\delta_{AB}$ denotes the Kronecker delta, and $Y_1$ and $Y_2$ are
two vector fields along $\lambda$ with
\begin{eqnarray}
  \nabla_K \nabla_K Y_A &=& R(K,Y_A,K),
  \label{eq:Jacobi} \\
  g(K,Y_A) &=& 0,
  \label{eq:YA1}
\end{eqnarray}%
such that $Y_1 (s)$, $Y_2 (s)$, and $K(s) $ are linearly independent
for almost all $s$. As usual, $R$ denotes the curvature tensor, defined by
\begin{equation}
  \label{eq:defR}
  R(X,Y,Z) = \nabla_X \nabla_Y Z - \nabla_Y \nabla_X Z - \nabla_{[X,Y]}Z.
\end{equation}
Equation~(\ref{eq:Jacobi}) is the Jacobi equation. It is a precise mathematical formulation
of the statement that ``the arrow-head of $Y_A$ traces an infinitesimally
neighboring geodesic''. Equation~(\ref{eq:YA1}) guarantees that this neighboring
geodesic is, again, lightlike and spatially related to $\lambda$. 
Vector fields
$Y_A$ that satisfy Equation~(\ref{eq:Jacobi}) are known as \emph{Jacobi 
vector fields}.
\\

\noindent
{\bf Sachs basis.} \\
\noindent
For discussing the geometry of infinitesimally thin bundles it
is usual to introduce a \emph{Sachs basis}, i.e., two vector fields
$E_1$ and $E_2$ along $\lambda$ that are orthonormal, orthogonal to
$K= \dot{\lambda}$, and parallelly transported,
\begin{equation}
  \label{eq:EA}
  g(E_A,E_B) = \delta_{AB},
  \qquad
  g(K, E_A) = 0,
  \qquad
  \nabla_K E_A = 0.
\end{equation}
Apart from the possibility to interchange them, $E_1$ and $E_2$ are
unique up to transformations
\begin{eqnarray}
  \tilde{E}_1 &=& \cos \alpha \, E_1 + \sin \alpha \, E_2 + a_1 K,
  \label{eq:Etrafo1} \\
  \tilde{E}_2 &=& - \sin \alpha \, E_1 + \cos \alpha \, E_2 + a_2 K,
  \label{eq:Etrafo2}
\end{eqnarray}%
where $\alpha$, $a_1$, and $a_2$ are constant along $\lambda$. A Sachs basis
determines a unique vector field $U$ with $g(U,U)=-1$ and $g(U,K)=1$ along
$\lambda$ that is perpendicular to $E_1$, and $E_2$. As $K$ is assumed
past-oriented, $U$ is future-oriented. In the rest system of the observer
field $U$, the Sachs basis spans the 2-space perpendicular to the ray. It is
helpful to interpret this 2-space as a ``screen''; correspondingly,
linear combinations of $E_1$ and $E_2$ are often refered to as ``screen
vectors''. \\

\noindent
{\bf Jacobi matrix.} \\
\noindent
With respect to a Sachs basis, the basis vector fields $Y_1$ and
$Y_2$ of an infinitesimally thin bundle can be represented as
\begin{equation}
  \label{eq:D}
  Y_A = D_A^B E_B + y_A K.
\end{equation}
The \emph{Jacobi matrix} $\boldsymbol{D} = (D_A^B)$ relates the shape
of the cross-section of the infinitesimally
thin bundle to the Sachs basis (see Figure~\ref{fig:shape}).
Equation~(\ref{eq:Jacobi}) implies that $\boldsymbol{D}$ satisfies
the \emph{matrix Jacobi equation}
\begin{equation}
  \label{eq:JacobiD}
  \ddot{\boldsymbol{D}} = \boldsymbol{D} \boldsymbol{R},
\end{equation}
where an overdot means derivative with respect to the affine parameter
$s$, and
\begin{equation}
  \label{eq:tidalR}
  \boldsymbol{R} =
  \begin{pmatrix}
    \Phi_{00} & 0 \\
    0 & \Phi_{00}
  \end{pmatrix}
  +
  \begin{pmatrix}
    -\real (\psi_0) ~ & \imag (\psi_0) \\
    \imag (\psi_0) ~  & \real (\psi_0)
  \end{pmatrix}
\end{equation}
is the \emph{optical tidal matrix}, with
\begin{equation}
  \label{eq:Phipsi}
  \Phi_{00} = - \frac{1}{2} \Ric (K,K),
  \qquad
  \psi_0 = - \frac{1}{2} C \left( E_1-iE_2,K,E_1-iE_2,K \right).
\end{equation}
Here $\Ric$ denotes the Ricci tensor, defined by
$\Ric (X,Y) = \tr \left( R (\cdot,X,Y) \right)$,
and $C$ denotes the conformal curvature tensor (=\,Weyl tensor). The
notation in Equation~(\ref{eq:Phipsi}) is chosen in agreement with the
Newman--Penrose formalism (cf., e.g., \cite{chandrasekhar-83}). As $Y_1$,
$Y_2$, and $K$ are not everywhere linearly dependent,
$\det ( \boldsymbol{D})$ does not vanish
identically. Linearity of the matrix Jacobi equation
implies that $\det (\boldsymbol{D})$ has only isolated
zeros. These are the ``caustic points'' of the bundle (see below). \\

\noindent
{\bf Shape parameters.} \\
\noindent
The Jacobi matrix $\boldsymbol{D}$ can be parametrized
according to
\begin{equation}
  \label{eq:shape}
  \boldsymbol{D} =
  \begin{pmatrix}
    \cos \psi ~ & - \sin \psi \\
    \sin \psi ~ & \cos \psi
  \end{pmatrix}
  \begin{pmatrix}
    D_+ & 0 \\
    0 & D_-
  \end{pmatrix}
  \begin{pmatrix}
    \cos \chi   ~ & \sin \chi \\
    - \sin \chi ~ & \cos \chi
  \end{pmatrix}.
\end{equation}
Here we made use of the well-known facts that any matrix can be written as the
product of an orthogonal and a symmetric matrix and that any symmetric
matrix can be diagonalized by an orthogonal transformation. Our definition of
infinitesimally thin bundles implies that $D_+$ and $D_-$ are non-zero almost
everywhere. In the representation of Equation~(\ref{eq:shape}), the
extremal points of the bundle's elliptical cross-section are given by the
position vectors
\begin{eqnarray}
  Y_+ &=& \cos \psi \, Y_1 + \sin \psi \, Y_2 \simeq
  D_+ \left( \cos \chi \, E_1 + \sin \chi \, E_2 \right),
  \label{eq:Yp} \\
  Y_- &=& - \sin \psi \, Y_1 + \cos \psi \, Y_2 \simeq
  D_- \left( - \sin \chi \, E_1 + \cos \chi \, E_2 \right),
  \label{eq:Ym}
\end{eqnarray}%
where $\simeq$ means equality up to multiples of $K$.
Hence, $|D_+|$ and $|D_-|$ give the semi-axes of the elliptical
cross-section and $\chi$ gives the angle by which the ellipse is
rotated with respect to the Sachs basis (see Figure~\ref{fig:shape}).
We call $D_+$, $D_-$, and $\chi$ the \emph{shape
parameters} of the bundle. This name is taken from Frittelli, Kling, and
Newman~\cite{frittelli-kling-newman-2001a, frittelli-kling-newman-2001b}
who actually use, instead of $D_+$ and $D_-$, the equivalent 
quantities $D_+D_-$ and $D_+/D_-$. 
For the case that the infinitesimally thin bundle can be embedded in 
a wave front, the shape parameters $D_+$ and $D_-$ have the following
interesting property (see Kantowski et al.~\cite{kantowski-68,
dwivedi-kantowski-72}).
$\dot{D}_+ / D_+$ and $\dot{D}_- / D_-$ give the principal curvatures
of the wave front in the rest system of the observer field $U$ which is
perpendicular to the Sachs basis. The notation $D_+$ and
$D_-$, which is taken from~\cite{dwivedi-kantowski-72},
is convenient because it often allows to write two equations in the
form of one equation with a $\pm$ sign (see, e.g., Equation~(\ref{eq:dotshape})
or Equation~(\ref{eq:ssscalars}) below). The angle $\chi$ can be directly linked
with observations if a light source emits linearly polarized light (see
Section~\ref{ssec:distortion}). 

For any infinitesimally thin bundle, given in terms of $Y_1$ and $Y_2$, we can 
choose the Sachs basis as we like. This freedom leads to two ambiguities in the 
definition of $D_+$ and $D_-$. Firstly, the transformation $(E_1,E_2) \mapsto 
(-E_1,E_2)$ results in $(D_+,D_-, \chi , \psi ) \mapsto 
(-D_+,D_-,-\chi , \psi)$, and the analogous transformation  
$(E_1,E_2) \mapsto (E_1,-E_2)$ results in $(D_+,D_-, \chi , \psi ) 
\mapsto (D_+,-D_-,-\chi , \psi)$; this shows that the signs of $D_+$ and $D_-$ 
are ambiguous. Secondly, the transformation 
$(E_1,E_2) \mapsto (E_2,-E_1)$ results in  
$(D_+,D_-, \chi , \psi ) \mapsto (D_-,D_+,\chi ,\psi + \pi /2)$; this shows that $D_+$ 
and $D_-$ can be interchanged. The most interesting case for us is that of an 
infinitesimally thin bundle that issues from a vertex at an observation event 
$p_\mathrm{O} = \lambda (0)$ into the past. For such bundles we can remove
the sign ambiguity in the definition of $D_+(s)$ and $D_-(s)$ by requiring that 
they are positive for small positive values of $s$. The freedom of interchanging
them can be removed, e.g., by requiring that $D_+(s) \ge D_-(s)$ for small
positive values of $s$; for spherically symmetric and static spacetimes, however,
another convention is more convenient, see Section \ref{ssec:ss} below. If we 
have chosen a convention that makes $D_+$ and $D_-$ unique along the bundle, 
the Sachs basis can still be changed by a transformation~(\ref{eq:Etrafo1},
\ref{eq:Etrafo2}). Under such a transformation 
the shape parameters change according to $\tilde{D}_{\pm} =
D_{\pm}$, $\tilde{\chi} = \chi - \alpha$, $\tilde{\psi} = \psi$.
This demonstrates the important fact that the shape and the
size of the cross-section of an infinitesimally thin bundle
have an invariant (observer-independent) meaning~\cite{sachs-61}. \\
\epubtkImage{figure03.png}
{\begin{figure}[hptb]
   \def\epsfsize#1#2{0.55#1}
   \centerline{\epsfbox{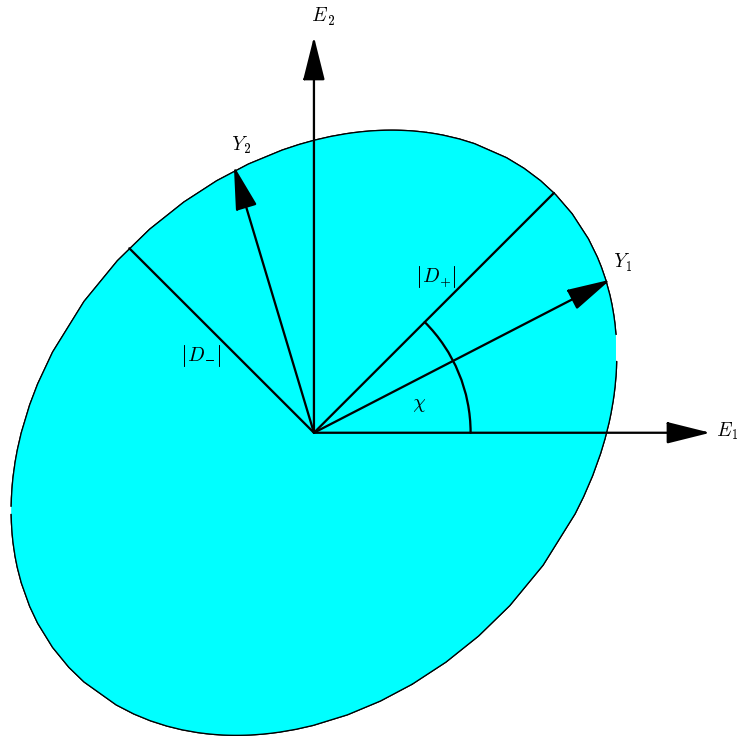}}
   \caption{Cross-section of an infinitesimally thin bundle. The
     Jacobi matrix~(\ref{eq:shape}) relates the Jacobi fields $Y_1$ and
     $Y_2$ that span the bundle to the Sachs basis vectors $E_1$ and $E_2$.
     The shape parameters $D_+$, $D_-$, and $\chi$ determine the outline of
     the cross-section; the angle $\psi$ that appears in Equation~(\ref{eq:shape})
     does not show in the outline. The picture shows the projection into the
     2-space (``screen'') spanned by $E_1$ and $E_2$; note that, in general,
     $Y_1$ and $Y_2$ have components perpendicular to the screen.}
   \label{fig:shape}
 \end{figure}
}

\noindent
{\bf Optical scalars.} \\
\noindent
Along each infinitesimally thin bundle one defines the \emph{deformation
matrix} $\boldsymbol{S}$ by
\begin{equation}
  \label{eq:dotD}
  \dot{\boldsymbol{D}} = \boldsymbol{D} \boldsymbol{S}.
\end{equation}
This reduces the second-order linear differential equation~(\ref{eq:JacobiD})
for $\boldsymbol{D}$ to a first-order non-linear differential equation for
$\boldsymbol{S}$,
\begin{equation}
  \label{eq:JacobiS}
  \dot{\boldsymbol{S}} + \boldsymbol{S} \boldsymbol{S} = \boldsymbol{R}.
\end{equation}
It is usual to decompose $\boldsymbol{S}$ into antisymmetric,
symmetric-tracefree, and trace parts,
\begin{equation}
  \label{eq:scalars}
  \boldsymbol{S} =
  \begin{pmatrix}
    0        ~ & \omega \\
    - \omega ~ & 0
  \end{pmatrix}
  +
  \begin{pmatrix}
    \sigma_1 ~ & \sigma_2 \\
    \sigma_2 ~ & - \sigma_1
  \end{pmatrix}
  +
  \begin{pmatrix}
    \theta ~ & 0 \\
    0      ~ & \theta
  \end{pmatrix}.
\end{equation}
This defines the \emph{optical scalars} $\omega$ (\emph{twist}),
$\theta$ (\emph{expansion}), and $(\sigma_1, \sigma_2)$
(\emph{shear}). One usually combines them into two complex
scalars $\varrho=\theta + i \omega$ and $\sigma = \sigma_1 + i
\sigma_2$. A change~(\ref{eq:Etrafo1}, \ref{eq:Etrafo2})
of the Sachs basis affects the optical scalars according to
$\tilde{\varrho} =\varrho$ and $\tilde{\sigma} = e^{-2i\alpha}
\sigma$. Thus, $\varrho$ and $|\sigma|$ are invariant. If
rewritten in terms of the optical scalars, Equation~(\ref{eq:JacobiS})
gives the \emph{Sachs equations}
\begin{eqnarray}
  \dot{\varrho} &=& - \varrho^2 - | \sigma |^2 + \Phi_{00},
  \label{eq:rhoSachs} \\
  \dot{\sigma} &=&
  - \sigma \left( \varrho + \overline{\varrho} \right) + \psi_0.
  \label{eq:sigmaSachs}
\end{eqnarray}%
One sees that the Ricci curvature term $\Phi_{00}$ directly
produces expansion (focusing) and that the conformal curvature
term $\psi_0$ directly produces shear. However, as the shear
appears in Equation~(\ref{eq:rhoSachs}), conformal curvature indirectly
influences focusing (cf.\ Penrose~\cite{penrose-66}). With $\boldsymbol{D}$
written in terms of the shape parameters and $\boldsymbol{S}$
written in terms of the optical scalars, Equation~(\ref{eq:dotD})
results in
\begin{equation}
  \label{eq:dotshape}
  \dot{D}_{\pm} + i \dot{\chi} D_{\pm} - i \dot{\psi} D_{\mp} =
  \left( \rho \pm e^{-2i\chi} \sigma \right) D_{\pm}.
\end{equation}
Along $\lambda$, Equations~(\ref{eq:rhoSachs}, \ref{eq:sigmaSachs})
give a system of 4 real first-order differential equations for the 4 real
variables $\varrho$ and $\sigma$; if $\varrho$ and $\sigma$ are known,
Equation~(\ref{eq:dotshape}) gives a system of 4 real first-order differential
equations for the 4 real variables $D_{\pm}$, $\chi$, and $\psi$.
The twist-free solutions ($\varrho$ real) to Equations~(\ref{eq:rhoSachs},
\ref{eq:sigmaSachs}) constitute a 3-dimensional linear subspace of the
4-dimensional space of all solutions. This subspace carries a natural
metric of Lorentzian signature, unique up to a conformal factor, and was
nicknamed \emph{Minikowski space}
in~\cite{barraco-kozameh-newman-tod-90}. \\

\noindent
{\bf Conservation law.} \\
\noindent
As the optical tidal matrix $\boldsymbol{R}$ is symmetric, for any two
solutions $\boldsymbol{D}_1$ and $\boldsymbol{D}_2$ of the matrix
Jacobi equation~(\ref{eq:JacobiD}) we have
\begin{equation}
  \label{eq:conservationD}
  \dot{\boldsymbol{D}}_1 \boldsymbol{D}_2^T -
  \boldsymbol{D}_1 \dot{\boldsymbol{D}}_2^T = \mbox{constant},
\end{equation}
where $(~)^T$ means transposition.
Evaluating the case $\boldsymbol{D}_1=\boldsymbol{D}_2$ shows that
for every infinitesimally thin bundle
\begin{equation}
  \label{eq:conservationomega}
  \omega D_+ D_- = \mbox{constant}.
\end{equation}
Thus, there are two types of infinitesimally thin bundles: those
for which this constant is non-zero and those for which it is zero. In
the first case the bundle is twisting ($\omega \neq 0$ everywhere)
and its cross-section nowhere collapses to a line or to a
point ($D_+ \neq 0$ and $D_- \neq 0$ everywhere). In the second
case the bundle must be non-twisting ($\omega = 0$ everywhere), because
our definition of infinitesimally thin bundles implies that
$D_+ \neq 0$ and $D_- \neq 0$ almost everywhere. A quick calculation 
shows that $\omega = 0$ is 
exactly the integrability condition that makes sure that the 
infinitesimally thin bundle can be embedded in a wave front. (For
the definition of wave fronts see Section~\ref{ssec:front}.)
In other words, an infinitesimally thin bundle is twist-free if and
only if we can find a wave front such that $\lambda$ is one of the 
generators and the vector fields $Y_1$ and $Y_2$ connect 
$\lambda$ with infinitesimally neighboring generators.
For a (necessarily twist-free) infinitesimally thin bundle, points
where one of the two shape parameters $D_+$ and $D_-$ vanishes
are called \emph{caustic points} of \emph{multiplicity}
one, and points where both shape parameters $D_+$ and $D_-$ vanish
are called \emph{caustic points} of
\emph{multiplicity} two. This notion coincides exactly with the
notion of caustic points, or conjugate points, of
wave fronts as introduced in Section~\ref{ssec:front}. The behavior
of the optical scalars near caustic points can be deduced
from Equation~(\ref{eq:dotshape}) with Equations~(\ref{eq:rhoSachs},
\ref{eq:sigmaSachs}). For a caustic point of multiplicty one
at $s=s_0$ one finds
\begin{eqnarray}
  \theta (s) &=& \frac{1}{2(s-s_0)} \left( 1 + {\cal O}(s-s_0) \right),
  \label{eq:caust1} \\
  \left| \sigma (s) \right| &=&
  \frac{1}{2(s-s_0)} \left( 1 + {\cal O}(s-s_0) \right).
  \label{eq:caust2}
\end{eqnarray}%
By contrast, for a caustic point of multiplicity two at $s=s_0$
the equations read (cf.~\cite{seitz-schneider-ehlers-94})
\begin{eqnarray}
  \theta (s) &=& \frac{1}{s-s_0} + {\cal O}(s-s_0),
  \label{eq:caust3} \\
  \sigma (s) &=& \frac{1}{3} \psi_0 (s_0) (s-s_0) +
  {\cal O} \left( (s-s_0)^2 \right).
  \label{eq:caust4}
\end{eqnarray}%

\noindent
{\bf Infinitesimally thin bundles with vertex.} \\
\noindent
We say that an infinitesimally thin bundle has a \emph{vertex}
at $s=s_0$ if the Jacobi matrix satisfies
\begin{equation}
  \label{eq:vertex}
  \boldsymbol{D} (s_0) = \boldsymbol{0},
  \qquad
  \dot{\boldsymbol{D}} (s_0) = \boldsymbol{1}.
\end{equation}
A vertex is, in particular, a caustic point of multiplicity two.
An infinitesimally thin bundle with a vertex must be non-twisting.
While any non-twisting infinitesimally thin bundle can be
embedded in a wave front, an infinitesimally thin bundle with
a vertex can be embedded in a light cone. Near the vertex, to
within a first-order approximation with respect to $s-s_0$, it
has a circular cross-section.
If $\boldsymbol{D}_1$ has a vertex
at $s_1$ and $\boldsymbol{D}_2$ has a vertex at $s_2$, the
conservation law~(\ref{eq:conservationD}) implies
\begin{equation}
  \label{eq:etherington}
  \boldsymbol{D}_2^T (s_1) = - \boldsymbol{D}_1 (s_2).
\end{equation}
This is Etherington's~\cite{etherington-33} reciprocity law.
The method by which this law was proven here follows
Ellis~\cite{ellis-72} (cf.\ Schneider, Ehlers, and
Falco~\cite{schneider-ehlers-falco-92}). Etherington's
reciprocity law is of relevance, in particular in view of
cosmology, because it relates the luminosity distance to the
area distance (see Equation~(\ref{eq:reci})). It was independently
rediscovered in the 1960s by Sachs and Penrose (see~\cite{penrose-66,
  kristian-sachs-66}).

The results of this section are the basis for Sections~\ref{ssec:distance},
\ref{ssec:distortion}, and~\ref{ssec:brightness}.


\subsection{Distance measures}
\label{ssec:distance}

In this section we summarize various distance measures that are
defined in an arbitrary spacetime. Some of them are directly related
to observable quantities with relevance for lensing. The material
of this section makes use of the results on infinitesimally thin bundles
which are summarized in Section~\ref{ssec:Sachs}. All of the distance
measures to be discussed refer to a past-oriented lightlike geodesic
$\lambda$ from an observation event $p_\mathrm{O}$ to an emission
event $p_\mathrm{S}$ (see Figure~\ref{fig:distance}).
Some of them depend on the 4-velocity $U_\mathrm{O}$ of the observer at $p_\mathrm{O}$
and/or on the 4-velocity $U_\mathrm{S}$ of the light source at $p_\mathrm{S}$. If a vector
field $U$ with $g(U,U)=-1$ is distinguished on $\mathcal{M}$, we can choose
for the observer an integral curve of $U$ and for the light sources all
other integral curves of $U$. Then each of the distance measures becomes
a function of the observational coordinates $(s, \Psi, \Theta, \tau)$
(recall Section~\ref{ssec:cone}). \\

\epubtkImage{figure04.png}
{\begin{figure}[hptb]
   \def\epsfsize#1#2{0.47#1}
   \centerline{\epsfbox{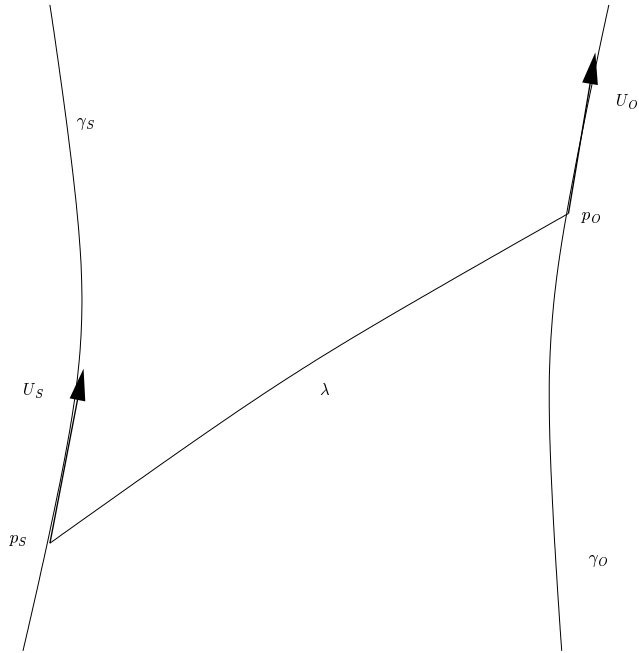}}
   \caption{Past-oriented lightlike geodesic $\lambda$ from an
     observation event $p_\mathrm{O}$ to an emission event
     $p_\mathrm{S}$. $\gamma_\mathrm{O}$ is the worldline of the
     observer, $\gamma_\mathrm{S}$ is the worldline of the light
     source. $U_\mathrm{O}$ is the 4-velocity of the observer at
     $p_\mathrm{O}$ and $U_\mathrm{S}$ is the 4-velocity of the
     light source at $p_\mathrm{S}$.}
   \label{fig:distance}
 \end{figure}
}

\noindent
{\bf Affine distance.} \\
\noindent
There is a unique affine parametrization $s \longmapsto \lambda (s)$ for
each lightlike geodesic through the observation event $p_\mathrm{O}$ such that
$\lambda (0) = p_\mathrm{O}$ and $g \left( \dot{\lambda} (0), U_\mathrm{O} \right) = 1$.
Then the affine parameter $s$ itself can be viewed as a distance measure.
This \emph{affine distance} has the desirable features that it increases
monotonously along each ray and that it coincides in an infinitesimal
neighborhood of $p_\mathrm{O}$ with Euclidean distance in the rest system of
$U_\mathrm{O}$. The affine distance depends on the 4-velocity $U_\mathrm{O}$ of the observer
but not on the 4-velocity $U_\mathrm{S}$ of the light source. It is a
mathematically very convenient notion, but it is not an observable.
(It can be operationally realized in terms of an observer field whose
4-velocities are parallel along the ray. Then the affine distance
results by integration if each observer measures the length of an
infinitesimally short part of the ray in his rest system. However, in
view of astronomical situations this is a purely theoretical construction.)
The notion of affine distance was introduced by Kermack, McCrea, and
Whittaker~\cite{kermack-mccrea-whittaker-32}. \\

\noindent
{\bf Travel time.} \\
\noindent
As an alternative distance measure one can use the \emph{travel time}. This
requires the choice of a time function, i.e., of a function $t$ that slices
the spacetime into spacelike hypersurfaces $t = \mbox{constant}$. (Such a
time function globally exists if and only if the spacetime is stably
causal; see, e.g., \cite{hawking-ellis-73}, p.~198.) The travel time is equal to
$t(p_\mathrm{O})-t(p_\mathrm{S})$, for each $p_\mathrm{S}$ on the past
light cone of $p_\mathrm{O}$. In other
words, the intersection of the light cone with a hypersurface
$t=\mbox{constant}$ determines events of equal travel time; we call
these intersections ``instantaneous wave fronts'' (recall
Section~\ref{ssec:front}). Examples of instantaneous wave fronts are
shown in Figures~\ref{fig:schwfrt}, \ref{fig:monofrt1}, \ref{fig:monofrt2},
\ref{fig:strfrt1}, and~\ref{fig:strfrt2}. The travel time increases
monotonously along each ray. Clearly, it depends neither on the 4-velocity
$U_\mathrm{O}$ of the observer nor on the 4-velocity $U_\mathrm{S}$ of the light source. Note
that the travel time has a unique value at each point of $p_\mathrm{O}$'s past light
cone, even at events that can be reached by two different rays from $p_\mathrm{O}$.
Near $p_\mathrm{O}$ the travel time coincides with Euclidean distance in the
observer's rest system only if $U_\mathrm{O}$ is perpendicular to the hypersurface
$t = \mbox{constant}$ with $dt(U_\mathrm{O})=1$. (The latter equation is true if
along the observer's world line the time function $t$ coincides with proper
time.) The travel time is not directly observable. However, travel time
differences are observable in multiple-imaging situations if the intrinsic
luminosity of the light source is time-dependent. To illustrate this, think of
a light source that flashes at a particular instant. If the flash reaches
the observer's wordline along two different rays, the proper time difference
$\Delta \tau_\mathrm{O}$ of the two arrival events is directly measurable. For a time
function $t$ that along the observer's worldline coincides with proper time,
this observed \emph{time delay} $\Delta \tau_\mathrm{O}$ gives the difference in
travel time for the two rays. In view of applications, the measurement of
time delays is of great relevance for quasar lensing. For the double
quasar 0957+561 the observed time delay $\Delta \tau_\mathrm{O}$ is
about 417~days (see, e.g., \cite{petters-levine-wambsganss-2001},
p.~149). \\

\noindent
{\bf Redshift.} \\
\noindent
In cosmology it is common to use the \emph{redshift} as a distance
measure. For assigning a redshift to a lightlike geodesic $\lambda$
that connects the observation event $p_\mathrm{O}$ on the worldline $\gamma_\mathrm{O}$
of the observer with the emission event $p_\mathrm{S}$ on the worldline $\gamma_\mathrm{S}$
of the light source, one considers a neighboring lightlike geodesic that
meets $\gamma_\mathrm{O}$ at a proper time interval $\Delta \tau_\mathrm{O}$ from $p_\mathrm{O}$
and $\gamma_\mathrm{S}$ at a proper time interval $\Delta \tau_\mathrm{S}$ from $p_\mathrm{S}$.
The redshift $z$ is defined as
\begin{equation}
  \label{eq:z}
  z = \!\!\underset{\Delta \tau_\mathrm{S} \to 0}{\lim}
  \frac{\Delta \tau_\mathrm{O} -
  \Delta \tau_\mathrm{S}}{\Delta \tau_\mathrm{S}}.
\end{equation}
If $\lambda$ is affinely parametrized with $\lambda (0) = p_\mathrm{O}$
and $\lambda (s) = p_\mathrm{S}$, one finds that $z$ is given by
\begin{equation}
  \label{eq:redshift}
  1 + z =
  \frac{g\left( \dot{\lambda} (0), U_\mathrm{O} \right)}
  {g\left( \dot{\lambda} (s), U_\mathrm{S} \right)}.
\end{equation}
This general \emph{redshift formula} is due to Kermack, McCrea, and
Whittaker~\cite{kermack-mccrea-whittaker-32}. Their proof is based
on the fact that $g(\dot{\lambda},Y)$ is a constant for
all Jacobi fields $Y$ that connect $\lambda$ with an infinitesimally
neighboring lightlike geodesic. The
same proof can be found, in a more elegant form, in~\cite{brill-72}
and in~\cite{straumann-84}, p.~109. An alternative proof, based on
variational methods, was given by Schr\"odinger~\cite{schroedinger-56}.
Equation~(\ref{eq:redshift}) is in agreement with the Hamilton formalism for
photons. Clearly, the redshift depends on the 4-velocity $U_\mathrm{O}$ of the
observer and on the 4-velocity $U_\mathrm{S}$ of the light source. If a vector
field $U$ with $g(U,U)=-1$ has been distinguished on $\mathcal{M}$, we
may choose one integral curve of $U$ as the observer
and all other integral curves of $U$ as the light sources.
Then the redshift becomes a function of the
observational coordinates $(s, \Psi, \Theta, \tau)$.
For $s \to 0$, the redshift goes to 0,
\begin{equation}
  \label{eq:hubble}
  z (s, \Psi, \Theta, \tau) = h(\Psi, \Theta, \tau) s + {\cal O}(s^2),
\end{equation}
with a (generalized) \emph{Hubble parameter} $h( \Psi, \Theta, \tau)$
that depends on spatial direction and on time. For criteria that $h$
and the higher-order coefficients are independent of $\Psi$ and
$\Theta$ see~\cite{hasse-perlick-99}. If the redshift
is known for one observer field $U$, it can be calculated for any
other $U$, according to Equation~(\ref{eq:redshift}), just by adding the usual
special-relativistic Doppler factors. Note that if $U_\mathrm{O}$ is given, the
redshift can be made to zero along any one ray $\lambda$ from $p_\mathrm{O}$ by
choosing the 4-velocities $U_{\lambda (s)}$
appropriately. This shows that $z$ is a reasonable distance measure only
for special situations, e.g., in cosmological models with $U$ denoting
the mean flow of luminous matter (``Hubble flow''). In any case, the
redshift is directly observable if the light source emits identifiable
spectral lines. For the calculation of Sagnac-like effects, the redshift
formula~(\ref{eq:redshift}) can be evaluated piecewise along broken lightlike
geodesics~\cite{bazanski-98}. \\

\noindent
{\bf Angular diameter distances.} \\
\noindent
The notion of \emph{angular diameter distance} is based on the intuitive
idea that the farther an object is away the smaller it looks, according to
the rule
\begin{equation}
  \label{eq:angular}
  \mbox{object diameter} = \mbox{angle} \times \mbox{distance}.
\end{equation}
The formal definition needs the results of Section~\ref{ssec:Sachs} on
infinitesimally thin bundles. One considers a past-oriented lightlike
geodesic $s \longrightarrow \lambda (s)$ parametrized by affine distance,
i.e., $\lambda (0) = p_\mathrm{O}$ and $g \bigl( \dot{\lambda} (0), U_\mathrm{O} \bigr) = 1$,
and along $\lambda$ an infinitesimally thin bundle with vertex at the
observer, i.e., at $s = 0$. Then the shape parameters $D_+(s)$ and $D_-(s)$
(recall Figure~\ref{fig:shape}) satisfy the initial conditions $D_{\pm}(0)=0$
and $\dot{D}_{\pm}(0)=1$. They have the following physical
meaning. If the observer sees a circular image of (small) angular diameter
$\alpha$ on his or her sky, the (small but extended) light source at
affine distance $s$ actually has an elliptical cross-section with
extremal diameters $\alpha | D_{\pm} (s)|$.
It is therefore reasonable to call $D_+$ and $D_-$ the
\emph{extremal angular diameter distances}. Near the vertex,
$D_+$ and $D_-$ are monotonously increasing functions of the affine
distance, $D_{\pm} (s) = s + {\cal O}(s^2)$. Farther away from the vertex, however,
they may become decreasing, so the functions $s \mapsto D_+ (s)$ and $s
\mapsto D_-(s)$ need not be invertible. At a caustic point of multiplicity
one, one of the two functions $D_+$ and $D_-$ changes sign; at a caustic
point of multiplicity two, both change sign (recall Section~\ref{ssec:Sachs}).
The image of a light source at affine distance $s$ is said to have
\emph{even parity} if $D_+(s)D_-(s)>0$ and \emph{odd parity} if
$D_+(s)D_-(s)<0$. Images with odd parity show the neighborhood of the
light source side-inverted in comparison to images with even parity.
Clearly, $D_+$ and $D_-$ are reasonable distance measures
only in a neighborhood of the vertex where they are monotonously
increasing. However, the physical relevance of $D_+$ and $D_-$ lies
in the fact that they relate cross-sectional diameters at the
source to angular diameters at the observer, and this is always true,
even beyond caustic points. $D_+$ and $D_-$ depend on the 4-velocity
$U_\mathrm{O}$ of the observer but not on the 4-velocity $U_\mathrm{S}$ of the source.
This reflects the fact that the angular diameter of an image on the
observer's sky is subject to aberration whereas the cross-sectional
diameter of an infinitesimally thin bundle has an invariant meaning
(recall Section~\ref{ssec:Sachs}). Hence, if the observer's worldline
$\gamma_\mathrm{O}$ has been specified, $D_+$ and $D_-$ are well-defined
functions of the observational coordinates $(s, \Psi, \Theta, \tau)$. \\

\noindent
{\bf Area distance.} \\
\noindent
The \emph{area distance} $D_{\mathrm{area}}$ is defined
according to the idea
\begin{equation}
  \label{eq:area}
  \mbox{object area} = \mbox{solid angle} \times \mbox{distance}^2.
\end{equation}
As a formal definition for $D_{\mathrm{area}}$, in terms of the extremal
angular diameter distances $D_+$ and $D_-$ as functions of affine distance
$s$, we use the equation
\begin{equation}
  \label{eq:Darea}
  D_{\mathrm{area}} (s) = \sqrt{\left| D_+(s) \, D_-(s) \right|}.
\end{equation}
$D_{\mathrm{area}} (s)^2$ indeed relates, for a bundle with vertex at the
observer, the cross-sectional area at the source to the opening solid angle
at the observer. Such a bundle has a caustic point exactly at those points
where $D_{\mathrm{area}} (s) = 0$. The area distance is often called
``angular diameter distance'' although, as indicated by Equation~(\ref{eq:Darea}),
the name ``averaged angular diameter distance'' would be more appropriate.
Just as $D_+$ and $D_-$, the area distance depends on the 4-velocity $U_\mathrm{O}$
of the observer but not on the 4-velocity $U_\mathrm{S}$ of the light source. The
area distance is observable for a light source whose true size is known
(or can be reasonably estimated). It is sometimes convenient to introduce
the \emph{magnification} or \emph{amplification factor}
\begin{equation}
  \label{eq:magnification}
  \mu (s) = \frac{s^2}{D_+(s) \, D_-(s)}.
\end{equation}
The absolute value of $\mu$ determines the area distance, and
the sign of $\mu$ determines the parity. In Minkowski spacetime,
$D_{\pm}(s)=s$ and, thus, $\mu (s) =1$. Hence, $|\mu (s)| >1$ means that a
(small but extended) light source at affine distance $s$ subtends a
larger solid angle on the observer's sky than a light source of the
same size at the same affine distance in Minkowski spacetime. Note
that in a multiple-imaging situation the individual images may have
different affine distances. Thus, the relative magnification factor of
two images is not directly observable. This is an important difference to
the magnification factor that is used in the quasi-Newtonian approximation
formalism of lensing. The latter is defined by comparison with an ``unlensed
image'' (see, e.g., \cite{schneider-ehlers-falco-92}), a notion that makes
sense only if the metric is viewed as a perturbation of some ``background''
metric. One can derive a differential equation for the area distance (or,
equivalently, for the magnification factor) as a function of affine
distance in the following way. On every parameter interval where
$D_+D_-$ has no zeros, the real part of Equation~(\ref{eq:dotshape}) shows
that the area distance is related to the expansion by
\begin{equation}
  \label{eq:thetaarea}
  \dot{D}_{\mathrm{area}} = \theta D_{\mathrm{area}}.
\end{equation}
Insertion into the Sachs equation~(\ref{eq:rhoSachs}) for $\theta = \varrho$
gives the \emph{focusing equation}
\begin{equation}
  \label{eq:focus}
  \ddot{D}_{\mathrm{area}} =
  - \left( | \sigma |^2 + \frac{1}{2}
  \Ric (\dot{\lambda}, \dot{\lambda}) \right)
  D_{\mathrm{area}}.
\end{equation}
Between the vertex at $s=0$ and the first conjugate point (caustic
point), $D_{\mathrm{area}}$ is determined by Equation~(\ref{eq:focus}) and the
initial conditions
\begin{equation}
  \label{eq:iniDarea}
  D_{\mathrm{area}} (0) = 0,
  \qquad
  \dot{D}_{\mathrm{area}} (0) = 1.
\end{equation}
The Ricci term in Equation~(\ref{eq:focus}) is non-negative if
Einstein's field equation holds and if the energy density is non-negative
for all observers (``weak energy condition''). Then Equations~(\ref{eq:focus},
\ref{eq:iniDarea}) imply that
\begin{equation}
  \label{eq:foctheo}
  D_{\mathrm{area}} (s) \le s,
\end{equation}
i.e., $1 \le \mu (s)$, for all $s$ between the vertex at $s=0$ and
the first conjugate point. In Minkowski spacetime, the equation
$D_{\mathrm{area}} (s) = s$ holds. 
Hence, the 
inequality~(\ref{eq:foctheo}) says that a gravitational
field has a focusing, as opposed to a defocusing, effect. This is
sometimes called the \emph{focusing theorem}. \\

\noindent
{\bf Corrected luminosity distance.} \\
\noindent
The idea of defining distance measures in terms of bundle cross-sections
dates back to Tolman~\cite{tolman-30} and Whittaker~\cite{whittaker-31}.
Originally, this idea was applied not to bundles with vertex at the
observer but rather to bundles with vertex at the light source. The
resulting analogue of the area distance is the so-called \emph{corrected
luminosity distance} $D'_{\mathrm{lum}}$. It relates, for a bundle with vertex
at the light source, the cross-sectional area at the observer to the opening
solid angle at the light source. Owing to Etherington's reciprocity law~(\ref{eq:etherington}), area distance and corrected
luminosity distance are related by
\begin{equation}
  \label{eq:reci}
  D'_{\mathrm{lum}} = (1+z) D_{\mathrm{area}}.
\end{equation}
The redshift factor has its origin in the fact that the definition of
$D'_{\mathrm{lum}}$ refers to an affine parametrization adapted to $U_\mathrm{S}$,
and the definition of $D_{\mathrm{area}}$ refers to an affine
parametrization adapted to $U_\mathrm{O}$. While $D_{\mathrm{area}}$ depends
on $U_\mathrm{O}$ but not on $U_\mathrm{S}$, $D'_{\mathrm{lum}}$ depends on $U_\mathrm{S}$ but not
on $U_\mathrm{O}$. \\

\noindent
{\bf Luminosity distance.} \\
\noindent
The physical meaning of the corrected luminosity distance is most easily
understood in the photon picture. For photons isotropically emitted from a
light source, the percentage that hit a prescribed area at the observer
is proportional to $1/(D'_{\mathrm{lum}})^2$. As the
energy of each photon undergoes a redshift, the \emph{energy flux} at the
observer is proportional to $1/(D_{\mathrm{lum}})^2$, where
\begin{equation}
  \label{eq:Dlum}
  D_{\mathrm{lum}} = (1+z) D'_{\mathrm{lum}} =
  (1+z)^2 D_{\mathrm{area}}.
\end{equation}
Thus, $D_{\mathrm{lum}}$ is the relevant quantity for calculating the luminosity
(apparent brightness) of pointlike light sources (see Equation~(\ref{eq:flux})).
For this reason $D_{\mathrm{lum}}$ is called the (uncorrected) \emph{luminosity
distance}. The observation that the purely geometric quantity $D'_{\mathrm{lum}}$
must be modified by an additional redshift factor to give the energy flux is
due to Walker~\cite{walker-34}. $D_{\mathrm{lum}}$ depends on the 4-velocity
$U_\mathrm{O}$ of the observer and of the 4-velocity $U_\mathrm{S}$ of the light source.
$D_{\mathrm{lum}}$ and $D'_{\mathrm{lum}}$ can be viewed as functions of
the observational coordinates $(s, \Psi, \Theta, \tau)$ if a vector
field $U$ with $g(U,U)=-1$ has been distinguished, one integral curve
of $U$ is chosen as the observer, and the other integral curves of $U$
are chosen as the light sources. In that case Equation~(\ref{eq:hubble}) implies
that not only $D_{\mathrm{area}}(s)$ but also $D_{\mathrm{lum}}(s)$ and
$D'_{\mathrm{lum}}(s)$ are of the form $s + {\cal O}(s^2)$. Thus, near the observer
all three distance measures coincide with Euclidean distance in the observer's
rest space. \\

\noindent
{\bf Parallax distance.} \\
\noindent
In an arbitrary spacetime, we fix an
observation event $p_\mathrm{O}$ and the observer's 4-velocity $U_\mathrm{O}$. We consider
a past-oriented lightlike geodesic $\lambda$ parametrized by affine distance,
$\lambda (0)=p_\mathrm{O}$ and $g \left( \dot{\lambda} (0), U_\mathrm{O} \right) = 1$. To a
light source passing through the event $\lambda (s)$ we assign the
(averaged) \emph{parallax distance} $D_{\mathrm{par}} (s)=-\theta (0)^{-1}$,
where $\theta$ is the expansion of an infinitesimally thin bundle
with vertex at $\lambda (s)$. This definition follows~\cite{jordan-ehlers-sachs-61}. Its relevance in view
of cosmology was discussed in detail by Rosquist~\cite{rosquist-88}.
$D_{\mathrm{par}}$ can be measured by performing
the standard trigonometric parallax method of elementary Euclidean
geometry, with the observer at $p_\mathrm{O}$ and an assistant observer at the
perimeter of the bundle, and then averaging over all possible positions
of the assistant. Note that the method refers to a bundle with vertex
at the light source, i.e., to light rays that leave the
light source simultaneously. (Averaging is not necessary if this
bundle is circular.) $D_{\mathrm{par}}$ depends on the 4-velocity
of the observer but not on the 4-velocity of the light source. To within
first-order approximation near the observer it coincides with
affine distance (recall Equation~(\ref{eq:caust3})). For the potential obervational
relevance of $D_{\mathrm{par}}$ see~\cite{rosquist-88},
and~\cite{schneider-ehlers-falco-92}, p.~509.

In view of lensing, $D_+$, $D_-$, and $D_{\mathrm{lum}}$
are the most important distance measures because they are related to
image distortion (see Section~\ref{ssec:distortion}) and to the brightness
of images (see Section~\ref{ssec:brightness}). In spacetimes with
many symmetries, these quantities can be explicitly calculated (see
Section~\ref{ssec:flat} for conformally flat spacetimes, and
Section~\ref{ssec:ss} for spherically symmetric static spacetimes).
This is impossible in a spacetime without symmetries, in particular
in a realistic cosmological model with inhomogeneities (``clumpy universe'').
Following Kristian and Sachs~\cite{kristian-sachs-66}, one often uses
series expansions with respect to $s$. For statistical considerations one
may work with the focusing equation in a Friedmann--Robertson--Walker spacetime
with average density (see Section~\ref{ssec:flat}), or with a heuristically
modified focusing equation taking clumps into account. The latter leads to
the so-called \emph{Dyer--Roeder distance}~\cite{dyer-roeder-72, dyer-roeder-73}
which is discussed in several text-books (see, e.g., \cite{schneider-ehlers-falco-92}).
(For pre-Dyer--Roeder papers on optics in cosmological models with inhomogeneities,
see the historical notes in~\cite{kantowski-98}.) As overdensities have a focusing
and underdensities have a defocusing effect, it is widely believed (following~\cite{weinberg-76}) that after averaging over sufficiently large angular scales
the Friedmann--Robertson--Walker calculation gives the correct distance-redshift relation.
However, it was argued by Ellis, Bassett, and Dunsby~\cite{ellis-bassett-dunsby-98} that
caustics produced by the lensing effect of overdensities lead to a systematic bias
towards smaller angular sizes (``shrinking''). For a spherically symmetric inhomogeneity,
the effect on the distance-redshift relation can be calculated analytically~\cite{mustapha-bassett-hellaby-ellis-98}. For thorough discussions of light propagation
in a clumpy universe also see Pyne and Birkinshaw~\cite{pyne-birkinshaw-96}, and Holz
and Wald~\cite{holz-wald-98}.


\subsection{Image distortion}
\label{ssec:distortion}

In special relativity, a spherical object always shows a circular
outline on the observer's sky, independent of its state of motion~\cite{penrose-59, terrell-59}. In general relativity, this is no longer
true; a small sphere usually shows an elliptic outline on the
observer's sky. This \emph{distortion} is caused by the shearing effect
of the spacetime geometry on light bundles. For the calculation of
image distortion we need the material of Sections~\ref{ssec:Sachs}
and~\ref{ssec:distance}. For an observer with 4-velocity $U_\mathrm{O}$ at an
event $p_\mathrm{O}$, there is a unique affine parametrization $s \longmapsto
\lambda (s)$ for each lightlike geodesic through $p_\mathrm{O}$ such that
$\lambda (0) = p_\mathrm{O}$ and $g \bigl( \dot{\lambda} (0), U_\mathrm{O} \bigr) = 1$.
Around each of these $\lambda$ we can consider an infinitesimally
thin bundle with vertex at $s=0$. The elliptical cross-section of
this bundle can be characterized by the shape parameters $D_+ (s)$,
$D_- (s) $ and $\chi (s)$ (recall Figure~\ref{fig:shape}). 
As outlined in 
Section~\ref{ssec:Sachs}, we choose the convention of having 
$D_+ (s)$ and $D_- (s)$ positive for small positive $s$. 
In the terminology
of Section~\ref{ssec:distance}, $s$ is the affine distance, and $D_+ (s)$
and $D_- (s)$ are the extremal angular diameter distances. The complex
quantity
\begin{equation}
  \label{eq:ellipticity}
  \epsilon (s) =
  \left( \frac{D_+(s)}{D_-(s)} - \frac{D_-(s)}{D_+(s)} \right)
  e^{2i \chi (s)}
\end{equation}
is called the \emph{ellipticity} of the bundle. The phase of $\epsilon$
determines the position angle of the elliptical cross-section of the bundle
with respect to the Sachs basis. The absolute value of $\epsilon (s)$
determines the eccentricity of this cross-section; $\epsilon (s) =0$
indicates a circular cross-section and $\vert \epsilon (s) \vert = \infty$
indicates a caustic point of multiplicity one. (It is also common to use
other measures for the eccentricity, e.g., $\vert D_+ - D_- \vert /
\vert D_+ + D_- \vert$.) From Equation~(\ref{eq:dotshape}) with
$\varrho = \theta$ we get the derivative of $\epsilon$ with respect to the
affine distance $s$,
\begin{equation}
  \label{eq:dotepsilon}
  \dot{\epsilon} = 2 \sigma \sqrt{| \epsilon |^2 + 4}.
\end{equation}
The initial conditions $D_{\pm}(0)=0$, $\dot{D}_{\pm}(0)=1$ imply
\begin{equation}
  \label{eq:iniepsilon}
 \epsilon (0) = 0.
\end{equation}
Equation~(\ref{eq:dotepsilon}) and Equation~(\ref{eq:iniepsilon}) determine $\epsilon$
if the shear $\sigma$ is known. The shear, in turn, is determined
by the Sachs equations~(\ref{eq:rhoSachs}, \ref{eq:sigmaSachs})
and the initial conditions~(\ref{eq:caust3}, \ref{eq:caust4}) with
$s_0 = 0$ for $\theta ( = \varrho)$ and $\sigma$.

It is recommendable to change from the $\epsilon$ determined this way
to $\varepsilon = -\overline{\epsilon}$. This transformation corresponds
to replacing the Jacobi matrix $\boldsymbol{D}$ by its inverse. The original
quantity $\epsilon (s)$ gives the true shape of objects at affine distance
$s$ that show a circular image on the observer's sky. The new quantity
$\varepsilon (s)$ gives the observed shape for objects at affine distance
$s$ that actually have a circular cross-section. In other words, if a
(small) spherical body at affine distance $s$ is observed, the ellipticity
of its image on the observer's sky is given by $\varepsilon (s)$.

By Equations~(\ref{eq:dotepsilon}, \ref{eq:iniepsilon}), $\epsilon$ vanishes along
the entire ray if and only if the shear $\sigma$ vanishes along the entire ray.
By Equations~(\ref{eq:sigmaSachs}, \ref{eq:caust4}), the shear vanishes along the
entire ray if and only if the conformal curvature term $\psi_0$ vanishes along
the entire ray. The latter condition means that $K= \dot{\lambda}$ is tangent
to a \emph{principal null direction} of the conformal curvature tensor
(see, e.g., Chandrasekhar~\cite{chandrasekhar-83}). At a point where
the conformal curvature tensor is not zero, there are at most four
different principal null directions. Hence, the distortion effect
vanishes along all light rays if and only if the conformal curvature
vanishes everywhere, i.e., if and only if the spacetime is
conformally flat. This result is due to Sachs~\cite{sachs-61}.
An alternative proof, based on expressions for image distortions
in terms of the exponential map, was given by Hasse~\cite{hasse-87}.

For any observer, the distortion measure $\varepsilon = - \overline{\epsilon}$
is defined along every light ray from every point of the observer's worldline.
This gives $\varepsilon$ as a function of the observational coordinates
$(s, \Psi, \Theta, \tau)$ (recall Section~\ref{ssec:cone}, in
particular Equation~(\ref{eq:obsco})). If we fix $\tau$ and $s$, $\varepsilon$ is
a function on the observer's sky. (Instead of $s$, one may choose any of
the distance measures discussed in Section~\ref{ssec:distance}, provided
it is a unique function of $s$.) In spacetimes with
sufficiently many symmetries, this function can be explicitly determined in
terms of integrals over the metric function. This will be worked out
for spherically symmetric static spacetimes in Section~\ref{ssec:ss}.
A general consideration of image distortion and example calculations
can also be found in papers by Frittelli, Kling and Newman~\cite{frittelli-kling-newman-2001a, frittelli-kling-newman-2001b}.
Frittelli and Oberst~\cite{frittelli-oberst-2001} calculate
image distortion by a ``thick gravitational lens'' model within
a spacetime setting.

In cases where it is not possible to determine $\varepsilon$ by explicitly
integrating the relevant differential equations, one may consider series
expansions with respect to the affine parameter $s$. This technique, which
is of particular relevance in view of cosmology, dates back to Kristian
and Sachs~\cite{kristian-sachs-66} who introduced image distortion as an
observable in cosmology. In lowest non-vanishing order,
$\varepsilon (s, \Psi, \Theta, \tau_\mathrm{O})$ is quadratic with respect to
$s$ and completely determined by the conformal curvature tensor at the
observation event $p_\mathrm{O} = \gamma (\tau_\mathrm{O})$, as can be read
from Equations~(\ref{eq:dotepsilon}, \ref{eq:iniepsilon}, \ref{eq:caust4}).
One can classify all possible distortion patterns
on the observer's sky in terms of the Petrov type of the Weyl tensor~\cite{chrobok-perlick-2001}. As outlined in~\cite{chrobok-perlick-2001},
these patterns are closely related to what Penrose and Rindler~\cite{penrose-rindler-86} call the \emph{fingerprint} of the Weyl tensor.
At all observation events where the Weyl tensor is non-zero, the following
is true. There are at most four points
on the observer's sky where the distortion vanishes, corresponding to
the four (not necessarily distinct) principal null directions of the
Weyl tensor. For type $N$, where all four principal null directions
coincide, the distortion pattern is shown in Figure~\ref{fig:distortion}.

\epubtkImage{figure05.png}
{\begin{figure}[hptb]
   \def\epsfsize#1#2{0.8#1}
   \centerline{\epsfbox{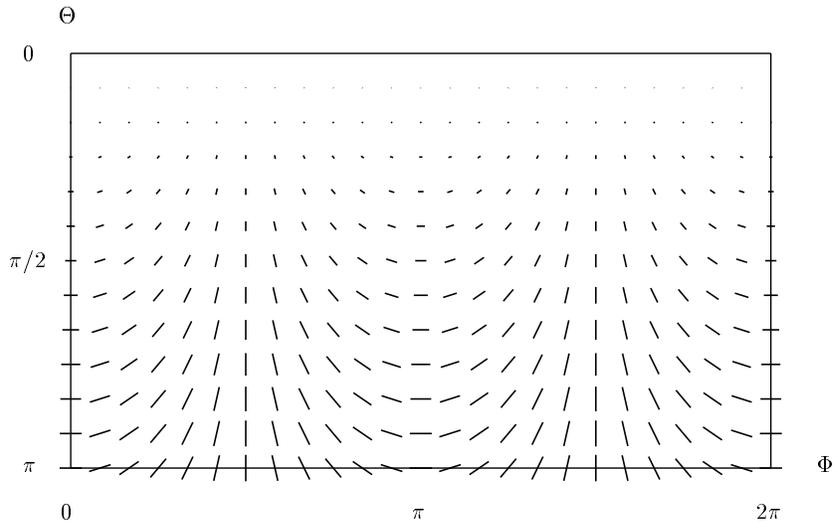}}
   \caption{Distortion pattern. The picture shows, in a Mercator
     projection with $\Phi$ as the horizontal and $\Theta$ as the
     vertical coordinate, the celestial sphere of an observer at a
     spacetime point where the Weyl tensor is of Petrov type $N$. The
     pattern indicates the elliptical images of spherical objects to
     within lowest non-trivial order with respect to distance. The
     length of each line segment is a measure for the eccentricity of
     the elliptical image, the direction of the line segment indicates
     its major axis. The distortion effect vanishes at the north pole
     $\Theta = 0$ which corresponds to the fourfold principal null
     direction. Contrary to the other Petrov types, for type N the
     pattern is universal up to an overall scaling factor. The picture
     is taken from~\cite{chrobok-perlick-2001} where the distortion
     patterns for the other Petrov types are given as well.}
   \label{fig:distortion}
 \end{figure}
}

The distortion effect is routinely observed since the mid-1980s
in the form of arcs and (radio) rings
(see~\cite{schneider-ehlers-falco-92, petters-levine-wambsganss-2001,
  wambsganss-98} for an overview).
In these cases a distant galaxy appears strongly elongated in one direction. Such
strong elongations occur near a caustic point of multiplicity one where
$| \varepsilon | \to \infty$. In the case
of rings and (long) arcs, the entire bundle cannot be treated as
infinitesimally thin, i.e., a theoretical description of the effect
requires an integration. For the idealized case of a point source,
images in the form of (1-dimensional) rings on the observer's sky
occur in cases of rotational symmetry and are usually called ``Einstein
rings'' (see Section~\ref{ssec:ss}). The rings that are actually observed
show extended sources in situations close to rotational symmetry.

For the majority of galaxies that are not distorted into arcs or rings,
there is a ``weak lensing'' effect on the apparent shape that can be
investigated statistically. The method is based on the assumption that
there is no prefered direction in the universe, i.e., that the axes
of (approximately spheroidal) galaxies are randomly distributed. So,
without a distortion effect, the axes of galaxy images should make a
randomly distributed angle with the $(\Psi, \Theta)$ grid on the
observer's sky. Any deviation from a random distribution is to be attributed
to a distortion effect, produced by the gravitational field of intervening
masses. With the help of the quasi-Newtonian approximation, this
method has been elaborated into a sophisticated formalism for
determining mass distributions, projected onto the plane perpendicular
to the line of sight, from observed image distortions. This is one
of the most important astrophysical tools for detecting (dark)
matter. It has been used to determine the mass distribution in
galaxies and galaxy clusters, and to probe the large-scale structure 
of the universe (see~\cite{bartelmann-schneider-2001, huterer-2010} 
for reviews). 
From a methodological point of view, it would be desirable 
to analyse this important line of astronomical research within a 
spacetime setting. This should give prominence to the role of the 
conformal curvature tensor.

Another interesting way of observing weak image distortions is possible
for sources that emit linearly polarized radiation. This is true
for many radio galaxies. (Polarization measurements are also relevant
for strong-lensing situations; see Schneider, Ehlers, and
Falco~\cite{schneider-ehlers-falco-92}, p.~82 for an example.)
The method is based on the geometric optics approximation of Maxwell's
theory. In this approximation, the polarization direction is parallel
along each ray between source and observer~\cite{ehlers-67}
(cf., e.g., \cite{misner-thorne-wheeler-73}, p.~577). We may, thus, 
choose the Sachs basis $(E_1, E_2)$ such that the plane spanned by
$K$ and $E_1$ gives the polarization direction. This fixes the Sachs
basis up to transformations~(\ref{eq:Etrafo2}) with $\alpha = 0$, 
i.e., it gives an unambiguous (observer-independent) meaning to the
angle $\chi$ in Figure~\ref{fig:shape}. If a light source (e.g., a galaxy) 
shows an approximately elliptic shape on the observer's sky, it is 
reasonable to assume that at the light source the polarization direction 
is aligned with one of the axes, i.e., $2 \chi (s) / \pi \in \mathbb{Z}$. 
A distortion effect is verified if the \emph{observed} polarization direction 
is not aligned with an axis of the image, $2 \chi (0) / \pi \notin
\mathbb{Z}$. It is to be emphasized that such a change of the angle
$\chi$ along the ray \emph{cannot} be the result of a rotation; the bundles 
under consideration have a vertex and are, thus, twist-free. It can only 
be the result of successive shearing processes, governed by the behaviour
of the conformal curvature tensor along the ray. Also, the effect has nothing 
to do with the rotation of an observer field; we have already stressed that
the angle $\chi$ is observer-independent. Related misunderstandings have 
been clarified by Panov and Sbytov~\cite{panov-sbytov-92, panov-sbytov-98}. 
So far, this distortion effect has not been observed. (Panov and
Sbytov~\cite{panov-sbytov-92} have clearly shown that an anisotropy
observed by Birch~\cite{birch-82}, even if real, cannot be interpreted in this way.)
Its future detectability is estimated, for distant radio
sources, in~\cite{surpi-harari-99}.

The effect of a gravitatational field on the polarization direction of light
was first discussed by Skrotskii~\cite{skrotskii-57} in 1957 and is therefore 
sometimes called the ``Skrotskii effect''. If the spacetime is
conformally stationary, and if the worldlines of observers and light sources 
are integral curves of the conformal Killing vector field, the effect can be 
expressed in terms of the ``Fermat geometry'' of 3-space~\cite{hasse-perlick-93},
see Section~\ref{ssec:conf} below for the definition of the Fermat geometry.
(Note that Figure~1 in~\cite{hasse-perlick-93} is erroneous because it
ignores the fact that, in general, the principal shear directions of a bundle
are not parallel along the central ray.) Relative to a frame that is parallel
with respect to the Fermat metric, one finds a rotation of the polarization
direction that is analogous to the well-known Faraday rotation in a magnetic field.
In this analogy, the magnetic field corresponds to the rotation (twist) of 
the conformal Killing vector field. Because of this analogy, the Skrotskii
effect is also known as the ``gravitational Faraday effect''. It has been
quite extensively discussed for stationary spacetimes and, in particular,
for the Kerr metric (see, e.g.,~\cite{godfrey-70, su-mallet-80, fayos-llosa-82,
ishihara-takahashi-tomimatsu-88, nourizonoz-99, sereno-2005}).
All these articles give formulas for the rotation of the
polarization direction relative to a frame distinguished by the symmetry
assumptions. This rotation should \emph{not}  be confused with 
the above-mentioned motion of the polarization direction relative to the
orientation of the image. The latter is a distortion effect, governed by
the conformal curvature tensor; the former is a gravitomagnetic effect,
governed by the rotation of a distinguished observer field.


\subsection{Brightness of images}
\label{ssec:brightness}

For calculating the brightness of images we need the definitions and
results of Section~\ref{ssec:distance}. In particular we need the
luminosity distance $D_{\mathrm{lum}}$ and its relation to other distance
measures. We begin by considering a point source (worldline) that emits
isotropically with (bolometric, i.e., integrated over all frequencies)
luminosity $L$. By definition of $D_{\mathrm{lum}}$, in this case the
energy flux at the observer is
\begin{equation}
  \label{eq:flux}
  F = \frac{L}{4 \pi D_{\mathrm{lum}}{}^2}.
\end{equation}
$F$ is a measure for the brightness of the image on the
observer's sky. The \emph{magnitude} $m$ used by astronomers is essentially the
negative logarithm of $F$,
\begin{equation}
  \label{eq:mag}
  m = 2.5 \log_{10} \left( D_{\mathrm{lum}}{}^2 \right) -
  2.5 \log_{10} (L) + m_0,
\end{equation}
with $m_0$ being a universal constant. In Equation~(\ref{eq:flux}), $D_{\mathrm{lum}}$
can be expresed in terms of the area distance $D_{\mathrm{area}}$ and
the redshift $z$ with the help of the general relation~(\ref{eq:Dlum}).
This demonstrates that the magnification factor $\mu$, which is
defined by Equation~(\ref{eq:magnification}), admits the following reinterpretation.
$| \mu (s) |$ relates the flux from a point source at affine distance $s$
to the flux from a point source with the same luminosity at the same affine
distance and at the same redshift in Minkowski spacetime.

$D_{\mathrm{lum}}$ can be explicitly calculated in spacetimes
where the Jacobi fields along lightlike geodesics can be explicitly
determined. This is true, e.g., in spherically symmetric and
static spacetimes where the extremal angular diameter distances
$D_+$ and $D_-$ can be calculated in terms of integrals over the
metric coefficients. The resulting formulas are given in
Section~\ref{ssec:ss} below. Knowledge of $D_+$ and $D_-$ immediately
gives the area distance $D_{\mathrm{area}}$ via Equation~(\ref{eq:Darea}).
$D_{\mathrm{area}}$ together with the redshift determines $D_{\mathrm{lum}}$
via Equation~(\ref{eq:Dlum}). Such an explicit calculation is, of course, possible
only for spacetimes with many symmetries.

By Equation~(\ref{eq:Dlum}), the zeros of $D_{\mathrm{lum}}$ coincide with the
zeros of $D_{\mathrm{area}}$, i.e., with the caustic points. Hence, in
the ray-optical treatment a point source is infinitely bright
(magnitude $m = - \infty$) if it passes through the caustic of
the observer's past light cone. A wave-optical treatment shows
that the energy flux at the observer is actually bounded by
diffraction. In the quasi-Newtonian approximation formalism, this
was demonstrated by an explicit calculation for light rays deflected
by a spheroidal mass by Ohanian~\cite{ohanian-83}
(cf.~\cite{schneider-ehlers-falco-92}, p.~220). Quite generally, the
ray-optical calculation of the energy flux gives incorrect results
if, for two different light paths from the source worldline to the
observation event, the time delay is smaller than or approximately equal to the
coherence time. Then interference effects give rise to frequency-dependent
corrections to the energy flux that have to be calculated with the help
of wave optics. In multiple-imaging situations, the time delay decreases
with decreasing mass of the deflector. If the deflector is a cluster of
galaxies, a galaxy, or a star, interference effects can be ignored.
Gould~\cite{gould-92} suggested that they could be observable if a
deflector of about $10^ {-15}$ Solar masses happens to be close to the
line of sight to a gamma-ray burster. In this case, the angle-separation
between the (unresolvable) images would be of the order $10^ {-15}$
arcseconds (``femtolensing''). Interference effects could make a
frequency-dependent imprint on the total intensity. Ulmer and
Goodman~\cite{ulmer-goodman-95} discussed related effects for
deflectors of up to $10^{-11}$ Solar masses. Femtolensing has not been
observed so far. However, it is an interesting future perspective for
lensing effects where wave optics has to be taken into account. This would
give practical relevance to the theoretical work of Herlt and
Stephani~\cite{herlt-stephani-76, herlt-stephani-78} who
calculated gravitational lensing on the basis of wave optics in the
Schwarzschild spacetime. Wave-optical aspects of gravitational lensing
are also discussed in~\cite{nakamura-deguchi-99}.

We now turn to the case of an extended source, whose surface makes up
a 3-dimensional timelike submanifold $\mathcal{T}$ of the spacetime. In
this case the radiation is characterized by the \emph{surface
brightness} $B$ (=\,luminosity $L$ per area) at the source
and by the \emph{intensity} $I$ (=\,energy flux $F$ per solid angle) at
the observer. For each past-oriented light ray from an observation
event $p_\mathrm{O}$ and to an event $p_\mathrm{S}$ on $\mathcal{T}$, we can
relate $B$ and $I$ in the following way. By definition, the area distance
$D_{\mathrm{area}}$ relates the area at the source to the solid angle at
the observer, so we get from Equation~(\ref{eq:flux}) $I = B
D_{\mathrm{area}}{}^2 / (4 \pi D_{\mathrm{lum}}{}^2)$. 
As area distance and luminosity distance are related by a
redshift factor, according to the general law~(\ref{eq:Dlum}), this gives
the relation
\begin{equation}
  \label{eq:intensity}
  I = \frac{B}{4 \pi (1+z)^4}.
\end{equation}
This result is, of course, valid only if the radiation from
different parts of the emitting surface is incoherent; otherwise
interference effects have to be taken into account. The most
remarkable feature of Equation~(\ref{eq:intensity}) is that all distance
measures have dropped out. Save for a redshift factor, the
(observed) intensity of a radiating surface is the same for
all observers.

The law for point sources~(\ref{eq:flux}) and the law for extended
sources~(\ref{eq:intensity}) refer to bolometric quantities, i.e., to integration
over all frequencies. As every astronomical observation is restricted to a
certain frequency range, it is actually necessary to consider
frequency-specific quantities. For a point source, one writes
$L=\int_0^{\infty} \ell (\omega_\mathrm{S}) d \omega_\mathrm{S}$ and
$F=\int_0^{\infty} f (\omega_\mathrm{O}) d \omega_\mathrm{O}$, where the
specific luminosity $\ell$ is a function of the emitted frequency
$\omega_\mathrm{S}$ and the specific flux $f$ is a function of the received
frequency $\omega_\mathrm{O}$. As $\omega_\mathrm{S}$ and $\omega_\mathrm{O}$ are related
by a redshift factor, the frequency-specific version of
Equation~(\ref{eq:flux}) reads
\begin{equation}
  \label{eq:specificflux}
  f(\omega_\mathrm{O}) =
  \frac{\ell \left( \omega_\mathrm{O} (1+z) \right)
  (1+z)}{4 \pi D_{\mathrm{lum}}{}^2}.
\end{equation}
Similarly, for an extended source one introduces a specific surface
brightness $b$ and a specific intensity $i$ such that $B=\int_0^{\infty}
b (\omega_\mathrm{S}) d \omega_\mathrm{S}$ and $I=\int_0^{\infty} i (\omega_\mathrm{O}) d \omega_\mathrm{O}$.
Then one gets the following frequency-specific version of Equation~(\ref{eq:intensity}).
\begin{equation}
  \label{eq:specificintensity}
  i(\omega_\mathrm{O}) =
  \frac{b \left( \omega_\mathrm{O} (1+z) \right)}{4 \pi (1+z)^3}.
\end{equation}
The results summarized in this section can also be derived from
the kinetic theory of photons (see, e.g., \cite{ehlers-73}). In the
photon picture, the three redshift factors in Equation~(\ref{eq:specificintensity})
are easily understood: The first reflects the fact that each photon
undergoes a redshift; the second relates the rate of emission (with
respect to proper time at the source) to the rate of reception (with
respect to proper time at the obsever); the third reflects the aberration
effect on the angular size of the source in dependence of the motion of
the observer.

As an example for the calculation of the brightness of images we consider
the Schwarzschild spactime (see Figure~\ref{fig:schwDe}).


\subsection{Conjugate points and cut points}
\label{ssec:cut}

In general, the past light cone of an event forms caustics
and transverse self-intersections, i.e., it is neither an
embedded nor an immersed submanifold. The relevance of this
fact in view of lensing was emphasized already in
Section~\ref{ssec:cone}. In the following we demonstrate that
caustics and transverse self-intersections of the light cone
are related to extremizing properties of lightlike geodesics.
A light cone with a caustic and a transverse self-intersection
is shown in Figure~\ref{fig:strcon2}.

In this section and in Section~\ref{ssec:crit} we
use mathematical techniques which are related to the
Penrose--Hawking singularity theorems. For background material,
see Penrose~\cite{penrose-72}, Hawking and Ellis~\cite{hawking-ellis-73},
O'Neill~\cite{oneill-83}, and Wald~\cite{wald-84}.

Recall from Section~\ref{ssec:front} that the caustic
of the past light cone of $p_\mathrm{O}$ is the set of all points
where this light cone is not an immersed submanifold. A point
$p_\mathrm{S}$ is in the caustic if a generator $\lambda$ of the light
cone intersects at $p_\mathrm{S}$ an infinitesimally neighboring generator.
In this situation $p_\mathrm{S}$ is said to be conjugate to $p_\mathrm{O}$ along
$\lambda$. The caustic of the past light cone of $p_\mathrm{O}$ is also
called the ``past lightlike conjugate locus'' of $p_\mathrm{O}$.

The notion of conjugate points is related to the extremizing
properties of lightlike geodesics in the following way.
Let $\lambda$ be a past-oriented lightlike geodesic with
$\lambda (0) = p_\mathrm{O}$. Assume that $p_\mathrm{S}=\lambda (s_0)$ is the
first conjugate point along this geodesic. This means that
$p_\mathrm{S}$ is in the caustic of the past light cone of $p_\mathrm{O}$
and that $\lambda$ does not meet the caustic at parameter values
between 0 and $s_0$. Then a well-known theorem says that
all points $\lambda (s)$ with $0<s<s_0$ cannot be reached from
$p_\mathrm{O}$ along a timelike curve arbitrarily close to $\lambda$,
and all points $\lambda (s)$ with $s>s_0$ can. For a proof we
refer to Hawking and Ellis~\cite{hawking-ellis-73},
Proposition 4.5.11 and Proposition 4.5.12. It might be helpful
to consult O'Neill~\cite{oneill-83}, Chapter 10, Proposition 48,
in addition.

Here we have considered a past-oriented lightlike geodesic
because this is the situation with relevance to lensing.
Actually, Hawking and Ellis consider the time-reversed situation,
i.e., with $\lambda$ future-oriented. Then the result can be
phrased in the following way. A material particle may catch up
with a light ray $\lambda$ after the latter has passed through
a conjugate point and, for particles staying close to $\lambda$,
this is impossible otherwise. The restriction to particles staying
close to $\lambda$ is essential. Particles ``taking a short cut''
may very well catch up with a lightlike geodesic even if the latter
is free of conjugate points.

For a discussion of the extremizing property in the global sense,
not restricted to timelike curves close to $\lambda$, we need the
notion of \emph{cut points}. The precise definition of cut points
reads as follows.

As ususal, let $I^- (p_\mathrm{O})$ denote the chronological past of $p_\mathrm{O}$, i.e.,
the set of all $q \in \mathcal{M}$ that can be reached from $p_\mathrm{O}$ along
a past-pointing timelike curve. In Minkowski spacetime, the boundary
$\partial I^- (p_\mathrm{O})$ of $I^-(p_\mathrm{O})$ is just the past light cone of $p_\mathrm{O}$
united with $\{p_\mathrm{O}\}$. In an arbitrary spacetime, this is not true. A
lightlike geodesic $\lambda$ that issues from $p_\mathrm{O}$ into the past is
always confined to the closure of $I^-(p_\mathrm{O})$, but it need not stay on
the boundary. The last point on $\lambda$ that is on the boundary
is by definition~\cite{budic-sachs-76} the \emph{cut point} of $\lambda$.
In other words, it is exactly the part of $\lambda$ beyond the cut point
that can be reached from $p_\mathrm{O}$ along a timelike curve. The union of all
cut points, along any past-pointing lightlike geodesic $\lambda$ from
$p_\mathrm{O}$, is called the \emph{cut locus} of the past light cone (or the
past lightlike cut locus of $p_\mathrm{O}$). For the
light cone in Figure~\ref{fig:strcon1} this is the curve (actually
2-dimensional) where the two sheets of the light cone intersect. For
the light cone in Figure~\ref{fig:strcon2} the cut locus is the same set
plus the swallow-tail point (actually 1-dimensional). For a detailed
discussion of cut points in manifolds with metrics of Lorentzian signature,
see~\cite{beem-ehrlich-easley-96}. For positive definite metrics, the notion
of cut points dates back to Poincar\'e~\cite{poincare-05} and
Whitehead~\cite{whitehead-35}.

For a generator $\lambda$ of the past light cone of $p_\mathrm{O}$, the cut point of
$\lambda$ does not exist in either of the two following cases:
\begin{enumerate}
\item $\lambda$ always stays on the boundary $\partial I^-
  (p_\mathrm{O})$, i.e., it never loses its extremizing property.
  \label{item_1}
\item $\lambda$ is always in $I^-(p_\mathrm{O})$, i.e., it fails to be
  extremizing from the very beginning.
  \label{item_2}
\end{enumerate}
Case~\ref{item_2} occurs, e.g., if there is a closed timelike curve through $p_\mathrm{O}$.
More precisely, Case~\ref{item_2} is excluded if the \emph{past distinguishing condition}
is satisfied at $p_\mathrm{O}$, i.e., if for $q \in \mathcal{M}$ the implication
\begin{equation}
  \label{eq:distinguish}
  I^-(q) = I^-(p_\mathrm{O}) \quad \Longrightarrow \quad q = p_\mathrm{O}
\end{equation}
holds. If the implication~(\ref{eq:distinguish}) is true, the following 
can be shown:

\begin{penumerate}%
  If, along $\lambda$, the point $\lambda (s)$ is conjugate to
  $\lambda (0)$, the cut point of $\lambda$ exists and
  it comes on or before $\lambda (s)$.
  \label{item_p1}
\end{penumerate}

\begin{penumerate}%
  Assume that a point $q$ can be reached from $p_\mathrm{O}$ along
  two different lightlike geodesics $\lambda_1$ and $\lambda_2$ from
  $p_\mathrm{O}$. Then the cut point of $\lambda_1$ and of $\lambda_2$
  exists and it comes on or before $q$.
  \label{item_p2}
\end{penumerate}

\begin{penumerate}%
  If the cut locus of a past light cone is empty, this past light cone
  is an embedded submanifold of $\mathcal{M}$.
  \label{item_p3}
\end{penumerate}
\vspace{0.8 em}

\noindent
For proofs see~\cite{perlick-2000}; The proofs can also be found in
or easily deduced
from~\cite{beem-ehrlich-easley-96}. Statement~\ref{item_p1} says that
conjugate points and cut points are related by the easily remembered
rule ``the cut point comes first''. Statement~\ref{item_p2} says that
a ``cut'' between two geodesics is indicated by the occurrence of a
cut point. However, it does \emph{not} say that exactly at the cut
point a second geodesic is met. Such a stronger statement, which truly
justifies the name ``cut point'', holds in globally hyperbolic
spacetimes (see Section~\ref{ssec:hypcrit}). Statement~\ref{item_p3} implies
that the occurrence of transverse self-intersections of a light cone
are always indicated by cut points. Note, however, that transverse
self-intersections of the past light cone of $p_\mathrm{O}$ may occur
inside $I^- (p_\mathrm{O})$ and, thus, far away from the cut locus.

Statement~\ref{item_p1} implies that $\partial I^- (p_\mathrm{O})$ is
an immersed submanifold everywhere
except at the cut locus and, of course, at the vertex $p_\mathrm{O}$. It is known
(see~\cite{hawking-ellis-73}, Proposition 6.3.1) that $\partial I^- (p_\mathrm{O})$
is achronal (i.e., it is impossible to connect any two of its points by a
timelike curve) and thus a 3-dimensional Lipschitz topological submanifold.
By a general theorem of Rademacher (see~\cite{federer-69}, Theorem 3.6.1),
this implies that $\partial I^- (p_\mathrm{O})$ is differentiable almost
everywhere, i.e., that the cut locus has measure zero in $\partial I^- (p_\mathrm{O})$.
Note that this argument does not necessarily imply that the cut locus is
a ``small'' subset of $\partial I^- (p_\mathrm{O})$. Chru{\'s}ciel
and Galloway~\cite{chrusciel-galloway-98} have demonstrated, by way of
example, that an achronal subset $\mathcal{A}$ of a spacetime may fail to
be differentiable on a set that is dense in $\mathcal{A}$. So our reasoning
so far does not even exclude the possibility that the cut locus is dense
in an open subset of $\partial I^- (p_\mathrm{O})$. This possibility can be excluded in
globally hyperbolic spacetimes where the cut locus is always a closed subset of
$\mathcal{M}$ (see Section~\ref{ssec:hypcrit}). In general, the cut locus need
not be closed as is exemplified by Figure~\ref{fig:strcon1}.

In Section~\ref{ssec:crit} we investigate the relevance of
cut points (and conjugate points) for multiple imaging.


\subsection{Criteria for multiple imaging}
\label{ssec:crit}

To investigate whether multiple imaging occurs in a spacetime
$(\mathcal{M},g)$, we choose any point $p_\mathrm{O}$ (observation
event) and any timelike curve $\gamma_\mathrm{S}$ (wordline of light
source) in $\mathcal{M}$. The following cases are possible:
\begin{enumerate}
\item {\em There is no past-pointing lightlike geodesic from $p_\mathrm{O}$ to
  $\gamma_\mathrm{S}$}. Then the observer at $p_\mathrm{O}$ does not see
  any image of the light source ${\gamma_\mathrm{S}}$. For instance, this occurs
  in Minkowski spacetime for an inextendible worldline $\gamma_\mathrm{S}$
  that asymptotically approaches the past light cone of
  $p_\mathrm{O}$.
  \label{case_1}
\item {\em There is exactly one past-pointing lightlike geodesic from
  $p_\mathrm{O}$ to ${\gamma}_\mathrm{S}$}. Then the observer at $p_\mathrm{S}$ sees
  exactly one image of the light source ${\gamma}_\mathrm{S}$. This
  is the situation naively taken for granted in pre-relativistic
  astronomy.
  \label{case_2}
\item {\em There are at least two but not more than denumerably
  many past-pointing lightlike geo\-desics from $p_\mathrm{O}$ to
  ${\gamma}_\mathrm{S}$}. Then the observer at $p_\mathrm{O}$ sees finitely
  or infinitely many distinct images of ${\gamma}_\mathrm{O}$ at
  his or her celestial sphere.
  \label{case_3}
\item {\em There are more than denumerably many past-pointing lightlike
  geo\-desics from $p$ to ${\gamma}$}. This happens, e.g., in rotationally
  symmetric situations where it gives rise to the so-called ``Einstein
  rings'' (see Section~\ref{ssec:ss}). It also happens, e.g., in
  plane-wave spacetimes (see Section~\ref{ssec:wave}).
  \label{case_4}
\end{enumerate}
If Case~\ref{case_3} or~\ref{case_4} occurs, astronomers speak of
\emph{multiple imaging}. We first demonstrate that Case~\ref{case_4}
is exceptional. It is easy to prove (see,
e.g., \cite{perlick-2000}, Proposition 12) that no finite segment of
the timelike curve $\gamma_\mathrm{S}$ can be contained in the past light cone
of $p_\mathrm{O}$. Thus, if there is a continuous one-parameter family of
lightlike geodesics that connect $p_\mathrm{O}$ and $\gamma_\mathrm{O}$, then all family members
meet $\gamma_\mathrm{S}$ at the same point, say $p_\mathrm{S}$. This point must be in the
caustic of the light cone because through all non-caustic points there is
only a discrete number of generators. One can always find a point $p_\mathrm{O}'$ arbitrarily
close to $p_\mathrm{O}$ such that $\gamma_\mathrm{S}$ does not meet the caustic
of the past light cone of $p_\mathrm{O}'$ (see, e.g., \cite{perlick-2000},
Proposition 10). Hence, by an arbitrarily small perturbation of $p_\mathrm{O}$
one can always destroy a Case~\ref{case_4} situation. One may interpret this result
as saying that Case~\ref{case_4} situations have zero probability. This is, indeed,
true as long as we consider point sources (worldlines). The observed
rings and arcs refer to extended sources (worldtubes) which are
close to the caustic (recall Section~\ref{ssec:distortion}). Such situations
occur with non-zero probability.

We will now show how multiple imaging is related to the notion of
cut points (recall Section~\ref{ssec:cut}). For any point $p_\mathrm{O}$ in an arbitrary
spacetime, the following criteria for multiple imaging hold:

\begin{cenumerate}%
  Let $\lambda$ be a past-pointing lightlike geodesic from
  $p_\mathrm{O}$ and let $p_\mathrm{S}$ be a point on $\lambda$
  beyond the cut point or beyond the first conjugate point. Then there
  is a timelike curve $\gamma_\mathrm{S}$ through $p_\mathrm{S}$ that
  can be reached from $p_\mathrm{O}$ along a second past-pointing
  lightlike geodesic.
  \label{item_c1}
\end{cenumerate}

\begin{cenumerate}%
   Assume that at $p_\mathrm{O}$ the past-distinguishing
   condition~(\ref{eq:distinguish}) is satisfied. If a timelike curve
   $\gamma_\mathrm{S}$ can be reached from $p_\mathrm{O}$ along two
   different past-pointing lightlike geodesics, at least one of them
   passes through the cut locus of the past light cone of
   $p_\mathrm{O}$ on or before arriving at $\gamma_\mathrm{S}$.
  \label{item_c2}
\end{cenumerate}
\vspace{0.8 em}

\noindent
For proofs see~\cite{perlick-96}
or~\cite{perlick-2000}. (In~\cite{perlick-96} Criterion~\ref{item_c2}
is formulated with the strong causality condition, although the
past-distinguishing condition is sufficient.) Criteria~\ref{item_c1}
and~\ref{item_c2}
say that the occurrence of cut points is sufficient and, in
past-distinguishing spacetimes, also necessary for multiple imaging.
The occurrence of conjugate points is sufficient but, in general, not
necessary for multiple imaging (see Figure~\ref{fig:strcon1} for an
example without conjugate points where multiple imaging occurs). 
So we have the following diagram:

\renewcommand{\arraystretch}{1.3}
\begin{center}
  \begin{tabular}{c|c|c}
    \hline \hline
    Occurrence of: &
    Sufficient for multiple imaging in: &
    Necessary for multiple imaging in: \\
    \hline
    cut point &
    arbitrary spacetime &
    past-distinguishing spacetime \\
    conjugate point &
    arbitrary spacetime &
    -- \\
    \hline \hline
  \end{tabular}
\end{center}
\renewcommand{\arraystretch}{1.0}

It is well known (see~\cite{hawking-ellis-73}, in particular
Proposition 4.4.5) that, under conditions which are to be
considered as fairly general from a physical point of view, a
lightlike geodesic must either be incomplete or contain a pair
of conjugate points. These ``fairly general conditions'' are, e.g.,
the weak energy condition and the so-called generic condition
(see~\cite{hawking-ellis-73} for details). This result implies the
occurrence of conjugate points and, thus, of multiple imaging, for a
large class of spacetimes.

The occurrence of conjugate points has an important consequence
in view of the focusing equation for the area distance $D_{\mathrm{area}}$
(recall Section~\ref{ssec:distance} and, in particular,
Equation~(\ref{eq:focus})).
As $D_{\mathrm{area}}$ vanishes at the vertex $s=0$ and at each
conjugate point, there must be a parameter value $s_m$ with
$\dot{D}_{\mathrm{area}} (s_m) = 0$ between the vertex and the
first conjugate point. An elementary evaluation of the focusing
equation~(\ref{eq:focus}) then implies
\begin{equation}
  \label{eq:focusest}
  1 \le \int_0^{s_m} \!\!\! s \left( | \sigma (s) |^2 +
  \left| \frac{1}{2} \Ric \left( \dot{\lambda} (s),
  \dot{\lambda} (s)\right) \right| \right) ds.
\end{equation}
As the Ricci term is related to the energy density via
Einstein's field equation,~(\ref{eq:focusest}) gives an estimate of
energy-density-plus-shear along the ray. If we observe a
multiple imaging situation, and if we know (or assume) that we are
in a situation where conjugate points are necessary for multiple
imaging, we have thus an estimate on energy-density-plus-shear
along the ray. This line of thought was worked out, under additional
assumptions on the spacetime, in~\cite{padmanabhan-subramanian-88}.


\subsection{Fermat's principle for light rays}
\label{ssec:fermat}

It is often advantageous to characterize light rays by a variational
principle, rather than by a differential equation. This is particularly
true in view of applications to lensing. If we have chosen a point
$p_\mathrm{O}$ (observation event) and a timelike curve $\gamma_\mathrm{S}$ (worldline of
light source) in spacetime $\mathcal{M}$, we want to determine all
past-pointing lightlike geodesics from $p_\mathrm{O}$ to $\gamma_\mathrm{S}$. When working
with a differential equation for light rays, we have to calculate
\emph{all} light rays issuing from $p_\mathrm{O}$ into the past, and to see
which of them meet $\gamma_\mathrm{S}$. If we work with a variational principle,
we can restrict to curves from $p_\mathrm{O}$ to $\gamma_\mathrm{S}$ at the outset.

To set up a variational principle, we have to choose the trial curves
among which the solution curves are to be determined and the functional
that has to be extremized. Let $\mathcal{L}_{p_\mathrm{O}, \gamma_\mathrm{S}}$ denote the
set of all past-pointing lightlike curves from $p_\mathrm{O}$ to $\gamma_\mathrm{S}$. This
is the set of trial curves from which the lightlike \emph{geodesics}
are to be singled out by the variational principle. Choose a past-oriented
but otherwise arbitrary parametrization for the timelike curve $\gamma_\mathrm{S}$
and assign to each trial curve the parameter at which it arrives. This
gives the \emph{arrival time functional} $T : \mathcal{L}_{p_\mathrm{O}, \gamma_\mathrm{S}}
\longrightarrow \mathbb{R}$ that is to be extremized. With respect to an
appropriate differentiability notion for $T$, it turns out that the critical points
(i.e., the points where the differential of $T$ vanishes) are exactly the
geodesics in $\mathcal{L}_{p_\mathrm{O}, \gamma_\mathrm{S}}$. This result (or its
time-reversed version) can be viewed as a general-relativistic Fermat
principle:
\begin{quote}
  Among all ways to move from $p_\mathrm{O}$ to $\gamma_\mathrm{S}$ in
  the past-pointing (or future-pointing) direction at the speed of
  light, the actual light rays choose those paths that make the
  arrival time stationary.
\end{quote}
This formulation of Fermat's principle was suggested in 1990 by
Kovner~\cite{kovner-90}, a local version (restricted to a convex
normal neighborhood) can be found already in a 1938 paper by
Temple~\cite{temple-38}. 
The crucial idea is to refer to the arrival
time which is given only along the curve $\gamma_\mathrm{S}$, and not to some kind of
global time which in an arbitrary spacetime does not even exist. The proof
that the solution curves of Kovner's variational principle are, indeed,
exactly the lightlike geodesics was given in~\cite{perlick-90a}. The proof can
also be found, with a slight restriction on the spacetime that simplifies
matters considerably, in~\cite{schneider-ehlers-falco-92}. An alternative
version, based on making $\mathcal{L}_{p_\mathrm{O}, \gamma_\mathrm{S}}$ into a Hilbert
manifold, is given in~\cite{perlick-95}.

As in ordinary optics, the light rays make the arrival time stationary
but not necessarily minimal. A more detailed investigation shows
that for a geodesic $\lambda \in \mathcal{L}_{p_\mathrm{O} \gamma_\mathrm{S}}$ the following
holds. (For the notion of conjugate points see
Sections~\ref{ssec:front} and~\ref{ssec:cut}.)

\begin{aenumerate}%
  If along $\lambda$ there is no point conjugate to $p_\mathrm{O}$,
  $\lambda$ is a strict local minimum of $T$.
  \label{item_a1}
\end{aenumerate}

\begin{aenumerate}%
  If $\lambda$ passes through a point conjugate to $p_\mathrm{O}$
  before arriving at $\gamma$, it is a saddle of $T$.
  \label{item_a2}
\end{aenumerate}

\begin{aenumerate}%
  If $\lambda$ reaches the first point conjugate to $p_\mathrm{O}$
  exactly on its arrival at $\gamma_\mathrm{S}$, it may be a local
  minimum or a saddle but not a local maximum.
  \label{item_a3}
\end{aenumerate}
\vspace{0.8 em}

\noindent
For a proof see~\cite{perlick-90a} or~\cite{perlick-95}.
The fact that local maxima cannot occur is easily understood from the
geometry of the situation: For every trial curve we can find a 
neighboring trial curve with a larger $T$ by putting ``wiggles'' 
into it, preserving the lightlike character of the curve. Also for 
Fermat's principle in ordinary optics, where light propagation is 
characterized by a positive index of refraction on Euclidean 3-space, 
the extremum is never a local maximum, as is mentioned, e.g., in 
Born and Wolf~\cite{born-wolf-2002}, p.~137. Note, however, that 
in the quasi-Newtonian lensing approximation with one or more
deflector planes, where only broken straight lines are allowed
as trial paths, local maxima \emph{do} occur, see, 
e.g., \cite{petters-levine-wambsganss-2001}. Also, in the general
formalism of ray optics, where the rays are the solutions of Hamilton's
equations with an unspecified Hamiltonian, local maxima \emph{do}
occur, unless a certain regularity condition is imposed on the 
Hamiltonian~\cite{perlick-2000b}.   

The advantage of Kovner's version of Fermat's principle is that it
works in an \emph{arbitrary} spacetime. In particular, the spacetime
need not be stationary and the light source may arbitrarily move
around (at subluminal velocity, of course). This allows applications
to dynamical situations, e.g., to lensing by gravitational
waves (see Section~\ref{ssec:wave}). If the spacetime is stationary or conformally
stationary, and if the light source is at rest, a purely spatial
reformulation of Fermat's principle is possible.
This more specific version of Femat's principle is known since
decades and has found various applications to lensing (see
Section~\ref{ssec:conf}). A more sophisticated application of Fermat's
principle to lensing theory is to put up a Morse theory in order to
prove theorems on the possible number of images. In its strongest
version, this approach has to presuppose a globally hyperbolic
spacetime and will be reviewed in Section~\ref{ssec:morse}.

For a generalization of Kovner's version of Fermat's principle to the
case that observer and light source have a spatial extension
see~\cite{perlick-piccione-98}.

An alternative variational principle was introduced by Frittelli
and Newman~\cite{frittelli-newman-99} and evaluated
in~\cite{frittelli-newman-2002, frittelli-kling-newman-2002}. While
Kovner's principle, like the classical Fermat principle, is a
varional principle for rays, the Frittelli--Newman principle is
a variational principle for wave fronts. (For the definition of
wave fronts see Section~\ref{ssec:front}.) Although Frittelli and
Newman call their variational principle a version of Fermat's
principle, it is actually closer to the classical Huygens principle
than to the classical Fermat principle. Again, one
fixes $p_\mathrm{O}$ and $\gamma_\mathrm{S}$ as above. To define the trial maps, one
chooses a set $\mathcal{W}(p_\mathrm{O})$ of wave fronts, such that for each
lightlike geodesic through $p_\mathrm{O}$ there is exactly one wave front in
$\mathcal{W} (p_\mathrm{O})$ that contains this geodesic. Hence, $\mathcal{W} (p_\mathrm{O})$
is in one-to-one correspondence to the lightlike directions at $p_\mathrm{O}$
and thus to the 2-sphere. Now let $\mathcal{W} (p_\mathrm{O}, \gamma_\mathrm{S})$
denote the set of all wave fronts in $\mathcal{W} (p_\mathrm{O})$ that meet
$\gamma_\mathrm{S}$. We can then define the arrival time functional
$T : \mathcal{W} (p_\mathrm{O}, \gamma_\mathrm{S}) \longrightarrow \mathbb{R}$ by
assigning to each wave front the parameter value at which it
intersects $\gamma_\mathrm{S}$. There are some cases to be excluded
to make sure that $T$ is defined on an open subset of
$\mathcal{W} (p_\mathrm{O}) \simeq S^2$, single-valued and differentiable.
If this is the case, one finds that $T$ is stationary at $W \in
\mathcal{W} (p_\mathrm{O})$ if and only if $W$ contains a lightlike geodesic
from $p_\mathrm{O}$ to $\gamma_\mathrm{S}$. Thus, to each image of $\gamma_\mathrm{S}$ on
the sky of $p_\mathrm{O}$ there corresponds a critical point of $T$. The great
technical advantage of the Frittelli--Newman principle over the Kovner
principle is that $T$ is defined on a \emph{finite dimensional}
manifold, directly to be identified with (part of) the observer's
celestial sphere. The arrival time $T$ in the Frittelli--Newman approach
is directly analogous to the ``Fermat potential'' in the quasi-Newtonian
formalism which is discussed, e.g., in~\cite{schneider-ehlers-falco-92}. In
view of applications, a crucial point is that the space $\mathcal{W} (p_\mathrm{O})$
is a matter of choice; there are many wave fronts which have one light ray
in common. There is a natural choice, e.g., in asymptotically simple
spacetimes (see Section~\ref{ssec:asy}).

Frittelli, Newman, and collaborators have used their variational
principle in combination with the exact lens map (recall
Section~\ref{ssec:cone}) to discuss thick and thin lens models from a
spacetime perspective~\cite{frittelli-newman-2002,
  frittelli-kling-newman-2002}. Methods from differential topology or
global analysis, e.g., Morse theory, have not yet been applied to the
Frittelli--Newman principle.

\newpage


\section{Lensing in Globally Hyperbolic Spacetimes}
\label{sec:hyp}

In a globally hyperbolic spacetime, considerably stronger
statements on qualitative lensing features can be made
than in an arbitrary spacetime. This includes, e.g.,
multiple imaging criteria in terms of cut points or
conjugate points, and also applications of Morse theory.
The value of these results lies in the fact that they
hold in globally hyperbolic spacetimes without
symmetries, where lensing cannot be studied by explicitly
integrating the lightlike geodesic equation.

The most convenient formal definition of global hyperbolicity
is the following. In a spacetime $(\mathcal{M},g)$, a subset
$\mathcal{C}$ of $\mathcal{M}$ is called a \emph{Cauchy surface}
if every inextendible causal (i.e., timelike or lightlike) curve
intersects $\mathcal{C}$ exactly once. A spacetime is globally
hyperbolic if and only if it admits a Cauchy surface. The name
globally hyperbolic refers to the fact that for hyperbolic
differential equations, like the wave equation, existence and
uniqueness of a global solution is guaranteed for initial
data given on a Cauchy surface. For details on globally
hyperbolic spacetimes see, e.g., \cite{hawking-ellis-73,
beem-ehrlich-easley-96}.
It was demonstrated by Geroch~\cite{geroch-70} that every gobally
hyperbolic spacetime admits a continuous function $t : \mathcal{M}
\longrightarrow \mathbb{R}$ such that $t^{-1} (t_0)$ is a Cauchy
surface for every $t_0 \in \mathbb{R}$. A complete proof of the
fact that such a Cauchy time function can be chosen differentiable
was given much later by Bernal and S{\'a}nchez~\cite{bernal-sanchez-2003, bernal-sanchez-2005, bernal-sanchez-2006}.

The topology of a globally hyperbolic spacetime is determined by the
topology of any of its Cauchy surfaces, $\mathcal{M} \simeq \mathcal{C}
\times \mathbb{R}$. Note, however, that the converse is not true because
$\mathcal{C}_1 \times \mathbb{R}$ may be homeomorphic (and even
diffeomorphic) to $\mathcal{C}_2 \times \mathbb{R}$ without
$\mathcal{C}_1$ being homeomorphic to $\mathcal{C}_2$. For instance, one
can construct a globally hyperbolic spacetime with topology
$\mathbb{R}^4$ that admits a Cauchy surface which is not
homeomorphic to $\mathbb{R}^3$~\cite{newman-clarke-87}.

In view of applications to lensing the following observation
is crucial. If one removes a point, a worldline (timelike curve),
or a world tube (region with timelike boundary) from an arbitrary
spacetime, the resulting spacetime cannot be globally hyperbolic.
Thus, restricting to globally hyperbolic spacetimes excludes all
cases where a deflector is treated as non-transparent by cutting
its world tube from spacetime (see Figure~\ref{fig:strcon1} for an
example). Note, however, that this does not mean that globally hyperbolic
spacetimes can serve as models only for transparent deflectors. First,
a globally hyperbolic spacetime may contain ``non-transparent''
regions in the sense that a light ray may be trapped in a spatially
compact set. Second, the region outside the horizon of a (Schwarzschild,
Kerr, \ldots) black hole is globally hyperbolic.


\subsection{Criteria for multiple imaging in globally hyperbolic spacetimes}
\label{ssec:hypcrit}

In Section~\ref{ssec:cut} we have considered the past light cone
of an event $p_\mathrm{O}$ in an arbitrary spacetime. We have seen that
conjugate points (=\,caustic points) indicate that the
past light cone fails to be an immersed submanifold and that
cut points indicate that it fails to be an embedded submanifold.
In a globally hyperbolic spacetime $(\mathcal{M},g)$, the following
additional statements are true.

\begin{henumerate}%
  The past light cone of any event $p_\mathrm{O}$, together with the
  vertex $\{p_\mathrm{O} \}$, is closed in $\mathcal{M}$.
  \label{item_h1}
\end{henumerate}

\begin{henumerate}%
  The cut locus of the past light cone of $p_\mathrm{O}$ is closed in
  $\mathcal{M}$.
  \label{item_h2}
\end{henumerate}

\begin{henumerate}%
  Let $p_\mathrm{S}$ be in the cut locus of the past light cone of
  $p_\mathrm{O}$ but not in the conjugate locus (=\,caustic). Then
  $p_\mathrm{S}$ can be reached from $p_\mathrm{O}$ along two
  different lightlike geodesics. The past light cone of $p_\mathrm{O}$
  has a transverse self-intersection at $p_\mathrm{S}$.
  \label{item_h3}
\end{henumerate}

\begin{henumerate}%
  The past light cone of $p_\mathrm{O}$ is an embedded submanifold if
  and only if its cut locus is empty.
  \label{item_h4}
\end{henumerate}
\vspace{0.8 em}

\noindent
Analogous results hold, of course, for the future light cone, but
the past version is the one that has relevance for lensing. For proofs
of these statements see~\cite{beem-ehrlich-easley-96},
Propositions~9.35 and~9.29 and Theorem~9.15, and~\cite{perlick-2000},
Propositions~13, 14, and~15. According to Statement~\ref{item_h3}, a
``cut point'' indicates a
``cut'' of two lightlike geodesics. For geodesics in Riemannian manifolds
(i.e., in the positive definite case), an analogous statement holds if
the Riemannian metric is complete and is known as \emph{Poincar\'e theorem}~\cite{poincare-05, whitehead-35}. It was this theorem that motivated
the name ``cut point''. Note that Statement~\ref{item_h3} is not true without the
assumption that $p_\mathrm{S}$ is not in the caustic. This is exemplified by the
swallow-tail point in Figure~\ref{fig:strcon2}. However, as points
in the caustic of the past light cone of $p_\mathrm{O}$ can be reached from $p_\mathrm{O}$
along two ``infinitesimally close'' lightlike geodesics, the name ``cut
point'' may be considered as justified also in this case.

In addition to Statements~\ref{item_h1} and~\ref{item_h2} one would
like to know whether in
globally hyperbolic spactimes the caustic of the past light cone
of $p_\mathrm{O}$ (also known as the past lightlike conjugate locus of $p_\mathrm{O}$)
is closed. 

This question is closely related to the question
of whether in a complete Riemannian manifold the conjugate locus
of a point is closed. For both questions, the answer was widely
believed to be `yes' although actually it is `no'. To the surprise
of many, Margerin~\cite{margerin-93} constructed Riemannian
metrics on the 2-sphere such that the conjugate locus of a
point is not closed. Taking the product of such a
Riemannian manifold with 2-dimensional Minkowski space
gives a globally hyperbolic spacetime in which the caustic
of the past light cone of an event is not closed.

In Section~\ref{ssec:crit} we gave criteria for the number of past-oriented
lightlike geodesics from a point $p_\mathrm{O}$ (observation event) to a timelike
curve $\gamma_\mathrm{S}$ (worldline of a light source) in an arbitrary spacetime.
With Statements~\ref{item_h1}, \ref{item_h2}, \ref{item_h3},
and~\ref{item_h4} at hand, the following stronger criteria can be given.

Let $(\mathcal{M},g)$ be globally hyperbolic, fix a point $p_\mathrm{O}$ and an
inextendible timelike curve $\gamma_\mathrm{S}$ in $\mathcal{M}$. Then the following
is true:

\begin{henumerate}%
  Assume that $\gamma_\mathrm{S}$ enters into the chronological past
  $I^- (p_\mathrm{O})$ of $p_\mathrm{O}$. Then there is a
  past-oriented lightlike geodesic $\lambda$ from $p_\mathrm{O}$ to
  $\gamma_\mathrm{S}$ that is completely contained in the boundary of
  $I^- (p_\mathrm{O})$. This geodesic does not pass through a cut
  point or through a conjugate point before arriving at
  $\gamma_\mathrm{S}$.
  \label{item_h5}
\end{henumerate}

\begin{henumerate}%
  Assume that $\gamma_\mathrm{S}$ can be reached from $p_\mathrm{O}$
  along a past-oriented lightlike geodesic that passes through a
  conjugate point or through a cut point before arriving at
  $\gamma_\mathrm{S}$. Then $\gamma_\mathrm{S}$ can be reached from
  $p_\mathrm{O}$ along a second past-oriented lightlike geodesic.
  \label{item_h6}
\end{henumerate}
\vspace{0.8 em}

\noindent
Statement~\ref{item_h5} was proven in~\cite{uhlenbeck-75} with the
help of Morse theory. For a more
elementary proof see~\cite{perlick-2000}, Proposition~16. Statement~\ref{item_h5} gives
a characterization of the \emph{primary image} in globally hyperbolic
spacetimes. (By definition, an image is ``primary'' if no other image 
shows the light source at an older age.) 
The condition of $\gamma_\mathrm{S}$ entering
into the chronological past of $p_\mathrm{O}$ is necessary to exclude the case that
$p_\mathrm{O}$ sees no image of
$\gamma_\mathrm{S}$. Statement~\ref{item_h5} implies that there is a
unique primary image unless $\gamma_\mathrm{S}$ passes through the cut locus of the
past light cone of $p_\mathrm{O}$. The primary image has even parity. If the weak
energy condition is satisfied, the focusing theorem implies that the
primary image has magnification factor $\ge 1$, i.e., that it appears
brighter than a source of the same luminosity at the same affine distance
and at the same redshift in Minkowski spacetime (recall
Sections~\ref{ssec:distance} and~\ref{ssec:brightness}, in
particular the inequality~(\ref{eq:foctheo})).

For a proof of Statement~\ref{item_h6} see~\cite{perlick-2000}, Proposition~17.


\subsection{Wave fronts in globally hyperbolic spacetimes}
\label{ssec:hypfront}

In Section~\ref{ssec:front} the notion of wave fronts was
discussed in an arbitrary spacetime $(\mathcal{M},g)$. It was
mentioned that a wave front can be viewed as a subset of the
space $\mathcal{N}$ of all lightlike geodesics in $(\mathcal{M},g)$.
This approach is particularly useful in globally hyperbolic
spacetimes, as was demonstrated by Low~\cite{low-93, low-98}. The
construction is based on the observations that, if $(\mathcal{M},g)$ is
globally hyperbolic and $\mathcal{C}$ is a smooth Cauchy surface,
the following is true:

\begin{nenumerate}%
  $\mathcal{N}$ can be identified with a sphere bundle over
  $\mathcal{C}$. The identification is made by assigning to each
  lightlike geodesic its tangent line at the point where it intersects
  $\mathcal{C}$. As every sphere bundle over an orientable 3-manifold
  is trivializable, $\mathcal{N}$ is diffeomorphic to
  $\mathcal{C} \times S^2$.
  \label{item_n1}
\end{nenumerate}

\begin{nenumerate}%
  $\mathcal{N}$ carries a natural contact structure. (This contact
  structure is also discussed, in twistor language,
  in~\cite{penrose-rindler-86}, volume~II.)
  \label{item_n2}
\end{nenumerate}

\begin{nenumerate}%
  The wave fronts are exactly the Legendre submanifolds of $\mathcal{N}$.
  \label{item_n3}
\end{nenumerate}
\vspace{0.8 em}

\noindent
Using Statement~\ref{item_n1}, the projection from $\mathcal{N}$ to $\mathcal{C}$ assigns
to each wave front its intersection with $\mathcal{C}$, i.e., an
``instantaneous wave front'' or ``small wave front'' (cf.\
Section~\ref{ssec:front} for terminology). The points where this
projection has non-maximal rank give the
caustic of the small wave front. According to the general stability results of
Arnold (see~\cite{arnold-gusein-varchenko-85}), the only caustic points that
are stable with respect to local perturbations within the class of
Legendre submanifolds are cusps and swallow-tails. By
Statement~\ref{item_n3}, perturbing
within the class of Legendre submanifolds is the same as perturbing
within the class of wave fronts. For this local stability result the
assumption of global hyperbolicity is irrelevant because every
spacelike hypersurface is a Cauchy surface for an appropriately chosen
neighborhood of any of its points. So we get the
result that was already mentioned in Section~\ref{ssec:front}: In an arbitrary
spacetime, a caustic point of an instantaneous wave front is stable if and only if
it is a cusp or a swallow-tail. Here stability refers to perturbations
that keep the metric and the hypersurface fixed and perturb
the wave front within the class of wave fronts. For a picture of an
instantaneous wave front with cusps and a swallow-tail point, see
Figure~\ref{fig:strfrt2}. In Figure~\ref{fig:schwfrt}, the caustic points
are neither cusps nor swallow-tails, so the caustic is unstable.


\subsection{Fermat's principle and Morse theory in globally hyperbolic spacetimes}
\label{ssec:morse}

In an arbitrary spacetime, the past-oriented lightlike geodesics from
a point $p_\mathrm{O}$ (observation event) to a timelike curve $\gamma_\mathrm{S}$
(worldline of light source) are the solutions of a variational
principle (Kovner's version of Fermat's principle; see
Section~\ref{ssec:fermat}). Every solution of this variational
principle corresponds to an image on $p_\mathrm{O}$'s sky of $\gamma_\mathrm{S}$.
Determining the number of images is the same as determining the
number of solutions to the variational problem. If the variational
functional satisfies some technical conditions, the number of solutions
to the variational principle can be related to the topology of the
space of trial paths. This is the content of Morse theory. In the
case at hand, the ``technical conditions'' turn out to be satisfied
in globally hyperbolic spacetimes.

To briefly review Morse theory, we consider a differentiable function
$F : \mathcal{X} \longrightarrow \mathbb{R}$ on a real manifold
$\mathcal{X}$. Points where the differential of $F$ vanishes are
called \emph{critical points} of $F$. A critical point is called
\emph{non-degenerate} if the Hessian of $F$ is non-degenerate at
this point. $F$ is called a \emph{Morse function} if all its critical
points are non-degenerate. In applications to variational problems,
$\mathcal{X}$ is the space of trial maps, $F$ is the functional to
be varied, and the critical points of $F$ are the solutions to the
variational problem. The non-degeneracy condition guarantees that
the character of each critical point -- local minimum, local maximum,
or saddle -- is determined by the Hessian of $F$ at this point. The
index of the Hessian is called the \emph{Morse index} of the critical
point. It is defined as the maximal dimension of a subspace on which the
Hessian is negative definite. At a local minimum the Morse index is zero, at
a local maximum it is equal to the dimension of $\mathcal{X}$.

Morse theory was first worked out by Morse~\cite{morse-34} for the case
that $\mathcal{X}$ is finite-dimensional and compact (see
Milnor~\cite{milnor-63} for a detailed exposition). The main result is
the following. On a compact manifold $\mathcal{X}$, for every Morse function
the \emph{Morse inequalities}
\begin{equation}
  \label{eq:morsein}
  N_k \ge B_k,
  \qquad
  k = 0,1,2,\dots,
\end{equation}
and the \emph{Morse relation}
\begin{equation}
  \label{eq:morserel}
  \sum_{k=0}^{\infty} (-1)^k N_k = \sum_{k=0}^{\infty} (-1)^k B_k
\end{equation}
hold true. Here $N_k$ denotes the number of critical points with Morse index
$k$ and $B_k$ denotes the $k$th \emph{Betti number} of $\mathcal{X}$.
Formally, $B_k$ is defined for each topological space $\mathcal{X}$
in terms of the $k$th singular homology space $H_k ( \mathcal{X})$
with coefficients in a field $\mathbb{F}$ (see, e.g., \cite{dold-80},
p.~32). (The results of Morse theory hold for
any choice of $\mathbb{F}$.) Geometrically, $B_0$ counts the connected
components of $\mathcal{X}$ and, for $k \ge 1$, $B_k$ counts the ``holes''
in $\mathcal{X}$ that prevent a $k$-cycle with coefficients in $\mathbb{F}$
from being a boundary. In particular, if $\mathcal{X}$ is contractible
to a point, then $B_k = 0$ for $k \ge 1$. The right-hand side of
Equation~(\ref{eq:morserel}) is, by definition, the \emph{Euler characteristic} of
$\mathcal{X}$. By compactness of $\mathcal{X}$, all $N_k$ and $B_k$ are
finite and in both sums of Equation~(\ref{eq:morserel}) only finitely many
summands are different from zero.

Palais and Smale~\cite{palais-63, palais-smale-64} realized that the
Morse inequalities and the Morse relations are also true for a Morse
function $F$ on a non-compact and possibly infinite-dimensional Hilbert
manifold, provided that $F$ is bounded below and satisfies a
technical condition known as \emph{Condition C} or \emph{Palais--Smale
condition}. In that case, the $N_k$ and $B_k$ need not be finite.

The standard application of Morse theory is the geodesic problem for
Riemannian (i.e., positive definite) metrics: given two
points in a Riemannian manifold, to find the geodesics that join
them. In this case $F$ is the ``energy functional'' (squared-length
functional). Varying the energy functional is related to varying
the length functional like Hamilton's principle is related to
Maupertuis' principle in classical mechanics. For the space
$\mathcal{X}$ one chooses, in the Palais--Smale approach~\cite{palais-63},
the $H^1$-curves between the given two points. (An $H^n$-curve is a
curve with locally square-integrable $n$th derivative). This is
an infinite-dimensional Hilbert manifold. It has the same homotopy type
(and thus the same Betti numbers) as the \emph{loop space} of the
Riemannian manifold. (The loop space of a connected topological
space is the space of all continuous curves joining any two
fixed points.) On this Hilbert manifold, the energy functional is
always bounded from below, and its critical points are exactly the
geodesics between the given end-points. A critical point (geodesic)
is non-degenerate if the two end-points are not conjugate to each
other, and its Morse index is the number of conjugate points in the interior,
counted with multiplicity (``Morse index theorem''). The Palais--Smale
condition is satisfied if the Riemannian manifold is complete. So
one has the following result: Fix any two points in a complete Riemannian
manifold that are not conjugate to each other along any geodesic. Then
the Morse inequalities~(\ref{eq:morsein}) and the Morse
relation~(\ref{eq:morserel}) are true, with $N_k$ denoting the
number of geodesics with Morse index $k$ between the two points and $B_k$
denoting the $k$th Betti number of the loop space of the Riemannian
manifold. The same result is achieved in the original version of
Morse theory~\cite{morse-34} (cf.~\cite{milnor-63}) by choosing
for $\mathcal{X}$ the space of broken geodesics between the two
given points, with $N$ break points, and sending $N \to \infty$
at the end.

Using this standard example of Morse theory as a pattern, one can
prove an analogous result for Kovner's version of Fermat's principle.
The following hypotheses have to be satisfied:

\begin{menumerate}%
  $p_\mathrm{O}$ is a point and $\gamma_\mathrm{S}$ is a timelike
  curve in a globally hyperbolic spacetime $(\mathcal{M},g)$.
  \label{item_m1}
\end{menumerate}

\begin{menumerate}%
  $\gamma_\mathrm{S}$ does not meet the caustic of the past light cone
  of $p_\mathrm{O}$.
  \label{item_m2}
\end{menumerate}

\begin{menumerate}%
  Every continuous curve from $p_\mathrm{O}$ to $\gamma_\mathrm{S}$
  can be continuously deformed into a past-oriented lightlike curve,
  with all intermediary curves starting at $p_\mathrm{O}$ and
  terminating on $\gamma_\mathrm{S}$.
  \label{item_m3}
\end{menumerate}
\vspace{0.8 em}

\noindent
The global hyperbolicity assumption in Statement~\ref{item_m1} is analogous
to the completeness assumption in the Riemannian case.
Statement~\ref{item_m2} is the direct analogue of the non-conjugacy condition
in the Riemannian case. 
Statement~\ref{item_m3} is necessary for relating the
space of trial paths (i.e., of past-oriented lightlike
curves from $p_\mathrm{O}$ to $\gamma_\mathrm{S}$) to the loop space of
the spacetime manifold or, equivalently, to the loop
space of a Cauchy surface. If Statements~\ref{item_m1}, \ref{item_m2},
and~\ref{item_m3} are valid,
the Morse inequalities~(\ref{eq:morsein}) and the Morse relation~(\ref{eq:morserel}) are true, with $N_k$ denoting the number
of past-oriented lightlike geodesics from $p_\mathrm{O}$ to $\gamma_\mathrm{S}$
that have $k$ conjugate points in its interior, counted with
muliplicity, and $B_k$ denoting the $k$th Betti number of the
loop space of $\mathcal{M}$ or, equivalently, of a Cauchy surface.
This result was proven by Uhlenbeck~\cite{uhlenbeck-75}
\`a la Morse and Milnor, and by
Giannoni and Masiello~\cite{giannoni-masiello-96} in an
infinite-dimensional Hilbert manifold setting \`a la Palais
and Smale. A more general version, applying to spacetime
regions with boundaries, was worked out by Giannoni, Masiello,
and Piccione~\cite{giannoni-masiello-piccione-97,
giannoni-masiello-piccione-98}. In the work of Giannoni et al.,
the proofs are given in greater detail than in the work
of Uhlenbeck.

If Statements~\ref{item_m1}, \ref{item_m2}, and~\ref{item_m3} are
satisfied, Morse theory gives us the following
results about the number of images of $\gamma_\mathrm{S}$ on the sky
of $p_\mathrm{O}$ (cf.~\cite{mckenzie-85}):

\begin{renumerate}%
  If $\mathcal{M}$ is not contractible to a point, there are
  infinitely many images. This follows from
  Equation~(\ref{eq:morsein}) because for the loop space of a
  non-contractible space either $B_0$ is infinite or almost all $B_k$
  are different from zero~\cite{serre-51}.
  \label{item_r1}
\end{renumerate}

\begin{renumerate}%
  If $\mathcal{M}$ is contractible to a point, the total number of
  images is infinite or odd. This follows from
  Equation~(\ref{eq:morserel}) because in this case the loop space of
  $\mathcal{M}$ is contractible to a point, so all Betti numbers $B_k$
  vanish with the exception of $B_0=1$. As a consequence,
  Equation~(\ref{eq:morserel}) can be written as $N_+ - N_- = 1$,
  where $N_+$ is the number of images with even parity (geodesics with
  even Morse index) and $N_-$ is the number of images with odd parity
  (geodesics with odd Morse index), hence $N_+ + N_- = 2 N_- + 1$.
  \label{item_r2}
\end{renumerate}
\vspace{0.8 em}

\noindent
These results apply, in particular, to the following situations
of physical interest: \\

\noindent
{\bf Black hole spacetimes.} \\
\noindent
Let $(\mathcal{M},g)$ be the domain of outer communication
of the Kerr spacetime, i.e., the region between the (outer)
horizon and infinity (see Section~\ref{ssec:kerr}). Then
the assumption of global hyperbolicity is satisfied and
$\mathcal{M}$ is not contractible to a point. Statement~\ref{item_m3}
is satisfied if
$\gamma_\mathrm{S}$ is inextendible and approaches neither the
horizon nor (past lightlike) infinity for $t \to - \infty$.
(This can be checked with the help of an analytical criterion
that is called the ``metric growth condition'' in~\cite{uhlenbeck-75}.)
If, in addition Statement~\ref{item_m2} is satisfied, the reasoning of
Statement~\ref{item_r1} applies.
Hence, a Kerr black hole produces infinitely many images, under fairly
generic conditions on the motion of the light source. The details of this 
argument are worked out, for the more general case of a Kerr-Newman 
black hole, in \cite{hasse-perlick-2006}. 
\\

\noindent
{\bf Asymptotically simple and empty spacetimes.} \\
\noindent
As discussed in Section~\ref{ssec:asy}, asymptotically simple
and empty spacetimes are globally hyperbolic and contractible
to a point. They can be viewed as models of isolated transparent
gravitational lenses. Statement~\ref{item_m3} is satisfied if
$\gamma_\mathrm{S}$ is inextendible
and bounded away from past lightlike infinity $\J^{\,-}$. If, in addition,
Statement~\ref{item_m2} is satisfied, Statement~\ref{item_r2}
guarantees that the number of images is
infinite or odd. If it were infinite, we had as the limit curve a
past-inextendible lightlike geodesic that would not go out to
$\J^{\,-}$, in contradiction to the definition of asymptotic simplicity.
So the number of images must be finite and odd. The same odd-number
theorem can also be proven with other methods (see
Section~\ref{ssec:asy}). \\

In this way Morse theory provides us with precise mathematical versions
of the statements ``A black hole produces infinitely many images''
and ``An isolated transparent gravitational lens produces an odd number
of images''. When comparing this theoretical result with observations
one has to be aware of the fact that some images might be hidden
behind the deflecting mass, some might be too faint for being
detected, and some might be too close together for being resolved.

In conformally stationary spacetimes, with $\gamma_\mathrm{S}$ being
an integral curve
of the conformal Killing vector field, a simpler version of Fermat's
principle and Morse theory can be used (see Section~\ref{ssec:conf}).


\subsection{Lensing in asymptotically simple and empty spacetimes}
\label{ssec:asy}

In elementary optics one often considers ``light sources at infinity"
which are characterized by the fact that all light rays emitted from
such a source are parallel to each other. In general relativity, ``light
sources at infinity" can be defined if one restricts to a special class
of spacetimes. These spacetimes, known as ``asymptotically simple and
empty'' are, in particular, globally hyperbolic. Their formal definition,
which is due to Penrose~\cite{penrose-64}, reads as follows
(cf.~\cite{hawking-ellis-73}, p.~222., and~\cite{frauendiener-2004},
Section~2.3).
(Recall that a spacetime is called ``strongly causal'' if each neighborhood
of an event $p$ admits a smaller neighborhood that is intersected by
any non-spacelike curve at most once.)

A spacetime $(\mathcal{M},g,)$ is called {\em asymptotically simple and empty} if
there is a strongly causal spacetime $({\tilde{\mathcal{M}}},{\tilde{g}})$ with the
following properties:

\begin{senumerate}%
  $\mathcal{M}$ is an open submanifold of ${\tilde{\mathcal{M}}}$ with
  a non-empty boundary $\partial \mathcal{M} $.
  \label{item_s1}
\end{senumerate}

\begin{senumerate}%
  There is a smooth function
  $\Omega : {\tilde{\mathcal{M}}} \longrightarrow {\mathbb{R}}$ such
  that $\mathcal{M} = \{ p \in {\tilde{\mathcal{M}}} | \Omega (p) > 0 \}$,
  $\partial \mathcal{M} = \{ p \in {\tilde{\mathcal{M}}} | \Omega (p) = 0 \}$,
  $d \Omega \neq 0$ everywhere on $\partial \mathcal{M}$ and
  ${\tilde{g}} = \Omega^2 g $ on $\mathcal{M}$.
  \label{item_s2}
\end{senumerate}

\begin{senumerate}%
  Every inextendible lightlike geodesic in ${\mathcal{M}}$ has past
  and future end-point on $\partial \mathcal{M} $.
  \label{item_s3}
\end{senumerate}

\begin{senumerate}%
  There is a neighborhood $\mathcal{V}$ of $\partial \mathcal{M}$ such
  that the Ricci tensor of $g$ vanishes on $\mathcal{V} \cap
  \mathcal{M}$.
  \label{item_s4}
\end{senumerate}
\vspace{0.8 em}

\noindent
Asymptotically simple and empty spacetimes are mathematical models of transparent
uncharged gravitating bodies that are isolated from all other gravitational sources.
In view of lensing, the transparency condition~\ref{item_s3} is particularly important.

We now summarize some well-known facts about asymptotically simple
and empty spacetimes (cf.\ again~\cite{hawking-ellis-73}, p.~222,
and~\cite{frauendiener-2004}, 
Section~2.3). Every asymptotically
simple and empty spacetime is globally hyperbolic. $\partial \mathcal{M}$
is a ${\tilde{g}}$-lightlike hypersurface of ${\tilde{\mathcal{M}}}$.
It has two connected components, denoted $\J^{\,+}$ and $\J^{\,-}$. Each
lightlike geodesic in $(\mathcal{M},g)$ has past end-point on $\J^{\,-}$ and
future end-point on $\J^{\,+}$. Geroch~\cite{geroch-71} gave a proof that
every Cauchy surface $\mathcal{C}$ of an asymptotically simple and empty
spacetime has topology $\mathbb{R}^3$ and that $\J^{\,\pm}$ has topology
$S^2 \times \mathbb{R}$. The original proof, which is repeated in~\cite{hawking-ellis-73}, is incomplete. A complete proof that $\mathcal{C}$
must be contractible and that $\J^{\,\pm}$ has topology $S^2 \times \mathbb{R}$
was given by Newman and Clarke~\cite{newman-clarke-87}
(cf.~\cite{newman-89}); the stronger statement that $\mathcal{C}$ must
have topology $\mathbb{R}^3$ needs the assumption that the Poincar\'e
conjecture is true (i.e., that every compact and simply connected
3-manifold is a 3-sphere). In~\cite{newman-clarke-87} the authors
believed that the Poincar{\'e} conjecture was proven, but the proof
they are refering to was actually based on an error.  As the more recent
proof of the Poincar\'e conjecture by Perelman~\cite{perelman-2003}
(cf.~\cite{morgan-tian-2007}) has been generally accepted as being correct, the matter is now settled.

As $\J^{\,\pm}$ is a lightlike hypersurface in $\tilde{\mathcal{M}}$,
it is in particular a wave front in the sense of Section~\ref{ssec:front}.
The generators of $\J^{\,\pm}$ are the integral curves of the gradient of
$\Omega$. The generators of $\J^{\,-}$ can be interpreted as the ``worldlines''
of light sources at infinity that send light into $\mathcal{M}$. The
generators of $\J^{\,+}$ can be interpreted as the ``worldlines'' of observers
at infinity that receive light from $\mathcal{M}$. This interpretation is
justified by the observation that each generator of $\J^{\,\pm}$ is the limit
curve for a sequence of timelike curves in $\mathcal{M}$.

For an observation event $p_\mathrm{O}$ inside $\mathcal{M}$ and light sources at
infinity, lensing can be investigated in terms of the exact lens map
(recall Section~\ref{ssec:cone}), with the role of the source surface
$\mathcal{T}$ played by $\J^{\,-}$. (For the mathematical properties of the
lens map it is rather irrelevant whether the source surface is timelike,
lightlike or even spacelike. What matters is that the arriving light rays
meet the source surface transversely.) In this case the lens map
is a map $S^2 \rightarrow S^2$, namely from the celestial sphere
of the observer to the set of all generators of $\J^{\,-}$. One can
construct it in two steps: First determine the intersection of
the past light cone of $p_\mathrm{O}$ with $\J^{\,-}$, then project along the
generators. The intersections of light cones with $\J^{\,\pm}$
(``light cone cuts of null infinity'') have been studied
in~\cite{kozameh-newman-83, kozameh-lamberti-reula-91}.

One can assign a mapping degree (=\,Brouwer degree\,=\,winding number)
to the lens map $S^2 \rightarrow S^2$ and prove that it must be
$\pm 1$~\cite{perlick-2001}.
(The proof is based on ideas of~\cite{newman-clarke-87, newman-89}.
Earlier proofs of similar statements -- \cite{kozameh-lamberti-reula-91},
Lemma 1, and~\cite{perlick-2000}, Theorem 6 -- are incorrect, as
outlined in~\cite{perlick-2001}.) Based on this result, the following
odd-number theorem can be proven for observer and light source inside
$\mathcal{M}$~\cite{perlick-2001}: Fix a point $p_\mathrm{O}$ and a timelike curve
$\gamma_\mathrm{S}$ in an asymptotically simple and empty spacetime $(\mathcal{M},g)$.
Assume that the image of $\gamma_\mathrm{S}$ is a closed subset of
${\tilde{\mathcal{M}}} \setminus \J^{\,+}$ and that $\gamma_\mathrm{S}$ meets
neither the point $p_\mathrm{O}$ nor the caustic of the past light cone of
$p_\mathrm{O}$. Then the number of past-pointing lightlike geodesics
from $p_\mathrm{O}$ to $\gamma_\mathrm{S}$ in $\mathcal{M}$ is finite and odd.
The same result can be proven with the help of Morse theory (see
Section~\ref{ssec:morse}).

We will now give an argument to the effect that in an asymptotically
simple and empty spacetime the non-occurrence of multiple imaging is
rather exceptional. The argument starts from a standard result that
is used in the Penrose--Hawking singularity theorems. This standard
result, given as Proposition 4.4.5 in~\cite{hawking-ellis-73}, says that
along a lightlike geodesic that starts at a point $p_\mathrm{O}$ there must
be a point conjugate to $p_\mathrm{O}$, provided that
\begin{enumerate}
\item the so-called generic condition is satisfied at $p_\mathrm{O}$,
\item the weak energy condition is satisfied along the geodesic, and
\item the geodesic can be extended sufficiently far.
\end{enumerate}
The last assumption is certainly
true in an asymptotically simple and empty spacetime because
there all lightlike geodesics are complete. Hence, the generic condition
and the weak energy condition guarantee that every past light cone must
have a caustic point. We know from Section~\ref{ssec:hypcrit} that
this implies multiple imaging for every observer. In other words,
the only asymptotically simple and empty spacetimes in which multiple
imaging does \emph{not} occur are non-generic cases (like Minkowski
spacetime) and cases where the gravitating bodies have negative energy.

The result that, under the aforementioned conditions, light cones
in an asymptotically simple and empty spacetime must have caustic
points is due to~\cite{iriondo-kozameh-rojas-99}. This paper investigates
the past light cones of points on $\J^{\,+}$ and their caustics.
These light cones are the generalizations, to an arbitrary
asymptotically simple and empty spacetime, of the lightlike
hyperplanes in Minkowski spacetime. With their help, the eikonal
equation (Hamilton--Jacobi equation) $g^{ij} \partial_i S \partial_j S =0$
in an asymptotically simple and empty spacetime
can be studied in analogy to Minkowski
spacetime~\cite{frittelli-newman-silva-99a, frittelli-newman-silva-99b}.
In Minkowski spacetime the lightlike hyperplanes are associated
with a two-parameter family of solutions to the eikonal equation.
In the terminology of classical mechanics such a family is called
a \emph{complete integral}. Knowing a complete integral allows constructing
all solutions to the Hamilton--Jacobi equation. In
an asymptotically simple and empty spacetime the past light
cones of points on $\J^{\,+}$ give us, again, a complete integral
for the eikonal equation, but now in a generalized sense,
allowing for caustics. These past light cones are wave fronts,
in the sense of Section~\ref{ssec:front}, and cannot be represented
as surfaces $S = \mbox{constant}$ near caustic points. The way
in which all other wave fronts can be determined from knowledge
of this distinguished family of wave fronts is detailed
in~\cite{frittelli-newman-silva-99b}. The distinguished family
of wave fronts gives a natural choice for the space of
trial maps in the Frittelli--Newman variational principle which
was discussed in Section~\ref{ssec:fermat}.

\newpage


\section{Lensing in Spacetimes with Symmetry}
\label{sec:symmetry}


\subsection{Lensing in conformally flat spacetimes}
\label{ssec:flat}

By definition, a spacetime is conformally flat if the conformal curvature
tensor (=\,Weyl tensor) vanishes. An equivalent condition is that every point
admits a neighborhood that is conformal to an open subset of Minkowski
spacetime. As a consequence, conformally flat spacetimes
have the same local conformal symmetry as Minkowski spacetime, that is they
admit 15 independent conformal Killing vector fields. The global topology,
however, may be different from the topology of Minkowski spacetime. The class
of conformally flat spacetimes includes all (kinematic) Robertson--Walker
spacetimes. Other physically interesting examples are some (generalized)
interior Schwarzschild solutions and some pure radiation spacetimes. All
conformally flat solutions to Einstein's field equation with a perfect fluid
or an electromagnetic field are known
(see~\cite{stephani-kramer-maccallum-hoenselaers-herlt-2003},
Section~37.5.3).

If a spacetime is globally conformal to an open subset of Minkowski
spacetime, the past light cone of every event is an embedded submanifold.
Hence, multiple imaging cannot occur (recall Section~\ref{ssec:crit}).
For instance, multiple imaging occurs in spatially closed but not
in spatially open Robertson--Walker spacetimes. In any conformally
flat spacetime, there is no image distortion, i.e., a sufficiently small
sphere always shows a circular outline on the observer's sky (recall
Section~\ref{ssec:distortion}). Correspondingly, every infinitesimally
thin bundle of light rays with a vertex is circular, i.e., the extremal
angular diameter distances $D_+$ and $D_-$ coincide (recall
Section~\ref{ssec:distance}). In addition, $D_+=D_-$ also coincides with the
area distance $D_{\mathrm{area}}$, at least up to sign. $D_+=D_-$ changes
sign at every caustic point. As $D_+$ has a zero if and only if $D_-$ has a zero,
all caustic points of an infinitesimally thin bundle with vertex are of
multiplicity two (\emph{anastigmatic focusing}), so all images have
even parity.

The geometry of light bundles can be studied directly in terms of the
Jacobi equation (=\,equation of geodesic deviation) along lightlike
geodesics. For a detailed investigation of the latter in conformally flat
spacetimes, see~\cite{peters-76a}. The more special case of
Friedmann--Lema{\^ \i}tre-Robertson--Walker spacetimes (with dust, radiation,
and cosmological constant) is treated in~\cite{ellis-vanelst-99}.
For bundles with vertex, one is left with one scalar equation for
$D_+=D_-=\pm D_{\mathrm{area}}$, that is the focusing equation~(\ref{eq:focus}) with $\sigma = 0$. This equation can be explicitly
integrated for Friedmann--Robertson--Walker spacetimes (dust without
cosmological constant). In this way one gets, for the standard observer
field in such a spacetime, relations between redshift and (area or
luminosity) distance in closed form~\cite{mattig-57}. There are generalizations
for a Robertson--Walker universe with dust plus cosmological constant~\cite{kaufman-71} and dust plus radiation plus cosmological constant~\cite{dabrowski-stelmach-86}. Similar formulas can be written for the
relation between age and redshift~\cite{thomas-kantowski-2000}.


\subsection{Lensing in conformally stationary spacetimes}
\label{ssec:conf}

Conformally stationary spacetimes are models for gravitational fields that are
time-independent up to an overall conformal factor. (The time-dependence of
the conformal factor is important, e.g., if cosmic expansion is
to be taken into account.) This is a reasonable model assumption for many, though
not all, lensing situations of interest. It allows describing light rays
in a 3-dimensional (spatial) formalism that will be outlined in this section.
The class of conformally stationary spacetimes includes spherically symmetric
and static spacetimes (see Sections~\ref{ssec:ss}) and axisymmetric
stationary spacetimes (see Section~\ref{ssec:axistat}). Also, conformally flat
spacetimes (see Section~\ref{ssec:flat}) are conformally stationary, at least
locally. A physically relevant example where the conformal-stationarity
assumption is \emph{not} satisfied is lensing by a gravitational wave (see
Section~\ref{ssec:wave}).

By definition, a spacetime is conformally stationary if it admits a timelike
conformal Killing vector field $W$. If $W$ is complete and if there are no
closed timelike curves, the spacetime must be a product, $\mathcal{M} \simeq
\mathbb{R} \times \widehat{\mathcal{M}}$ with a (Hausdorff and
paracompact) 3-manifold $\widehat{\mathcal{M}}$ and $W$ parallel to the
$\mathbb{R}$-lines~\cite{harris-92}. If we denote the projection from
$\mathcal{M}$ to
$\mathbb{R}$ by $t$ and choose local coordinates $x=(x^1,x^2,x^3)$ on
$\widehat{\mathcal{M}}$, the metric takes the form
\begin{equation}
  \label{eq:confg}
  g = e^{2f(t,x)} \left( - (dt + \hat{\phi}_{\mu}(x) \, dx^{\mu})^2 +
  \hat{g}_{\mu \nu}(x) \, dx^{\mu} \, dx^{\nu} \right)
\end{equation}
with $\mu, \nu, \ldots = 1,2,3$.
The conformal factor $e^{2f}$ does not affect the lightlike geodesics
apart from their parametrization. So the paths of light rays are completely
determined by the metric $\hat{g} = \hat{g}_{\mu \nu} (x) dx^{\mu} dx^{\nu}$
and the one-form $\hat{\phi} = \hat{\phi}_{\mu} (x) dx^{\mu}$ which live
on $\widehat{\mathcal{M}}$. The metric $\hat{g}$ must be positive definite to give
a spacetime metric of Lorentzian signature. We call $f$ the \emph{redshift
potential}, $\hat{g}$ the \emph{Fermat metric} and $\hat{\phi}$ the
\emph{Fermat one-form}. The motivation for these names will become clear
from the discussion below.

If $\hat{\phi}_{\mu} = \partial_{\mu} h$, where $h$ is a function of
$x=(x^1,x^2,x^3)$, we can change the time coordinate according to $t \longmapsto t +
h(x)$, thereby transforming $\hat{\phi}_{\mu} dx^{\mu}$ to
zero, i.e., making the surfaces $t = \mbox{constant}$ orthogonal to the
$t$-lines. This is the \emph{conformally static} case. Also, Equation~(\ref{eq:confg}) includes
the stationary case ($f$ independent of $t$) and the static case ($\hat{\phi}_{\mu}
= \partial_{\mu} h$ and $f$ independent of $t$).

In Section~\ref{ssec:fermat} we have discussed Kovner's version of Fermat's
principle which characterizes the lightlike geodesics between a point
(observation event) $p_\mathrm{O}$ and a timelike curve (worldline of light source)
$\gamma_\mathrm{S}$. In a conformally stationary spacetime we may specialize to the
case that $\gamma_\mathrm{S}$ is an integral curve of the conformal Killing vector
field, parametrized by the ``conformal time'' coordinate $t$ (in the
past-pointing sense, to be in agreement with Section~\ref{ssec:fermat}).
Without loss of generality, we may assume that the observation event $p_\mathrm{O}$
takes place at $t=0$. Then for each trial path (past-oriented lightlike curve)
$\lambda$ from $p_\mathrm{O}$ to $\gamma_\mathrm{S}$ the arrival time is equal to the travel
time in terms of the time function $t$. By Equation~(\ref{eq:confg}) this puts the
arrival time functional into the following coordinate form
\begin{equation}
  \label{eq:confT}
  T(\lambda) = \int_{\ell_1}^{\ell_2}
  \left( \sqrt{\hat{g}_{\mu \nu}(x) \frac{d x^{\mu}}{d \ell}
  \frac{dx^{\nu}}{d \ell}} - \hat{\phi}_{\mu}(x)
  \frac{dx^{\mu}}{d \ell} \right) d\ell,
\end{equation}
where $\ell$ is any parameter along the trial path, ranging over an interval
$[\ell_1, \ell_2]$ that depends on the individual curve. The right-hand
side of Equation~(\ref{eq:confT}) is a functional for curves in $\widehat{\mathcal{M}}$ with
fixed end-points. The projections to $\widehat{\mathcal{M}}$ of light rays are the
stationary points of this functional. In general, the right-hand side
of Equation~(\ref{eq:confT}) is the length functional of a Finsler metric. In the
conformally static case $\hat{\phi}_{\mu} = \partial_{\mu} h$, the integral
over $\hat{\phi}_{\mu} (x) dx^{\mu} /d \ell$ is the same for all trial paths,
so we are left with the length functional of the Fermat metric $\hat{g}$. In
this case the light rays, if projected to $\widehat{\mathcal{M}}$, are the geodesics of
$\hat{g}$. Note that the travel time functional~(\ref{eq:confT}) is invariant
under reparametrization; in the terminology of classical mechanics, it is a
special case of \emph{Maupertuis' principle}. It is often convenient to
switch to a parametrization-dependent variational principle which, in the
terminology of classical mechanics, is called \emph{Hamilton's principle}.
The Maupertuis principle with action functional~(\ref{eq:confT}) corresponds
to Hamilton's principle with a Lagrangian
\begin{equation}
  \label{eq:confLag}
  \mathcal{L} = \frac{1}{2} \hat{g}_{\mu \nu} (x)
  \frac{dx^{\mu}}{d \ell} \frac{dx^{\nu}}{d \ell} -
  \hat{\phi}_{\mu} \frac{dx^{\mu}}{d \ell},
\end{equation}
(see, e.g., Carath{\'e}odory~\cite{caratheodory-82}, Sections~304\,--\,307).
The pertaining Euler--Lagrange equations read
\begin{equation}
  \label{eq:confLor}
  \hat{g}_{\mu \nu} \left( \frac{d^2 x^{\nu}}{d \ell^2} +
  \hat{\Gamma}^{\nu}_{\sigma \tau} \frac{d x^{\sigma}}{d \ell}
  \frac{d x^{\tau}}{d \ell} \right) =
  \left( \partial_{\nu} \hat{\phi}_{\mu} -
  \partial_{\mu} \hat{\phi}_{\nu} \right) \frac{dx^{\mu}}{d \ell}
\end{equation}
where $\hat{\Gamma}^{\nu}_{\sigma \tau}$ are the Christoffel symbols of
the Fermat metric $\hat{g}$. The solutions admit the constant of motion
\begin{equation}
  \label{eq:confE}
  \hat{g}_{\mu \nu} (x) \frac{dx^{\mu}}{d \ell}
  \frac{dx^{\nu}}{d \ell} = \mbox{constant},
\end{equation}
which can be chosen equal to 1 for each ray, such that $\ell$ gives the
$\hat{g}$-arclength. By Equation~(\ref{eq:confT}), the latter
gives the travel time if $\hat{\phi} = 0$. According to Equation~(\ref{eq:confLor}),
the \emph{Fermat two-form}
\begin{equation}
  \label{eq:omega}
  \hat{\omega} = d \hat{\phi}
\end{equation}
exerts a kind of Coriolis force on the light rays. This force has the same
mathematical structure as the \emph{Lorentz force} in a magnetostatic field.
In this analogy, $\hat{\phi}$ corresponds to the magnetic (vector) potential.
In other words, light rays in a conformally stationary spacetime behave like
charged particles, with fixed charge-to-mass ratio, in a magnetostatic field
$\hat{\omega}$ on a Riemannian manifold $(\hat{M},\hat{g})$.
For linearly polarized light, the Fermat geometry can also be used for 
describing the propagation of the polarization plane~\cite{hasse-perlick-93}.
One finds that the polarization plane undergoes a rotation similar to
the Faraday rotation in a magnetic field. This observation corroborates
the formal analogy between $\hat{\omega}$ and a magnetic field.
The gravitational analogue of the Faraday rotation was already 
discussed briefly in Section~\ref{ssec:distortion} above. 

Fermat's principle in static spacetimes dates back to
Weyl~\cite{weyl-17} (cf.~\cite{levicivita-18, synge-25}). The stationary
case was treated by Pham Mau Quan~\cite{pham-57}, who even took an isotropic
medium into account, and later, in a more elegant presentation, by
Brill~\cite{brill-73}. These versions of Fermat's principle are discussed
in several text-books on general relativity (see, e.g., \cite{misner-thorne-wheeler-73, frankel-79, straumann-84} for the
static and~\cite{landau-lifshitz-62} for the stationary case). A detailed
discussion of the conformally stationary case can be found in~\cite{perlick-90b}. Fermat's principle in conformally stationary spacetimes
was used as the starting point for deriving the lens equation of the
quasi-Newtonian apporoximation formalism by
Schneider~\cite{schneider-85}
(cf.~\cite{schneider-ehlers-falco-92}). As an alternative to the name
``Fermat metric'' (used, e.g., in~\cite{frankel-79, straumann-84, perlick-90b}),
the names ``optical metric'' (see, e.g., \cite{gibbons-perry-78,
dowker-kennedy-78, gibbons-werner-2008, gibbons-warnick-werner-2008}) 
and ``optical (reference) geometry'' (see, e.g.,
\cite{abramowicz-carter-lasota-88, kristiansson-sonego-abramowicz-98,
stuchlik-hledik-99a, stuchlik-hledik-juran-2000, hledik-2001, 
abramowicz-bengtsson-karas-rosquist-2002}) are also used.

In the conformally static case, one can apply the standard Morse theory for
Riemannian geodesics to the Fermat metric $\hat{g}$ to get results on the number
of $\hat{g}$-geodesics joining two points in space. This immediately gives results
on the number of lightlike geodesics joining a point in spacetime to an integral
curve of $W = \partial_t$. Completeness of the
Fermat metric corresponds to global hyperbolicity of the spacetime metric. The
relevant techniques, and their generalization to (conformally) stationary
spacetimes, are detailed in a book by Masiello~\cite{masiello-94}. (Note that,
in contrast to standard terminology, Masiello's definition of a stationary
spacetime includes the assumption that the hypersurfaces $t=\mbox{constant}$
are spacelike.) The resulting Morse theory is a special case of the Morse
theory for Fermat's principle in globally hyperbolic spacetimes (see
Section~\ref{ssec:morse}). In addition to Morse theory, other standard
methods from Riemannian geometry have been applied to the Fermat metric,
e.g., convexity techniques~\cite{giannoni-masiello-piccione-2001a,
giannoni-masiello-piccione-2001b}.

If the metric~(\ref{eq:confg}) is conformally static, $\hat{\phi}_{\mu} (x)=
\partial_{\mu} h (x)$, and if the Fermat metric is conformal to the
Euclidean metric, $\hat{g}_{\mu \nu} (x) = n(x)^2
\delta_{\mu \nu}$, the arrival time functional~(\ref{eq:confT}) can be written as
\begin{equation}
  \label{eq:confTn}
  T(\lambda) = \int_{\ell_1=0}^{\ell_2} n(x) \, d \ell + \mbox{constant},
\end{equation}
where $\ell$ is Euclidean arclength. Hence, Fermat's principle reduces to its
standard optics form for an isotropic medium with index of refraction $n$ on
Euclidean space. As a consequence, light propagation in a spacetime with the
assumed properties can be mimicked by a medium with an
appropriately chosen index of refraction. This remark applies, e.g., to
spherically symmetric and static spacetimes (see Section~\ref{ssec:ss}) and,
in particular, to the Schwarzschild spacetime (see Section~\ref{ssec:schw}).
The analogy with ordinary optics in media has been used for
constructing, in the laboratory, \emph{analogue models} for light propagation
in general-relativistic spacetimes (see~\cite{novello-visser-volovik-2002}).

Extremizing the functional~(\ref{eq:confTn})
is formally analogous to Maupertuis' principle for a particle in a scalar potential
on flat space, which is discussed in any book on classical mechanics. Dropping the
assumption that the Fermat one-form is a differential, but still requiring the Fermat
metric to be conformal to the Euclidean metric, corresponds to introducing an additional
vector potential. This form of the optical-mechanical analogy, for light rays in
stationary spacetimes whose Fermat metric is conformal to the Euclidean metric, is
discussed, e.g., in~\cite{alsing-98}.

The conformal factor $e^{2f}$ in Equation~(\ref{eq:confg}) does not affect
the paths of light rays. However, it does affect redshifts and distance
measures (recall Section~\ref{ssec:distance}). If $g$ is of
the form~(\ref{eq:confg}), for every lightlike geodesic $\lambda$ the
quantity $g(\dot{\lambda}, \partial_t)$ is a constant of motion. This leads
to a particularly simple form of the general redshift formula~(\ref{eq:z}).
We consider an arbitrary lightlike geodesic $s \mapsto \lambda (s)$ in terms of
its coordinate representation $s \mapsto \left( t(s),x^1(s),x^2(s),x^3(s) \right)$.
If both observer and emitter are at rest in the sense that their 4-velocities
$U_\mathrm{O}$ and $U_\mathrm{S}$ are parallel to $W = \partial_t$, Equation~(\ref{eq:z}) can be
rewritten as
\begin{equation}
  \label{eq:zpot}
  \log \left( 1 + z(s) \right) =
  f \left( t(s),x(s) \right) - f \left( t(0),x(0) \right).
\end{equation}
This justifies calling $f$ the redshift potential. It is shown in~\cite{hasse-perlick-88}
that there is a redshift potential for a congruence of timelike curves in a
spacetime if and only if the timelike curves are the integral curves of a conformal
Killing vector field. The notion of a redshift potential or redshift function
is also discussed in~\cite{dautcourt-87}. Note that Equation~(\ref{eq:zpot}) immediately
determines the redshift in conformally stationary spacetimes for \emph{any} pair of
observer and emitter. If the 4-velocity of the observer or of the emitter is not
parallel to $W = \partial_t$, one just has to add the usual special-relativistic
Doppler factor.

Conformally stationary spacetimes can be characterized by another interesting
property. Let $W$ be a timelike vector field in a spacetime and fix
three observers whose worldlines are integral curves of $W$. Then the angle
under which two of them are seen by the third one remains constant in the
course of time, for any choice of the observers, if and only if $W$ is
proportional to a conformal Killing vector field. For a proof see~\cite{hasse-perlick-88}.


\subsection{Lensing in spherically symmetric and static spacetimes}
\label{ssec:ss}

The class of spherically symmetric and static spacetimes is of particular relevance
in view of lensing, because it includes models for non-rotating stars and black
holes (see Sections~\ref{ssec:schw}, \ref{ssec:kottler}, \ref{ssec:reissner}),
but also for more exotic objects such as wormholes (see Section~\ref{ssec:worm}),
monopoles (see Section~\ref{ssec:mono}), naked singularities (see
Section~\ref{ssec:JNW}), and Boson or Fermion stars (see 
Section~\ref{ssec:boson}). A spherically symmetric and static 
spacetime can also be used, as a rough approximation, to model 
a star cluster, a galaxy or a cluster of galaxies.
Here we collect the relevant formulas for an unspecified spherically symmetric
and static metric. We find it convenient to write the metric in the form
\begin{equation}
  \label{eq:ssg}
  g = e^{2f(r)} \left( - dt^2 + S(r)^2 \, dr^2 +
  R(r)^2 \left( d\vartheta^2 + \sin^2 \vartheta \, d\varphi^2 \right) \right).
\end{equation}
As Equation~(\ref{eq:ssg}) is a special case of Equation~(\ref{eq:confg}), all results of
Section~\ref{ssec:conf} for conformally stationary metrics apply. However, much
stronger results are possible because for metrics of the form~(\ref{eq:ssg}) the
geodesic equation is completely integrable. Hence, all relevant quantities can be
determined explicitly in terms of integrals over the metric
coefficients. \\

\noindent
{\bf Redshift and Fermat geometry.} \\
\noindent
Comparison of Equation~(\ref{eq:ssg}) with the general form~(\ref{eq:confg}) of a
conformally stationary spacetime shows that here the redshift potential
$f$ is a function of $r$ only, the Fermat one-form $\hat{\phi}$ vanishes,
and the Fermat metric $\hat{g}$ is of the special form
\begin{equation}
  \label{eq:ssgF}
  \hat{g} = S(r)^2 \, dr^2 + R(r)^2 \left( d\vartheta^2 +
  \sin^2 \vartheta \, d\varphi^2 \right).
\end{equation}
This Fermat metric has several interesting applications. E.g., 
Gibbons and Werner~\cite{gibbons-werner-2008} have derived some 
lensing features of a spherically symmetric static fluid ball by applying 
the Gauss-Bonnet theorem to the corresponding Fermat metric (or 
optical metric). 
By Fermat's principle, the geodesics of $\hat{g}$ coincide with the
projection to 3-space of light rays. The travel time (in terms of the
time coordinate $t$) of a lightlike curve coincides with the
$\hat{g}$-arclength of its projection. By symmetry, every
$\hat{g}$-geodesic stays in a plane through the origin.
From Equation~(\ref{eq:ssgF}) we read that the sphere of radius $r$ has
area $4 \pi R(r)^2$ with respect to the Fermat metric. Also, Equation~(\ref{eq:ssgF})
implies that the second fundamental form of this sphere is a multiple
of its first fundamental form, with a factor
$-R'(r) \left( R(r) \, S(r) \right)^{-1}$. If
\begin{equation}
  \label{eq:rL}
  R'(r_\mathrm{p}) = 0,
\end{equation}
the sphere $r=r_\mathrm{p}$ is totally geodesic, i.e., a $\hat{g}$-geodesic
that starts tangent to this sphere remains in it. The best known
example for such a \emph{light sphere} or \emph{photon sphere} is
the sphere $r = 3m$ in the Schwarzschild spacetime (see
Section~\ref{ssec:schw}). Light spheres also occur in the spacetimes of
wormholes (see Section~\ref{ssec:worm}). If $R''(r_\mathrm{p}) < 0$, the circular
light rays in a light sphere are stable with respect to radial perturbations,
and if $R''(r_\mathrm{p}) > 0$, they are unstable like in the Schwarschild case.
The condition under which a spherically symmetric static spacetime
admits a light sphere was first given by Atkinson~\cite{atkinson-65}.
Abramowicz~\cite{abramowicz-90} has shown that for an observer traveling
along a circular light orbit (with subluminal velocity) there is no
centrifugal force and no gyroscopic precession. Claudel, Virbhadra, and
Ellis~\cite{claudel-virbhadra-ellis-2001} investigated, with the help
of Einstein's field equation and energy conditions, the amount of
matter surrounded by a light sphere. Among other things, they found
an energy condition under which a spherically symmetric static black
hole must be surrounded by a light sphere. A purely kinematical argument
shows that any spherically symmetric and static spacetime that has a
horizon at $r = r_\mathrm{H}$ and is asymptotically flat for $r \to \infty$
must contain a light sphere at some radius between $r_\mathrm{H}$ and $\infty$
(see Hasse and Perlick~\cite{hasse-perlick-2002}). In the same article,
it is shown that in any spherically symmetric static spacetime with a
light sphere there is gravitational lensing with infinitely many images.
Bozza~\cite{bozza-2002} investigated a \emph{strong-field limit} of lensing
in spherically symmetric static spacetimes, as opposed to the well-known
weak-field limit, which applies to light rays that come close to an unstable
light sphere. (In later papers, the term ``strong-field limit"
was replaced with ``strong-deflection limit". This is, indeed, more
appropriate because the gravitational field, measured in terms of tidal
forces, need not be particularly strong near an unstable light sphere.
The characteristic feature is that the bending angle goes to infinity, i.e.,
that light rays make arbitrarily many turns around the center if they
approach an unstable light sphere.)
This limit applies, in particular, to light rays that approach
the sphere $r=3m$ in the Schwarzschild spacetime
(see~\cite{bozza-capozziello-iovane-scarpetta-2001} and, for
illustrations, Figures~\ref{fig:schwlen}, \ref{fig:schwDT},
and~\ref{fig:schwDe}). The strong-deflection limit has also been 
applied to many other spherically symmetric and static metrics;
several examples are discussed in Section~\ref{sec:examples}
below. As demonstrated in the original article by
Bozza~\cite{bozza-2002}, the parameters that characterize
the strong-deflection limit can be used to distinguish between
different black-hole metrics. These parameters were related to 
quasi-normal modes in~\cite{stefanov-yazadjiev-gyulchev-2010}.
\\

\noindent
{\bf Index of refraction and embedding diagrams.} \\
\noindent
Transformation to an \emph{isotropic} radius coordinate ${\tilde{r}}$ via
\begin{equation}
  \label{eq:ssiso}
  \frac{S(r) \, dr}{R(r)} = \frac{d \tilde{r}}{\tilde{r}}
\end{equation}
takes the Fermat metric~(\ref{eq:ssgF}) to the form
\begin{equation}
  \label{eq:ssgFiso}
  \hat{g} = n(\tilde{r})^2 \left( d\tilde{r}^2 +
  \tilde{r}^2 (d\vartheta^2 + \sin^2 \vartheta \, d\varphi^2) \right)
\end{equation}
where
\begin{equation}
  \label{eq:ssn}
  n (\tilde{r}) = \frac{R(r)}{\tilde{r}}.
\end{equation}
On the right-hand side $r$ has to be expressed by $\tilde{r}$ with the
help of Equation~(\ref{eq:ssiso}).
The results of Section~\ref{ssec:conf} imply that the lightlike geodesics
in a spherically symmetric static spacetime are equivalent to the light
rays in a medium with index of refraction~(\ref{eq:ssn}) on Euclidean 3-space.
For arbitrary metrics of the form~(\ref{eq:ssg}), this result is due to
Atkinson~\cite{atkinson-65}. It reduces the lightlike geodesic problem in a
spherically symmetric static spacetime to a standard problem in ordinary optics,
as treated, e.g., in~\cite{luneburg-64}, \S 27,
and~\cite{lakshminarayanan-ghatak-thyagarajan-2001},
Section~4. One can combine this result with our earlier observation that the integral in
Equation~(\ref{eq:confTn}) has the same form as the functional in Maupertuis' principle in
classical mechanics. This demonstrates that light rays in spherically symmetric and
static spacetimes behave like particles in a spherically symmetric potential
on Euclidean 3-space (cf., e.g., \cite{evans-nandi-islam-96a}).

If the \emph{embeddability condition}
\begin{equation}
  \label{eq:ssemb}
  S(r)^2 \ge R'(r)^2
\end{equation}
is satisfied, we define a function $Z(r)$ by
\begin{equation}
  \label{eq:ssprofile}
  Z'(r) = \sqrt{ S(r)^2 - R'(r)^2 }.
\end{equation}
Then the Fermat metric~(\ref{eq:ssgF}) reads
\begin{equation}
  \label{eq:ssgFemb}
  \hat{g} = (d R(r))^2 + R(r)^2
  \left( d \vartheta^2 + \sin^2 \vartheta \, d\varphi^2 \right) +
  (dZ(r))^2.
\end{equation}
If restricted to the equatorial plane $\vartheta = \pi /2$, the
metric~(\ref{eq:ssgFemb}) describes a surface of revolution, embedded
into Euclidean 3-space as
\begin{equation}
  \label{eq:embrev}
  (r,\varphi) \mapsto
  \left( R(r) \cos \varphi, R(r) \, \sin \varphi, Z(r) \right).
\end{equation}
Such embeddings of the Fermat geometry have been visualized for 
several spacetimes of interest (see Figure~\ref{fig:schwemb} for 
the Schwarzschild case 
and~\cite{hledik-2001, hledik-stuchlik-cipko-2006} 
for other examples). 
This is quite
instructive because from a picture of a surface of revolution one can read the
qualitative features of its geodesics without calculating them. Note that
Equation~(\ref{eq:ssiso}) defines the isotropic radius coordinate uniquely up to
a multiplicative constant. Hence, the straight lines in this coordinate
representation give us an unambiguously defined reference grid for every
spherically symmetric and static spacetime. These straight
lines have been called \emph{triangulation lines} in~\cite{cowling-83, cowling-84},
where their use for calculating bending angles, exactly or approximately,
is outlined. \\

\noindent
{\bf Light cone.} \\
\noindent
In a spherically symmetric static spacetime, the
(past) light cone of an event $p_\mathrm{O}$ can be written in terms of integrals
over the metric coefficients. We first restrict to the equatorial plane
$\vartheta = \pi /2$. The $\hat{g}$-geodesics are then determined
by the Lagrangian
\begin{equation}
  \label{eq:LF}
  \mathcal{L} =
  \frac{1}{2} \left( S(r)^2 \left( \frac{dr}{d \ell} \right)^2 +
  R(r)^2 \left( \frac{d \varphi}{d \ell} \right)^2 \right).
\end{equation}
The Euler-Lagrange equations read
\begin{equation}
  \label{eq:ELF1}
  \frac{d}{d \ell} \left( S(r)^2\frac{dr}{d \ell} \right) =
   S(r) S'(r) \left(\frac{dr}{d \ell} \right)^2 + R(r) R'(r)
  \left( \frac{d \varphi}{d \ell} \right)^2 .
\end{equation}
\begin{equation}
  \label{eq:ELF2}
  \frac{d}{d \ell} \left( R(r)^2 \frac{d \varphi}{d \ell} \right) =
  0 .
\end{equation}
After dividing the first equation by $R(r)^2 (d \varphi / d \ell )^2$,
and using the second equation, we find
\begin{equation}
  \label{eq:ELrp}
  \frac{d}{d \varphi} \left( \frac{S(r)^2}{R(r)^2}
  \frac{dr}{d \varphi} \right) =
   \frac{S(r) S'(r)}{R(r)^2} \left(\frac{dr}{d \varphi} \right)^2 
   + \frac{R'(r)}{R(r)} .
\end{equation}
Equations (\ref{eq:ELF1}) and (\ref{eq:ELF2}) give the
light rays parametrized by $\hat{g}$-arclength (which equals 
travel time) $\ell$,  Equation (\ref{eq:ELrp}) can be used for 
determining the orbits of light rays if the parametrization 
plays no role.    

For fixed radius value $r_\mathrm{O}$, initial conditions
\begin{equation}
  \begin{array}{rclrcl}
    r(0) &=& r_\mathrm{O}, &
    \qquad
    \displaystyle \frac{dr}{d \ell} (0) &=&
    \displaystyle \frac{\cos \Theta}{S(r_\mathrm{O})}, \\ [1.0 em]
    \varphi (0) &=& 0, &
    \qquad
    \displaystyle \frac{d \varphi}{d \ell} (0) &=&
    \displaystyle \frac{\sin \Theta}{R(r_\mathrm{O})}
  \end{array}
\end{equation}
determine a unique solution $r= \mathsf{r} (\ell, \Theta)$, $\varphi
= \phi (\ell, \Theta)$ of the Euler--Lagrange equations (\ref{eq:ELF1}) and (\ref{eq:ELF2}). 
$\Theta$ measures the
initial direction with respect to the symmetry axis (see
Figure~\ref{fig:sslens}). We get all light rays issuing from the event
$r=r_\mathrm{O}$, $\varphi = 0$, $\vartheta =\pi /2$, $t=t_\mathrm{O}$ into the past
by letting $\Theta$ range from 0 to $\pi$ and applying rotations around
the symmetry axis. This gives us the past light cone of this event
in the form
\begin{equation}
  \label{eq:sscone}
  (\ell, \Psi, \Theta) \longmapsto
  \begin{pmatrix}
    t_\mathrm{O} - \ell \\
    \mathsf{r} (\ell, \Theta) \, \sin \phi (\ell, \Theta) \, \cos \Psi \\
    \mathsf{r} (\ell, \Theta) \, \sin \phi (\ell, \Theta) \, \sin \Psi \\
    \mathsf{r} (\ell, \Theta) \, \cos \phi (\ell, \Theta)
  \end{pmatrix}.
\end{equation}
$\Psi$ and $\Theta$ are spherical coordinates on the observer's sky.
If we let $t_\mathrm{O}$ float over $\mathbb{R}$, we get the observational
coordinates~(\ref{eq:obsco}) for an observer on a $t$-line, up to
two modifications. First, $t_\mathrm{O}$ is not the same as proper time $\tau$;
however, along each $t$-line they are related just by a constant,
\begin{equation}
  \label{eq:taut}
  \frac{d \tau}{d t_\mathrm{O}} = e^{-f(r_\mathrm{O})}.
\end{equation}
Second, $\ell$ is not the same as the affine parameter $s$; along a
ray with initial direction $\Theta$, they are related by
\begin{equation}
  \label{eq:ells}
  \frac{ds}{d \ell} = e^{f (\mathrm{r} (\ell, \Theta))}.
\end{equation}
The constants of motion
\begin{equation}
  \label{eq:EL}
  R(r)^2 \frac{d \varphi}{d \ell} = R(r_\mathrm{O}) \sin \Theta,
  \qquad
  S(r)^2 \left( \frac{d r}{d \ell} \right)^2 +
  R(r)^2 \left( \frac{d \varphi}{d \ell} \right)^2 = 1
\end{equation}
give us the functions $\mathsf{r} (\ell, \Theta)$, $\phi (\ell, \Theta)$
in terms of integrals,
\begin{eqnarray}
  \ell &=& \int_{r_\mathrm{O} \dots}^{\dots \mathsf{r} (\ell, \Theta)}
  \frac{R(r) \, S(r) \, dr}{\sqrt{R(r)^2-R(r_\mathrm{O})^2 \, \sin^2 \Theta}},
  \label{eq:rell} \\
  \phi (\ell, \Theta) &=& R(r_\mathrm{O}) \, \sin \Theta
  \int_{r_\mathrm{O} \dots}^{\dots \mathsf{r} (\ell, \Theta)}
  \frac{S(r) \, dr}{R(r) \sqrt{R(r)^2-R(r_\mathrm{O})^2 \, \sin^2 \Theta}}.
  \label{eq:rphi}
\end{eqnarray}%
Here the notation with the dots is a short-hand; it means that the
integral is to be decomposed into sections where $\mathsf{r}(\ell, \Theta)$
is a monotonous function of $\ell$, and that the absolute value of
the integrals over all sections have to be added up. Turning points
occur at radius values where $R(r) = R(r_\mathrm{O}) \sin \Theta$
and $R'(r) \neq 0$ (see Figure~\ref{fig:schwpot}). If the metric coefficients
$S$ and $R$ have been specified, these integrals can be calculated and
give us the light cone (see Figure~\ref{fig:schwcon} for an example).
Having parametrized the rays with $\hat{g}$-arclength (=\,travel time),
we immediately get the intersections of the light cone with hypersurfaces
$t= \mbox{constant}$ (``instantaneous wave fronts''); see
Figures~\ref{fig:schwfrt}, \ref{fig:monofrt1},
and~\ref{fig:monofrt2}. \\

\noindent
{\bf Exact lens map and various approximation methods.} \\
\noindent

Recall from Section~\ref{ssec:cone} that the exact lens
map~\cite{frittelli-newman-99} refers to a chosen observation event
$p_\mathrm{O}$ and a chosen ``source surface'' $\mathcal{T}$. In
general, for $\mathcal{T}$ we
may choose any 3-dimensional submanifold that is ruled by timelike curves.
The latter are to be interpreted as wordlines of light sources. In a
spherically symmetric and static spacetime, we may take advantage of the
symmetry by choosing for $\mathcal{T}$ a sphere $r = r_\mathrm{S}$ with its ruling
by the $t$-lines. This restricts the consideration to lensing for static
light sources. Note that, for an observer at $r_\mathrm{O}$, 
all static light sources at radius $r_\mathrm{S}$
undergo the same redshift, $\log (1+z) = f(r_\mathrm{S}) - f(r_\mathrm{O})$.
Without loss of generality, we place the observation event
$p_\mathrm{O}$ on the 3-axis. This gives us the past light cone in
the representation~(\ref{eq:sscone}). To each ray from the observer,
with initial direction characterized by $\Theta$, we can assign the
total angle $\Phi (\Theta)$ the ray sweeps out on its way from $r_\mathrm{O}$
to $r_\mathrm{S}$ (see Figure~\ref{fig:sslens}). $\Phi (\Theta)$ is given by
Equation~(\ref{eq:rphi}),
\begin{equation}
  \label{eq:ssPhi}
  \Phi (\Theta) = R(r_\mathrm{O}) \, \sin \Theta
  \int_{r_\mathrm{O} \dots}^{\dots r_\mathrm{S}}
  \frac{S(r) \, dr}{R(r) \sqrt{R(r)^2 - R(r_\mathrm{O})^2 \, \sin^2 \Theta}},
\end{equation}
where the same short-hand notation is used as in Equation~(\ref{eq:rphi}).
$\Phi (\Theta)$ is not necessarily defined for all $\Theta$ because some 
light rays that start at $r_\mathrm{O}$ may not reach $r_\mathrm{S}$. Also, 
$\Phi (\Theta)$ may be multi-valued because a light ray may intersect the 
sphere $r=r_\mathrm{S}$ several times. Equation~(\ref{eq:sscone}) gives us 
the (possibly multi-valued) lens map
\begin{equation}
  \label{eq:sslens}
  (\Psi, \Theta) \longmapsto
  \begin{pmatrix}
    r_\mathrm{S} \, \sin \Phi (\Theta) \, \cos \Psi \\
    r_\mathrm{S} \, \sin \Phi (\Theta) \, \sin \Psi \\
    r_\mathrm{S} \, \cos \Phi (\Theta)
  \end{pmatrix}.
\end{equation}
This version of the exact lens map in spherically symmetric
and static spacetimes was first considered in~\cite{perlick-2004}. 
It is interesting to compare it with the standard lens map (or 
lens equation) in the quasi-Newtonian approximation formalism, 
see e.g. Wambsganss~\cite{wambsganss-98}, Section~3.1. In both cases, 
rotational symmetry about the axis through the 
observer has the effect that in essence the lens map reduces to 
a map from an angle to another angle; the first angle, here $\Theta$, 
determines the position of the image on the observer's sky, the 
second angle, here $\Phi ( \Theta )$, gives the actual position of 
the light source.  If the metric coefficients $R(r)$ and $S(r)$ are 
given, the integrals in Equation (\ref{eq:ssPhi}) can be numerically
calculated and from the result all lensing features can be determined 
with arbitrary accuracy. As an example, the exact lens map will be
evaluated for the Schwarzschild metric in Section~\ref{ssec:schw}
below. In~\cite{perlick-2004}), the examples of an Ellis
wormhole (cf. Section~\ref{ssec:worm}) and of a Barriola-Vilenkin
monopole (cf. Section~\ref{ssec:mono}) were treated. 
-- Note that $\Phi (\Theta)$ may take any value between 0 and infinity. 
A value $\Phi (\Theta) > 2 \pi$ occurs whenever a light ray makes more 
than one full turn around the center. For each image we can define the 
\emph{order}
\begin{equation}
  \label{eq:k}
  i (\Theta) =
  \min \left\{ m \in \mathbb{N} \,\big|\, \Phi (\Theta) < m \pi \right\},
\end{equation}
which counts how often the ray has crossed the axis. (If the lens map is
multi-valued, one should introduce an index to label different images 
that correspond to the same angle $\Theta$). In accordance with 
the terminology introduced in Section~\ref{ssec:hypcrit}, an image 
of order 0 is called  \emph{primary}, an image of order 1 is called 
\emph{secondary}, and so on. The standard example where images 
of arbitrarily high order occur is the Schwarzschild spacetime (see 
Section~\ref{ssec:schw}). For a light source which is 
not perfectly aligned with the observer and the center, images of 
even order have even parity and line up on one side of the direction 
towards the center; images of odd order have odd parity and line up 
on the other side of the direction towards the center.  In the case of 
perfect alignment, a sequence of Einstein rings is seen. An Einstein 
ring of order 0 is called \emph{primary}, an Einstein ring of
order 1 is called \emph{secondary}, and so on. -- 
We can rewrite the exact lens map in a spherically symmetric and static
spacetime in a form more similar to the standard quasi-Newtonian 
lens map if we make two additional assumptions which are satisfied
in many, though not all, situations of interest:
\begin{itemize}
\item The spacetime is asymptotically flat and both $r_\mathrm{O}$ 
and $r_\mathrm{S}$ are very large. 
\item The source is almost exactly opposite to the observer, i.e., $\Phi (\Theta)$
is close to an odd multiple of $\pi$.  
\end{itemize}
The first assumption makes sure that the lens map is single-valued, and both 
assumptions together imply that along each (past-oriented) light ray from
the observer to the source the radius coordinate has precisely one turning-point. 
For a light ray with turning point at $r_m(\Theta)$, the asymptotic
assumption allows to approximate Equation~(\ref{eq:ssPhi}) by
\begin{equation}
  \label{eq:Phiasy}
  \Phi (\Theta) = 2 \int_{r_m (\Theta)}^{\infty}
  \frac{S(r) \, dr}{R(r) \sqrt{R(r)^2 - R(r_\mathrm{O})^2 \, \sin^2 \Theta}}.
\end{equation}
To link up with the notation of the standard lens map, we introduce distances $D_{d}$ and $D_{ds}$ 
and angles $\theta = \pi - \Theta$, $\beta ( \theta )$ and $\hat{\alpha} ( \theta )$ according to 
Figure~\ref{fig:vele}. The alignment assumption implies that $\beta (\theta )$ is small, and the
asymptotic condition implies that the bending angle $\hat{\alpha}$ can be approximated as
\begin{equation}
  \label{eq:bendasy}
  \hat{\alpha} ( \theta ) = \Phi ( \Theta ) - \pi 
\end{equation}
After some elementary geometry, one finds that
\begin{equation}
  \label{eq:vele}
  \mathrm{tan} \, \beta ( \theta ) = \mathrm{tan} \, \theta -
  \frac{D_{ds}}{D_d + D_{ds}} \Big( \mathrm{tan} \, \theta
  + \mathrm{tan} \big( \hat{\alpha} ( \theta ) - \theta \big) \Big).
\end{equation}
This is the lens equation of Virbhadra and 
Ellis~\cite{virbhadra-ellis-2000} 
(cf.~\cite{virbhadra-narasimha-chitre-98} for an 
earlier version). Equation (\ref{eq:vele}) gives a well-defined 
(single-valued) lens map $\theta \mapsto \beta ( \theta )$ if we 
insert Equations~(\ref{eq:bendasy}) and (\ref{eq:Phiasy}). 
The Virbhadra-Ellis lens map may be called ``almost exact''. 
It is based on approximations as to the positions of source 
and observer, but it is not restricted to the case that the 
bending angle is small. As a matter of fact, the bending angle 
may be arbitrarily large; $\hat{\alpha} ( \theta )$ diverges 
to infinity if the turning point $r_m (\Theta)$ approaches an 
unstable light sphere. (It was already mentioned that an
unstable light sphere occurs at a radius value $r_\mathrm{p}$ 
if and only if $R'(r_\mathrm{p})=0$ and $R''(r_\mathrm{p})>0$;
the standard example is the sphere at $r_p=3m$ in 
the Schwarzschild spacetime, see Section~\ref{ssec:schw}). 
It was shown by 
Bozza~\cite{bozza-2002} that, whenever an unstable
light sphere is approached, the divergence of the bending
angle is logarithmic. The Virbhadra-Ellis lens equation
was originally introduced for the Schwarzschild 
metric~\cite{virbhadra-ellis-2000} where it approximates 
the exact treatment remarkably well within a wide range of 
validity~\cite{frittelli-kling-newman-2000}. In comparison to
the exact lens map, the Virbhadra-Ellis lens map has the 
appealing property of resembling the standard quasi-Newtonian
lens map as much as possible. On the other hand, neither
analytical nor numerical evaluation of the ``almost exact 
lens map'' is significantly easier than that of the exact 
lens map. The Virbhadra-Ellis lens map was successfully applied
to many spherically symmetric and static spacetimes, several
examples are considered in Section~\ref{sec:examples} below. 
Bozza~\cite{bozza-2008b} compared the Virbhadra-Ellis lens
equation with other approximate lens equations  that had been
proposed for spherically symmetric and static spacetimes and,
in particular, for the Schwarzschild metric: 
\begin{itemize}
\item the Ohanian lens equation, which was implicitly contained 
in Ohanian's pioneering work~\cite{ohanian-87} on Schwarzschild 
lensing, 
\item  a modification of the Virbhadra-Ellis lens equation, 
introduced by Da{\c b}rowski and 
Schunck~\cite{dabrowski-schunck-2000} in their treatment of 
lensing by a boson star,  
\item a lens equation introduced by Bozza and 
Sereno~\cite{bozza-sereno-2006} that is essentially 
equivalent to the Ohanian lens equation but replaces 
an angle centered at the lens by an
angle centered at the observer, 
\item a new lens equation that is, again, a slight modification 
of the Ohanian lens equation. 
\end{itemize}
All these lens equations relax the alignment condition but retain 
some kind of asymptotic assumption. After discussing the accuracy 
of these various lens equations in realistic situations, Bozza argues 
in favour of the Ohanian lens equation and its modifications. 
-- In addition to approximative lens equations, several 
other approximation techniques have
been developed for lensing in spherically symmetric and
static spacetimes. Amore and Arceo~\cite{amore-arceo-2006}
expressed the bending angle analytically as a rapidly 
convergent series; this approach was further developed
in~\cite{amore-arceo-fernandez-2006, 
amore-cervantes-pace-fernandez-2007}.
Keeton and Petters~\cite{keeton-petters-2005}
expanded corrections to the weak-deflection limit as a 
Taylor series in the gravitational radius of the lens. In two
follow-up papers, they applied this formalism to post-Newtonian
metrics~\cite{keeton-petters-2006a} and to braneworld
black holes~\cite{keeton-petters-2006b}.  A major purpose of 
all approximation methods mentioned is to test general relativity 
by comparing Schwarzschild lensing to lensing in alternative 
theories of gravity, see Section~\ref{ssec:schw}.

\epubtkImage{figure06.png}
{\begin{figure}[hptb]
   \def\epsfsize#1#2{0.8#1}
   \centerline{\epsfbox{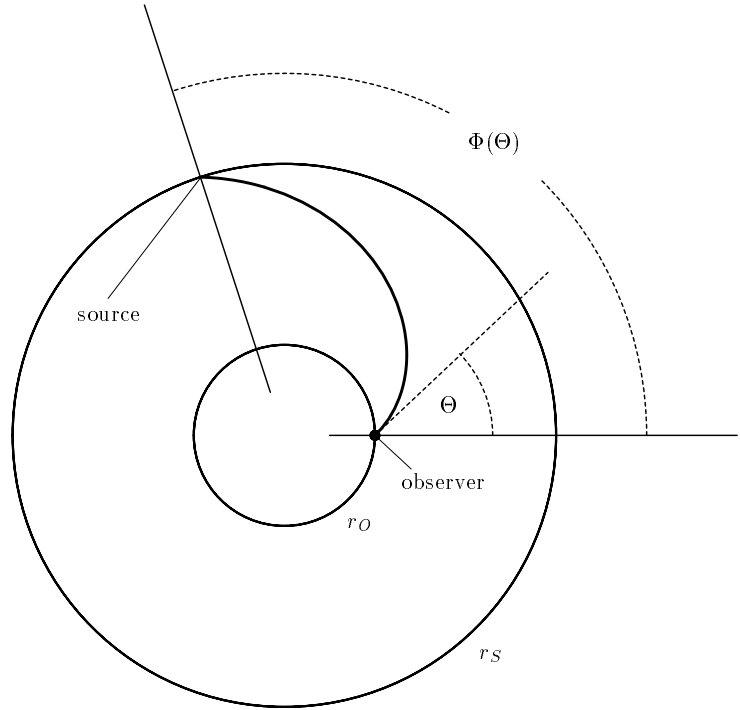}}
   \caption{Illustration of the exact lens map in spherically
     symmetric static spacetimes. The picture shows a spatial
     plane. The observation event (dot) is at $r=r_\mathrm{O}$, static
     light sources are distributed at $r = r_\mathrm{S}$. $ \Theta$ is
     the colatitude coordinate on the observer's sky. It takes values
     between 0 and $\pi$. $\Phi (\Theta)$ is the angle swept out by
     the ray with initial direction $\Theta$ on its way from
     $r_\mathrm{O}$ to $r_\mathrm{S}$. It takes values between 0 and
     $\infty$. In general, neither existence nor uniqueness of
     $\Phi (\Theta)$ is guaranteed for given $\Theta$. A similar
     picture is in~\cite{perlick-2004}.}
   \label{fig:sslens}
 \end{figure}
}

\epubtkImage{figure06a.png}
{\begin{figure}[hptb]
   \def\epsfsize#1#2{0.5#1}
   \centerline{\epsfbox{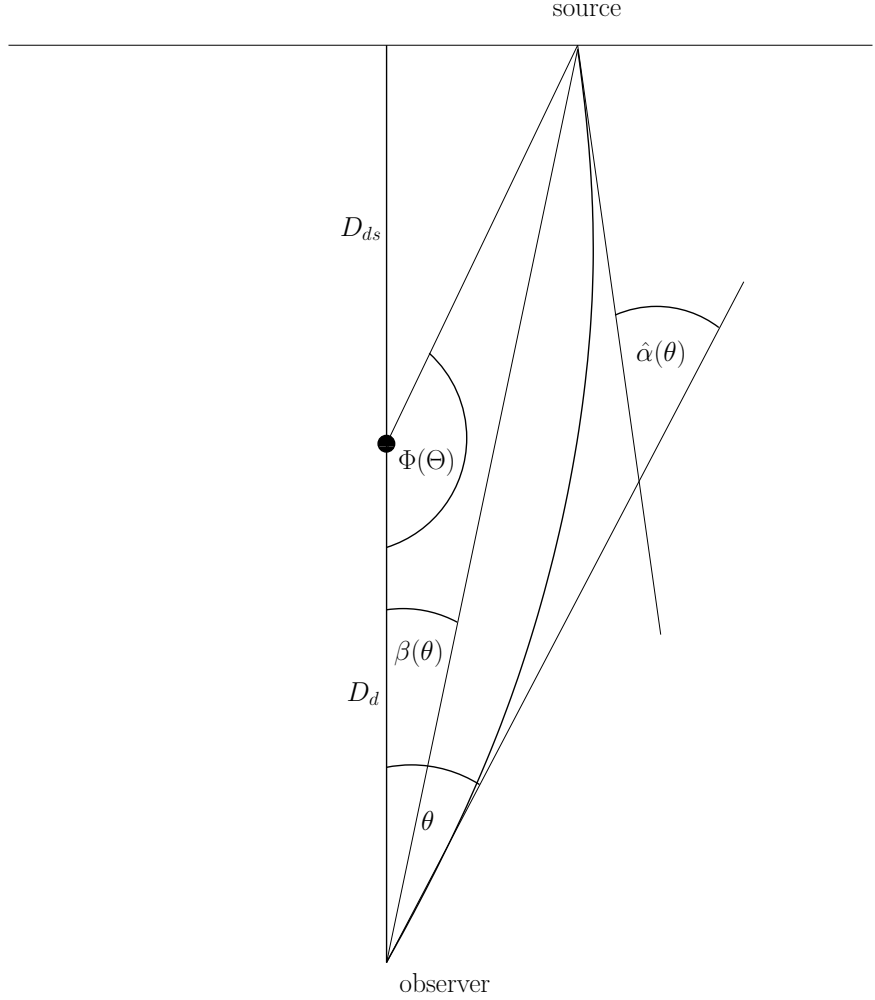}}
   \caption{Illustration of the ``almost exact'' lens map 
     of Virbhadra and Ellis~\cite{virbhadra-ellis-2000}. As almost 
     perfect alignment is assumed, we may think of the light sources 
     as being distributed on a plane, rather than on a sphere as for
     the exact lens map. $D_d$ and $D_{ds}$ are measured 
     in terms of the radial coordinate. The difference between $\Theta$ 
     and $\theta = \pi - \Theta$ must be observed when comparing with
     the exact lens map. $\beta (\theta )$ is the angle between the axis 
     and a straight line, in the coordinate picture, that connects observer 
     and source. In contrast to the angle $\Phi ( \Theta )$, the angle 
     $\beta (\theta )$ is not invariant under a transformation of the 
     radial coordinate; neither $\Phi ( \Theta )$ nor $\beta (\theta )$ is
     observable. Note that the Virbhadra-Ellis lens map also applies to 
     cases where $\Phi ( \Theta )$ is bigger than $2 \pi$.}
   \label{fig:vele}
 \end{figure}
}

\paragraph{Distance measures, image distortion and brightness of
 images.} ~\\
For calculating image distortion (see Section~\ref{ssec:distortion})
and the brightness of images (see Section~\ref{ssec:brightness})
we have to consider infinitesimally thin bundles with vertex at
the observer. In a spherically symmetric and static spacetime,
we can apply the orthonormal derivative operators $\partial_{\Theta}$
and $\sin \Theta \, \partial_{\Psi}$
to the representation~(\ref{eq:sscone}) of the past light cone. Along
each ray, this gives us two Jacobi fields $Y_1$ and $Y_2$ which span
an infinitesimally thin bundle with vertex at the observer. $Y_1$
points in the \emph{radial} direction and $Y_2$ points in the
\emph{tangential} direction (see Figure~\ref{fig:radtang1}). The radial
and the tangential direction are orthogonal to each other and, by
symmetry, parallel-transported along each ray. Thus, we can choose the
Sachs basis $(E_1 ,E_2)$ such that $Y_1=D_+E_1$ and $Y_2=D_-E_2$.
The coefficients $D_+$ and $D_-$ are unique if we require them to
be positive near the vertex. $D_+$ and $D_-$ are the extremal angular 
diameter distances of Section~\ref{ssec:distance} with respect to a static
observer (because the $(\Psi, \Theta)$-grid refers to a static observer).
In the case at hand, they are called the \emph{radial} and \emph{tangential}
angular diameter distances. They can be calculated by normalizing $Y_1$
and $Y_2$,
\begin{eqnarray}
  D_+ (\ell, \Theta) &=&
  e^{f (\mathsf{r} (\ell, \Theta))} \, R(r_\mathrm{O}) \, \cos \Theta \,
  \sqrt{R \left( \mathsf{r}(\ell, \Theta) \right)^2 -
  R(r_\mathrm{O}^2) \, \sin^2\Theta} 
  \nonumber \\
  &&\times \int_{r_\mathrm{O} \dots}^{\dots \mathsf{r} (\Theta, \ell)} \!\!\!\!\!
  \frac{ S(r) \, R(r) dr}{\sqrt{R(r)^2 - R(r_\mathrm{O})^2 \, \sin^2 \Theta}^{\,3}},
  \label{eq:ssD+} \\
  D_- (\ell, \Theta) &=& e^{f ( \mathsf{r} (\ell, \Theta))} \,
  R \left( \mathsf{r} (\ell, \Theta) \right)
  \frac{\sin \phi (\ell, \Theta)}{\sin \Theta}.
  \label{eq:ssD-}
\end{eqnarray}%
These formulas have been derived first for the special case of the Schwarz\-schild
metric by Dwivedi and Kantowski~\cite{dwivedi-kantowski-72} and then for
arbitrary spherically symmetric static spacetimes by Dyer~\cite{dyer-77}.
(In~\cite{dyer-77}, Equation~(\ref{eq:ssD-}) is erroneously given only for the case
that, in our notation, $e^{f(r)}R(r)=r$.) From these formulas we immediately
get the area distance $D_{\mathrm{area}} = \sqrt{| D_+ D_- |}$
for a static observer and, with the help of the redshift $z$, the luminosity
distance $D_{\mathrm{lum}} = (1+z)^2 D_{\mathrm{area}}$
(recall Section~\ref{ssec:distance}). In this way, Equation~(\ref{eq:ssD+}) and
Equation~(\ref{eq:ssD-}) allow to calculate the brightness of images according to
the formulas of Section~\ref{ssec:brightness}. Similarly, Equation~(\ref{eq:ssD+}) and
Equation~(\ref{eq:ssD-}) allow to calculate image distortion in terms of the ellipticity
$\varepsilon$ (recall Section~\ref{ssec:distortion}). In general, $\varepsilon$
is a complex quantity, defined by Equation~(\ref{eq:ellipticity}). In the case at
hand, it reduces to the real quantity $\varepsilon = D_- / D_+ - D_+ / D_-$.
The expansion $\theta$ and the shear $\sigma$ of the bundles under consideration
can be calculated from Kantowski's formula~\cite{kantowski-68,
dwivedi-kantowski-72},
\begin{equation}
  \label{eq:ssscalars}
  \dot{D}_{\pm} = \left( \theta \pm \sigma \right) D_{\pm},
\end{equation}
to which Equation~(\ref{eq:dotshape}) reduces in the case at hand. The dot
(= derivative with respect to the affine parameter $s$) is related to the
derivative with respect to $\ell$ by Equation~(\ref{eq:ells}).
Evaluating Equations~(\ref{eq:ssD+}, \ref{eq:ssD-}) in connection
with the exact lens map leads to quite convenient formulas, for static light
sources at $r=r_\mathrm{S}$. Setting $\mathsf{r} (\ell, \Theta) = r_\mathrm{S}$ and
$\phi (\ell, \Theta) = \Phi (\Theta)$ and comparing with Equation~(\ref{eq:ssPhi})
yields (cf.~\cite{perlick-2004})
\begin{eqnarray}
  D_+ (\Theta) &=& e^{f (r_\mathrm{S})} \,
  \sqrt{R(r_\mathrm{S})^2-R(r_\mathrm{O})^2 \, \sin^2 \Theta} \; \Phi' (\Theta),
  \label{eq:ssPhiD+} \\
  D_- (\Theta) &=& e^{f ( r_\mathrm{S})} \, R(r_\mathrm{S}) \, \sin \Phi (\Theta).
  \label{eq:ssPhiD-}
\end{eqnarray}%
These formulas immediately give image distortion
and the brightness of images if the map $\Theta \mapsto
\Phi (\Theta)$ is known. \\

\epubtkImage{figure07.png}
{\begin{figure}[hptb]
   \def\epsfsize#1#2{1.0#1}
   \centerline{\epsfbox{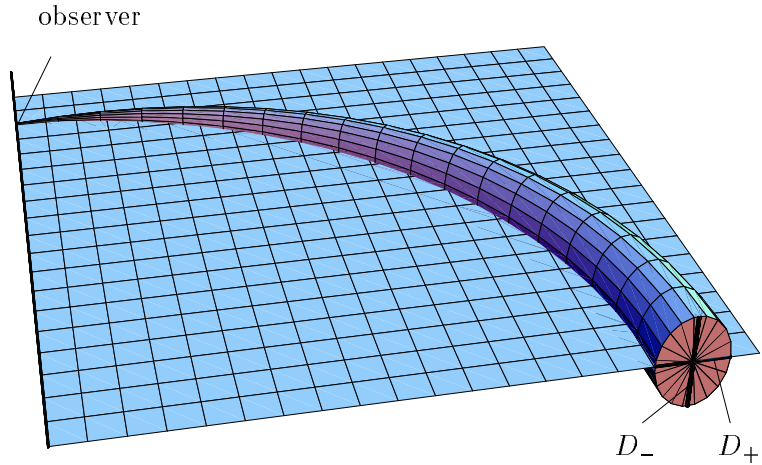}}
   \caption{Thin bundle around a ray in a spherically symmetric
     static spacetime. The picture is purely spatial, i.e., the time
     coordinate $t$ is not shown. The ray is contained in a plane, so
     there are two distinguished spatial directions orthogonal to the
     ray: the ``radial'' direction (in the plane) and the
     ``tangential'' direction (orthogonal to the plane). For a bundle
     with vertex at the observer, the radial diameter of the
     cross-section equals $2|D_+|$, and the tangential diameter 
     of the cross-section equals $2|D_-|$. Up to the first caustic
     point, $D_+$ and $D_-$ are positive. In contrast to the
     general situation of Figure~\ref{fig:shape}, here the angle
     $\chi$ is always zero (if the Sachs basis $(E_1,E_2)$ is chosen
     appropriately). }
   \label{fig:radtang1}
 \end{figure}
}

\noindent
{\bf Caustics of light cones.} \\
\noindent
Quite generally, the past light cone has a caustic point exactly
where at least one of the extremal angular diameter distances $D_+$,
$D_-$ vanishes (see Sections~\ref{ssec:front}, \ref{ssec:Sachs},
and~\ref{ssec:distance}). In the case at hand, zeros of $D_+$ are
called \emph{radial caustic points} and zeros of $D_-$ are called \emph{tangential
caustic points} (see Figure~\ref{fig:radtang2}). By Equation~(\ref{eq:ssD-}),
tangential caustic points occur if $\phi (\ell, \Theta)$ is a multiple of
$\pi$, i.e., whenever a light ray crosses the axis of symmetry through
the observer (see Figure~\ref{fig:radtang2}). Symmetry implies that a point source
is seen as a ring (``Einstein ring'') if its worldline crosses a tangential
caustic point. By contrast, a point source whose wordline crosses a radial
caustic point is seen infinitesimally extended in the radial direction.
The set of all tangential caustic points of the past light cone is called
the tangential caustic for short. In general, it has several connected
components. In accordance with the order of images, as defined in
Equation (\ref{eq:k}), these connected components can be labeled
as \emph{primary}, \emph{secondary}, etc.\ tangential caustics. 
Each connected component 
is a spacelike curve in spacetime which projects to (part of) the axis of 
symmetry through the observer. The radial caustic is a lightlike surface in 
spacetime unless at points where it meets the axis; its projection to space is 
rotationally symmetric around the axis.
The best known example for a tangential caustic, with infinitely many
connected components, occurs in the Schwarzschild spacetime (see
Figure~\ref{fig:schwcon}). It is also instructive to visualize
radial and tangential caustics in terms of instantaneous wave fronts, i.e.,
intersections of the light cone with hypersurfaces $t = \mbox{constant}$.
Examples are shown in Figures~\ref{fig:schwfrt}, \ref{fig:monofrt1},
and~\ref{fig:monofrt2}. By symmetry, a tangential caustic point
of an instantaneous wave front can be neither a cusp nor a swallow-tail.
Hence, the general result of Section~\ref{ssec:front} implies that
the tangential caustic is always unstable. The radial caustic in
Figure~\ref{fig:monofrt2} consists of cusps and is, thus, stable.

\epubtkImage{figure08.png}
{\begin{figure}[hptb]
   \def\epsfsize#1#2{1.0#1}
   \centerline{\epsfbox{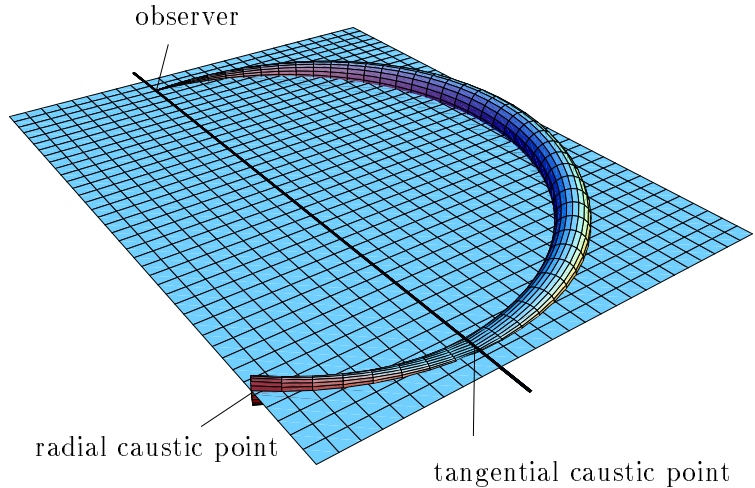}}
   \caption{Tangential and radial caustic points. Tangential
     caustic points, $D_-=0$, occur on the axis of symmetry through
     the observer. A (point) source at a tangential caustic point is
     seen as a (1-dimensional) Einstein ring on the observer's
     sky. A point source at a radial caustic point, $D_+=0$, appears
     ``infinitesimally extended'' in the radial direction.}
   \label{fig:radtang2}
 \end{figure}
}


\subsection{Lensing in axisymmetric stationary spacetimes}
\label{ssec:axistat}

Axisymmetric stationary spacetimes are of interest in view of lensing
as general-relativistic models for rotating deflectors. The best known
and most important example is the Kerr metric which describes a rotating
black hole (see Section~\ref{ssec:kerr}). For non-collapsed rotating objects,
exact solutions of Einstein's field equation are known only for the idealized
cases of infinitely long cylinders (including string models; see
Section~\ref{ssec:str}) and disks (see Section~\ref{ssec:disk}).
Here we collect, as a preparation for these examples, some formulas for an
unspecified axisymmetric stationary metric. The latter can be written
in coordinates $(y^1,y^2,\varphi,t)$, with capital indices $A,B, \dots$ taking
the values 1 and 2, as
\begin{equation}
  \label{eq:asg}
  g = g_{tt} (y) \, dt^2 + 2 g_{t\varphi} (y) \, dt \, d \varphi +
  g_{\varphi \varphi} (y) \, d \varphi^2 + g_{AB} (y) \, dy^A \, dy^B,
\end{equation}
where all metric coefficients depend on $y=(y^1,y^2)$ only. We assume
that the integral curves of $\partial_{\varphi}$ are closed, with the usual
$(2 \pi)$-periodicity, and that the 2-dimensional orbits spanned by
$\partial_{\varphi}$ and $\partial_t$ are timelike. Then the Lorentzian
signature of $g$ implies that $g_{AB} (y)$ is positive definite. In general,
the vector field $\partial_t$ need not be timelike and the hypersurfaces
$t=\mbox{constant}$ need not be spacelike.
Our assumptions allow for transformations $(\varphi, t) \mapsto
(\varphi + \Omega t, t)$ with a constant $\Omega$. If, by such
a transformation, we can achieve that $g_{tt} < 0$ everywhere,
we can use the purely spatial formalism for light rays in terms
of the Fermat geometry (recall Section~\ref{ssec:conf}). Comparison
of Equation~(\ref{eq:asg}) with Equation~(\ref{eq:confg}) shows that the redshift potential $f$,
the Fermat metric $\hat{g}$, and the Fermat one-form $\hat{\phi}$ are
\begin{eqnarray}
  e^{2f} &=& - g_{tt},
  \label{eq:asf} \\
  \hat{g} &=& - \frac{g_{AB}}{g_{tt}} \, dx^A \, dx^B +
  \frac{g_{t \varphi}^2 - g_{tt} \, g_{\varphi \varphi}}{g_{tt}^2} \, d\varphi^2,
  \label{eq:asgF} \\
  \hat{\phi} &=& - \frac{g_{t \varphi}}{g_{tt}} \, d\varphi,
  \label{eq:asphiF}
\end{eqnarray}%
respectively. If it is not possible to make $g_{tt}$ negative on the
entire spacetime domain under consideration, the Fermat geometry is
defined only locally and, therefore, of limited usefulness. This is
the case, e.g., for the Kerr metric where, in Boyer--Lindquist
coordinates, $g_{tt}$ is positive in the ergosphere (see
Section~\ref{ssec:kerr}).

Variational techniques related to Fermat's principal in stationary
spacetimes are detailed in a book by Masiello~\cite{masiello-94}.
Note that, in contrast to standard terminology, Masiello's definition
of stationarity includes the assumption that the surfaces
$t=\mbox{constant}$ are spacelike.

For a rotating body with an equatorial plane (i.e., with reflectional
symmetry), the Fermat metric of the equatorial plane can be represented by
an embedding diagram, in analogy to the spherically symmetric static
case (recall Figure~\ref{fig:schwemb}). However, one should keep in mind that in the
non-static case the lightlike geodesics do \emph{not} correspond to
the geodesics of $\hat{g}$ but are affected, in addition, by a sort of
Coriolis force produced by $\hat{\phi}$. For a review on embedding diagrams,
including several examples (see~\cite{hledik-2001}).

\newpage


\section{Examples}
\label{sec:examples}


\subsection{Schwarzschild spacetime}
\label{ssec:schw}

The (exterior) Schwarzschild metric
\begin{equation}
  \label{eq:schwarzschild}
  g = - \left( 1 - \frac{2m}{r} \right) dt^2 +
  \left( 1 - \frac{2m}{r} \right)^{-1} dr^2 +
  r^2 \left( d \vartheta^2 + \sin^2 \vartheta \, d\varphi^2 \right)
\end{equation}
has the form~(\ref{eq:ssg}) with
\begin{equation}
  \label{eq:schwRS}
  e^{2f(r)} = S(r)^{-1} = 1 - \frac{2m}{r},
  \qquad
  R(r) = \frac{r}{\sqrt{1 - \frac{2m}{r}}}.
\end{equation}
It is the unique spherically symmetric vacuum solution of Einstein's
field equation. At the same time, it is the most important and best
understood spacetime in which lensing can be explicitly studied without
approximations. Schwarzschild lensing beyond the weak-field approximation
has astrophysical relevance in view of black holes and neutron stars.
The increasing evidence that there is a supermassive black hole at
the center of our Galaxy (see~\cite{falcke-hehl-2003} for background
material) is a major motivation for a detailed study of Schwarzschild
lensing (and of Kerr lensing; see Section~\ref{ssec:kerr}).
In the following we consider the Schwarzschild metric with a constant
$m>0$ and we ignore the region $r<0$ for which the singularity at $r=0$
is naked. The Schwarzschild metric is static on the region $2m<r<\infty$.
(The region $r<0$ for $m>0$ is equivalent to the region $r>0$ for $m<0$.
It is usually considered as unphysical but has found some recent interest
in connection with lensing by wormholes; see Section~\ref{ssec:worm}.) \\

\noindent
{\bf Historical notes.} \\
\noindent
Shortly after the discovery of the Schwarzschild metric
by Schwarzschild~\cite{schwarzschild-16} and independently by
Droste~\cite{droste-16}, basic features of its lightlike geodesics
were found by Flamm~\cite{flamm-16}, Hilbert~\cite{hilbert-17}, and
Weyl~\cite{weyl-17}. Detailed studies of its timelike and lightlike
geodesics were made by Hagihara~\cite{hagihara-31} and
Darwin~\cite{darwin-59, darwin-61}. 
For a fairly complete list of the
pre-1979 literature on Schwarzschild geodesics see Sharp~\cite{sharp-79}.
All modern text-books on general relativity include a section on
Schwarzschild geodesics, but not all of them go beyond the
weak-field approximation. For a particularly detailed exposition see
Chandrasekhar~\cite{chandrasekhar-83}. \\

\noindent
{\bf Redshift and Fermat geometry.} \\
\noindent
The redshift potential $f$ for the Schwarz\-schild metric is given
in Equation~(\ref{eq:schwRS}). With the help of $f$ we can directly calculate
the redshift via Equation~(\ref{eq:zpot}) if observer and light source
are static (i.e., $t$-lines). If the light source or the observer
does not follow a $t$-line, a Doppler factor has to be added. Independent
of the velocity of observer and light source, the redshift becomes
arbitrarily large if the light source is sufficiently close to the
horizon. For light source and observer freely falling, the redshift
formula was discussed by Ba{\.z}a{\'n}ski and
Jaranowski~\cite{bazanski-jaranowski-89}. If projected to 3-space, the
light rays
in the Schwarzschild spacetime are the geodesics of the Fermat metric which
can be read from Equation~(\ref{eq:ssgF}) (cf.\ Frankel~\cite{frankel-79}), 
\begin{equation}
  \label{eq:schwgF}
  \hat{g} = \frac{dr^2}{(1-\frac{2m}{r})^2} +
  \frac{r^2( d \vartheta^2 + \sin \vartheta \, d \varphi^2)}{1-\frac{2m}{r}}.
\end{equation}
The metric coefficient $R(r)$, as given by
Equation~(\ref{eq:schwRS}), has a strict minimum at $r=3m$ and no
other extrema (see Figure~\ref{fig:schwpot}). Hence, there is an unstable light sphere at
this radius (recall Equation~(\ref{eq:rL})). The existence of circular light rays
at $r=3m$ was noted already by Hilbert~\cite{hilbert-17}. The relevance of
these circular light rays in view of lensing was clearly seen by
Darwin~\cite{darwin-59,darwin-61} 
and Atkinson~\cite{atkinson-65}. They
realized, in particular, that a Schwarzschild black hole produces infinitely
many images of each light source, corresponding to an infinite sequence of
light rays whose limit curve asymptotically spirals towards a circular light ray. 
The
circular light rays at $r=3m$ are also associated with other physical effects
such as centrifugal force reversal and ``locking'' of gyroscopes. These
effects have been discussed with the help of the Fermat geometry (=\,optical
reference geometry) in various articles by Abramowicz and
collaborators (see, e.g., \cite{abramowicz-lasota-74,
  abramowicz-carter-lasota-88, abramowicz-prasanna-90,
  abramowicz-92}). \\

\epubtkImage{figure09.png}
{\begin{figure}[hptb]
   \def\epsfsize#1#2{0.7#1}
   \centerline{\epsfbox{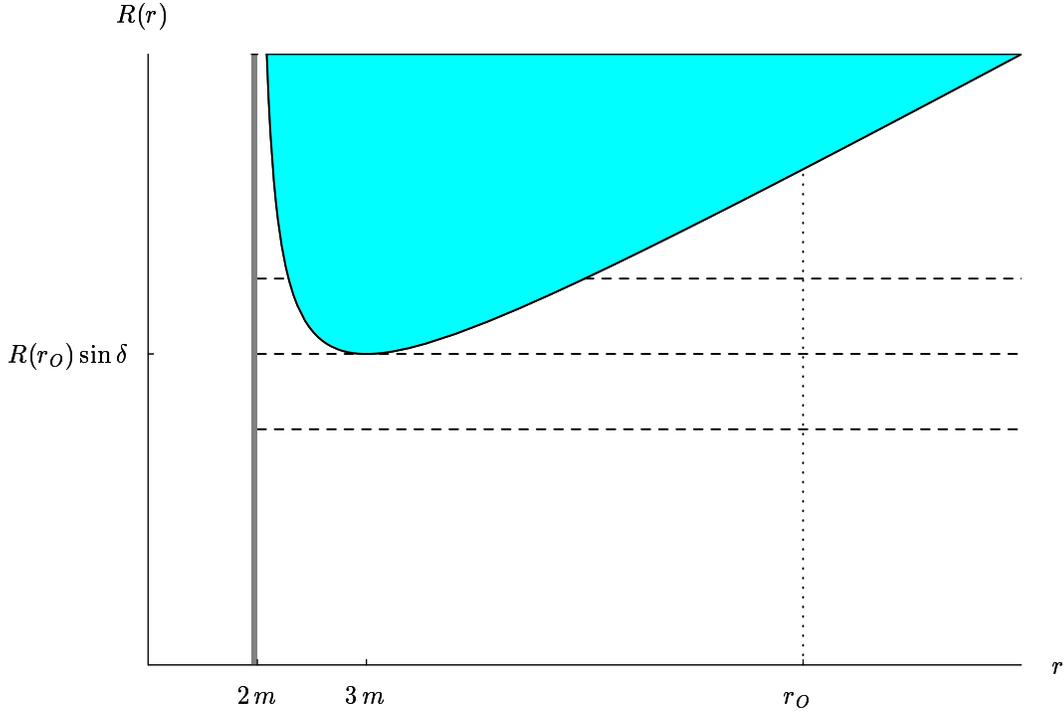}}
   \caption{The function $R(r)$ for the Schwarzschild
     metric. Light rays that start at $r_\mathrm{O}$ with initial
     direction $\Theta$ are confined to the region where
     $R(r) \ge R(r_\mathrm{O}) \, \sin \Theta$. The equation
     $R(3m) = R(r_\mathrm{O}) \, \sin \delta$ defines for each
     $r_\mathrm{O}$ a critical value $\delta$. A light ray from
     $r_\mathrm{O}$ with $\Theta =\delta$ asymptotically approaches
     $r=3m$.}
   \label{fig:schwpot}
 \end{figure}
}

\noindent
{\bf Index of refraction and embedding diagrams.} \\
\noindent
We know from Section~\ref{ssec:ss} that light rays in any spherically
symmetric and static spacetime can be characterized by an index of
refraction. This requires introducing an isotropic radius coordinate
$\tilde{r}$ via Equation~(\ref{eq:ssiso}). In the Schwarzschild case, $\tilde{r}$
is related to the Schwarzschild radius coordinate $r$ by
\begin{equation}
  \label{eq:schwiso}
  {\tilde{r}} = \frac{1}{2} \left( \sqrt{r^2-2mr} + r - m \right),
  \qquad
  r = \frac{\left( 2 \tilde{r} + m \right)^2 }{4 \tilde{r}}.
\end{equation}
$\tilde{r}$ ranges from $m/2$ to infinity if $r$ ranges from $2m$
to infinity. In terms of the isotropic coordinate, the Fermat
metric~(\ref{eq:schwgF}) takes the form
\begin{equation}
  \label{eq:ssgFiso2}
  \hat{g} = n({\tilde{r}})^2
  \left( d{\tilde{r}}^2 + {\tilde{r}}^2 \left( d\vartheta^2 +
  \sin^2 \vartheta \, d\varphi^2 \right) \right)
\end{equation}
with
\begin{equation}
  \label{eq:schwn}
  n (\tilde{r}) = \left( 1 + \frac{m}{2 \tilde{r}} \right)^3
  \left( 1 - \frac{m}{2 \tilde{r}} \right)^{-1}.
\end{equation}
Hence, light propagation in the Schwarzschild metric can be mimicked
by the index of refraction~(\ref{eq:schwn}); see Figure~\ref{fig:schwn}.
The index of refraction~(\ref{eq:schwn}) is known since Weyl~\cite{weyl-23}.
It was employed for calculating lightlike Schwarzschild geodesics, exactly
or approximately, e.g., in~\cite{atkinson-65, nandi-islam-95,
evans-nandi-islam-96b, lerner-97}. This index of refraction can be
modeled by a fluid flow~\cite{rosquist-2003}.
The embeddability condition~(\ref{eq:ssemb}) is satisfied for $r > 2.25 m$
(which coincides with the so-called \emph{Buchdahl limit}). On this domain the
Fermat geometry, if restricted to the equatorial plane $\vartheta = \pi /2$,
can be represented as a surface of revolution in Euclidean 3-space (see
Figure~\ref{fig:schwemb}). The entire region $r>2m$ can be 
isometrically embedded into a space of constant negative
curvature~\cite{abramowicz-bengtsson-karas-rosquist-2002}. 
\\

\epubtkImage{figure10.png}
{\begin{figure}[hptb]
   \def\epsfsize#1#2{0.5#1}
   \centerline{\epsfbox{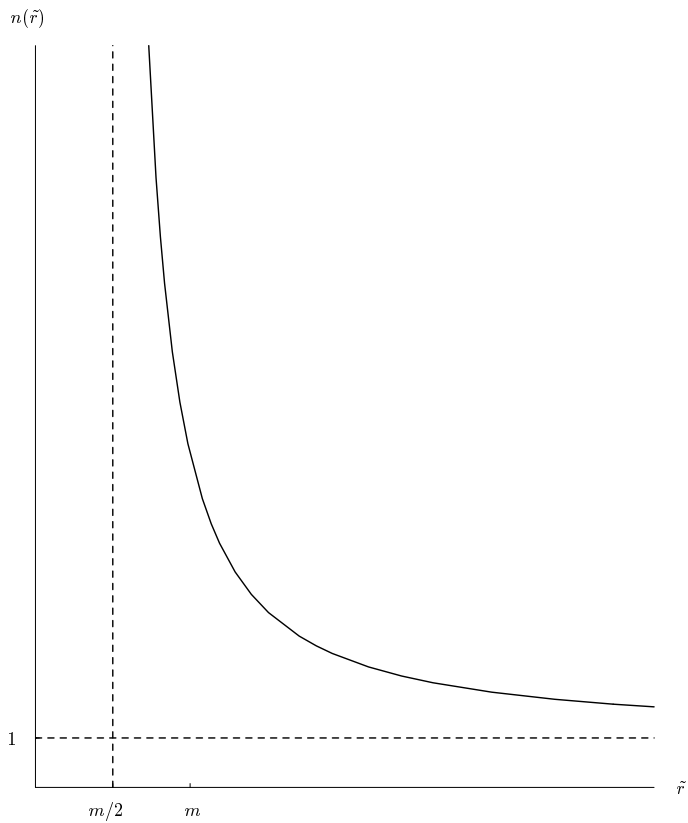}}
   \caption{Index of refraction $n (\tilde{r})$, given by
     Equation~(\ref{eq:schwn}), for the Schwarzschild metric as a
     function of the isotropic coordinate $\tilde{r}$.}
   \label{fig:schwn}
 \end{figure}
}

\epubtkImage{figure11.png}
{\begin{figure}[hptb]
   \def\epsfsize#1#2{1.0#1}
   \centerline{\epsfbox{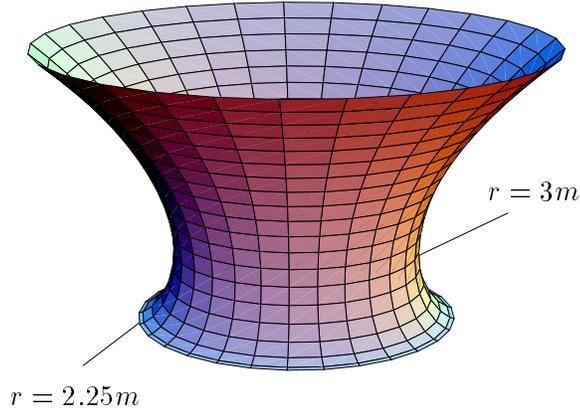}}
   \caption{Fermat geometry of the equatorial plane of the
     Schwarzschild spacetime, embedded as a surface of revolution into
     Euclidean 3-space. The neck is at $r = 3m$ (i.e., $\tilde{r}
     \approx 1.87 m$), the boundary of the embeddable part at
     $r=2.25 m$ (i.e., $\tilde{r}=m$). The geodesics of this surface
     of revolution give the light rays in the Schwarzschild spacetime. A
     similar figure can be found
     in~\cite{abramowicz-carter-lasota-88} (also cf.~\cite{hledik-2001}).}
   \label{fig:schwemb}
 \end{figure}
}

\noindent
{\bf Lensing by a Schwarzschild black hole.} \\
\noindent
To get a Schwarzschild black hole, one joins at $r=2m$ the static Schwarzschild
region $2m<r<\infty$ to the non-static Schwarzschild region $0<r<2m$ in
such a way that \emph{ingoing} light rays can cross this surface but
\emph{outgoing} cannot. If the observation event $p_\mathrm{O}$ is at $r_\mathrm{O}>2m$,
only the region $r>2m$ is of relevance for lensing, because the past light
cone of such an event does not intersect the black-hole horizon at $r=2m$.
(For a Schwarzschild white hole see below.) Such a light cone is depicted in
Figure~\ref{fig:schwcon} (cf.~\cite{kling-newman-99}).
The picture was produced with the help of the representation~(\ref{eq:sscone})
which requires integrating Equation~(\ref{eq:rell}) and Equation~(\ref{eq:rphi}). For
the Schwarzschild case, these are elliptical integrals. Their numerical
evaluation is an exercise for students (see~\cite{bruckman-esteban-93} for
a MATHEMATICA program). Note that the evaluation of Equation~(\ref{eq:rell}) and
Equation~(\ref{eq:rphi}) requires knowledge of the turning points. In the Schwarzschild
case, there is at most one turning point $r_m (\Theta)$ along each ray
(see Figure~\ref{fig:schwpot}), and it is given by the cubic equation
\begin{equation}
  \label{eq:schwrmin}
  r_m (\Theta)^3 \, (r_\mathrm{O}-2m) -
  r_m (\Theta) \, r_\mathrm{O}^3 \, \sin^2 \Theta +
  2 m r_\mathrm{O}^3 \, \sin^2 \Theta = 0.
\end{equation}
The representation~(\ref{eq:sscone}) in terms of Fermat arclength $\ell$
(= travel time) gives us the intersections of the light cone
with hypersurfaces $t= \mbox{constant}$. These ``instantaneous
wave fronts'' are depicted in Figure~\ref{fig:schwfrt} (cf.~\cite{hanni-77}). With
the light cone explicitly known, one can analytically verify that
every inextendible timelike curve in the region $r>2m$ intersects the
light cone infinitely many times, provided it is bounded away from the
horizon and from (past lightlike) infinity. This shows that the observer
sees infinitely many images of a light source with this worldline.
The same result can be proven with the help of Morse theory (see
Section~\ref{ssec:morse}), where one has to exclude the case that the
worldline meets the caustic of the light cone. In the latter case the 
light source is seen as an Einstein ring. Note that a moving source 
might appear simultaneously as a point image and as an Einstein ring 
on the observer's sky. For static light sources (i.e., $t$-lines), however,
either all images are Einstein rings or none. 
For such light sources
we can study lensing in the exact-lens-map formulation of
Section~\ref{ssec:ss} (see in particular Figure~\ref{fig:sslens}). Also,
Section~\ref{ssec:ss} provides us with formulas for distance measures,
brightness, and image distortion which we just have to specialize to
the Schwarzschild case. For another treatment of Schwarzschild lensing
with the help of the exact lens map, see~\cite{frittelli-kling-newman-2000}.
We place our static light sources at radius $r_\mathrm{S}$. If $r_\mathrm{O}<r_\mathrm{S}$
and $3m<r_\mathrm{S}$, only light rays with $\Theta < \delta$,
\begin{equation}
  \label{eq:schwesc}
  \sin \delta := \frac{R(3m)}{R(r_\mathrm{O})} =
  \sqrt{\frac{27 m^2 (r_\mathrm{O}-2m)}{r_\mathrm{O}^3}},
\end{equation}
can reach the radius value $r_\mathrm{S}$ (see Figure~\ref{fig:schwpot}). Rays with
$\Theta = \delta$ asymptotically spiral towards the light sphere at $r=3m$.
$\delta$ lies between 0 and $\pi /2$ for $r_\mathrm{O} < 3m$ and between $\pi /2$ and
$\pi$ for $r_\mathrm{O} > 3 m$. The \emph{escape cone} defined by Equation~(\ref{eq:schwesc})
is depicted, for different values of $r_\mathrm{O}$, in Figure~\ref{fig:schwesc}.
It gives the domain of definition for the lens map.
The lens map is graphically discussed in Figure~\ref{fig:schwlen}. The
pictures are valid for $r_\mathrm{O} = 5m$ and $r_\mathrm{S} =10m$. Qualitatively, however,
they look the same for all cases with $r_\mathrm{S}>r_\mathrm{O}$ and $r_\mathrm{S}>3m$. From the
diagram one can read the position of the infinitely many images for each
light source which, for the two light sources on the axis, degenerate into
infinitely many Einstein rings. For each fixed source, the images are
ordered by the number $i$ ($=0,1,2,3,\dots$) which counts how often
the ray has crossed the axis. This coincides with ordering according to travel
time. With increasing order $i$, the images come closer and closer to the rim at
$\Theta = \delta$ (see Figure~\ref{fig:schwlen}) and their brightness decreases
rapidly (see Figure~\ref{fig:schwDe}). For a light source not on the axis, images of even 
order are upright and line up on one side of the direction towards the center, images of odd 
order are side-inverted (see Figure~\ref{fig:schwDT}) and line up on the other side of the 
direction towards the center (see Figure~\ref{fig:schwlen}).
These basic features of Schwarzschild lensing are known since pioneering
papers by Darwin~\cite{darwin-59} 
and Atkinson~\cite{atkinson-65}
(cf.~\cite{luminet-79, ohanian-87, lano-89}).
Various methods of how multiple imaging by
a black hole could be discovered, directly or indirectly, have been
discussed~\cite{luminet-79, lano-89, bao-hadrava-ostgaard-94a,
  bao-hadrava-ostgaard-94b,
petters-2002, paolis-geralico-ingrosso-nucita-2003}. Related work has also
been done for Kerr black holes (see Section~\ref{ssec:kerr}). An interesting
suggestion was made in~\cite{holz-wheeler-2002}. A Schwarzschild black hole,
somewhere in the universe, would send photons originating from our Sun back
to the vicinity of our Sun (``boomerang photons''~\cite{stuckey-93}). If the
black hole is sufficiently close to our Solar system, this would produce
images of our own Sun on the sky that could be detectable.
Quite generally,
one speaks of \emph{retrolensing} when a gravitating mass sends light 
back into approximately the same direction from which it has come in.
Retrolensed images have not been observed so far, the perspectives are 
discussed, e.g., in~\cite{paolis-geralico-ingrosso-nucita-2003, 
eiroa-torres-2004}.
-- The lensing effect of a
Schwarzschild black hole has been visualized in two ways:
\begin{enumerate}
\item by showing the visual appearance of some background pattern as
  distorted by the black hole~\cite{cunningham-75,
    schastok-soffel-ruder-schneider-87, nemiroff-93, 
    bakala-cermak-hledik-stuchlik-truparova-2007, mueller-weiskopf-2010},
  also cf.~\cite{mueller-2008, mueller-2009}.
\item by showing the visual appearence of an accretion disk around the
  black hole~\cite{luminet-79, fukue-yokoyama-88,
    bao-hadrava-ostgaard-94a, bao-hadrava-ostgaard-94b}, 
  also cf.~\cite{broderick-blandford-2003, broderick-blandford-2004, 
  cadez-kostic-2005}.
\end{enumerate}
In the course of time the ray tracing programs on which these 
visualizations are based have become more and more advanced,
taking not only redshift and magnification (including
higher-order images) but also Fraunhofer diffraction (due to the
finite aperture of the observer's eye) or scattering into account. 
Ray tracing programs have also been developed for 
the more general case of the Kerr metric, see Section~\ref{ssec:kerr}. -- 
Interest in Schwarzschild lensing
(and Kerr lensing) beyond the weak-field approximation has
greatly increased with the growing evidence that there is a 
supermassive black hole at the center of our galaxy, and 
probably at the center of most galaxies. Higher-order images,
where a light ray makes at least one full turn around the 
center, have not been observed so far, but they are thought to 
be relevant for future observations. It was already emphasized
that, even if the bending angles are arbitrarily large, 
all lensing properties of a Schwarzschild black hole can
be calculated exactly, in terms of elliptic integrals; 
then these integrals can be evaluated numerically with
arbitrary accuracy and the results can be discussed 
graphically, as exemplified in  Figures~\ref{fig:schwlen},
\ref{fig:schwDT} and~\ref{fig:schwDe}. However, for
practical purposes many authors found it convenient
to develop approximation methods that go beyond,
or are complementary to, the weak-field approximation,
rather than to work with the exact formulas. 
Two approximation methods have proven particularly useful:
Virbhadra and Ellis~\cite{virbhadra-ellis-2000} developed 
a lens equation that applies to the case that source and
observer are in the asymptotic region and approximately
aligned with the center, but is not restricted to light rays
that remain in the asymptotic region. Bozza et 
al.~\cite{bozza-capozziello-iovane-scarpetta-2001,
bozza-2002}
introduced a strong-field limit (or strong-deflection
limit) that describes light rays the better the more turns
they make around the center. Both methods, along with
other approximation techniques~\cite{amore-arceo-2006, 
amore-arceo-fernandez-2006, amore-cervantes-pace-fernandez-2007, 
keeton-petters-2005, keeton-petters-2006a, keeton-petters-2006b}
for light bending in spherically-symmetric and static spacetimes, 
have already been discussed in Section~\ref{ssec:ss} above. 
Especially for the Schwarzschild metric, Iyer and 
Petters~\cite{iyer-petters-2007} have demonstrated that, by combining 
a strong-deflection series expansion and a weak-deflection series 
expansion, one gets an approximation that is within $1 \,$\% of 
the exact bending angle value for light rays traversing 
anywhere between the photon sphere and infinity. A main goal
of all these endeavours is to provide a new test of general
relativity with the help of higher-order images, once they have
been observed. As shown by Bozza~\cite{bozza-2002}, 
the separation of higher-order images and their decrease
in magnitude can be used for discriminating between different
black holes. Hence, the observation of higher-order images
would reveal if the bending object can be modeled 
as a Schwarzschild black hole, or if an alternative model 
has to be used. An example for such an alternative model
is the Reissner-Nordstr{\"o}m black hole (see 
Section~\ref{ssec:reissner}). Other spherically-symmetric 
and static black hole models have found some interest 
because their existence is predicted by alternative theories
of gravity. E.g., the bending properties have been worked out 
for black holes from string theory~\cite{bhadra-2003}, from
braneworld gravity~\cite{kar-sinha-2003, whisker-2005, eiroa-2005a, 
eiroa-2005b, binnun-2010}, from Einstein-Born-Infeld theory~\cite{eiroa-2006},
from dilaton theory~\cite{mukherjee-majumdar-2007, ghosh-sengupta-2010} 
and from Ho{\v r}ava-Lifshitz gravity~\cite{chen-jing-2009}.
Up to now there is no observational indication that any
of these black holes exist in nature. The future observation of
higher-order images could help to find out if they exist. For 
the time being, all observations are in agreement with the 
assumption that the existing black holes are Schwarzschild
or Kerr black holes, as predicted by standard general relativity. 
Schwarzschild lensing as a tool for probing the supermassive 
objects at the center of galaxies is
discussed in detail by Virbhadra~\cite{virbhadra-2009}. For a 
recent review on black hole lensing in general see 
Bozza~\cite{bozza-2010}.\\

\epubtkImage{figure12.png}
{\begin{figure}[hptb]
   \def\epsfsize#1#2{1.1#1}
   \centerline{\epsfbox{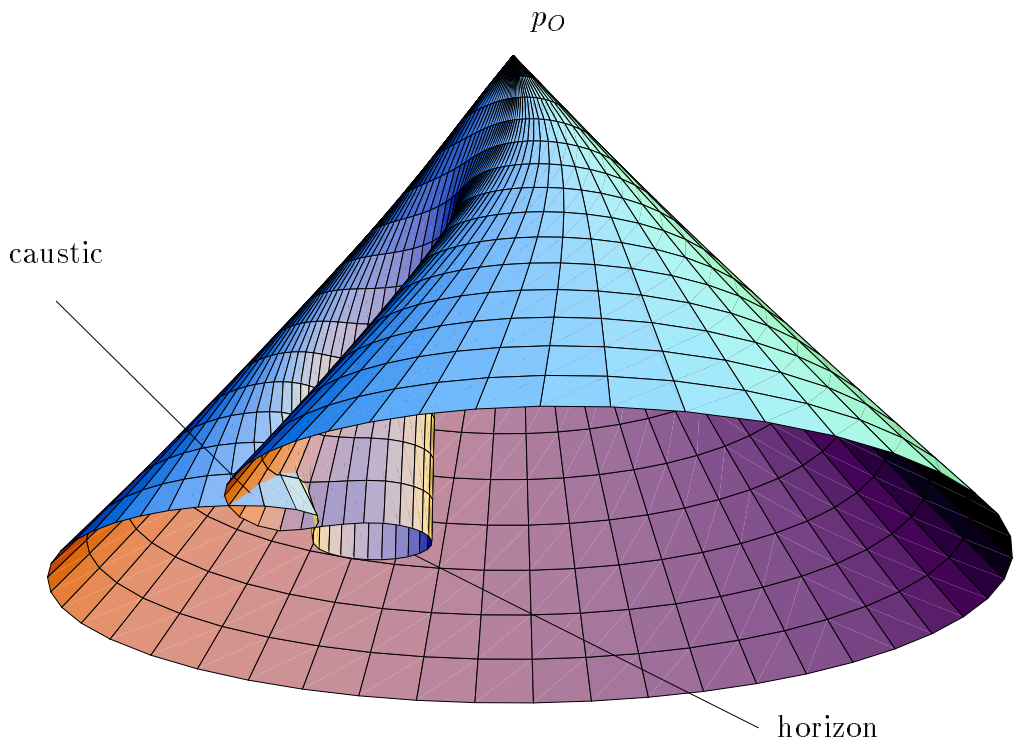}}
   \caption{Past light cone in the Schwarzschild spacetime. One
     sees that the light cone wraps around the horizon, then forms a
     tangential caustic. In the picture the caustic looks like a
     transverse self-intersection because one spatial dimension is
     suppressed. (Only the hyperplane $\vartheta = \pi /2$ is shown.)
     There is no radial caustic. If one follows the light rays further
     back in time, the light cone wraps around the horizon again and
     again, thereby forming infinitely many tangential caustics which
     alternately cover the radius line through the observer and the
     radius line opposite to the observer. In spacetime, each caustic
     is a spacelike curve along which $r$ ranges from $2m$ to $\infty$,
     whereas $t$ ranges from $-\infty$ to some maximal value and then
     back to $-\infty$. Equal-time sections of this light cone are
     shown in Figure~\ref{fig:schwfrt}.}
   \label{fig:schwcon}
 \end{figure}
}

\epubtkImage{figure13.png}
{\begin{figure}[hptb]
   \def\epsfsize#1#2{1.0#1}
   \centerline{\epsfbox{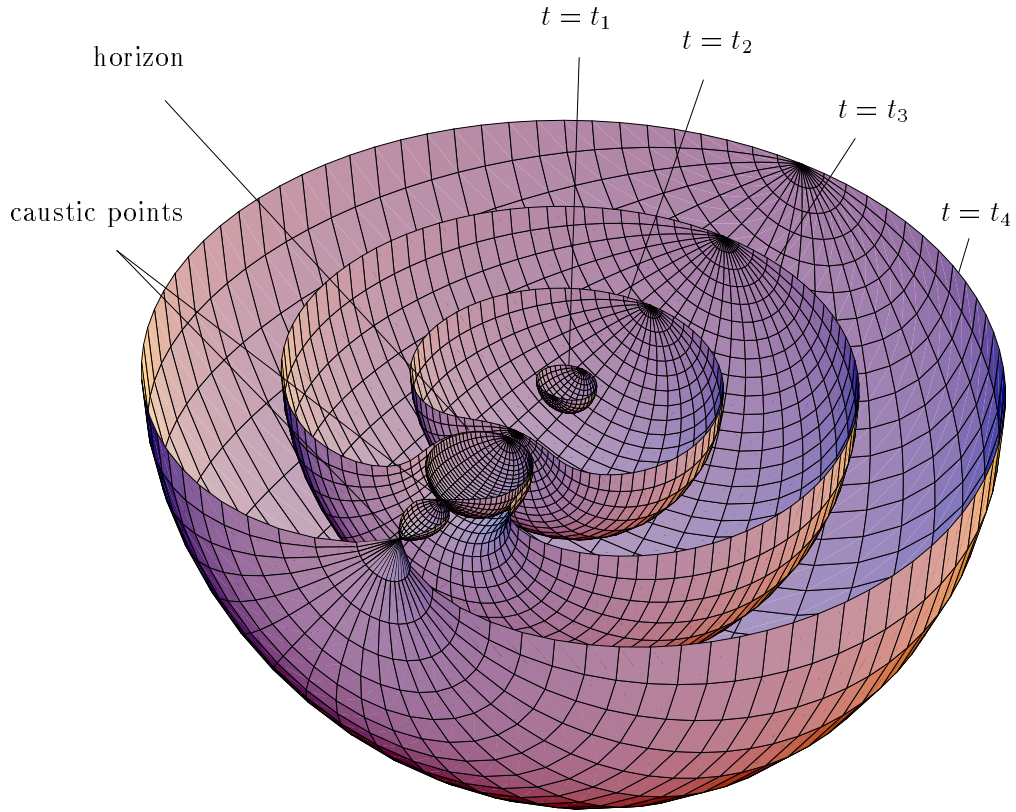}}
   \caption{Instantaneous wave fronts of the light cone in the
     Schwarzschild spacetime. This picture shows intersections of the
     light cone in Figure~\ref{fig:schwcon} with hypersurfaces
     $t= \mbox{constant}$ for four $t$-values, with $t_1>t_2>t_3>t_4$. The
     instantaneous wave fronts wrap around the horizon and, after
     reaching the first caustic, have two caustic points each. If one
     goes further back in time than shown in the picture, the wave
     fronts another time wrap around the horizon, reach the second
     caustic, and now have four caustic points each, and so on. In
     comparison to Figure~\ref{fig:schwcon}, the representation in
     terms of instantaneous wave fronts has the advantage that all
     three spatial dimensions are shown.}
   \label{fig:schwfrt}
 \end{figure}
}

\epubtkImage{figure14.png}
{\begin{figure}[hptb]
   \def\epsfsize#1#2{1.05#1}
   \centerline{\epsfbox{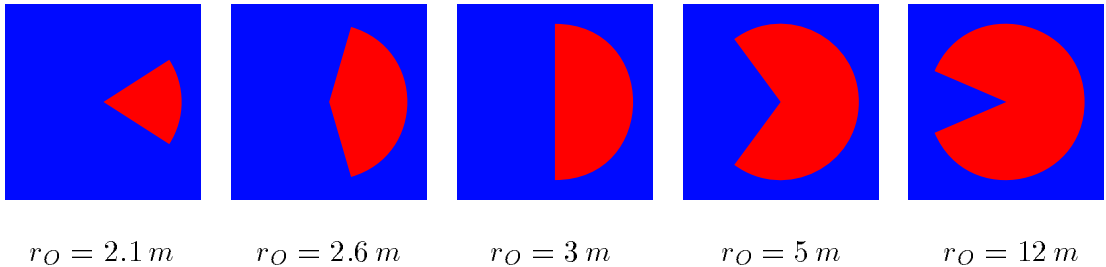}}
   \caption{Escape cones in the Schwarzschild metric, for five
     values of $r_\mathrm{O}$. For an observer at radius
     $r_\mathrm{O}$, light sources distributed at a radius
     $r_\mathrm{S}$ with $r_\mathrm{S} > r_\mathrm{O}$ and
     $r_\mathrm{S} > 3m$ illuminate a disk whose angular radius
     $\delta$ is given by Equation~(\ref{eq:schwesc}). The boundary of
     this disk corresponds to light rays that spiral towards the light
     sphere at $r=3m$. If $r_\mathrm{O}$ is big, the bright disk covers
     almost the whole sky, leaving a small dark disk that is called the
     ''shadow" of the black hole. With decreasing $r_\mathrm{O}$, the 
     shadow becomes bigger and bigger until, for $r_\mathrm{O} \to 2m$,
     it covers the whole sky. Figure~\ref{fig:schwpot} illustrates that
       the notion of escape cones is meaningful for any spherically
     symmetric and static spacetime where $R$ has one minimum and no
     other extrema~\cite{pande-durgapal-86}. For the Schwarzschild
     spacetime, the escape cones were first mentioned
     in~\cite{oppenheimer-snyder-39, metzner-63}, and explicitly
     calculated in~\cite{synge-66}. A picture similar to this one can
     be found, e.g., in~\cite{chandrasekhar-83}, p.130.}
   \label{fig:schwesc}
 \end{figure}
}

\epubtkImage{figure15.png}
{\begin{figure}[hptb]
   \def\epsfsize#1#2{0.52#1}
   \centerline{\epsfbox{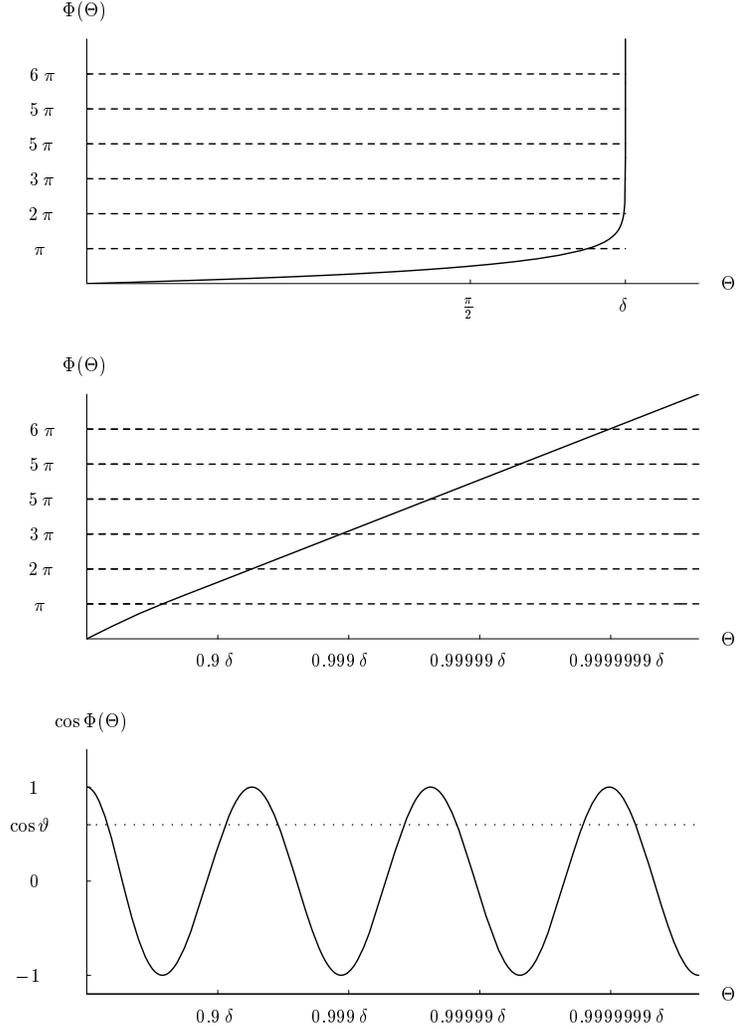}}
   \caption{Lens map for the Schwarzschild metric. The observer is
     at $r_\mathrm{O}=5m$, the light sources are at
     $r_\mathrm{S}=10m$. $\Theta$ is the colatitude on the observer's
     sky and $\Phi (\Theta)$ is the angle swept out by the ray (see
     Figure~\ref{fig:sslens}). $\Phi (\Theta)$ was calculated with the
     help of Equation~(\ref{eq:ssPhi}). $\Theta$ is restricted by the
     opening angle $\delta$ of the observer's escape cone (see
     Figure~\ref{fig:schwesc}). Rays with $\Theta = \delta$
     asymptotically spiral towards the light sphere at $r=3m$. The
     first diagram (cf.~\cite{frittelli-kling-newman-2000}, Figure 5)
     shows that $\Phi (\Theta)$ ranges from 0 to $\infty$ if $\Theta$
     ranges from 0 to $\delta$. So there are infinitely many Einstein
     rings (dashed lines) whose angular radius approaches
     $\delta$. One can analytically prove~\cite{luminet-79,
       ohanian-87,   bozza-capozziello-iovane-scarpetta-2001} that the
     divergence of $\Phi (\Theta)$ for $\Theta \to \delta$ is
     logarithmic. This is true whenever light rays approach an
     unstable light sphere~\cite{bozza-2002}. The second diagram shows
     $\Phi (\Theta)$ over a logarithmic $\Theta$-axis. The graph of
     $\Phi$ approaches a straight line which was called the
     ``strong-field limit'' by Bozza et
     al.~\cite{bozza-capozziello-iovane-scarpetta-2001,
       bozza-2002}. The picture illustrates that it is a good
     approximation for all light rays that make at least one full
     turn. The third diagram shows $\cos \Phi (\Theta)$ over a
     logarithmic $\Theta$-axis. For every source position
     $0 < \vartheta < \pi$ one can read the position of the images
     (dotted line). There are infinitely many, numbered by their
     \emph{order}~(\ref{eq:k}) that counts how often the light ray has
     crossed the axis. Images of odd order are on one side of the
     black hole, images of even order on the other. For the sources at
     $\vartheta = \pi$ and $\vartheta = 0$ one can read the positions
     of the Einstein rings.}
   \label{fig:schwlen}
 \end{figure}
}

\epubtkImage{figure16.png}
{\begin{figure}[hptb]
   \def\epsfsize#1#2{0.52#1}
   \centerline{\epsfbox{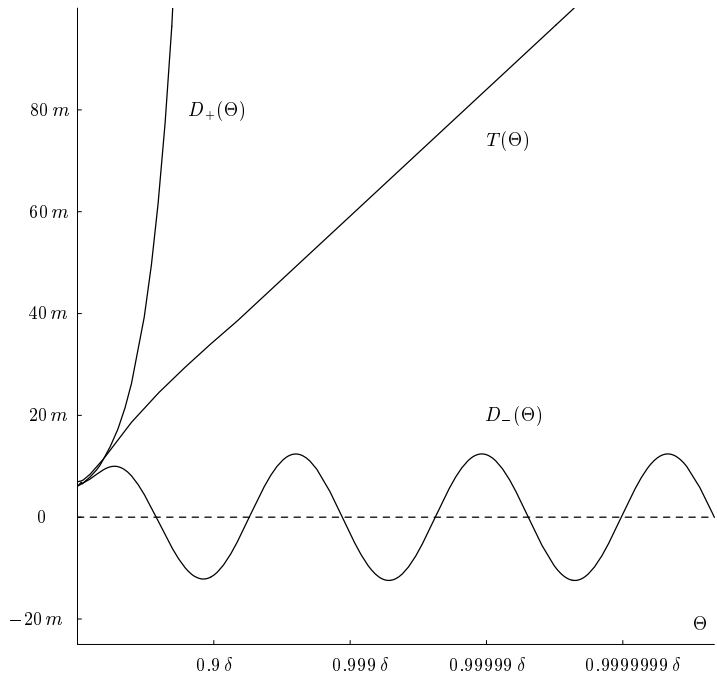}}
   \caption{Radial angular diameter distance $D_+ (\Theta)$,
     tangential angular diameter distance $D_- (\Theta)$ and travel
     time $T (\Theta)$ in the Schwarschild spacetime. The data are the
     same as in Figure~\ref{fig:schwlen}. For the definition of $D_+$
     and $D_-$ see Figure~\ref{fig:radtang1}. $D_{\pm} (\Theta)$ can
     be calculated from $\Phi (\Theta)$ with the help of
     Equation~(\ref{eq:ssPhiD+}) and Equation~(\ref{eq:ssPhiD-}). For
     the Schwarzschild case, the resulting formulas are due
     to~\cite{dwivedi-kantowski-72} (cf.~\cite{dyer-77,
       frittelli-kling-newman-2000}). Zeros of $D_-$ indicate Einstein
     rings. If $D_+$ and $D_-$ have different signs, the observer sees
     a side-inverted image. The travel time $T( \Theta)$ (= Fermat
     arclength) can be calculated from Equation~(\ref{eq:rell}). One
     sees that, over the logarithmic $\Theta$-axis used here, the
     graph of $T$ approaches a straight line. This illustrates that
     $T(\Theta)$ diverges logarithmically if $\Theta$ approaches its
     limiting value $\delta$. This can be verified analytically and is
     characteristic of all cases where light rays approach an unstable
     light sphere~\cite{bozza-mancini-2004}.}
   \label{fig:schwDT}
 \end{figure}
}

\epubtkImage{figure17.png}
{\begin{figure}[hptb]
   \def\epsfsize#1#2{0.52#1}
   \centerline{\epsfbox{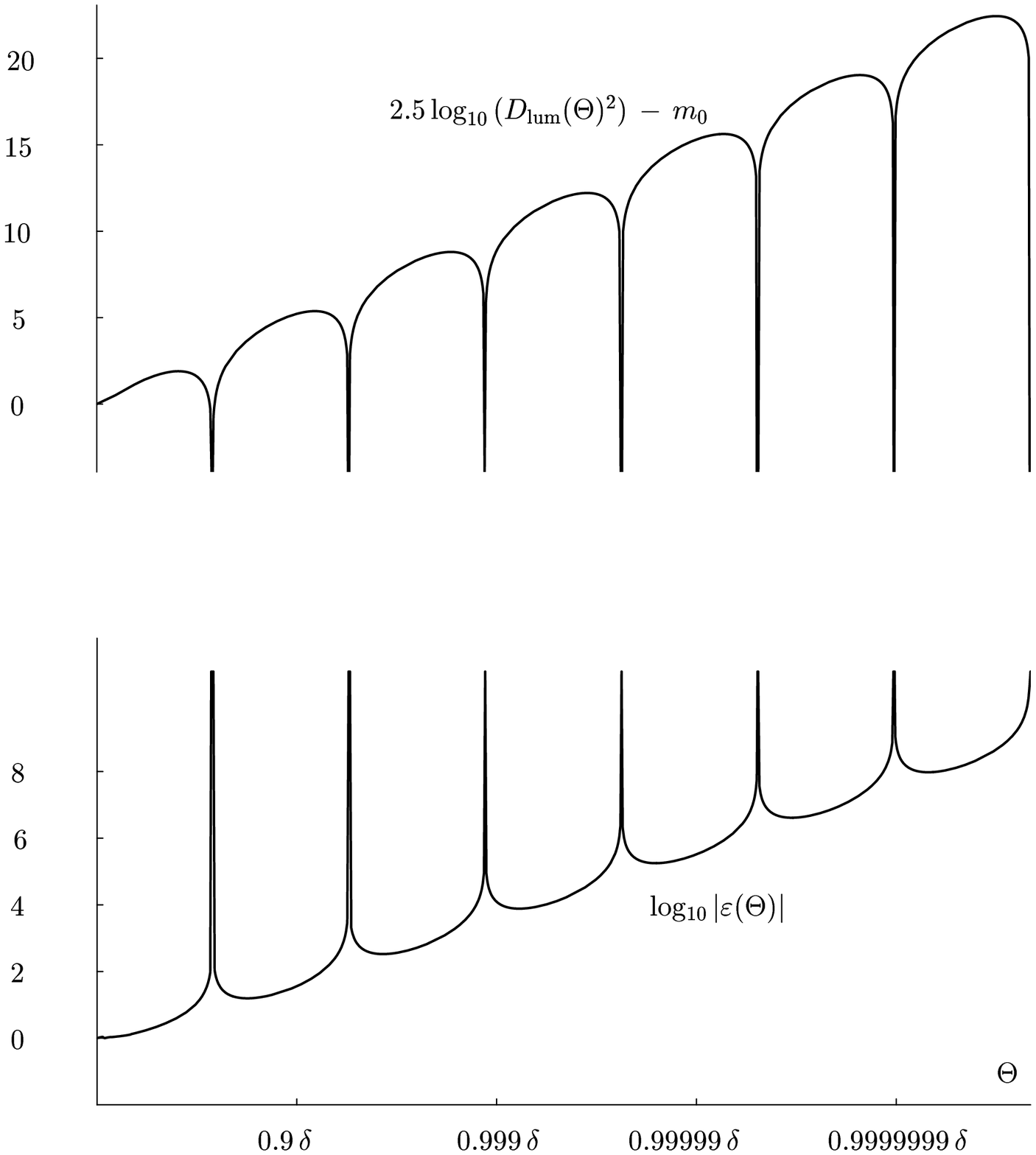}}
   \caption{Luminosity distance $D_{\mathrm{lum}} (\Theta)$ and
     ellipticity $\varepsilon (\Theta)$ (image distortion) in the
     Schwarzschild spacetime. The data are the same as in
     Figures~\ref{fig:schwlen} and~\ref{fig:schwDT}. If point sources
     of equal bolometric luminosity are distributed at
     $r=r_\mathrm{S}$, the plotted function
     $2.5 \log_{10} \left( D_{\mathrm{lum}} (\Theta)^2 \right)$
     gives their magnitude on the observer's sky, modulo an additive
     constant $m_0$. For the calculation of $D_{\mathrm{lum}}$ one
     needs $D_+$ and $D_-$ (see Figure~\ref{fig:schwDT}), and the
     general relations~(\ref{eq:Darea}) and~(\ref{eq:Dlum}). This
     procedure follows~\cite{dwivedi-kantowski-72} (cf.~\cite{dyer-77,
       frittelli-kling-newman-2000}). For source and observer at large
     radius, related calculations can also be found
     in~\cite{luminet-79, ohanian-87, lano-89,
       virbhadra-ellis-2000}. Einstein rings have magnitude $- \infty$
     in the ray-optical treatment. For a light source not on the axis,
     the image of order $i+2$ is fainter than the image of order $i$
     by $2.5 \log_{10}(e^{2 \pi}) \approx 6.8$ magnitudes,
     see~\cite{luminet-79, ohanian-87}. (This is strictly true in the
     ``strong-field limit'', or ``strong-deflection limit'', which is
     explained in the caption of Figure~\ref{fig:schwlen}.) The above
     picture is similar to Figure~6 in~\cite{ohanian-87}. Note that it
     refers to point sources and \emph{not} to a radiating spherical
     surface $r=r_\mathrm{S}$ of constant surface brightness; by
     Equation~(\ref{eq:intensity}), the latter would show a constant
     intensity. The lower part of the diagram illustrates image
     distortion in terms of $ \varepsilon = \frac{D_-}{D_+}-
     \frac{D_+}{D_-} $. Clearly, $| \varepsilon |$ is infinite at each
     Einstein ring. The double-logarithmic representation shows that
     beyond the second Einstein ring all images are extremely
     elongated in the tangential direction,
     $| \varepsilon | > 100$. Image distortion in the Schwarzschild
     spacetime is also treated in~\cite{dyer-77,
       frittelli-kling-newman-2001a, frittelli-kling-newman-2001b}, an
     approximation formula is derived in~\cite{noonan-83}.}
   \label{fig:schwDe}
 \end{figure}
}

\noindent
{\bf Lensing by a non-transparent Schwarzschild star.} \\
\noindent
To model a non-trans\-parent star of radius $r_*$ one has to restrict the exterior
Schwarzschild metric to the region $r>r_*$. Lightlike geodesics terminate when
they arrive at $r=r_*$. The star's radius cannot be smaller
than $2m$ unless it is allowed to be time-dependent. The qualitative
features of lensing depend on whether $r_*$ is bigger than $3m$.
Stars with $2m < r_* \le 3m$ are called
\emph{ultracompact}~\cite{iyer-vishveshwara-dhurandhar-85}. Their
existence is speculative.
The lensing properties of an ultracompact star are the same as
that of a Schwarzschild black hole of the same mass, for observer and
light source in the region $r>r_*$. In particular, the apparent angular
radius $\delta$ on the observer's sky of an ultracompact star is given by the
escape cone of Figure~\ref{fig:schwesc}. Also, an ultracompact star
produces the same infinite sequence of images of each light source as a
black hole. For $r_*>3m$, only finitely many of the images survive
because the other lightlike geodesics are blocked. A non-transparent star
has a finite focal length $r_\mathrm{f}>2m$ in the sense that parallel light
from infinity is focused along a line that extends from radius value
$r_\mathrm{f}$ to infinity. $r_\mathrm{f}$ depends on $m$ and on $r_*$. For the values of
our Sun one finds $r_\mathrm{f}=550$ au (1 au = 1 astronomical unit = average
distance from the Earth to the Sun). An
observer at $r \ge r_\mathrm{f}$ can observe strong lensing effects of the Sun
on distant light sources. The idea of sending a spacecraft to $r \ge r_\mathrm{f}$
was occasionally discussed in the literature~\cite{eshleman-79, nemiroff-ftaclas-97, turyshev-andersson-2003}.
The lensing properties of a non-transparent Schwarzschild star
have been illustrated by showing the appearance of the star's surface to a
distant observer. For $r_*$ bigger than but of the same order of magnitude as
$3m$, this has relevance for neutron stars (see~\cite{winterberg-phillips-73,
pechenick-ftaclas-cohen-83, ftaclas-kearney-pechenick-86, riffert-meszaros-88a,
riffert-meszaros-88b, nollert-ruder-herold-kraus-89}). $r_*$ may be chosen
time-dependent, e.g., to model a non-transparent collapsing star. A star starting
with $r_*>2m$ cannot reach $r=2m$ in finite Schwarzschild coordinate time $t$
(though in finite proper time of an observer at the star's surface), i.e., for
a collapsing star one has $r_*(t) \to 2m$ for $t \to \infty$. To a distant observer,
the total luminosity of a freely (geodesically) collapsing star is attenuated exponentially,
$L(t) \propto \exp \left( -t (3 \sqrt{3} m)^{-1} \right)$. This formula
was first derived by Podurets~\cite{podurets-65} with an incorrect factor 2
under the exponent and corrected by Ames and Thorne~\cite{ames-thorne-68}. Both
papers are based on kinetic photon theory (Liouville's equation). An alternative
derivation of the luminosity formula, based on the optical scalars, was given by
Dwivedi and Kantowski~\cite{dwivedi-kantowski-72}. Ames and Thorne also
calculated the spectral distribution of the radiation as a function of time
and position on the apparent disk of the star. All these analyses considered
radiation emitted at an angle $\le \pi /2$ against the normal of the star as
measured by a static (Killing) observer. Actually, one has to refer not to
a static observer but to an observer comoving with the star's surface. This
modification was worked out by Lake and Roeder~\cite{lake-roeder-79}. 
An interesting approximation formula was derived by 
Beloborodov~\cite{beloborodov-2002}. He showed that a light ray that is 
emitted at radius $r_{\mathrm{S}}$ at an angle $\alpha$ with respect 
to the radial direction escapes to infinity at an angle $\psi$, approximately 
given by $1-\mathrm{cos} \, \alpha  =  \big( 1- \mathrm{cos} \, \psi \big)
\big( 1-2m/r_{\mathrm{S}} \big)$. As an application, he discusses the 
light bending of pulsars. Another approximation formula for the bending
angle was found by Mutka and 
M{\" a}h{\" o}hnen~\cite{mutka-mahonen-2002a, mutka-mahonen-2002b}.
\\

\noindent
{\bf Lensing by a transparent Schwarzschild star.} \\
\noindent
To model a transparent star of radius $r_*$ one has to join the exterior
Schwarzschild metric at $r=r_*$ to an interior (e.g., perfect fluid) metric.
Lightlike geodesics of the exterior Schwarzschild metric are to be joined
to lightlike geodesics of the interior metric when they arrive at $r=r_*$.
The radius $r_*$ of the star can be time-independent only if $r_*>2m$.
For $2m < r_* \le 3m$ (\emph{ultracompact star}), the lensing properties for
observer and light source in the region $r>r_*$ differ from the black hole
case only by the possible occurrence of additional images, corresponding to
light rays that pass through the star. Inside such a transparent
ultracompact star, there is at least one stable photon sphere, in
addition to the unstable one at $r=3m$ outside the star
(cf.~\cite{hasse-perlick-2002}). In principle, there may be
arbitrarily many photon spheres~\cite{karlovini-rosquist-samuelsson-2002}.
For $r_* >3m$, the lensing properties depend on whether there are light
rays trapped inside the star. For a perfect fluid with constant density,
this is not the case; the resulting spacetime is then asymptotically
simple, i.e., all inextendible light rays come from infinity
and go to infinity. General results (see Section~\ref{ssec:asy})
imply that then the number of images must be finite and odd. The light
cone in this exterior-plus-interior Schwarzschild spacetime is discussed
in detail by Kling and Newman~\cite{kling-newman-99}. (In this paper the
authors constantly refer to their interior metric as to a ``dust'' where
obviously a perfect fluid with constant density is meant.) Effects on
light rays issuing from the star's interior have been discussed already
earlier by Lawrence~\cite{lawrence-79}. The ``escape cones'', which are
shown in Figure~\ref{fig:schwesc} for the exterior Schwarzschild metric
have been calculated by Jaffe~\cite{jaffe-70} for points inside the star.
The focal length of a transparent star with constant density is smaller than
that of a non-transparent star of the same mass and radius. For the mass and the
radius of our Sun, one finds 30 au for the transparent case, in contrast
to the above-mentioned 550 au for the non-transparent
case~\cite{nemiroff-ftaclas-97}. Radiation from a spherically symmetric
homogeneous dust star that collapses to a black hole is calculated
in~\cite{shapiro-96}, using kinetic theory. A collapsing inhomogeneous 
spherically symmetric dust configuration may form a naked singularity. 
Its visual appearance, and other observable features, are discussed 
in~\cite{dwivedi-98, deshingkar-joshi-dwivedi-2002, nolan-mena-2002, 
nakao-kobayashi-ishihara-2003}. This analysis was generalized from the 
dust case to more general matter models in~\cite{giambo-2006, deshingkar-2009}. 
\\

\noindent
{\bf Lensing by a Schwarzschild white hole.} \\
\noindent
To get a Schwarzschild white hole one joins at $r=2m$ the static Schwarzschild
region $2m<r<\infty$ to the non-static Schwarzschild region $0<r<2m$ at
$r=2m$ in such a way that \emph{outgoing} light rays can cross this surface
but \emph{ingoing} cannot. In analogy to the gravitational collapse of a
spherically symmetric star into  a black hole, one can consider the outburst
of a white hole into a spherically symmetric star. The observable effects for
an observer in the region $r>2m$ are discussed
in~\cite{faulkner-hoyle-narlikar-64, narlikar-apparao-75, narlikar-kapoor-78,
  dultzin-hacyan-77, lake-roeder-78a, lake-roeder-78b}.


\subsection{Kottler spacetime}
\label{ssec:kottler}

The Kottler metric
\begin{equation}
  \label{eq:kottler}
  g = - \left( 1 - \frac{2m}{r} - \frac{\Lambda r^2}{3} \right) dt^2 +
  \frac{dr^2}{ 1 - \frac{2m}{r} - \frac{\Lambda r^2}{3}} +
  r^2 \left( d \vartheta^2 + \sin^2 \vartheta \, d\varphi^2 \right)
\end{equation}
is the unique spherically symmetric solution of Einstein's
vacuum field equation with a cosmological constant $\Lambda$. It has
the form~(\ref{eq:ssg}) with
\begin{equation}
  \label{eq:kottlerRS}
  e^{2f(r)} = S(r)^{-1} = 1 - \frac{2m}{r} - \frac{\Lambda r^2}{3},
  \qquad
  R(r) = \frac{r}{\sqrt{1 - \frac{2m}{r} - \frac{\Lambda r^2}{3}}}.
\end{equation}
It is also known as the Schwarzschild--deSitter metric for $\Lambda >0$
and as the Schwarzschild--anti-deSitter metric for $\Lambda <0$.
The Kottler metric was found independently by Kottler~\cite{kottler-18}
and by Weyl~\cite{weyl-19}. For $\Lambda = 0$, it reduces to the 
Schwarzschild metric (\ref{eq:schwarzschild}).

In the following we consider the Kottler metric with a constant $m>0$ and
we ignore the region $r<0$ for which the singularity at $r=0$ is naked, for any
value of $\Lambda$. For $\Lambda<0$, there is one horizon at a radius $r_\mathrm{H}$
with $0<r_\mathrm{H}<2m$; the staticity condition $e^{f(r)}>0$ is satisfied on the region
$r_\mathrm{H} < r < \infty$. For $0 <\Lambda < (3m)^{-2}$, there are two horizons at
radii $r_\mathrm{H1}$ and $r_\mathrm{H2}$ with $2m<r_\mathrm{H1}<3m<r_\mathrm{H2}$;
the staticity condition $e^{f(r)}>0$ is satisfied on the region
$r_\mathrm{H1} < r <r_\mathrm{H2}$.
For $\Lambda > (3m)^{-2}$ there is no horizon and no static region. At
the horizon(s), the Kottler metric can be analytically extended into
non-static regions. For $\Lambda <0$, the resulting global structure is
similar to the Schwarzschild case. For $0<\Lambda<(3m)^{-2}$, the resulting
global structure is more complex (see~\cite{lake-roeder-77}). The extreme
case $\Lambda = (3m)^{-2}$ is discussed in~\cite{podolsky-99}.

For any value of $\Lambda$, the Kottler metric has a light sphere at $r=3m$.
Escape cones and embedding diagrams for the Fermat geometry (optical geometry)
can be found in~\cite{stuchlik-hledik-99b, hledik-2001} (cf.\
Figures~\ref{fig:schwesc} and~\ref{fig:schwemb} for the Schwarzschild case).
Similarly to the Schwarzschild spacetime, the Kottler spacetime can
be joined to an interior perfect-fluid metric with constant density.
Embedding diagrams for the Fermat geometry (optical geometry) of the
exterior-plus-interior spacetime can be found 
in~\cite{stuchlik--hledik-soltes-ostgaard-2001}. 
For the optical appearance of a Kottler black hole 
see~\cite{bakala-cermak-hledik-stuchlik-truparova-2007},
and for the optical appearance of a Kottler white hole 
see~\cite{lake-roeder-78a}. The shape of infinitesimally thin 
light bundles in the Kottler spacetime is determined 
in~\cite{dyer-77}.

In view of gravitational lensing, the Kottler metric is of particular interest because 
it can be used to answer the question of how the bending angle of light is 
affected by a cosmological constant. To that end one has to consider
the orbits of light rays in the Kottler spacetime and to investigate how
they differ from the orbits of light rays in the Schwarzschild spacetime.
The first person who looked into this question was, surprisingly late, 
Islam in 1983~\cite{islam-83}. He found that the bending angle of 
light is not affected at all by a cosmological constant. His conclusion,
which eventually turned out to be erroneous, was based on the (correct) 
observation that the differential equation for the orbits of light rays in 
the Kottler spacetime is exactly the same as in the Schwarzschild spacetime. 
To verify this, it suffices to insert the metric coefficients $S(r)$ and $R(r)$ 
from Equation  (\ref{eq:kottlerRS}) into Eq. (\ref{eq:ELrp}). This results in
the differential equation
\begin{equation}
  \label{eq:kottlerrp}
  \frac{d^2r}{d \varphi} - \frac{2}{r} \left(\frac{dr}{d \varphi} \right) ^2 -r
    +3 m = 0 
\end{equation}
which is, indeed, independent of $\Lambda$. Hence, the orbits of light rays in 
the Kottler spacetime are given by exactly the same coordinate equations as in
the Schwarzschild spacetime. On the basis of this result, it was generally accepted
for more than two decades that the gravitational bending of light is unaffected
by a cosmological constant. (See, however, Lake~\cite{lake-2002} for an 
interesting caveat, as to the question of whether the constant $m$ has the same
physical meaning in the Kottler case as in the Schwarzschild case.) Only in 2007
was it shown, in a paper by Rindler and Ishak~\cite{rindler-ishak-2007},
that this conclusion was incorrect. The fact that the coordinate expressions for the
orbits of light rays in the Kottler spacetime are the same as in the Schwarzschild
spacetime does \emph{not} imply that the bending angles are the same. The reason
is that physically measured angles differ from coordinate angles; the physically
measured angles involve the metric, and the metric \emph{does} depend on $\Lambda$.
The analysis of Rindler and Ishak showed that, in contrast to earlier belief, a positive 
cosmological constant would have a diminishing effect on the bending angle. This is 
in perfect agreement with the intuitive idea that a positive cosmological constant 
has a repelling effect (i.e., that it tends to weaken the gravitational attraction).
In terms of the Fermat metric (or optical metric), recall Equation (\ref{eq:ssgF}), 
the Rindler-Ishak result can be rephrased in the following 
way~\cite{gibbons-warnick-werner-2008}: The Fermat metric of the
Kottler spacetime is projectively equivalent to the Fermat metric of the
Schwarzschild spacetime (i.e., the unparametrized geodesics are the same),
but not conformally equivalent (i.e., angles are different). Sereno supported
(and slightly modified) the results of Rindler and Ishak in two papers.
In the first one~\cite{sereno-2008} he analyzed the influence of a cosmological
constant on the bending of light in the weak deflection limit. He found 
that, in the case of a positive $\Lambda$, the radius of 
an Einstein ring decreases and, in a multiple 
imaging situation, the images are demagnified and the time delay increases. In the
second paper~\cite{sereno-2009} he demonstrated that the influence of a cosmological 
constant on the lens equation can be partly (but not completely) absorbed by an 
appropriate redefinition of the angular diameter distance. He argued that, for physical
reasons,  one should express all results in terms of angular diameter distances, rather
than in terms of the radial Schwarzschild-like coordinate, as in the Rindler-Ishak paper.
In the same paper, Sereno also calculated the influence of a cosmological constant on
the redshifts in a multiple imaging situation. Further contributions to the subject 
were made by Sch{\"u}cker~\cite{schuecker-2009}. In contrast to Sereno, Sch{\"u}cker 
deliberately avoided any reference to a lens equation; instead, he concentrated on the 
difference between coordinate angles and physical angles.

The Rindler-Ishak paper has caused a fairly large number of follow-up papers. Although 
some of them were critical, it sems fair to say that, by now, it is generally accepted that 
a positive cosmological constant has a diminishing effect on the bending angle of light. 
However, there is still a controversy about the question of whether this effect is actually 
observable, in realistic astrophysical situations. In order to answer this question,  it is 
not sufficient to analyze the light bending in the Kottler metric, which describes the 
gravitational field around an isolated mass in a world with a cosmologal constant. It 
is rather necessary to take the influence of a cosmological background spacetime into 
account. This has been done by applying the Einstein-Straus method with a cosmological 
constant, i.e., by matching a Kottler vacuole at an outer boundary to a Robertson-Walker 
spacetime. Calculations of light bending in such a composed spacetime were undertaken 
by Ishak et al.~\cite{ishak-rindler-dossett-moldenhauer-allison-2008}, and then, e.g., 
by Sch{\"u}cker~\cite{schuecker-2009b}. Whereas these papers come to the conclusion 
that the effect of $\Lambda$ on the light bending by some galaxy clusters could be 
observable, some other authors feel that this effect is negligibly small, see 
e.g.~\cite{simpson-peacock-heavens-2010}. For a recent review article on the 
topic the reader may consult Ishak and Rindler~\cite{ishak-rindler-2010}.


\subsection{Reissner--Nordstr\"om spacetime}
\label{ssec:reissner}

The Reissner--Nordstr\"om metric
\begin{equation}
  \label{eq:reissner}
  g = - \left( 1 - \frac{2m}{r} + \frac{e^2}{r^2} \right) dt^2 +
  \frac{dr^2}{ 1 - \frac{2m}{r} + \frac{e^2}{r^2}} +
  r^2 \left( d \vartheta^2 + \sin^2 \vartheta \, d\varphi^2 \right)
\end{equation}
is the unique spherically symmetric and asymptotically flat solution of the
Einstein--Maxwell equations. It has the form~(\ref{eq:ssg}) with
\begin{equation}
  \label{eq:reissnerRS}
  e^{2f(r)} = S(r)^{-1} = 1 - \frac{2m}{r} + \frac{e^2}{r^2},
  \qquad
  R(r) = \frac{r}{\sqrt{1 - \frac{2m}{r} + \frac{e^2}{r^2}}}.
\end{equation}
It describes the field around an isolated spherical object with mass $m$ and
charge $e$. The Reissner--Nordstr\"om metric was found independently by
Reissner~\cite{reissner-16}, Weyl~\cite{weyl-17}, and
Nordstr\"om~\cite{nordstrom-18}. A fairly complete list of the
pre-1979 literature
on Reissner--Nordstr\"om geodesics can be found in~\cite{sharp-79}.
A detailed account of Reissner--Nordstr\"om geodesics is given
in~\cite{chandrasekhar-83}. (The Reissner--Nordstr{\"o}m spacetime can
be modified by introducing a cosmological constant. This generalized
Reissner--Nordstr{\"o}m spacetime, whose global structure is investigated
in~\cite{laue-weiss-77}, will not be considered here.)

We assume $m>0$ and ignore the region $r<0$ for which the singularity at
$r=0$ is naked, for any value of $e$. Two cases are to be
distinguished:
\begin{enumerate}
\item $0 \le e^2 \le m^2$; in this case the staticity condition
  $e^{f(r)}>0$ is satisfied on the regions $0<r<m-\sqrt{m^2-e^2}$
  and $m+\sqrt{m^2-e^2}<r<\infty$, i.e., there are two horizons.
  \label{item_x1}
\item $m^2<e^2$; then the staticity condition $e^{f(r)}>0$ is
  satisfied on the entire region $0 < r < \infty$, i.e., there is no
  horizon and the singularity at $r=0$ is naked.
  \label{item_x2}
\end{enumerate}
By switching to isotropic coordinates, one can describe light propagation in
the Reissner--Nordstr\"om metric by an index of refraction (see, e.g.,
\cite{evans-nandi-islam-96a}). 
The resulting Fermat geometry (optical
geometry) is discussed, in terms of embedding diagrams for the black-hole case and for
the naked-singularity case, in~\cite{kristiansson-sonego-abramowicz-98,
abramowicz-bengtsson-karas-rosquist-2002} (cf.~\cite{hledik-2001}).
The visual appearance of a background, as distorted by a Reissner--Nordstr\"om
black hole, is calculated in~\cite{metzenthen-90}. Lensing by a charged neutron
star, whose exterior is modeled by the Reissner--Nordstr\"om metric, is
the subject of~\cite{dabrowski-osarczuk-92, dabrowski-osarczuk-95}.
The lensing properties of a Reissner--Nordstr\"om black hole are qualitatively
(though not quantitatively) the same as that of a Schwarzschild black hole.
The reason is the following. For a Reissner--Nordstr\"om black hole, the
metric coefficient $R(r)$ has one local minimum and no other extremum
between horizon and infinity, just as in the Schwarzschild case (recall
Figure~\ref{fig:schwpot}). The minimum of $R(r)$ indicates an unstable
light sphere towards which light rays can spiral asymptotically, thereby 
defining the ``shadow" of a Reissner-Nordstr{\"o}m black hole. 
The
existence of this minimum, and of no other extremum, was responsible for
all qualitative features of Schwarzschild lensing. Correspondingly,
Figures~\ref{fig:schwlen}, \ref{fig:schwDT}, and~\ref{fig:schwDe} also
qualitatively illustrate lensing by a Reissner--Nordstr\"om black hole.
In particular, there is an infinite sequence of images for each light source,
corresponding to an infinite sequence of light rays whose limit
curve asymptotically spirals towards the light sphere. One can consider
the ``strong-field limit''~\cite{bozza-capozziello-iovane-scarpetta-2001,
bozza-2002} of lensing for a Reissner--Nordstr\"om black hole, in analogy to
the Schwarzschild case which is indicated by the asymptotic straight line in
the middle graph of Figure~\ref{fig:schwlen}. Bozza~\cite{bozza-2002}
investigates whether quantitative features of the ``strong-field limit'',
e.g., the slope of the asymptotic straight line, can be used to distinguish
between different black holes. For the Reissner--Nordstr\"om black hole,
image positions and magnifications have been calculated
in~\cite{eiroa-romero-torres-2002}, and travel times have been
calculated in~\cite{rubio-2003}. In both cases, the authors use the
``almost exact lens map'' of Virbhadra and Ellis~\cite{virbhadra-ellis-2000}
(recall Section~\ref{ssec:ss}) and analytical methods of Bozza et
al.~\cite{bozza-capozziello-iovane-scarpetta-2001, bozza-2002,
bozza-mancini-2004}. The question of whether the ``shadow" of a Reissner-Nordstr{\"o}m
black hole can be observationally distinguished from that of a Schwarzschild
black hole is discussed in \cite{zakharov-paolis-ingrosso-nucita-2005a}. 


\subsection{Morris--Thorne wormholes}
\label{ssec:worm}

We consider a spacetime whose metric is of the form~(\ref{eq:ssg})
with $e^{f(r)} S(r) = 1$, i.e.,
\begin{equation}
  \label{eq:wormg}
  g = - e^{2f(r)} dt^2 + dr^2 + e^{2f(r)} R(r)^2
  \left( d\vartheta^2 + \sin^2 \vartheta \, d\varphi^2 \right),
\end{equation}
where $r$ ranges from $- \infty$ to $\infty$. We call such a spacetime a
\emph{Morris--Thorne wormhole} (see~\cite{morris-thorne-88}) if
\begin{equation}
  \label{eq:wormasy}
  f(r) \underset{r\to \pm \infty}{\longrightarrow} 0,
  \qquad
  r^{-2} R(r)^2 \underset{r\to \pm \infty}{\longrightarrow} 1,
\end{equation}
such that the metric~(\ref{eq:wormg}) is asymptotically flat for $r \to - \infty$
and for $r \to \infty$.

A particular example of a Morris--Thorne wormhole is the Ellis wormhole~\cite{ellis-73} where
\begin{equation}
  \label{eq:ellisg}
  f(r) = 0,
  \qquad
  R(r) = \sqrt{r^2 + a^2}
\end{equation}
with a constant $a$. The Ellis wormhole has an unstable light sphere at $r=0$, i.e.,
at the ``neck'' of the wormhole. It is easy to see that every Morris--Thorne wormhole
must have an unstable light sphere at some radius between $r=-\infty$ and $r = \infty$.
This has the consequence~\cite{hasse-perlick-2002} that every Morris--Thorne
wormhole produces an infinite sequence of images of an appropriately placed light
source. This infinite sequence corresponds to infinitely many light rays whose
limit curve asymptotically spirals towards the unstable light sphere.

Lensing by the Ellis wormhole was discussed in~\cite{chetouani-clement-84}; in this
paper the authors identified the region $r >0$ with the region $r<0$
and they developed a scattering formalism, assuming that observer and light
source are in the asymptotic region. Lensing by the Ellis wormhole was also
discussed in~\cite{perlick-2004} in terms of the exact lens map.
The resulting features are qualitatively very similar to the Schwarzschild
case, with the radius values $r= - \infty$, $r=0$, $r=\infty$ in the wormhole
case corresponding to the radius values $r= 2m$, $r=3m$, $r=\infty$ in the
Schwarzschild case. With this correspondence, Figures~\ref{fig:schwlen},
\ref{fig:schwDT}, and~\ref{fig:schwDe} qualitatively illustrate lensing by
the Ellis wormhole. More generally, the same qualitative features occur
whenever the metric function $R(r)$ has one minimum and no other extrema,
as in Figure~\ref{fig:schwpot}. Lensing by the Ellis wormhole (and other types
of wormholes) is also discussed in~\cite{nandi-zhang-zakharov-2006}. 
For a detailed discussion of lensing by Morris-Thorne wormholes, including 
visualizations, see~\cite{mueller-2004, mueller-2008}.

If observer and light source are on the same side of the wormhole's neck, and
if only light rays in the asymptotic region are considered, lensing by a
wormhole can be studied in terms of the quasi-Newtonian approximation
formalism~\cite{kim-cho-96}. However, as wormholes are typically associated
with negative energy densities~\cite{morris-thorne-88,
morris-thorne-yurtsever-88}, the usual assumption of the
quasi-Newtonian approximation formalism that the mass density is positive
cannot be maintained. This observation has raised some interest in lensing
by negative masses, in particular in the question of whether negative
masses can be detected by their (``microlensing'') effect on
the energy flux from sources passing behind them. So far, related
calculations~\cite{cramer-forward-morris-visser-benford-landis-96,
safonova-torres-romero-2002} have been done only in the quasi-Newtonian
approximation formalism.


\subsection{Barriola--Vilenkin monopole}
\label{ssec:mono}

The \emph{Barriola--Vilenkin monopole}~\cite{barriola-vilenkin-89} is
given by the metric
\begin{equation}
  \label{eq:monog}
  g = - dt^2 + dr^2 + k^2 r^2
  (d \vartheta^2 + \sin^2 \vartheta \, d\varphi^2),
\end{equation}
with a constant $k<1$. There is a deficit solid angle and a singularity
at $r=0 $; the plane $t={\mbox{constant}}$, $\vartheta = \pi /2$ has the
geometry of a cone. (Similarly, for $k>1$ one gets a surplus solid angle.)
The Einstein tensor has non-vanishing components $G_{tt} = -
G_{rr} = (1-k^2)/r^2$.

The metric~(\ref{eq:monog}) was briefly mentioned as an example for a conical
singularity by Sokolov and Starobinsky~\cite{sokolov-starobinsky-77}.
Barriola and Vilenkin~\cite{barriola-vilenkin-89} realized that this
metric can be used as a model for monopoles that might exist in the universe,
resulting from breaking a global ${\cal O}(3)$ symmetry. They also discussed the
question of whether such monopoles could be detected by their lensing
properties which were characterized on the basis of some
approximative assumptions (cf.~\cite{durrer-94}). However, such
approximative assumptions are actually not necessary. The
metric~(\ref{eq:monog}) has the nice property that the geodesics can be
written explicitly in terms of elementary functions. This allows
to write down explicit expressions for image positions and observables
such as angular diameter distances, luminosity distances, image
distortion, etc. (see~\cite{perlick-2004}).
Note that because of the
deficit angle the metric~(\ref{eq:monog}) is not asymptotically flat
in the usual sense. (It is ``quasi-asymptotically flat'' in the sense
of~\cite{nucamendi-sudarsky-97}.) For this reason, the ``almost exact
lens map'' of Virbhadra and Ellis~\cite{virbhadra-ellis-2000} (see
Section~\ref{ssec:ss}), is not applicable to this case, at least not
without modification.

The metric~(\ref{eq:monog}) is closely related to the metric of a static
string (see metric~(\ref{eq:strg}) with $a=0$). Restricting metric~(\ref{eq:monog})
to the hyperplane $\vartheta = \pi /2$ and restricting metric~(\ref{eq:strg})
with $a=0$ to the hyperplane $z = \mbox{constant}$ gives the same
(2\,+\,1)-dimensional metric. Thus, studying light rays in the equatorial
plane of a Barriola--Vilenkin monopole is the same as studying light rays
in a plane perpendicular to a static string. Hence, the multiple
imaging properties of a Barriola--Vilenkin monopole can be deduced
from the detailed discussion of the string example in
Section~\ref{ssec:str}. In particular Figures~\ref{fig:strcon1}
and~\ref{fig:strcon2} show the light cone of a non-transparent and
of a transparent monopole if we interpret the missing spatial dimension
as circular rather than linear. This makes an important difference.
While in the string case the cone of Figures~\ref{fig:strcon1} has
a 2-dimensional set of transverse self-intersection points, the
corresponding cone for the monopole has a 1-dimensional radial
caustic. The difference is difficult to visualize in spacetime
pictures; it is therefore recommendable to use a purely spatial
visualization in terms of instantaneous wave fronts (intersections
of the light cone with hypersurfaces $t= \mbox{constant}$) (compare
Figures~\ref{fig:monofrt1} and~\ref{fig:monofrt2} with
Figures~\ref{fig:strfrt1} and~\ref{fig:strfrt2}).

\epubtkImage{figure18.png}
{\begin{figure}[hptb]
   \def\epsfsize#1#2{1.0#1}
   \centerline{\epsfbox{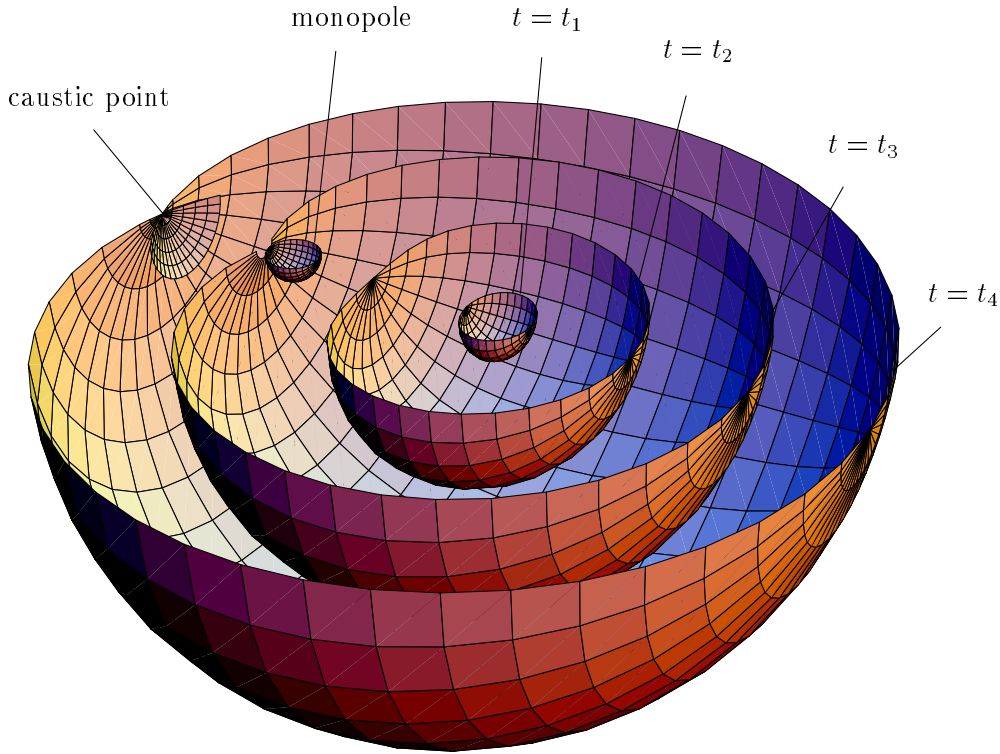}}
   \caption{Instantaneous wave fronts in the spacetime of a
     non-transparent Barriola--Vilenkin monopole with $k=0.8$. The
     picture shows in 3-dimensional space the intersections of the
     past light cone of some event with four hypersurfaces
     $t = \mbox{constant}$, at values $t_1>t_2>t_3>t_4$. Only one half
     of each instantaneous wave front and of the monopole is shown. When
     the wave front passes the monopole, a hole is pierced into it,
     then a tangential caustic develops. The caustic of each
     instantaneous wave front is a point, the caustic of the entire
     light cone is a spacelike curve in spacetime which projects to
     part of the axis in 3-space.}
   \label{fig:monofrt1}
 \end{figure}
}

\epubtkImage{figure19.png}
{\begin{figure}[hptb]
   \def\epsfsize#1#2{1.0#1}
   \centerline{\epsfbox{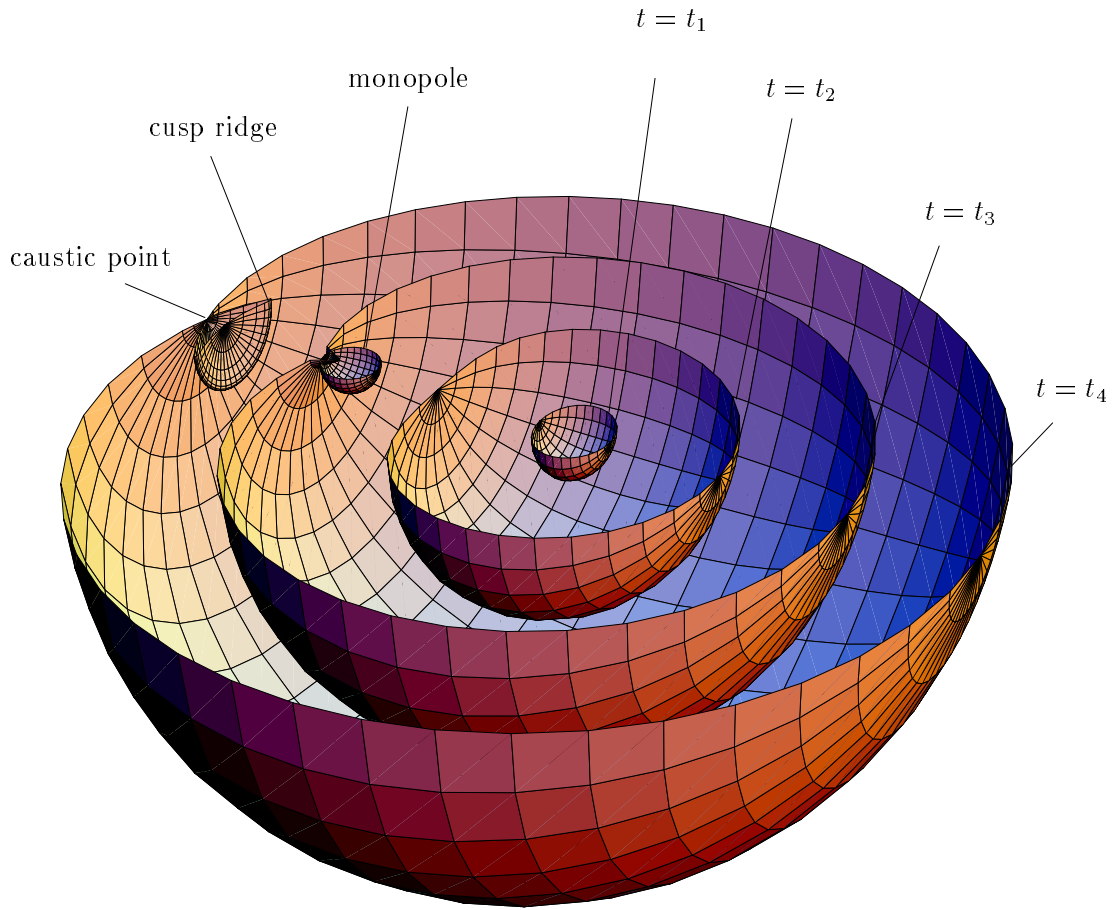}}
   \caption{Instantaneous wave fronts in the spacetime of a
     transparent Barriola--Vilenkin monopole with $k=0.8$. In addition
     to the tangential caustic of Figure~\ref{fig:monofrt1}, a radial
     caustic is formed. For each instantaneous wave front, the radial
     caustic is a cusp ridge. The radial caustic of the entire light
     cone is a lightlike 2-surface in spacetime which projects to a
     rotationally symmetric 2-surface in 3-space.}
   \label{fig:monofrt2}
 \end{figure}
}


\subsection{Janis--Newman--Winicour spacetime}
\label{ssec:JNW}

The Janis--Newman--Winicour metric~\cite{janis-newman-winicour-68} can
be brought into the form~\cite{virbhadra-97}
\begin{equation}
  \label{eq:jnwg}
  g = - \left( 1- \frac{2m}{\gamma r} \right)^{\gamma} dt^2 +
  \frac{dr^2}{\left( 1- \frac{2m}{\gamma r} \right)^{\gamma}} +
  \frac{r^2 \left( d\vartheta^2 + \sin^2 \vartheta \, d\varphi^2 \right)}
  {\left(1- \frac{2m}{\gamma r} \right)^{\gamma -1}},
\end{equation}
where $m$ and $\gamma$ are constants. It is the most general spherically
symmetric static and asymptotically flat solution of Einstein's field
equation coupled to a massless scalar field. For $\gamma = 1$ it reduces
to the Schwarzschild solution; in this case the scalar field vanishes. For
$m > 0$ and $\gamma \neq 1$, there is a naked curvature singularity at
$r = 2 m / \gamma$. Lensing in this spacetime was studied
in~\cite{virbhadra-narasimha-chitre-98, virbhadra-ellis-2002, 
virbhadra-keeton-2008}. The
main motivation was to find out whether the lensing characteristics
of such a naked singularity can be distinguished from lensing by a
Schwarschild black hole. The result is that the qualitative features of
lensing remain similar to the Schwarzschild case as long as $1/2
< \gamma < 1$. However, if $\gamma$ drops below $1/2$, they become
quite different. The reason is easily understood if we write
Equation~(\ref{eq:jnwg}) in the form~(\ref{eq:ssg}). The metric coefficient
\begin{equation}
  \label{eq:jnwR}
  R(r) = r \left( 1 - \frac{2 m}{\gamma r} \right)^{\frac{1}{2} - \gamma}
\end{equation}
has a minimum between the singularity and infinity as long as
$\frac{1}{2} < \gamma < 1$ (see Figure~\ref{fig:jnw}). This minimum
indicates an unstable light sphere (recall Equation~(\ref{eq:rL})), as in
the Schwarzschild case at $r=3m$. All qualitative features of lensing
carry over from the Schwarzschild case, i.e., Figures~\ref{fig:schwlen},
\ref{fig:schwDT}, and~\ref{fig:schwDe} remain qualitatively unchanged.
Clearly, the precise shape of the graph of $\Phi$ in
Figure~\ref{fig:schwlen} changes if $\gamma$ is changed. The question
of how the logarithmic asymptote (``strong-field limit'') depends on
$\gamma$ is dicussed in~\cite{bozza-2002}.
If $\gamma$ drops below $1/2$, $R(r)$ has no longer an extremum, i.e.,
there is no light sphere. Owing to a general result proven 
in~\cite{hasse-perlick-2002}, this implies that only finitely many 
images are possible. In~\cite{virbhadra-ellis-2002}
naked singularities of spherically symmetric spacetimes are called
\emph{weakly naked} if they are surrounded by a light sphere
(cf.~\cite{claudel-virbhadra-ellis-2001}). In a nutshell, weakly naked
singularities show the same qualitative lensing features as black holes. A
generalization of this result to spacetimes without spherical symmetry
has not been worked out so far.

\epubtkImage{figure20.png}
{\begin{figure}[hptb]
   \def\epsfsize#1#2{0.7#1}
   \centerline{\epsfbox{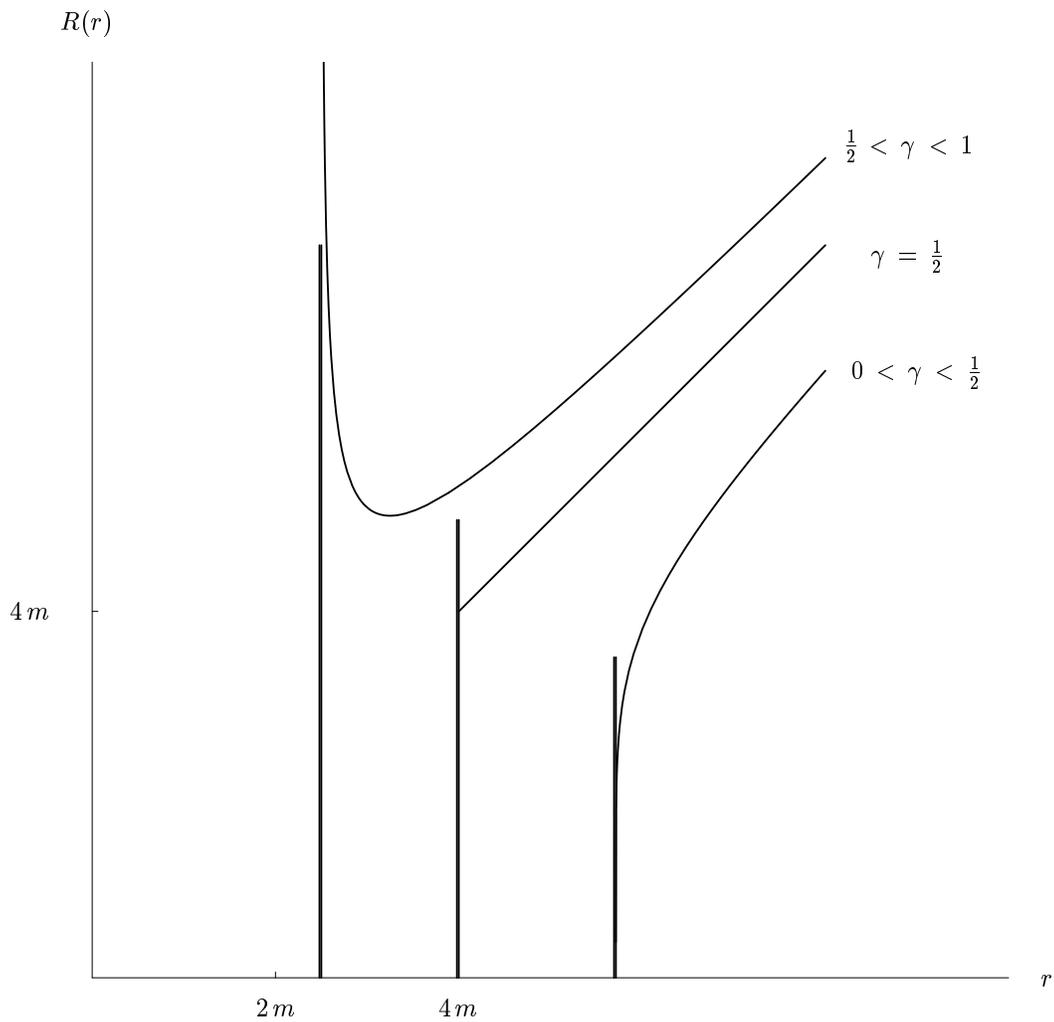}}
   \caption{Metric coefficient $R(r)$ for the
     Janis--Newman--Winicour metric. For $\frac{1}{2} < \gamma < 1$,
     $R(r)$ is similar to the Schwarzschild case $\gamma = 1$ (see
     Figure~\ref{fig:schwpot}). For $\gamma \le \frac{1}{2}$, $R(r)$
     has no longer a minimum, i.e., there is no longer a light sphere
     which can be asymptotically approached by light rays.}
   \label{fig:jnw}
 \end{figure}
}


\subsection{Boson and fermion stars}
\label{ssec:boson}

Spherically symmetric static solutions of Einstein's field equation
coupled to a scalar field may be interpreted as (uncharged,
non-rotating) \emph{boson stars} if they are free of singularities.
Because of the latter condition, the Janis--Newman--Winicour metric (see
Section~\ref{ssec:JNW}) does not describe a boson star.
The theoretical concept of boson stars goes back to~\cite{kaup-68,
ruffini-bonazzola-69}. The analogous idea of a \emph{fermion star},
with the scalar field replaced by a spin 1/2 (neutrino) field,
is even older~\cite{markov-64}. Until today there is no observational
evidence for the existence of either a boson or a fermion star.
However, they are considered, e.g., as hypothetical candidates
for supermassive objects at the center of galaxies
(see~\cite{schunck-liddle-97, torres-capozziello-lambiase-2000} for
the boson and~\cite{viollier-trautmann-tupper-93, tsiklauri-viollier-98} for
the fermion case). For the supermassive object at the center of our own
galaxy, evidence points towards a black hole, but the possibility that
it is a boson or fermion star cannot be completely excluded so far.

Exact solutions that describe boson or fermion stars have been found
only numerically (in 3\,+\,1 dimensions). For this reason there is no boson
star model for which the lightlike geodesics could be studied analytically.
Numerical studies of lensing have been carried out by D{\c a}browski and
Schunck~\cite{dabrowski-schunck-2000} for a transparent spherically symmetric
static maximal boson star, and by Bili{\'c}, Nikoli{\'c}, and
Viollier~\cite{bilic-nikolic-viollier-2000} for a transparent
spherically symmetric
static maximal fermion star. For the case of a fermion-fermion star (two
components) see~\cite{jin-zhang-zhu-2000}. In all three articles the authors
use the ``almost exact lens map'' of Virbhadra and Ellis (see
Section~\ref{ssec:ss}) which is valid for observer and light source in
the asymptotic region and almost aligned. D{\c a}browski and Schunck~\cite{dabrowski-schunck-2000} also discuss how the alignment assumption
can be dropped. The lensing features found in~\cite{dabrowski-schunck-2000}
for the boson star and in~\cite{bilic-nikolic-viollier-2000} for the fermion
star have several similarities. In both cases, there is a tangential
caustic and a radial caustic (recall Figure~\ref{fig:radtang2} for
terminology). A (point) source on the tangential caustic
(i.e., on the axis of symmetry through the observer) is seen as a
(1-dimenional) Einstein ring plus a (point) image in the center.
If the (point) source is moved away from the axis the Einstein ring
breaks into two (point) images, so there are three images altogether.
Two of them merge and vanish if the radial caustic is crossed.
So the qualitative lensing features are quite different from a
Schwarzschild black hole with (theoretically) infinitely many images
(see Section~\ref{ssec:schw}). The essential difference is
that in the case of a boson or fermion star there are no circular
lightlike geodesics towards which light rays could asymptotically
spiral.


\subsection{Kerr spacetime}
\label{ssec:kerr}

Next to the Schwarzschild spacetime, the Kerr spacetime is the physically
most relevant example of a spacetime in which lensing can be studied
explicitly in terms of the lightlike geodesics.
The Kerr metric is given in Boyer--Lindquist coordinates
$(r, \vartheta, \varphi,t)$ as
\begin{equation}
  \label{eq:kerrg}
  g = \varrho (r, \vartheta)^2
  \left( \frac{dr^2}{\Delta (r)} + d\vartheta^2 \right) +
  (r^2 + a^2) \sin^2 \vartheta \, d\varphi^2 - dt^2 +
  \frac{2mr}{\varrho (r, \vartheta)^2}
  \left( a \sin^2 \vartheta \, d\varphi - dt \right)^2,
\end{equation}
where $\varrho$ and $\Delta$ are defined by
\begin{equation}
  \label{eq:rhodelta}
  \varrho (r, \vartheta)^2 = r^2 + a^2 \cos^2 \vartheta,
  \qquad
  \Delta (r) = r^2 - 2mr + a^2,
\end{equation}
and $m$ and $a$ are two real constants. We assume $0<a<m$, with the Schwarzschild
case $a=0$ and the extreme Kerr case $a=m$ as limits. Then the Kerr metric
describes a rotating uncharged black hole of mass $m$ and specific angular
momentum $a$. (The case $a > m$, which describes a naked singularity, will be
briefly considered at the end of this section.) The \emph{domain of outer
communication} is the region between the (outer) horizon at
\begin{equation}
  \label{eq:kerrhor}
  r_+ = m + \sqrt{m^2 - a^2}
\end{equation}
and $r = \infty$. It is joined to the region $r<r_+$ in such a way that
past-oriented ingoing lightlike geodesics cannot cross the horizon. Thus,
for lensing by a Kerr black hole only the domain of outer communication is
of interest unless one wants to study the case of an observer who has fallen
into the black hole. The lensing properties of a Kerr black hole will be 
reviewed below. For the effect of a Kerr black hole on the propagation of 
the polarization plane of light (cf. Section~\ref{ssec:distortion}) see, 
e.g.,~\cite{godfrey-70, su-mallet-80, fayos-llosa-82, 
ishihara-takahashi-tomimatsu-88, nourizonoz-99, sereno-2005}. 
\\

\noindent
{\bf Historical notes.} \\
\noindent
The Kerr metric was found by Kerr~\cite{kerr-63}. The coordinate
representation~(\ref{eq:kerrg}) is due to Boyer and
Lindquist~\cite{boyer-lindquist-67}. The literature on lightlike
(and timelike) geodesics of the Kerr metric is abundant (for an overview
of the pre-1979 literature, see Sharp~\cite{sharp-79}). Detailed accounts
on Kerr geodesics can be found in the books by
Chandrasekhar~\cite{chandrasekhar-83} and O'Neill~\cite{oneill-95}. \\

\noindent
{\bf Fermat geometry.} \\
\noindent
The Killing vector field $\partial_t$ is not timelike on that part of
the domain of outer communication where $\varrho (r, \vartheta)^2 \le 2mr$.
This region is known as the \emph{ergosphere}. Thus, the general results
of Section~\ref{ssec:conf} on conformally stationary spacetimes apply
only to the region outside the ergosphere. On this region, the Kerr
metric is of the form~(\ref{eq:confg}), with redshift potential
\begin{equation}
  \label{eq:kerrf}
  e^{2f(r, \vartheta)} = 1- \frac{2mr}{\varrho (r, \vartheta)^2},
\end{equation}
Fermat metric
\begin{equation}
  \label{eq:kerrgfermat}
  \hat{g} = \frac{\varrho (r, \vartheta)^4}{\varrho (r, \vartheta)^2 - 2mr}
  \left( \frac{dr^2}{\Delta (r)} + d \vartheta^2 \right) +
  \frac{\varrho (r, \vartheta)^4 \, \Delta (r) \, \sin^2 \vartheta \, d\varphi^2}
  {\left( \varrho (r, \vartheta)^2 - 2 m r \right)^2},
\end{equation}
and Fermat one-form
\begin{equation}
  \label{eq:kerrphifermat}
  \hat{\phi} = \frac{2 m r a \, \sin^2 \vartheta \, d\varphi}
  {\varrho (r, \vartheta)^2 - 2mr}.
\end{equation}
(Equation~(\ref{eq:kerrgfermat}) corrects a misprint in~\cite{perlick-90b},
Equation~(66), where a square is missing.)
With the Fermat geometry at hand, the optical-mechanical analogy (Fermat's
principle versus Maupertuis' principle) allows to write the equation for
light rays in the form of Newtonian mechanics (cf.~\cite{alsing-98}).
Certain embedding diagrams for the Fermat geometry (optical reference
geometry) of the equatorial plane have been
constructed~\cite{stuchlik-hledik-99a, hledik-2001}. However, they are
less instructive than in the static case (recall
Figure~\ref{fig:schwemb}) because they do not represent the light rays
as geodesics of a Riemannian manifold. \\

\noindent
{\bf First integrals for lightlike geodesics.} \\
\noindent
Carter~\cite{carter-68} discovered that the geo\-de\-sic equation in the
Kerr metric admits another independent constant of motion $K$, in addition
to the constants of motion $L$ and $E$ associated with the Killing vector
fields $\partial_{\varphi}$ and $\partial_t$. This reduces the lightlike
geodesic equation to the following first-order form:
\begin{eqnarray}
  \varrho (r, \vartheta)^2 \dot{t} &=&
  a \left( L-E a \, \sin^2 \vartheta \right) +
  \frac{(r^2+a^2) \left((r^2+a^2)E-aL \right)}{\Delta (r)},
  \label{eq:kerrdt} \\
  \varrho (r, \vartheta)^2 \dot{\varphi} &=&
  \frac{L-E a \, \sin^2 \vartheta}{\sin^2 \vartheta} +
  \frac{(r^2+a^2) a E-a^2L}{\Delta (r)},
  \label{eq:kerrdphi} \\
  \varrho (r, \vartheta)^4 \dot{\vartheta}^2 &=&
  K - \frac{(L-E a \, \sin^2 \vartheta)^2}{\sin^2 \vartheta},
  \label{eq:kerrdtheta} \\
  \varrho (r, \vartheta)^4 \dot{r}^2 &=&
  - K \Delta (r) + \left((r^2+a^2)E - a L \right)^2.
  \label{eq:kerrdr}
\end{eqnarray}%
Here an overdot denotes differentiation with respect to an affine
parameter $s$. This set of equations allows writing the lightlike geodesics in terms
of elliptic integrals~\cite{bardeen-73}. Clearly, $\dot{\vartheta}$ and
$\dot{r}$ may change sign along a ray; thus, the integration of
Equation~(\ref{eq:kerrdtheta}) and Equation~(\ref{eq:kerrdr}) must be done piecewise.
The determination of the turning points where $\dot{\vartheta}$ and $\dot{r}$
change sign is crucial for numerical evaluation of these integrals and,
thus, for ray tracing in the Kerr spacetime (see, e.g., \cite{viergutz-93a,
rauch-blandford-94, fanton-calvani-felice-cadez-97}).
With the help of Equations~(\ref{eq:kerrdtheta}, \ref{eq:kerrdr}) one easily verifies
the following important fact (cf. \cite{hasse-perlick-2006}). 
Through each point of the region
\begin{equation}
  \label{eq:kerrK}
  \mathcal{K} \, : \,
  \left( 2 r \Delta (r) - (r-m) \, \varrho (r, \vartheta)^2 \right)^2 \le
  4 a^2 r^2 \Delta (r) \, \sin^2 \vartheta
\end{equation}
there is spherical light ray, i.e., a light ray along which $r$ is
constant (see Figure~\ref{fig:kerrK}). These spherical light rays are
unstable with respect to radial perturbations. For the spherical
light ray at radius $r_\mathrm{p}$ the constants of motion $E$, $L$, and $K$
satisfy
\begin{eqnarray}
  a \frac{L}{E} &=& r_\mathrm{p}^2 + a^2 -
  \frac{2 r_\mathrm{p} \Delta (r_\mathrm{p})}{r_\mathrm{p} - m},
  \label{eq:kerrLrL} \\
  \frac{K}{E^2} &=&
  \frac{4 r_\mathrm{p}^2 \Delta (r_\mathrm{p})}{(r_\mathrm{p} - m)^2}.
  \label{eq:kerrKrL}
\end{eqnarray}%
The region $\mathcal{K}$ is the Kerr analogue of the ``light sphere''
$r=3m$ in the Schwarzschild spacetime. \\

\epubtkImage{figure21.png}
{\begin{figure}[hptb]
   \def\epsfsize#1#2{0.7#1}
   \centerline{\epsfbox{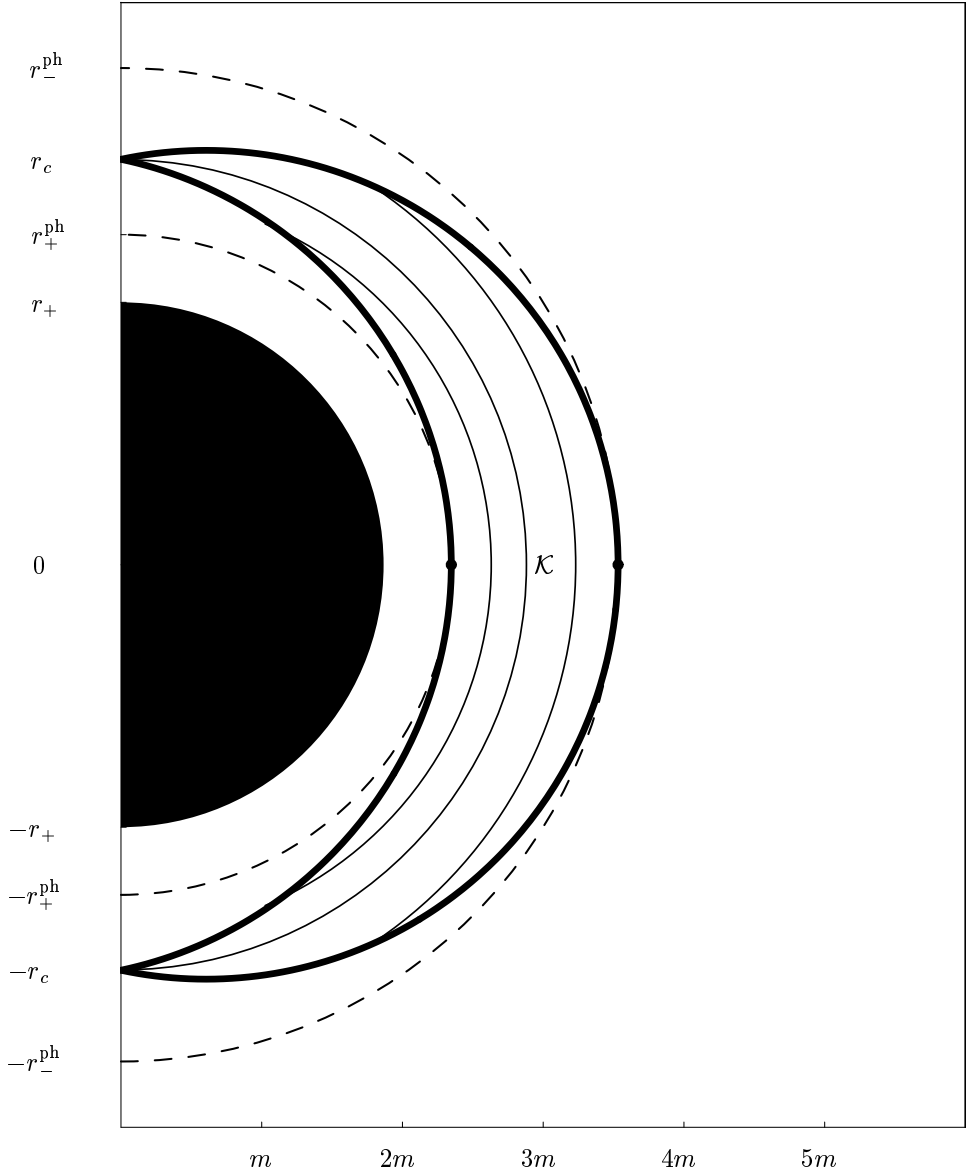}}
   \caption{The region $\mathcal{K}$, defined by
     Equation~(\ref{eq:kerrK}), in the Kerr spacetime. This picture,
     which can also be found in~\cite{hasse-perlick-2006}, is
     purely spatial and shows a meridional section
     $\varphi = \mbox{constant}$, with the axis of symmetry at the
     left-hand boundary. Through each point of $\mathcal{K}$ there is
     a spherical geodesic. Along each of these spherical geodesics,
     the coordinate $\vartheta$ oscillates between extremal values,
     corresponding to boundary points of $\mathcal{K}$, whereas the
     coordinate $\varphi$ proceeds according to Eq. (\ref{eq:kerrdphi}). 
     The region $\mathcal{K}$ meets the axis at radius $r_\mathrm{c}$, given by
     $r_\mathrm{c}^3-3 m r_\mathrm{c}^2 + a^2 r_\mathrm{c} + m a^2 = 0$.
     Its boundary intersects the equatorial plane in circles of radius
     $r_+^{\mathrm{ph}}$ (corotating circular light ray) and
     $r_-^{\mathrm{ph}}$ (counter-rotating circular light
     ray). $r_{\pm}^{\mathrm{ph}}$ are determined by
     $r_{\pm}^{\mathrm{ph}} (r_{\pm}^{\mathrm{ph}} - 3 m)^2 = 4 m a^2$
     and $r_+ < r_+^{\mathrm{ph}} < 3 m < r_-^{\mathrm{ph}} < 4 m $.
     In the Schwarzschild limit $a \to 0$ the region $\mathcal{K}$
     shrinks to the light sphere $r=3m$. In the extreme Kerr limit
     $a \to m$ the region $\mathcal{K}$ extends to the horizon because
     in this limit both $r_+^{\mathrm{ph}} \to m$ and $r_+ \to m$. For
     a caveat, as to geometric misinterpretations of this limit, see
     Figure~3 in~\cite{bardeen-73}.}
   \label{fig:kerrK}
 \end{figure}
}

\noindent
{\bf Light cone.} \\
\noindent
With the help of Equations~(\ref{eq:kerrdt}, \ref{eq:kerrdphi},
\ref{eq:kerrdtheta}, \ref{eq:kerrdr}), the past light cone of any
observation event $p_\mathrm{O}$
can be explicitly written in terms of elliptic integrals. In this representation
the light rays are labeled by the constants of motion $L/E$ and $K/E^2$.
In accordance with the general idea of observational
coordinates~(\ref{eq:obsco}), it is desirable to relabel them by
spherical coordinates
$(\Psi, \Theta)$ on the observer's celestial sphere. This requires choosing
an orthonormal tetrad $(e_0,e_1,e_2,e_3)$ at $p_\mathrm{O}$. It is convenient to
choose $e_1 \propto \partial_{\vartheta}$, $e_2 \propto \partial_{\varphi}$,
$e_3 \propto \partial_r$ and, thus, $e_0$ perpendicular to the hypersurface
$t=\mbox{constant}$ (``zero-angular-momentum observer''). For an
observation event in the equatorial plane, $\vartheta_\mathrm{O}= \pi /2$,
at radius $r_\mathrm{O}$, one finds
\begin{eqnarray}
  \frac{L}{E} &=& a + \frac{\left( (r_\mathrm{O} (r_\mathrm{O}^2+a^2) \,
  \sin \Theta \, \sin \Psi - a r_\mathrm{O}
  \sqrt{\Delta (r_\mathrm{O})} \right)}{r_\mathrm{O}
  \sqrt{\Delta (r_\mathrm{O})} + 2 m a \, \sin \Theta \, \sin \Psi},
  \label{eq:kerrLE} \\
  \frac{K}{E^2} &=& \frac{ r_\mathrm{O}^2
  \left( r_\mathrm{O}^2 + a^2 - a  \sqrt{\Delta (r_\mathrm{O})} \,
  \sin \Theta \, \sin \Psi \right)^2 -
  r_\mathrm{O}^3 \left( r_\mathrm{O} (r_\mathrm{O}^2 + a^2) +
  2 m a^2 \right) \cos^2 \Theta }{\left( r_\mathrm{O}
  \sqrt{\Delta (r_\mathrm{O})} + 2 m a \, \sin \Theta \, \sin \Psi \right)^2}.
  \label{eq:kerrKE}
\end{eqnarray}
As in the Schwarzschild case, some light rays from $p_\mathrm{O}$ go out to infinity
and some go to the horizon. In the Schwarzschild case, the borderline between
the two classes corresponds to light rays that asymptotically approach the
light sphere at $r=3m$. In the Kerr case, it corresponds to light rays
that asymptotically approach a spherical light ray in the region $\mathcal{K}$
of Figure~\ref{fig:kerrK}. The constants of motion for such light rays
are given by Equation~(\ref{eq:kerrLrL}, \ref{eq:kerrKrL}), with $r_\mathrm{p}$
varying between its extremal values $r_+^{\mathrm{ph}}$ and
$r_-^{\mathrm{ph}}$ (see again Figure~\ref{fig:kerrK}). Thereupon,
Equation~(\ref{eq:kerrLE}) and Equation~(\ref{eq:kerrKE})
determine the celestial coordinates $\Psi$ and $\Theta$ of those light rays that
approach a spherical light ray in $\mathcal{K}$. The resulting curve on the
observer's celestial sphere gives the apparent shape of the Kerr black
hole (see Figure~\ref{fig:kerresc}). For an observation event on the axis of
rotation, $\sin \vartheta_\mathrm{O} =0$, the Kerr light cone
is rotationally symmetric. The caustic consists of infinitely many spacelike
curves, as in the Schwarzschild case. A light source passing through  a
point of the caustic is seen as an Einstein ring. For observation events not on the
axis, the light cone has no rotational symmetry and the caustic structure is quite different
from the Schwarzschild case. The caustic still consists of infinitely many 
connected subsets (a primary caustic and infinitely many higher-order caustics),
but these are no longer spacelike curves. This fact is somewhat disguised if one 
restricts to light rays in the equatorial plane $\vartheta = \pi /2$ (which is possible,
of course, only if the observation event is in the equatorial plane). Then the
resulting 2-dimensional light cone looks indeed qualitatively similar to the
Schwarzschild cone of Figure~\ref{fig:schwcon} (cf.~\cite{hanni-77}), where
intersections of the light cone with hypersurfaces $t = \mbox{constant}$
are depicted. However, in the Kerr case the transverse self-intersection of
this 2-dimensional light cone does not occur on an axis of symmetry. Therefore,
the caustic of the full (3-dimensional) light cone is more involved than
in the Schwarzschild case. The primary caustic turns out to be not a spatially 
straight line, as in the Schwarzschild case, but rather a tube, with astroid
cross-section, that winds a certain number of times around the black hole;
for $a \to m$ it approaches the horizon in an infinite spiral motion. The
primary caustic of a Kerr light cone, with vertex in the equatorial plane far
away from the black hole, was numerically calculated and depicted, for 
$a=m$, by Rauch and Blandford~\cite{rauch-blandford-94}. A detailed
study of primary and higher-order caustics, for a Kerr light cone with vertex 
far away from the black hole but not necessarily in the equatorial plane, was
presented by Bozza~\cite{bozza-2008a}. This work, which contains several
pictures of Kerr caustics in 3+0 dimensions, is based on numerical calculations.
The results are in good agreement with analytical approximation methods
for studying the caustics. Two such methods exist which are complementary
to each other in the sense that the first is valid for light rays that come
close to a spherical light ray in the region $\mathcal{K}$ and the second 
is valid for light rays that stay far away from the black hole: The first
method is due to  Bozza, de Luca, Scarpetta and 
Sereno~\cite{bozza-luca-scarpetta-sereno-2005, bozza-scarpetta-2007} 
who analytically studied
higher-order caustics in the strong deflection limit; this approach is not
applicable to the primary caustic. The second method is due to Sereno and 
de Luca~\cite{sereno-luca-2008} who developed an analytic formula for the 
primary caustic that is valid up to fourth order in $m/b$ and $a/b$, where 
$b$ is the impact parameter. Taking all this together, a fairly clear picture
of the caustics of Kerr light cones has now emerged. Also, attempts have 
been made to visualize the Kerr light cones in terms of their  intersections 
with hypersurfaces $t=\mbox{constant}$, see Figure~1 
in~\cite{grave-frutos-mueller-adis-2008}. 
From the study of light cones one may switch to the study of arbitrary
wave fronts. (For the definition of wave fronts see Section~\ref{ssec:front}.)
Pretorius and Israel~\cite{pretorius-israel-98} determined all
axisymmetric wave fronts in the Kerr geometry. In
this class, they investigated in particular those members that
are asymptotic to Minkowski light cones at infinity (``quasi-spherical
light cones'') and they found, rather surprisingly, that they are free
of caustics. Special families of wave fronts in the Kerr spacetime 
are also considered, e.g., in~\cite{fletcher-lun-2003, hayward-2004,
bai-cao-gong-shang-wu-lau-2007}. 
\\

\epubtkImage{figure22.png}
{\begin{figure}[hptb]
   \def\epsfsize#1#2{0.7#1}
   \centerline{\epsfbox{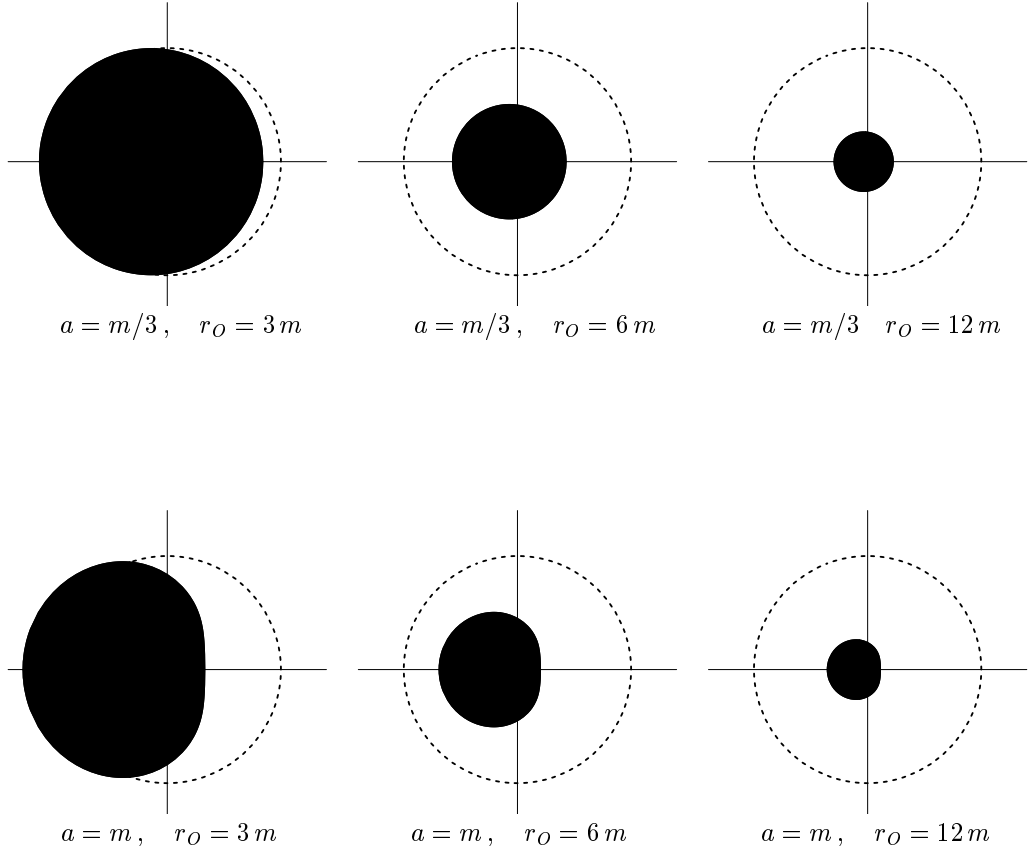}}
   \caption{Apparent shape of a Kerr black hole for an observer at
     radius $r_\mathrm{O}$ in the equatorial plane. (For the
     Schwarzschild analogue, see Figure~\ref{fig:schwesc}.) The
     spin vector of the black hole is pointing downwards. The
     pictures show the celestial sphere of an observer whose
     4-velocity is perpendicular to a hypersurface
     $t=\mbox{constant}$. (If the observer is moving one has to
     correct for aberration.) The dashed circle is the celestial
     equator, $\Theta = \pi /2$, and the crossing axes indicate the
     direction towards the center, $\Theta = \pi$. Past-oriented light
     rays go to the horizon if their initial direction is in the black
     disk and to infinity otherwise. Thus the black disk, known as the
     ``shadow" of the black hole, shows that part of the sky which is 
     not illuminated by light sources at a large radius. The boundary 
     of this disk corresponds to light rays that asymptotically approach 
     a spherical light ray in the region $\mathcal{K}$ of 
     Figure~\ref{fig:kerrK}. For an observer in the
     equatorial plane at infinity, the shadow of a Kerr black
     hole was correctly calculated and depicted by
     Bardeen~\cite{bardeen-73} (cf.~\cite{chandrasekhar-83},
     p.~358). Earlier work by Godfrey~\cite{godfrey-70} contains a
     mathematical error. Observability of the shadow of a Kerr black
     hole is discussed, e.g., in \cite{falcke-melia-agol-2000, 
     zakharov-paolis-ingrosso-nucita-2005b, hioki-maeda-2009}.}
   \label{fig:kerresc}
 \end{figure}
}

\noindent
{\bf Lensing by a Kerr black hole.} \\
\noindent
For an observation event $p_\mathrm{O}$ and a light source with worldline
$\gamma_\mathrm{S}$, both in the domain of outer communication of a Kerr black
hole, several qualitative features of lensing are unchanged in comparison
to the Schwarzschild case. If $\gamma_\mathrm{S}$ is past-inextendible, bounded
away from the horizon and from (past lightlike) infinity, and does not meet
the caustic of the past light cone of $p_\mathrm{O}$, the observer sees an infinite
sequence of images; for this sequence, the travel time (e.g., in terms
of the time coordinate $t$) goes to infinity. These statements have been 
proven in~\cite{hasse-perlick-2006} with the help of Morse theory 
(cf. Section~\ref{ssec:morse}). 
On the observer's sky the
sequence of images approaches the apparent boundary of the black hole
which is shown in Figure~\ref{fig:kerresc}. This follows from the fact
that
\begin{itemize}
\item the infinite sequence of images must have an accumulation point
  on the observer's sky, by compactness, and
\item the lightlike geodesic with this initial direction cannot go to
  infinity or to the horizon, by assumption on $\gamma_\mathrm{S}$.
\end{itemize}
If $\gamma_\mathrm{S}$ meets the caustic of $p_\mathrm{O}$'s
past light cone, the image is not an Einstein ring, unless $p_\mathrm{O}$ is
on the axis of rotation. It has only an ``infinitesimal'' angular
extension on the observer's sky. As always when a point source meets
the caustic, the ray-optical calculation gives an infinitely bright
image. Numerical studies show that in the Kerr spacetime, where
the caustic is a tube with astroid cross-section, the image is
very bright whenever the light source is inside the tube~\cite{rauch-blandford-94}.
In principle, with the lightlike geodesics given
in terms of elliptic integrals, image positions on the observer's sky
can be calculated explicitly. This has been worked out for several
special wordlines $\gamma_\mathrm{S}$. The case that $\gamma_\mathrm{S}$ is a circular timelike
geodesic in the equatorial plane of the extreme Kerr metric, $a=m$,
was treated by Cunningham and Bardeen~\cite{cunningham-bardeen-72,
bardeen-cunningham-73}. This example is of relevance in view of accretion
disks. Viergutz~\cite{viergutz-93a} developed a formalism for the case that
$\gamma_\mathrm{S}$ has constant $r$ and $\vartheta$ coordinates, i.e., for
a light source that stays on a ring around the axis. One aim of this approach,
which could easily be rewritten in terms of the exact lens map (recall
Section~\ref{ssec:cone}), was to provide a basis for numerical studies.
The case of a stationary light source (i.e., the case that 
$\gamma_\mathrm{S}$ is an integral curve of $\partial_t$) 
was investigated in great detail in a series of papers by Bozza, de Luca,
Scarpetta and Sereno~\cite{bozza-2003, bozza-luca-scarpetta-sereno-2005,
bozza-luca-scarpetta-2006, bozza-scarpetta-2007}. In all these papers the 
authors derive analytic approximation formulas using the 
strong-deflection
limit, i.e., the approximation is good for light rays that undergo a 
deflection of $\pi$ or more. Such light rays come close to one of the 
spherical light rays in the region $\mathcal{K}$, recall Figure~\ref{fig:kerrK}.
The first two papers in the series make the additional assumption that 
the light source and the observer are in the equatorial plane and that
not only the observer but also the light source is far away from the 
black hole; in the last two papers these assumptions are relaxed. This
series of papers gives a fairly complete analysis of Kerr lensing for
stationary light sources under the strong deflection hypothesis. An alternative
approach to Kerr lensing with stationary light sources, partly based on
numerical results, was brought forward by Vazquez
and Esteban~\cite{vazquez-esteban-2003}. 
All
these articles also calculate the brightness of images. This requires determining
the cross-section of infinitesimally thin bundles with a vertex, e.g., in terms
of the shape parameters $D_+$ and $D_-$ (recall Figure~\ref{fig:shape}). For a
bundle around an arbitrary light ray in the Kerr metric, all relevant equations
were worked out analytically by Pineault and Roeder~\cite{pineault-roeder-77a}.
However, the equations are much more involved than for the Schwarzschild case
and will not be given here. Lensing by a Kerr black hole has been visualized
(i) by showing the apparent distortion of a background pattern~\cite{pineault-roeder-77b, sikora-79} and (ii) by showing the
visual appearence of an accretion disk~\cite{pineault-roeder-77b,
polnarev-turchaninov-79, sikora-79, beckwith-done-2005}. 
The main difference, in
comparison to the Schwarzschild case, is in the loss of the left-right
symmetry. In view of observations, Kerr black holes are considered
as candidates for active galactic nuclei (AGN) since many years.
In particular, the X-ray variability of AGN is interpreted
as coming from a ``hot spot'' in an accretion disk that circles around
a Kerr black hole. Starting with the pioneering work in~\cite{cunningham-bardeen-72, bardeen-cunningham-73}, many articles
have been written on calculating the light curves and the spectrum
of such ``hot spots'', as seen by a distant observer (see, e.g., \cite{felice-nobili-calvani-74, asaoka-89, karas-bao-92,
jaroszynski-kurpiewski-97, fanton-calvani-felice-cadez-97}).
The spectrum can be calculated in terms of a \emph{transfer
function} that was tabulated, for some values of $a$,
in~\cite{cunningham-75b} (cf.~\cite{viergutz-93a, viergutz-93b}).
A Kerr black hole is also considered as the most likely candidate
for the supermassive object at the center of our own galaxy. (For background
material see~\cite{falcke-hehl-2003}.) In this case, the predicted
angular diameter of the black hole on our sky, in the sense
of Figure~\ref{fig:kerresc}, is about 30 microarcseconds;
this is not too far from the reach of current VLBI technologies~\cite{falcke-melia-agol-2000}. Also, the fact that the radio emission
from our galactic center is linearly polarized gives a good
motivation for calculating polarimetric images as produced by
a Kerr black hole~\cite{bromley-melia-liu-2001}. The calculation
is based on the geometric-optics approximation according to
which the polarization vector is parallel along the light ray.
In the Kerr spacetime, this parallel-transport law can be
explicitly written with the help of constants of
motion~\cite{connors-stark-77, pineault-roeder-77a, su-mallet-80}
(cf.~\cite{chandrasekhar-83}, p.~358). As to the large number of numerical
codes that have been written for calculating imaging properties of
a Kerr black hole the reader may consult~\cite{karas-vokrouhlicky-polnarev-92,
viergutz-93a, rauch-blandford-94, fanton-calvani-felice-cadez-97}.

\paragraph{Notes on Kerr naked singularities.} ~\\
The Kerr metric with $a>m$ describes a naked singularity. Until now there
is no observational indication that such objects exist in nature. The 
lightlike geodesics in a Kerr spacetime with $a>m$ have been studied 
in~\cite{calvani-felice-78, calvani-nobili-felice-78} (cf.~\cite{chandrasekhar-83},
p.~375). Observable effects of accretion disks around a Kerr naked singularity,
in comparison to a Kerr black hole, were discussed in~\cite{takahashi-harada-2010}.
The ``shadow" of a Kerr naked singularity was calculated 
in~\cite{vries-2000, hioki-maeda-2009} and, under different assumptions, 
in~\cite{bambi-freese-2009}. 

\paragraph{Notes on the Kerr--Newman spacetime.} ~\\
The Kerr--Newman spacetime (charged Kerr spacetime) is usually
thought to be of little astrophysical relevance because the net charge
of celestial bodies is small. For the lightlike geodesics in this spacetime
the reader may consult~\cite{calvani-felice-nobili-80, calvani-turolla-81}.
Embeddability diagrams of the equatorial plane of a Kerr-Newman spacetime
can be found in~\cite{stuchlik-hledik-juran-2000}. The shadow of a 
Kerr-Newman black hole, and of a Kerr-Newman naked singularity, was 
discussed in \cite{vries-2000, takahashi-2005}. A Morse-theoretical analysis 
of lensing in the Kerr-Newman spacetime can be found in~\cite{hasse-perlick-2006}.


\subsection{Rotating disk of dust}
\label{ssec:disk}

The stationary axisymmetric spacetime around a rigidly rotating disk
of dust was first studied in terms of a numerical solution to Einstein's
field equation by Bardeen and Wagoner~\cite{bardeen-wagoner-69,
bardeen-wagoner-71}. The exact solution was found much later by Neugebauer
and Meinel~\cite{neugebauer-meinel-93}. It is discussed,
e.g., in~\cite{neugebauer-kleinwaechter-meinel-96}. The metric cannot
be written in terms of elementary functions because it
involves the solution to an ultraelliptic integral equation. It
depends on a parameter $\mu$ which varies between zero and $\mu_\mathrm{c} =
4.62966 \dots$. For small $\mu$ one gets the Newtonian approximation,
for $\mu \to \mu_\mathrm{c}$ the extreme Kerr metric ($a=m$) is approached.
The lightlike geodesics in this spacetime have been studied
numerically and the appearance of the disk to a distant
observer has been visualized~\cite{weiskopf-ansorg-2000}. It would
be desirable to support these numerical results with exact statements.
From the known properties of the metric, only a few qualitative
lensing features of the disk can be deduced. As Minkowski spacetime
is approached for $\mu \to 0$, the spacetime must be asymptotically
simple and empty as long as $\mu$ is sufficiently small. (This is true,
of course, only if the disk is treated as transparent.) The general
results of Section~\ref{ssec:asy} imply that in this case the gravitational field
of the disk produces finitely many images of each light source, and
that the number of images is odd, provided that the worldline
of the light source is past-inextendible and does not go out to
past lightlike infinity. For larger values of $\mu$, this is no
longer true. For $\mu > 0.5$ there are two counter-rotating circular
lightlike geodesics in the equatorial plane, a stable one at a radius
$\tilde{\rho}_1$ inside the disk and an unstable one at a radius
$\tilde{\rho}_2$ outside the disk. (This follows from~\cite{ansorg-98}
where it is shown that for $\mu > 0.5$ \emph{timelike} counter-rotating
circular geodesics do not exist in a radius interval $[\tilde{\rho}_1,
\tilde{\rho}_2]$. The boundary values of this interval give the
radii of lightlike circular geodesics.) The existence of circular
light rays has the consequence that the number of images must
be infinite; this is obviously true if light source and observer
are exactly on the spatial track of such a circular light ray and,
by continuity, also in a neighborhood. For a better understanding
of lensing by the disk of dust it is desirable to investigate,
for each value of $\mu$ and each event $p_\mathrm{O}$: Which past-oriented
lightlike geodesics that issue from $p_\mathrm{O}$ go out to infinity and
which are trapped? Also, it is desirable to study the light cones
and their caustics.


\subsection{Straight spinning string}
\label{ssec:str}

Cosmic strings (and other topological defects) are expected to exist
in the universe, resulting from a phase transition in the early
universe (see, e.g., \cite{vilenkin-shellard-94} for a detailed account). So far,
there is no direct observational evidence for the existence of strings.
In principle, they could be detected by their lensing effect. The general 
perspective is discussed in~\cite{huterer-vachaspati-2003, 
gasparini-marshall-treu-morganson-dubath-2008, 
morganson-marshall-treu-schrabback-blandford-2010}. The object
CSL-1, which consists of a pair of galaxies, was discussed as a candidate
for lensing by a string for some time~\cite{sazhin-etal-2003}. However, more 
recent observations by the Hubble Space Telescope led to the conclusion that 
it is not a lensed image~\cite{sazhin-etal-2007}.

Basic lensing features for various string configurations are briefly
summarized in~\cite{anderson-96}. Here we consider the simple case of a
straight string that is isolated from all other masses. This is one of the
most attractive examples for investigating lensing from the spacetime
perspective without approximations. In particular, studying the light
cones in this metric is an instructive exercise. The geodesic equation
is completely integrable, and the geodesics can even be written
explicitly in terms of elementary functions.

We consider the spacetime metric
\begin{equation}
  \label{eq:strg}
  g = - (dt - a \, d\varphi)^2 + dz^2 +
  d \rho^2 + k^2 \rho^2 \, d{\varphi}^2,
\end{equation}
with constants $a$ and $k>0$. As usual, the azimuthal
coordinate $\varphi$ is defined modulo $2 \pi$. For $a=0$ and $k=1$,
metric~(\ref{eq:strg}) is the Minkowski metric in cylindrical coordinates. For
any other values of $a$ and $k$, the metric is still (locally) flat but
not globally isometric to Minkowski spacetime; there is a singularity
along the $z$-axis. For $a=0$ and $0 < k <1$, the plane $t= \mbox{constant}$,
$z= \mbox{constant}$ has the geometry of a cone with a deficit angle
\begin{equation}
  \label{eq:deficit}
  \delta = (1 - k) 2 \pi.
\end{equation}
(see Figure~\ref{fig:deficit}); for $k>1$ there is a surplus angle. Note
that restricting the metric~(\ref{eq:strg}) with $a=0$ to the hyperplane
$z = \mbox{constant}$ gives the same result as restricting the
metric~(\ref{eq:monog}) of the Barriola--Vilenkin monopole to the
hyperplane $\vartheta = \pi /2$.

The metric~(\ref{eq:strg}) describes the spacetime around a straight
spinning string. The constant $k$ is related to the string's mass-per-length
$\mu$, in Planck units, via
\begin{equation}
  \label{eq:strmu}
  k = 1 - 4 \mu,
\end{equation}
whereas the constant $a$ is a measure for the string's spin. Equation~(\ref{eq:strmu})
shows that we have to restrict to the deficit-angle case $k<1$ to have $\mu$
positive. 
One may treat the string as a line singularity, i.e., consider
the metric~(\ref{eq:strg}) for all $\rho > 0$. (This ``wire approximation'',
where the energy-momentum tensor of the string is concentrated on a 2-dimensional
submanifold, is mathematically delicate; see~\cite{geroch-traschen-87}.) For
a string of finite radius $\rho_*$ one has to match the metric~(\ref{eq:strg}) at
$\rho =\rho_*$ to an interior solution, thereby getting a metric that is
regular on all of $\mathbb{R}^4$. In view of lensing it is important to 
distinguish between a transparent string, where light rays are allowed to
pass through the interior solution, and a non-transparent string, where
light rays are blocked at the boundary of the string. 
\\

\epubtkImage{figure23.png}
{\begin{figure}[hptb]
   \def\epsfsize#1#2{0.4#1}
   \centerline{\epsfbox{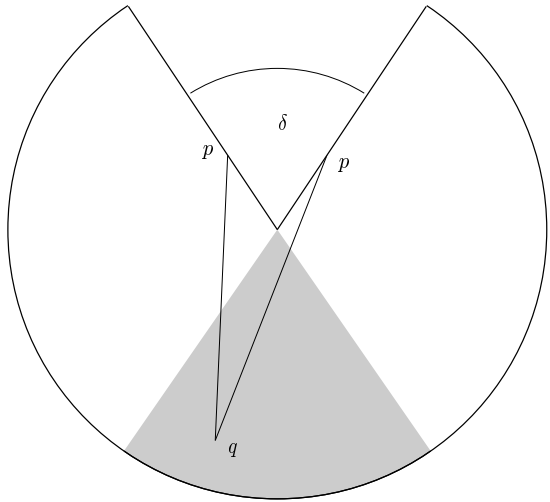}}
   \caption{On a cone with deficit angle $0 < \delta < \pi$, the
     point $p$ can be connected to every point $q$ in the
     double-imaging region (shaded) by two geodesics and to a point in
     the single-imaging region (non-shaded) by one geodesic.}
   \label{fig:deficit}
 \end{figure}
}

\noindent
{\bf Historical notes.} \\
\noindent
With $a=0$, the metric~(\ref{eq:strg}) and its geodesics were first
studied by Marder~\cite{marder-59,marder-62}. He also discussed the
matching to an interior solution, without, however, associating it with
strings (which were no issue at that time). The same metric was
investigated by Sokolov and Starobinsky~\cite{sokolov-starobinsky-77}
as an example for a conic singularity. Later Vilenkin~\cite{vilenkin-81,
vilenkin-84} showed that within the linearized Einstein
theory the metric~(\ref{eq:strg}) with $a=0$ describes the spacetime
outside a straight non-spinning string. Hiscock~\cite{hiscock-85},
Gott~\cite{gott-85}, and Linet~\cite{linet-85} realized that the same is
true in the full (non-linear) Einstein theory. Basic features of lensing
by a non-spinning string were found by Vilenkin~\cite{vilenkin-84}
and Gott~\cite{gott-85}. The matching to an interior solution for a spinning
string, $a \neq 0$, was worked out by Jensen and Soleng~\cite{jensen-soleng-92}.
Already earlier, the restriction of the metric~(\ref{eq:strg}) with
$a \neq 0$ to the hyperplane $z = 0$ was studied as the
spacetime of a spinning particle in 2\,+\,1 dimensions by Deser, Jackiw, and
't Hooft~\cite{deser-jackiw-thooft-84}. The geodesics in this
(2\,+\,1)-dimensional metric were first investigated by
Cl\'ement~\cite{clement-85} (cf.\ Krori, Goswami, and
Das~\cite{krori-goswami-das-93} for the (3\,+\,1)-dimensional
case). For geodesics in string metrics one may also consult Galtsov and
Masar~\cite{galtsov-masar-89}. The metric~(\ref{eq:strg}) can be generalized
to the case of several parallel strings (see Letelier~\cite{letelier-87}
for the non-spinning case, and Krori, Goswami, and Das~\cite{krori-goswami-das-93}
for the spinning case). Clarke, Ellis and Vickers~\cite{clarke-ellis-vickers-90} found obstructions against embedding
a string model close to metric~(\ref{eq:strg}) into an almost-Robertson--Walker
spacetime. This is a caveat, indicating that the lensing properties of ``real''
cosmic strings might be significantly different from the lensing
properties of the metric~(\ref{eq:strg}). \\

\noindent
{\bf Redshift and Fermat geometry.} \\
\noindent
The string metric~(\ref{eq:strg}) is stationary, so the results of
Section~\ref{ssec:conf} apply. Comparison of metric~(\ref{eq:strg}) with
metric~(\ref{eq:confg}) shows that the redshift potential vanishes, $f=0$.
Hence, observers on $t$-lines see each other without redshift. The Fermat
metric $\hat{g}$ and Fermat one-form $\hat{\phi}$ read
\begin{eqnarray}
  \hat{g} &=& dz^2 + d\rho^2 + k^2 \rho^2 \, d\varphi^2,
  \label{eq:strgF} \\
  \hat{\phi} &=& - a \, d\varphi.
  \label{eq:strphiF}
\end{eqnarray}%
As the Fermat one-form is closed, $d \hat{\phi} = 0$, the spatial
paths of light rays are the geodesics of the Fermat metric $\hat{g}$
(cf.\ Equation~(\ref{eq:confLor})), i.e., they are not affected by the spin of
the string. $\hat{\phi}$ can be transformed to zero by changing
from $t$ to the new time function $t - a \varphi$. Then the
influence of the string's spin on the travel time~(\ref{eq:confT})
vanishes as well. However, the new time function is not globally
well-behaved (if $a \neq 0$), because $\varphi$ is either
discontinuous or multi-valued on any region that contains a full
circle around the $z$-axis. As a consequence, $\hat{\phi}$ can
be transformed to zero on every region that does
not contain a full circle around the $z$-axis, but not globally.
This may be viewed as a gravitational analogue of the Aharonov--Bohm
effect (cf.~\cite{stachel-82}). The Fermat metric~(\ref{eq:strgF})
describes the product of a cone with the $z$-line. Its geodesics
(spatial paths of light rays) are straight lines if we cut the cone
open and flatten it out into a plane (see Figure~\ref{fig:deficit}).
The metric of a cone is (locally) flat but not (globally) Euclidean.
This gives rise to another analogue of the Aharonov--Bohm effect, to
be distinguished from the one mentioned above, which was discussed,
e.g., in~\cite{ford-vilenkin-81, bezerra-87, helliwell-konkowski-87}. \\

\noindent
{\bf Light cone.} \\
\noindent
For the metric~(\ref{eq:strg}), the lightlike geodesics can be explicitly
written in terms of elementary functions. One just has to apply the
coordinate transformation $(t, \varphi) \longmapsto (t
- a \varphi, k
\varphi)$ to the lightlike geodesics in Minkowski spacetime. As indicated above, the
new coordinates are not globally well-behaved on the entire spacetime. However, they
can be chosen as continuous and single-valued functions of the affine parameter $s$
along all lightlike geodesics through some chosen event, with $\varphi$
taking values in $\mathbb{R}$. In this way we get the following representation of
the lightlike geodesics that issue from the observation event
$(\rho = \rho_0, \varphi = 0, z=0, t=0)$ into the past:
\begin{eqnarray}
  \rho (s) &=&
  \sqrt{ s^2 \, \sin^2 \Theta + 2 s \rho_0 \, \sin \Theta \, \cos \Psi + \rho_0^2},
  \label{eq:strgeor} \\
  \mathrm{tan} \left( k \varphi (s) \right) &=&
  \frac{s \, \sin \Theta \, \sin \Psi}{ \rho_0 + s \, \sin \Theta \, \cos \Psi},
  \label{eq:strgeophi} \\
  z(s) &=& s \cos \Theta,
  \label{eq:strgeoz} \\
  t(s) &=& - s + a \varphi (s).
  \label{eq:strgeot}
\end{eqnarray}
The affine parameter $s$ coincides with $\hat{g}$-arclength $\ell$,
and $(\Psi, \Theta)$ parametrize the observer's celestial sphere,
\begin{equation}
  \label{eq:PsiTheta}
  \frac{d}{ds}
  \left.
    \begin{pmatrix}
      \rho (s) \, \cos \varphi (s) \\
      \rho (s) \, \sin \varphi (s) \\
      z(s)
    \end{pmatrix}
  \right|_{s=0} \!\!\!\!\! =
  \begin{pmatrix}
    \cos \Psi \, \sin \Theta \\
    \sin \Psi \, \sin \Theta \\
    \cos \Theta
  \end{pmatrix}.
\end{equation}
Equations~(\ref{eq:strgeor}, \ref{eq:strgeophi}, \ref{eq:strgeoz}, \ref{eq:strgeot}) give
us the light cone parametrized by $(s,\Theta,\Psi)$. The same equations
determine the intersection of the light cone with any timelike hypersurface
(source surface) and thereby the exact lens map in the sense of Frittelli and
Newman~\cite{frittelli-newman-99} (recall Section~\ref{ssec:cone}).
For $k=0.8$ and $a=0$, the light cone is depicted in Figure~\ref{fig:strcon1};
intersections of the light cone with hypersurfaces $t = \mbox{constant}$
(``instantaneous wave fronts'') are shown in Figure~\ref{fig:strfrt1}.
In both pictures we consider a non-transparent string of finite radius $\rho_*$,
i.e., the light rays terminate if they meet the boundary of the string.
Figures~\ref{fig:strcon2} and~\ref{fig:strfrt2} show how the light cone
is modified if the string is transparent. This requires matching the
metric~(\ref{eq:strg}) to an interior solution which is everywhere
regular and letting light rays pass through the interior. For the
non-transparent string, the light cone cannot form a caustic, because
the metric is flat. For the transparent string, light rays that pass
through the interior of the string do form a caustic. The special
form of the interior metric is not relevant. The caustic has the same
features for all interior metrics that monotonously interpolate between
a regular axis and the boundary of the string. Also, there is no
qualitative change of the light cone for a \emph{spinning} string as long
as the spin $a$ is small. Large values of $a$, however, change the picture
drastically. For $a^2>k^2 \rho_*^2$, where $\rho_*$ is the radius of
the string, the $\varphi$-lines become timelike on a neighborhood
of the string. As the $\varphi$-lines are closed, this indicates causality
violation. In this causality-violating region the hypersurfaces $t =
\mbox{constant}$ are not everywhere spacelike and, in particular,
not transverse to all lightlike geodesics. Thus, our notion of instantaneous
wave fronts becomes pathological. \\

\epubtkImage{figure24.png}
{\begin{figure}[hptb]
   \def\epsfsize#1#2{1.1#1}
   \centerline{\epsfbox{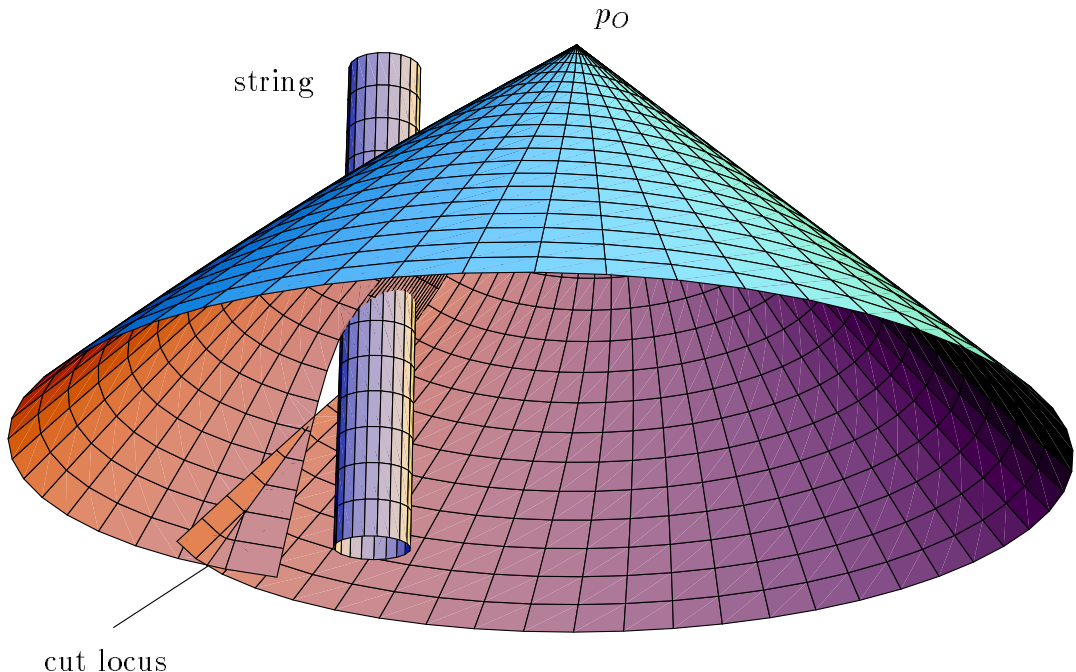}}
   \caption{Past light cone of an event $p_\mathrm{O}$ in the
     spacetime of a non-transparent string of finite radius $\rho_*$
     with $k = 0.8$ and $a=0$. The metric~(\ref{eq:strg}) is
     considered on the region $\rho > \rho_*$, and the light rays are
     cut if they meet the boundary of this region. The $z$ coordinate
     is not shown, the vertical coordinate is time $t$. The
     ``chimney'' indicates the region $\rho<\rho_*$ which is occupied
     by the string. The light cone has no caustic but a transverse
     self-intersection (cut locus). The cut locus, in the
     (2\,+\,1)-dimensional picture represented as a curve, is actually a
     2-dimensional spacelike submanifold. When passing through the
     cut locus, the lightlike geodesics leave the boundary of the
     chronological past $I^-(p_\mathrm{O})$. Note that the light cone
     is not a closed subset of the spacetime.}
   \label{fig:strcon1}
 \end{figure}
}

\epubtkImage{figure25.png}
{\begin{figure}[hptb]
   \def\epsfsize#1#2{1.0#1}
   \centerline{\epsfbox{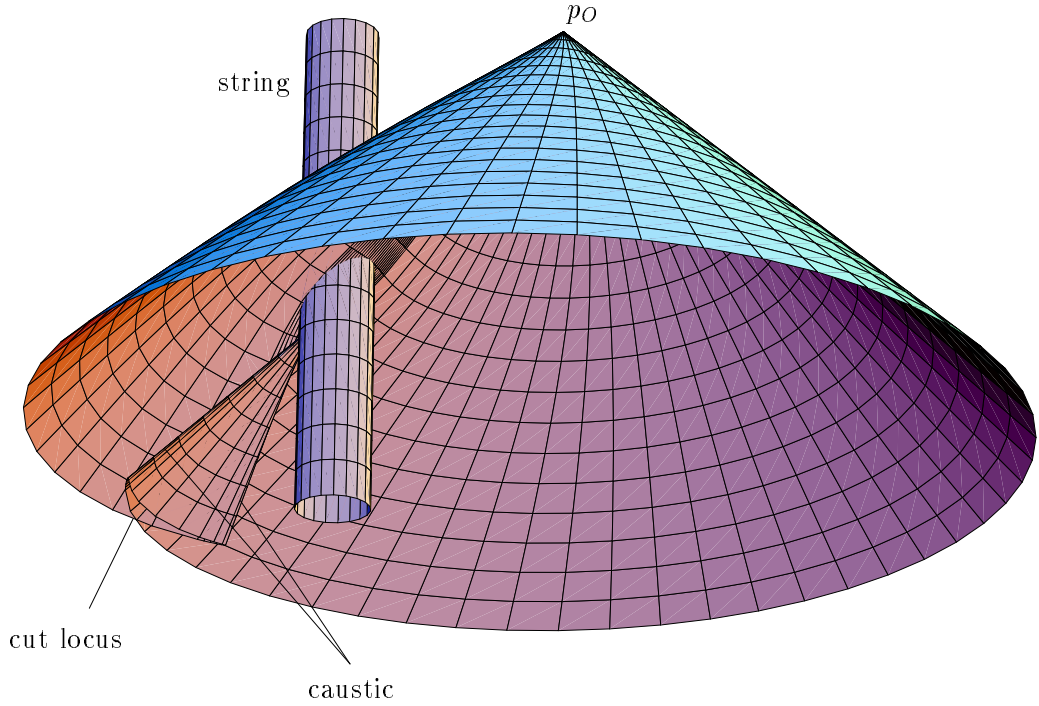}}
   \caption{Past light cone of an event $p_\mathrm{O}$ in the
     spacetime of a transparent string of finite radius $\rho_*$ with
     $k=0.8$ and $a=0$. The metric~(\ref{eq:strg}) is matched at
     $\rho = \rho_*$ to an interior metric, and light rays are allowed
     to pass through the interior region. The perspective is analogous
     to Figure~\ref{fig:strcon1}. The light rays which were blocked by
     the string in the non-transparent case now form a caustic. In the
     (2\,+\,1)-dimensional picture the caustic consists of two lightlike
     curves that meet in a swallow-tail point (see
     Figure~\ref{fig:swallow} for a close-up). Taking the $z$-dimension
     into account, the caustic actually consists of two lightlike
     2-manifolds (fold surfaces) that meet in a spacelike curve
     (cusp ridge). The third picture in Figure~\ref{fig:caustics}
     shows the situation projected to 3-space. Each of the
     past-oriented lightlike geodesics that form the caustic first
     passes through the cut locus (transverse self-intersection), then
     smoothly slips over one of the fold surfaces. The fold surfaces
     are inside the chronological past $I^-(p_\mathrm{O})$, the cusp
     ridge is on its boundary.}
   \label{fig:strcon2}
 \end{figure}
}

\epubtkImage{figure26.png}
{\begin{figure}[hptb]
   \def\epsfsize#1#2{1.0#1}
   \centerline{\epsfbox{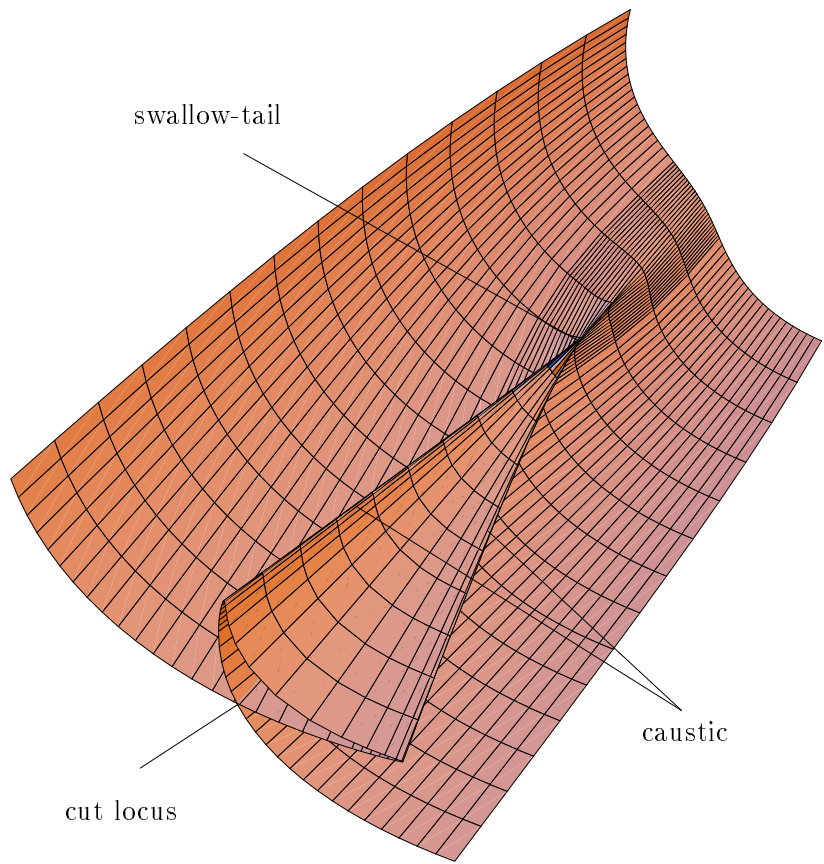}}
   \caption{Close-up of the caustic of
     Figure~\ref{fig:strcon2}. The string is not shown. Taking the
     $z$-dimension into account, the swallow-tail point is actually a
     spacelike curve (cusp ridge).}
   \label{fig:swallow}
 \end{figure}
}

\epubtkImage{figure27.png}
{\begin{figure}[hptb]
   \def\epsfsize#1#2{1.0#1}
   \centerline{\epsfbox{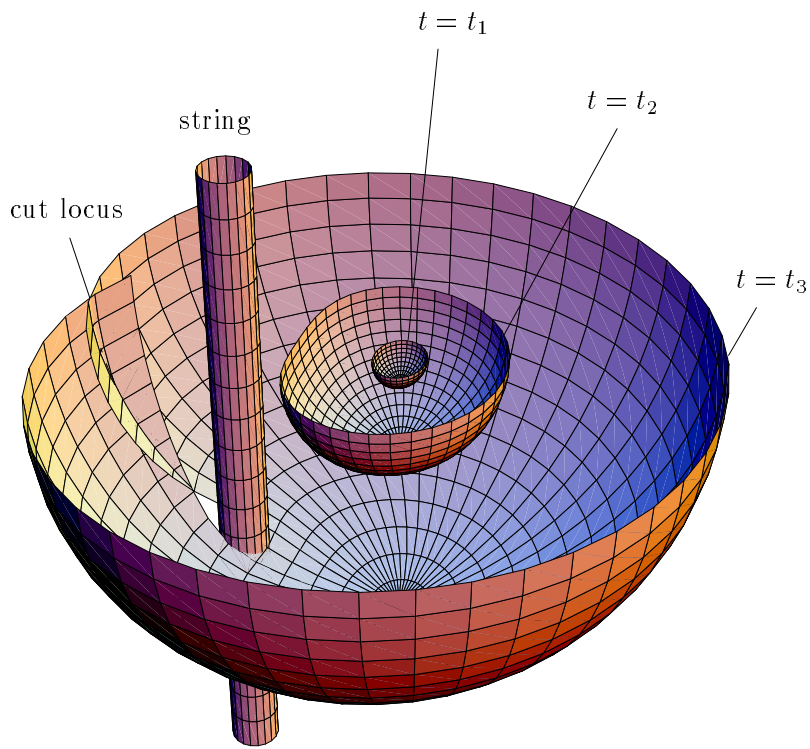}}
   \caption{Instantaneous wave fronts in the spacetime of a
     non-transparent string of finite radius $\rho_*$ with $k=0.8$ and
     $a=0$. The picture shows in 3-dimensional space the intersections
     of the light cone of Figure~\ref{fig:strcon1} with three
     hypersurfaces $t = \mbox{constant}$, at values $t_1>t_2>t_3$. The
     vertical coordinate is the $z$-coordinate which was suppressed in
     Figure~\ref{fig:strcon1}. Only one half of each instantaneous
     wave front is shown so that one can look into its interior. There
     is a transverse self-intersection (cut locus) but no caustic.}
   \label{fig:strfrt1}
 \end{figure}
}

\epubtkImage{figure28.png}
{\begin{figure}[hptb]
   \def\epsfsize#1#2{1.0#1}
   \centerline{\epsfbox{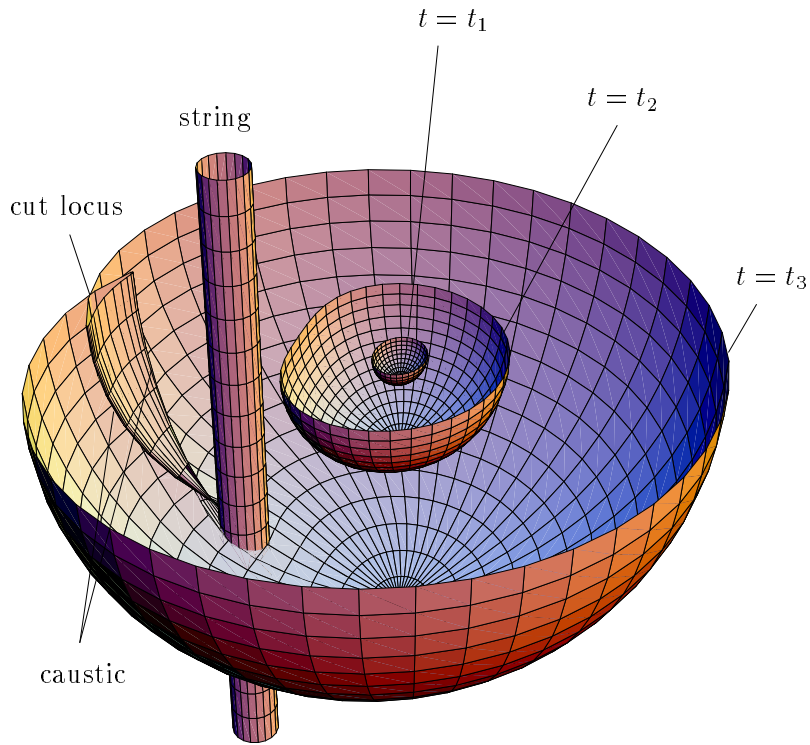}}
   \caption{Instantaneous wave fronts in the spacetime of a
     transparent string of finite radius $\rho_*$ with $k=0.8$ and
     $a=0$. The picture is related to Figure~\ref{fig:strcon2} as
     Figure~\ref{fig:strfrt1} is related to
     Figure~\ref{fig:strcon1}. Instantaneous wave fronts that have
     passed through the string have a caustic, consisting of two cusp
     ridges that meet in a swallow-tail point. This caustic is stable
     (see Section~\ref{ssec:front}). The caustic of the light cone in
     Figure~\ref{fig:strcon2} is the union of the caustics of its
     instantaneous wave fronts. It consists of two fold surfaces that
     meet in a cusp ridge, like in the third picture of
     Figure~\ref{fig:caustics}.}
   \label{fig:strfrt2}
 \end{figure}
}

\noindent
{\bf Lensing by a non-transparent string.} \\
\noindent
With the lightlike geodesics known in terms of elementary
functions, positions and properties of images can be explicitly
determined without approximation. We place the observation event
at $\rho = \rho_0$, $\varphi =0$, $z=0$, $t=0$, and we consider a
light source whose worldline is a $t$-line at $\rho = \rho_\mathrm{S}$,
$\varphi = \varphi_\mathrm{S}$, $z = z_\mathrm{S}$ with $0 \le \varphi_\mathrm{S} \le \pi$.
From Equations~(\ref{eq:strgeor}, \ref{eq:strgeophi}, \ref{eq:strgeoz})
we find that the images are in one-to-one correspondence with
integers $n$ such that
\begin{equation}
  \label{eq:strcrit}
  \left| \varphi_\mathrm{S} + 2 n \pi \right| < \pi /k.
\end{equation}
They can be numbered by the \emph{winding number} $n$ in the
order $n=0,-1,1,-2,2, \dots$ The total number of images depends on $k$.
Let $N_1(k)$ be the largest integer and $N_2(k)$ be the smallest integer
such that $N_1(k) \le 1/k < N_2(k)$. Of the two integers $N_1(k)$ and
$N_2(k)$, denote the odd one by $N_{\mathrm{odd}}(k)$ and the even one by
$N_{\mathrm{even}}(k)$. Then we find from Equation~(\ref{eq:strcrit})
\begin{eqnarray}
  0 \le \varphi_\mathrm{S} < \left| N_{\mathrm{even}}-1/k \right| \pi : &&
  ~ N_{\mathrm{odd}}(k) \mbox{ images},
  \label{eq:odd} \\
  \left| N_{\mathrm{even}}-1/k \right| \pi < \varphi_\mathrm{S} \le \pi : &&
  ~ N_{\mathrm{even}}(k) \mbox{ images}.
  \label{eq:even}
\end{eqnarray}%
Thus, the number of images is even in a wedge-shaped region behind the
string and odd everywhere else. If the light source approaches the
boundary between the two regions, one image vanishes behind the
string (see Figure~\ref{fig:deficit} for the case $1\le 1/k < 2$).
(If the non-transparent string has finite thickness, there
is also a region with no image at all, in the ``shadow'' of the string.)
The coordinates $(\Psi_n,\Theta_n)$ on the observer's sky of an
image with winding number $n$ and the affine parameter $s_n$ at which
the light source is met can be determined from Equations~(\ref{eq:strgeor},
\ref{eq:strgeophi}, \ref{eq:strgeoz}). We just have to insert
$\rho (s)= \rho_\mathrm{S}$, $\varphi (s) = \varphi_\mathrm{S} + 2 n \pi$, $z (s)= z_\mathrm{S}$
and to solve for $\mathrm{tan} \Psi = \mathrm{tan}
 \Psi_n$, $\mathrm{tan} \Theta = \mathrm{tan} \Theta_n$, $s=s_n$:
\begin{eqnarray}
  \mathrm{tan} \Psi_n &=&
  \frac{\rho_\mathrm{S} \, \sin \left( k (\varphi_\mathrm{S} + 2n \pi) \right)}
  {\rho_\mathrm{S} \, \cos \left( k (\varphi_\mathrm{S} + 2n \pi) \right) - \rho_0},
  \label{eq:strPhi} \\
  \mathrm{tan} \Theta_n &=& \frac{\sqrt{\rho_\mathrm{S}^2 +
  \rho_0^2 - 2 \rho_\mathrm{S} \rho_0 \,
  \cos \left( k (\varphi_\mathrm{S} + 2n \pi) \right)}}{z_\mathrm{S}},
  \label{eq:strTheta} \\
  s_n &=& \sqrt{z_\mathrm{S}^2 + \rho_\mathrm{S}^2 +
  \rho_0^2 - 2 \rho_\mathrm{S} \rho_0 \, \cos
  \left( k (\varphi_\mathrm{S} + 2n \pi) \right)}.
  \label{eq:strs}
\end{eqnarray}%
The travel time follows from Equation~(\ref{eq:strgeot}):
\begin{equation}
  \label{eq:strt}
  T_n = s_n - a (\varphi_\mathrm{S} + 2n \pi).
\end{equation}
It is the only relevant quantity that depends on the string's spin $a$. With the
observer on a $t$-line, the affine parameter $s$ coincides with the area distance,
$D_{\mathrm{area}} (s) = s$, because in the (locally) flat string spacetime the
focusing equation~(\ref{eq:focus}) reduces to
$\ddot{D}_{\mathrm{area}} = 0$.
For observer and light source on $t$-lines, the redshift vanishes, so $s$ also
coincides with the luminosity distance, $D_{\mathrm{lum}} (s) = s$, owing to
the general law~(\ref{eq:Dlum}). Hence, Equation~(\ref{eq:strs}) gives us the brightness
of images (see Section~\ref{ssec:brightness} for the relevant formulas). The
string metric produces no image distortion because the curvature tensor (and
thus, the Weyl tensor) vanishes (recall Section~\ref{ssec:distortion}).
Realistic string models yield a mass density $\mu$ that is smaller
than $10^{-4}$. So, by Equation~(\ref{eq:strmu}), only the case
$N_{\mathrm{odd}}(k)=1$ and $N_{\mathrm{even}}(k)=2$ is thought to
be of astrophysical relevance. In that case we have a
single-imaging region, $0 \le \varphi_\mathrm{S} < 2 \pi - \pi /k$, and a
double-imaging region, $2 \pi - \pi /k < \varphi_\mathrm{S} \le \pi$ (see
Figure~\ref{fig:deficit}). The occurrence of double-imaging and of
single imaging can also be read from Figure~\ref{fig:strcon1}. In
the double-imaging region we have a (``primary'') image with $n=0$
and a (``secondary'') image with $n=-1$. From Equations~(\ref{eq:strTheta},
\ref{eq:strs}) we read that the two images have different latitudes
and different brightnesses. However, for $k$ close to
1 the difference is small. If we express $k$ by Equation~(\ref{eq:deficit}) and
linearize Equations~(\ref{eq:strPhi}, \ref{eq:strTheta}, \ref{eq:strs},
\ref{eq:strt}) with respect to the deficit angle~(\ref{eq:deficit}),
we find
\begin{eqnarray}
  \Psi_0 &=& \frac{\rho_0 \pi - \rho_\mathrm{S} \varphi_\mathrm{S}}{\rho_\mathrm{S} + \rho_0} -
  \frac{\varphi_\mathrm{S} \rho_\mathrm{S} \delta}{(\rho_\mathrm{S} + \rho_0)2 \pi}
  \label{eq:strPhia1} \\
  \Psi_{-1} &=& \Psi_0 + \frac{\rho_\mathrm{S} \delta}{\rho_\mathrm{S} + \rho_0},
  \label{eq:strPhia2} \\
  \Theta_{-1} - \Theta_0 &=& 0,
  \label{eq:strThetaa} \\
  s_{-1} - s_0 &=& 0,
  \label{eq:strsa} \\
  T_{-1} - T_0 &=& 2 a \pi.
  \label{eq:strta}
\end{eqnarray}%
Hence, in this approximation the two images
have the same $\Theta-$coordinate; their angular distance $\Delta$ on
the sky is given by Vilenkin's~\cite{vilenkin-84} formula
\begin{equation}
  \label{eq:sep}
  \Delta = \frac{\rho_\mathrm{S} \delta \, \sin \Theta_0}{\rho_\mathrm{S} + \rho_0},
\end{equation}
and is thus independent of $\varphi_\mathrm{S}$; they have equal brightness
and their time delay is given by the string's spin $a$ via
Equation~(\ref{eq:strta}). All these results apply to the case that the
worldlines of the observer and of the light source are $t$-lines.
Otherwise redshift factors must be added. \\

\noindent
{\bf Lensing by a transparent string.} \\
\noindent
In comparison to a non-transparent string, a transparent string
produces additional images. These additional images correspond to
light rays that pass through the string. We consider the case
$a=0$ and $1 < 1/k < 2$, which is illustrated by Figures~\ref{fig:strcon1}
and~\ref{fig:strcon2}. The general features do not depend on
the form of the interior metric, as long as it monotonously interpolates
between a regular axis and the boundary of the string. In the
non-transparent case, there is a single-imaging region and a
double-imaging region. In the transparent case, the double-imaging
region becomes a triple-imaging region. The additional image corresponds to
a light ray that passes through the interior of the string and
then smoothly slips over one of the cusp ridges. The point where this
light ray meets the worldline of the light source is on the sheet of
the light cone between the two cusp ridges in Figure~\ref{fig:strcon2},
i.e., on the sheet that does not exist in the non-transparent case
of Figure~\ref{fig:strcon1}. From the picture it is obvious that the
additional image shows the light source at a younger age than the other
two images (so it is a ``tertiary image''). A light source whose worldline
meets the caustic of the observer's past light cone is on the borderline
between single-imaging and triple-imaging. In this case the tertiary
image coincides with the secondary image and it is particularly bright
(even infinitely bright according to the ray-optical treatment; recall
Section~\ref{ssec:brightness}). Under a small perturbation of the
worldline the bright image either splits into two or vanishes, so one
is left either with three images or with one image.


\subsection{Plane gravitational waves}
\label{ssec:wave}

A \emph{plane gravitational wave} is a spacetime with metric
\begin{equation}
  \label{eq:waveg}
  g = - 2 \, du \, dv -
  \left( f(u) (x^2 - y^2) + 2 g(u) x y \right) du^2 + dx^2 + dy^2,
\end{equation}
where $f(u)^2+g(u)^2$ is not identically zero. For any choice of $f(u)$ and
$g(u)$, the metric~(\ref{eq:waveg}) has vanishing Ricci tensor, i.e.,
Einstein's vacuum field equation is satisfied. The lightlike vector
field $\partial_v$ is covariantly constant. Non-flat spacetimes
with a covariantly constant lightlike vector field are called
\emph{plane-fronted waves with parallel rays} or \emph{pp-waves}
for short. They made their first appearance in a purely mathematical
study by Brinkmann~\cite{brinkmann-25}.

In spite of their high idealization, plane gravitational waves
are interesting mathematical models for studying the lensing
effect of gravitational waves. In particular, the focusing effect
of plane gravitational waves on light rays can be studied quite
explicitly, without any weak-field or small-angle approximations. This
focusing effect is reflected by an interesting light cone structure.

The basic features with relevance to lensing can be summarized in the
following way. If the profile functions $f$ and $g$ are differentiable,
and the coordinates $(x,y,u,v)$ range over $\mathbb{R}^4$, the
spacetime with the metric~(\ref{eq:waveg}) is geodesically
complete~\cite{ehlers-kundt-62}. With the exception of the integral
curves of $\partial_v$, all inextendible lightlike geodesics contain a
pair of conjugate points. Let $q$ be the first conjugate point along a
past-oriented lightlike geodesic from an observation event $p_\mathrm{O}$. Then
the first caustic of the past light cone of $p_\mathrm{O}$ is a parabola through
$q$. (It depends on the profile functions $f$ and $g$ whether or
not there are more caustics, i.e., second, third, etc.\ conjugate
points.) This parabola is completely contained in a hyperplane
$u = \mbox{constant}$. All light rays through $p_\mathrm{O}$, with the
exception of the integral curve of $\partial_v$, pass through this
parabola. In other words, the entire sky of $p_\mathrm{O}$, with the exception of
one point, is focused into a curve (see Figure~\ref{fig:wave1}).
This \emph{astigmatic focusing} effect of plane gravitational waves
was discovered by Penrose~\cite{penrose-65} who worked out the
details for ``sufficiently weak sandwich waves''. (The name ``sandwich
wave'' refers to the case that $f(u)$ and $g(u)$ are different from
zero only in a finite interval $u_1<u<u_2$.) Full proofs of the
above statements, for arbitrary profile functions $f$
and $g$, were given by Ehrlich and Emch~\cite{ehrlich-emch-92,
ehrlich-emch-93} (cf.~\cite{beem-ehrlich-easley-96}, Chapter~13). The
latter authors also demonstrate that plane gravitational wave spacetimes
are causally continuous but not causally simple. This strengthens Penrose's
observation~\cite{penrose-65} that they are not globally hyperbolic.
(For the hierarchy of causality notions see~\cite{beem-ehrlich-easley-96}.)
The generators of the light cone leave the boundary of the chronological
past $I^-(p_\mathrm{O})$ when they reach the caustic. Thus, the above-mentioned
parabola is also the cut locus of the past light cone. By the general results
of Section~\ref{ssec:crit}, the occurrence of a cut locus
implies that there is multiple imaging in the plane-wave spacetime. The
number of images depends on the profile functions. We may choose the
profile functions such that there is no second caustic. (The ``sufficiently
weak sandwich waves'' considered by Penrose~\cite{penrose-65} are of
this kind.) Then Figure~\ref{fig:wave1} demonstrates that an appropriately
placed worldline (close to the caustic) intersects the past light cone
exactly twice, so there is double-imaging. Thus, the plane waves demonstrate
that the number of images need not be odd, even in the case of a geodesically
complete spacetime with trivial topology.

\epubtkImage{figure29.png}
{\begin{figure}[hptb]
   \def\epsfsize#1#2{1.0#1}
   \centerline{\epsfbox{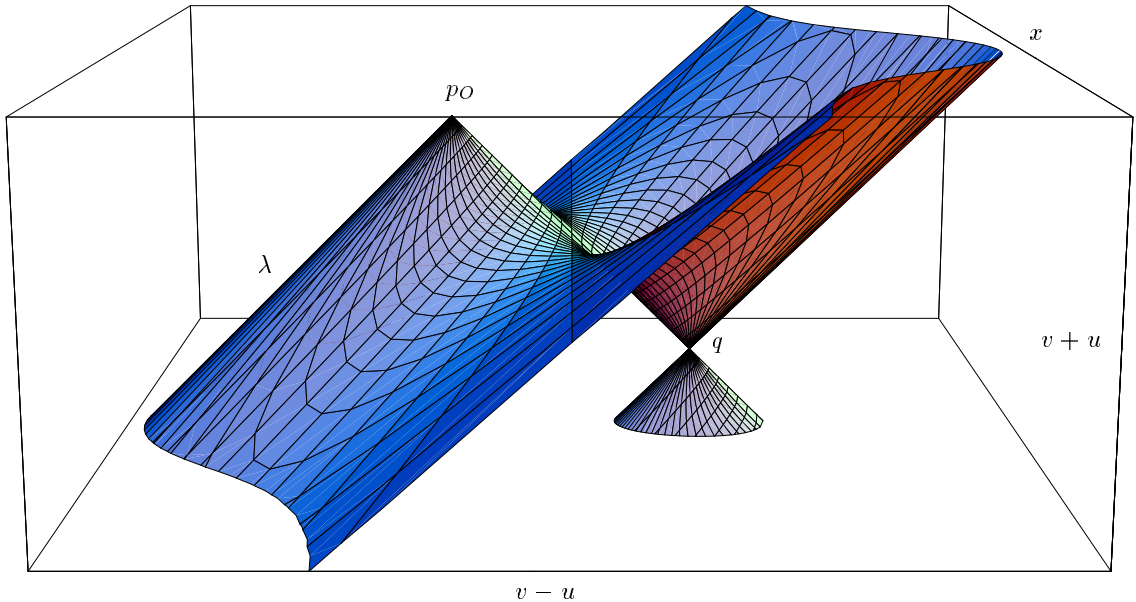}}
   \caption{Past light cone of an event $p_\mathrm{O}$ in the
     spacetime~(\ref{eq:waveg}) of a plane gravitational wave. The
     picture was produced with profile functions $f(u)>0$ and
     $g(u)=0$. Then there is focusing in the $x$-direction and
     defocusing in the $y$-direction. In the (2\,+\,1)-dimensional
     picture, with the $y$-coordinate not shown, the past light cone
     is completely refocused into a single point $q$, with the
     exception of one generator $\lambda$. It depends on the profile
     functions whether there is a second, third, and so on,
     caustic. In any case, the generators leave the boundary of the
     chronological past $I^-(p_\mathrm{O})$ when they pass through the
     first caustic. Taking the $y$-coordinate into account, the first
     caustic is not a point but a parabola (``astigmatic focusing'')
     (see Figure~\ref{fig:wave2}). An \emph{electromagnetic} plane wave
     (vanishing Weyl tensor rather than vanishing Ricci tensor) can
     refocus a light cone, with the exception of one generator, even
     into a point in 3\,+\,1 dimensions (``anastigmatic focusing'')
     (cf.\ Penrose~\cite{penrose-65} where a hand-drawing similar to the
     picture above can be found).}
   \label{fig:wave1}
 \end{figure}
}

\epubtkImage{figure30.png}
{\begin{figure}[hptb]
   \def\epsfsize#1#2{1.0#1}
   \centerline{\epsfbox{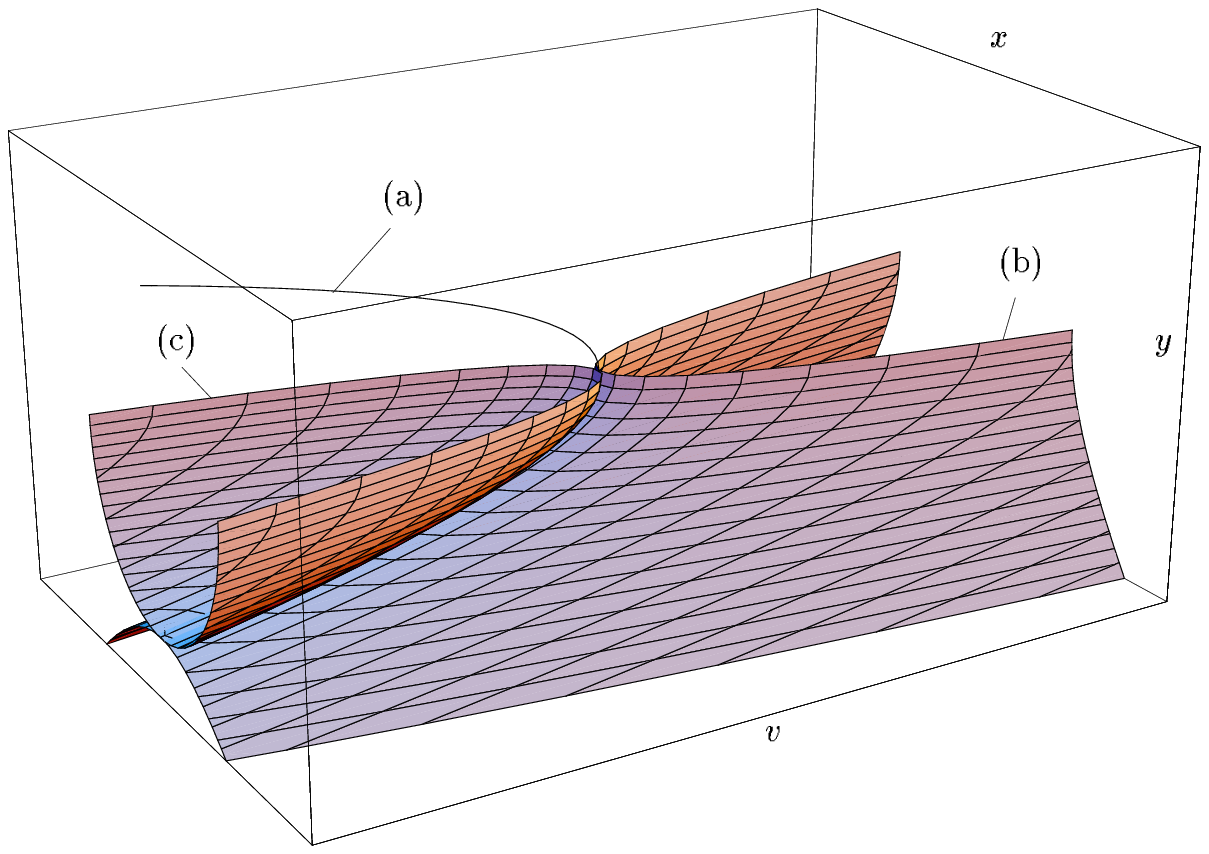}}
   \caption{``Small wave fronts'' of the light cone in the
     spacetime~(\ref{eq:waveg}) of a plane gravitational wave. The
     picture shows the intersection of the light cone of
     Figure~\ref{fig:wave1} with the lightlike hyperplane
     $u=\mbox{constant}$ for three different values of the constant:
     (a) exactly at the caustic (parabola), (b) at a larger value of
     $u$ (hyperbolic paraboloid), and (c) at a smaller value of $u$
     (elliptic paraboloid). In each case, the hyperplane
     $u=\mbox{constant}$ does not intersect the one generator
     $\lambda$ tangent to $\partial_v$; all other generators are
     intersected transversely and exactly once.}
   \label{fig:wave2}
 \end{figure}
}

The geodesic and causal structure of plane gravitational waves and,
more generally, of pp-waves is also studied
in~\cite{hubeny-rangamani-2002, candela-flores-sanchez-2003}.

One often considers profile functions $f$ and $g$ with Dirac-delta-like
singularities (``impulsive gravitational waves''). Then a mathematically
rigorous treatment of the geodesic equation, and of the geodesic deviation
equation, is delicate because it involves operations on distributions
which are not obviously well-defined. For a detailed mathematical
study of this situation see~\cite{steinbauer-98, kunzinger-steinbauer-99}.

Garfinkle~\cite{garfinkle-90} discovered an interesting example for
a pp-wave which is singular on a 2-dimensional worldsheet. This
exact solution of Einstein's vacuum field equation can be interpreted
as a wave that travels along a cosmic string. Lensing in this spacetime
was numerically discussed by Vollick and Unruh~\cite{vollick-unruh-91}.

The vast majority of work on lensing by gravitational waves is done
in the weak-field approximation. Both for the exact treatment and for the
weak-field approximation one may use Kovner's version of Fermat's
principle (see Section~\ref{ssec:fermat}), which has the advantage
that it allows for time-dependent situations. 
Applications of
this principle to gravitational waves have been worked out in the original
article by Kovner~\cite{kovner-90} and by Faraoni~\cite{faraoni-92,
faraoni-98}.

\newpage


\section{Acknowledgements}
\label{sec:acknowledgements}

I have profited very much from many suggestions and comments by J{\"u}rgen
Ehlers. Also, I wish to thank an anonymous referee for his detailed
and very helpful report.

\newpage


\bibliography{refs}

\end{document}